\documentclass[letterpaper,twocolumn,10pt]{article}
\usepackage{usenix-2020-09}

\usepackage{tikz}
\usepackage{amsmath}

\usepackage{colortbl}
\usepackage{graphicx,marvosym}
\newcommand{\platformname}{\textsc{SEC4SR}\xspace}

\newcommand{\defensename}{\textsc{Feature Compression}\xspace}
\newcommand{\defensenameabbr}{\textsc{FC}\xspace}
\newcommand{\chengk}[1]{\textcolor{cyan}{#1}}

\usepackage{amssymb}
\usepackage{amsmath}
\usepackage{pifont}
\newcommand{\cmark}{\ding{51}}%
\newcommand{\xmark}{\ding{55}}%
\usepackage{multirow}
\usepackage{url}
\usepackage{hyperref}
\usepackage{xcolor}
\usepackage{bm}
\usepackage{colortbl}
\usepackage{subfigure}
\usepackage{xspace}

\usepackage[many]{tcolorbox}

\begin{document}

\date{}

\title{\Large \bf \platformname: A Security Analysis Platform for Speaker Recognition}

 \author{
 {\rm Guangke Chen}\\
 ShanghaiTech University
 \and
 {\rm Zhe Zhao}\\
 ShanghaiTech University
 \and
  {\rm Fu Song}\\
  ShanghaiTech University
 \and
    {\rm Sen Chen}\\
 Tianjin  University
 \and
 {\rm Lingling Fan}\\
 Nankai  University
 \and
  {\rm Yang Liu}\\
Nanyang Technological University
 } 

\maketitle

\begin{abstract}
Adversarial attacks have been expanded to speaker recognition (SR). However, existing attacks are often
assessed using different SR models, recognition tasks and datasets,
and only few adversarial defenses borrowed from computer vision are considered. Yet, these defenses have not been thoroughly evaluated against
adaptive attacks. Thus, there is still a lack of quantitative understanding about the strengths and limitations
of adversarial attacks and defenses. More effective defenses are also required for securing 
SR systems.

To bridge this gap, we present \platformname,  the first platform enabling researchers to systematically and comprehensively evaluate adversarial
attacks and defenses in SR. \platformname incorporates 4 white-box and 2 black-box attacks,
24 defenses including our novel feature-level transformations. It also contains techniques for mounting adaptive attacks.
Using \platformname, we conduct thus far the largest-scale empirical
study on adversarial attacks and defenses in SR, involving 23 defenses, 15 attacks and 4 attack settings.
Our study provides lots of useful findings that may advance future research:
such as 
(1) all the transformations slightly degrade accuracy on benign examples and their effectiveness vary with attacks;
(2) most transformations become less effective under adaptive attacks, but some
transformations become more effective;
(3) few transformations combined with adversarial training yield stronger defenses over some but not all attacks, while
our feature-level transformation combined with adversarial training yields the strongest defense over all the attacks.
Extensive experiments demonstrate
capabilities and advantages of \platformname which can
benefit future research in SR.
\end{abstract}

\section{Introduction}\label{sec:introduction}
Speaker recognition (SR) is the process of
automatically verifying or identifying individual speakers 
by extracting and analyzing their unique acoustic characteristics~\cite{Homayoon11}.
SR has been supported by platforms such as Kaldi~\cite{kaldi} and MSR Identity~\cite{MSRIdentity},
adopted in commercial products such as Microsoft Azure~\cite{Azure}, Amazon Alexa~\cite{Alexa}, Google Home~\cite{googlehome}
and SpeechPro VoiceKey~\cite{SpeechPro},
and used in our daily life ranging from remote voice authentication in financial transaction~\cite{TD-Bank}, device access control in smart home~\cite{RenSYS16},
to voice control cars~\cite{vehicles}.

Machine learning including deep learning techniques are the mainstream
methods for implementing state-of-the-art speaker recognition systems (SRSs)~\cite{reynolds2000speaker,fortuna2005open,Homayoon11,SnyderGSMPK19,Sremath2016areview}.
However, machine learning models have been shown to be vulnerable to adversarial examples in various domains, e.g., computer vision~\cite{SzegedyZSBEGF13,goodfellow2014explaining,madry2017towards,madry2017towards}
and speech recognition~\cite{Carlini018,qin2019imperceptible,yuan2018commandersong,ijcai2019-741,247642,TaoriKCV19,LiW00020,yang2018characterizing}.
To systematically and comprehensively evaluate different adversarial attacks and defenses,
various platforms have been proposed such as Cleverhans~\cite{papernot2016technical}, Foolbox~\cite{rauber2017foolbox}, AdvBox~\cite{goodman2020advbox}, ART~\cite{nicolae2018adversarial}, advertorch~\cite{ding2019advertorch}, ARES~\cite{DDongFYPSXZ20}, DEEPSEC~\cite{LingJZWWLW19} and FenceBox~\cite{qiu2020fencebox}. 
These platforms facilitate research on adversarial examples.

Adversarial attacks have been expanded to SR for understanding security weaknesses in SR~\cite{DBLP:journals/corr/abs-1711-03280,abs-1801-03339,li2020adversarial,jati2021adversarial,zhang2021attack,xie2021real, LiZJXZWM020, xie2020enabling,WangGX20,shamsabadi2021foolhd,chen2019real,du2020sirenattack}. However, their evaluations vary in SR models, recognition tasks
and datasets, and only few adversarial defenses (i.e., adversarial training and input transformations) from computer vision are considered.
Yet, these defenses have not been thoroughly evaluated against adaptive attacks~\cite{CW17a,tramer2020adaptive} which are specific attacks
designed to circumvent a given defense.
Thus, there is still a lack of quantitative understanding about the strengths
and limitations of adversarial attacks and defenses.
More effective adversarial defenses are also required for securing SRSs.

To bridge this gap, we present the \emph{first} platform, named \platformname (\url{https://sec4sr.github.io}), for security analysis of SRSs.
\platformname incorporates all the adversarial attacks presented in
\cite{DBLP:journals/corr/abs-1711-03280,abs-1801-03339,li2020adversarial,zhang2021attack,jati2021adversarial,chen2019real,du2020sirenattack}, which are extended to more recognition tasks in \platformname.
Specifically, \platformname includes 4 representative white-box attacks and 2 black-box attacks.
\platformname also provides Backward Pass Differentiable
Approximation~\cite{athalye2018obfuscated} (BPDA), Expectation over Transformation (EOT)~\cite{athalye2018synthesizing} and Natural Evolution Strategy (NES)~\cite{wierstra2014natural}, enabling
to lift the attacks to adaptive versions for circumventing defenses. These adaptive attacks have never been considered
in SR except that NES was adopted to estimate gradients by the black-box attack FAKEBOB~\cite{chen2019real}.

\platformname also incorporates two promising adversarial training of \cite{jati2021adversarial},
all the input transformations of ~\cite{chen2019real,du2020sirenattack}.
Since these defenses are borrowed from computer vision,
they may become ineffective in the presence of adaptive attacks.
Thus, we investigate input transformations from speech
recognition, which are the first time used for securing SRSs, to our best knowledge.
We also propose transformations which manipulate \emph{features} of input voices.
Such feature-level transformations are novel and have never been studied either in speaker
or speech recognition. In total,  \platformname consists of 2 promising adversarial training and 22 diverse transformations (4 time domain and
3 frequency domain input transformations, 7 speech compressions and 8 feature-level transformations).
These transformations cover differentiable, non-differentiable, deterministic, randomized transformations.

\platformname further features 5 voice datasets and 3 mainstream SRSs which cover 4 different main recognition tasks.
We also implement 8 attack metrics and 3 defense metrics for quantitative understanding about the strengths and limitations
of adversarial attacks and defenses.

Using \platformname, we perform thus far the largest-scale
empirical study on adversarial attacks and defenses in SR, motivated by the following four research questions:
(RQ1) How weak SRSs are under adversarial attacks?
(RQ2) How effective are the transformations under non-adaptive attacks where the adversary is unaware of the defenses?
(RQ3) Do the transformations remain effective when the adversary owns complete knowledge of both the model and defense, i.e., adaptive attacks?
(RQ4) Can a combination of a transformation with adversarial training yields a stronger defense?
We obtain a set of interesting and insightful findings that may advance research on adversarial examples in SR.

To address RQ1, we perform both untargeted and targeted attacks using
5  attacks (15 attacks counting different attack parameters).
Most of attacks achieve 100\% attack success rate,
confirming that SRSs are vulnerable to adversarial attacks.
These results are served as baseline for evaluating defenses.

To address RQ2, we perform cross evaluation between 22 transformations
and 5 non-adaptive attacks (11 attacks counting different attack parameters).
The results show that all the transformations slightly degrade accuracy on benign examples and their effectiveness vary with attacks.

To address RQ3, we perform cross evaluation between 15 transformations (ineffective transformations in RQ2 are excluded)
and 5 adaptive attacks (7 attacks counting different attack parameters).
The results show that most transformations become less effective under adaptive attacks,
while some transformations become more effective. The latter reveals that
more powerful adaptive attacks are required.

To address RQ4, we perform cross evaluation between 6 transformations (ineffective transformations in RQ3 are excluded)
and 5 adaptive attacks on adversarially trained SRSs.
The results show that few transformations combined with adversarial training yield stronger defenses over some but not all attacks,
where our feature-level transformation combined with adversarial training yields the strongest defense over all the attacks
and alleviates the accuracy degradation induced by adversarial training.

To summarize, we make the following main contributions.
 \begin{itemize}\setlength{\itemsep}{0pt}
\setlength{\parskip}{0pt}
    \item We present \platformname, the \textit{first} platform for systematic and
comprehensive evaluation of different adversarial attacks and
defenses in SR. It features mainstream SRSs, proper voice datasets, white-box and black-box attacks, techniques for mounting
adaptive attacks, evaluation metrics and diverse defense solutions.

    \item Using \platformname, we perform the largest-scale empirical study on adversarial attacks and defenses in SR, involving 23 defenses, 15 attacks, and 4 attack settings.
    Our study provides lots of useful insights and findings that may advance research on adversarial examples in SR and assist the maintainers of SRSs to deploy proper defense solutions to enhance their systems.

    \item We propose a new type of feature-level transformations dedicated for SR, which is able to
    beat all the input transformations and results in the strongest defense against all the attacks when combined with adversarial training.
\end{itemize}

\section{Background}\label{sec:background}\vspace*{-1mm}

In this section, we introduce the preliminaries of mainstream and state-of-the-art SRSs and their recognition tasks.


\vspace*{-1mm}\subsection{Overview of Speaker Recognition}\vspace*{-1mm}
Modern and state-of-the-art SRSs use speaker embedding~\cite{wang2020simulation} to represent
acoustic characteristics of speakers as fixed-dimensional vectors.
The typical
speaker embedding is identity-vector (ivector)~\cite{DehakDKBOD09} based on the Gaussian Mixture Model (GMM)~\cite{ReynoldsR95,reynolds2000speaker}.
Recently, deep embedding was also proposed to compete with ivector.
It first uses deep learning to train a deep neural network, then
extracts acoustic characteristics from the deep neural network
and presents acoustic characteristics as
dvector~\cite{variani2014deep} or xvector~\cite{snyder2018x}, etc.

\begin{figure}[t]
    \centering
    \includegraphics[width=0.48\textwidth]{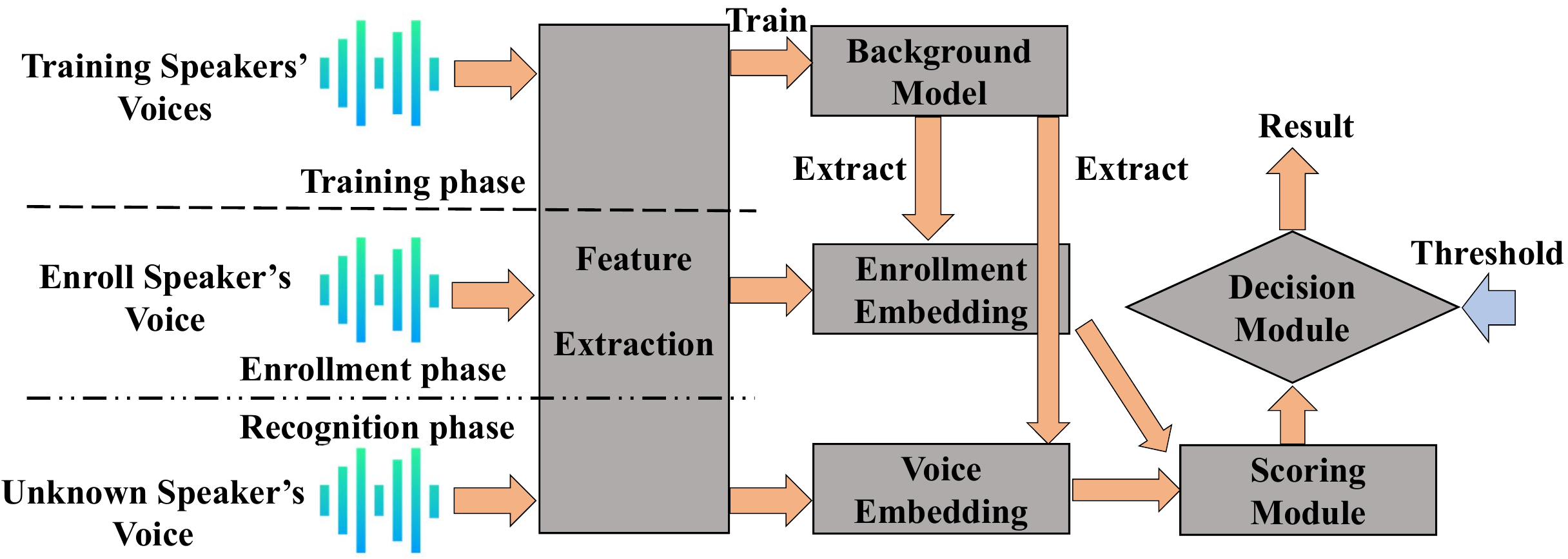}
    \vspace*{-6mm}
    \caption{Architecture of SRSs.}
    \label{fig:typical-SRSs}\vspace*{-4mm}
\end{figure}

A generic architecture of SRSs is shown in Figure~\ref{fig:typical-SRSs}, consisting of three phases: training, enrollment, and recognition phases.
In the training phase, a background model is trained using tens of thousands of voices from thousands of training speakers,
representing the speaker-independent distribution of voice features.
In the enrollment phase, the background model maps the voice uttered by each enrolling speaker to an \emph{enrollment embedding},
regarded as the unique identity of the enrolling speaker.
In the recognition phase, given a voice of an unknown speaker, the \emph{voice embedding} is extracted from the background model
for scoring. The scoring module measures the similarity between the \emph{enrollment embedding} and \emph{voice embedding} based on which the result is produced by the decision module.
There are two typical scoring approaches: Probabilistic Linear Discriminant Analysis (PLDA)~\cite{nandwana2019analysis}
and COSine Similarity (COSS)~\cite{dehak2010cosine}.
The former approach works well in most situations, but needs to be trained using voices~\cite{wang2020simulation}.
The latter approach is a reasonable substitution of PLDA
which does not need to be trained.


The feature extraction module converts raw speech signals to acoustic features
carrying characteristics of the raw signals.
Common feature extraction algorithms include Mel-Frequency Cepstral Coefficients (MFCC)~\cite{muda2010voice} and
Perceptual Linear Predictive (PLP)~\cite{hermansky1990perceptual}.

\vspace*{-2mm}\subsection{Recognition Tasks}\label{sec:SR-task}\vspace*{-1mm}
There are three main tasks in SR: close-set identification (CSI), speaker verification (SV), and open-set identification (OSI).
CSI identifies a speaker from a set of enrolled speakers, which is a multi-class classification problem.
SV verifies if an input voice is uttered by the unique enrolled speaker, according to a preset threshold, where
the input voice may be rejected by regarding the speaker as an imposter.
OSI utilizes the scores and a preset threshold to identify which enrolled speaker utters the input voice.
If the highest score is less than the threshold, the input voice is rejected by regarding the speaker as an imposter.
Furthermore, CSI could be classified into two sub-tasks:
CSI with enrollment (CSI-E) and  CSI without enrollment (CSI-NE).
CSI-E exactly follows the above description.
In contrast, CSI-NE does not have the enrollment phase
and the background model is directly utilized to identify speakers.
Thus, ideally,  
   a recognized speaker in CSI-NE task is involved in the training phase,
   while a recognized speaker in CSI-E task should have enrolled in the enrollment phase but may not involved in the training phase.

\vspace*{-1mm}\section{Attacks \& Defenses}\vspace*{-1mm}
In this section, we first briefly recap several typical adversarial attacks, including white-box and black-box attacks.
Then, we introduce two categories of defenses: {robust training and input transformation}. Finally, we introduce the techniques that are leveraged to mount adaptive attacks to bypass defenses~\cite{tramer2020adaptive}.

\subsection{Attacks}\label{sec:review-attacks}

A plethora of adversarial attacks have been proposed, most of which are primarily studied in computer vision~\cite{akhtar2018threat}.
It is largely unknown if they can successfully be ported to all the recognition tasks due to the unique property of SR~\cite{chen2019real}.
Hence, in this work, we present three representative white-box attacks
and two black-box attacks which have been demonstrated to be effective on at least one recognition task.

\subsubsection{White-box Attacks}

\noindent {\bf Fast Gradient Sign Method (FGSM)}~\cite{goodfellow2014explaining}. FGSM perturbs an input $x$ by performing one-step gradient ascent to maximize a loss
function. Formally, a potential adversarial example is:
\begin{center}
$\hat{x} = x + \epsilon \times \emph{sign}(\nabla_x f(x,y))$
\end{center}
where $\epsilon$ is the step size of gradient ascent and
$f(x,y)$ is the loss function describing the cost of classifying $x$ as label $y$.

\smallskip
\noindent {\bf Projected Gradient Descent (PGD)}~\cite{madry2017towards}. PGD is an iterative version of FGSM. In each iteration, PGD applies FGSM with a small step size $\alpha$ and clips the result to ensure that it stays within an $\epsilon$-neighborhood of the original input $x$. The intermediate example after the $i$-th iteration is:
\begin{center}
$x^i = \emph{clip}_{x,\epsilon}(x^{i-1} + \alpha \times \emph{sign}(\nabla_x f(x^{i-1},y)))$
\end{center}

Note that PGD attack starts from a randomly perturbed example rather than the original example, which helps the attack find a better local minimum.
We denote by PGD-$x$ the PGD attack with $x$ steps for iteration,
the larger $x$, the stronger the adversarial examples.

\smallskip
\noindent {\bf CW}~\cite{carlini2017towards}. CW is introduced to search for adversarial examples with small magnitude of perturbations.
It formulates finding adversarial examples as an optimization problem whose objective function is the trade-off (controlled by a factor $c$) between the effectiveness and imperceptibility of adversarial examples. The effectiveness is measured by a loss function $f(x,y)$ such that $f(x,y)\leq 0$ if and only if the attack succeeds. The imperceptibility can be instantiated by $L_0$, $L_2$, and $L_\infty$ distance between adversarial and original examples, leading to three versions of CW attack, denoted by CW$_0$, CW$_2$, and CW$_{\bm{\infty}}$, respectively. CW attack is equipped with a parameter $\kappa$,
the larger $\kappa$, the stronger the adversarial examples.
We denote by CW$_p$-$x$ the CW$_p$ attack with $\kappa=x$.

\subsubsection{Black-box Attacks}\label{sec:blackboxattacks}
\noindent {\bf FAKEBOB}~\cite{chen2019real}. FAKEBOB is similar to PGD except that it estimates gradients via Natural Evolution Strategy (cf. Section~\ref{sec:adaptive-attack-2-defense})
and the attack starts from the original input voice instead of a randomly perturbed
one. FAKEBOB adopts an early-stop strategy to reduce the number of queries, i.e., stop searching once an adversarial example is found.
Similar to the CW attack, FAKEBOB also provides an option to control the confidence of adversarial examples via
a parameter $\kappa$. Furthermore, FAKEBOB proposed the first algorithm to estimate the threshold for SV and OSI tasks.

\smallskip\noindent {\bf SirenAttack}~\cite{du2020sirenattack}.
SirenAttack searches for adversarial examples by leveraging the Particle Swarm Optimization (PSO)~\cite{PSO}, a gradient-free optimization method.
PSO first imitates the behaviour of a swarm of birds. Each particle is a candidate solution,
and is iteratively updated via the weighted linear combination of three parts, i.e., inertia, local best solution and global best solution.
When the algorithm terminates, an adversarial example is found.


\subsubsection{Attack Metrics}
In general, a good adversarial example should not only fool the model (effectiveness), but also be imperceptible to avoid being noticed by humans (stealthiness).
To evaluate the effectiveness, we define attack success rate ({ASR}).
For untargeted attack, ASR is the proportion of examples that are misclassified by the model.
For targeted attack, ASR is the proportion of examples that are recognized as the chosen targeted label.
To measure the stealthiness, we use the standard $L_2$ norm~\cite{carlini2017towards}, Signal-to-Noise Ratio (SNR)~\cite{yuan2018commandersong} and Perceptual Evaluation of Speech Quality (PESQ)~\cite{rix2001perceptual}. SNR is defined as $10\log_{10}\frac{P_x}{P_\delta}$, where $P_x$ is the power of the original example and $P_\delta$ is the power of the perturbation.
The calculation of PESQ is more involved.
{Intuitively, PESQ first applies an auditory transform to obtain the loudness spectra of the original and adversarial voices, and then compares these two loudness spectra to obtain a metric score whose value is in the range of -0.5 to 4.5~\cite{WangC18a}.}
We refer readers to \cite{rix2001perceptual} for more details. Smaller $L_2$, larger SNR and higher PESQ indicate better stealthiness.
Remark that our platform also provides Short-Time Objective Intelligibility score (STOI)~\cite{TaalHHJ11}, $L_0$, $L_1$, and $L_\infty$ norms
for measuring stealthiness.

\subsection{Defenses}\label{sec:review-defenses}
Though intensive defense methods against adversarial examples in computer vision
have been proposed, not all of them can be easily applied in SR.
Thus, in this section, we present potential effective
defense methods in SR.
These methods are grouped into robust training and input transformation. 

\subsubsection{Robust Training}
Robust training strengthens resistance of a model to adversarial examples.
Adversarial training is one of the most effective techniques, which augments the training data with adversarial examples.
Efficient adversarial attacks such as FGSM and PGD are widely-used for adversarial training~\cite{goodfellow2014explaining,madry2017towards,jati2021adversarial}.
In our platform, we implement both FGSM and PGD based adversarial training for SR.

\begin{table}[t]
    \centering
    \footnotesize\setlength\tabcolsep{4pt}
    \caption{Transformations, where D, R and Freq. denote Differentiable, Randomized and Frequency. 
    The transformations highlighted in \textcolor{blue}{blue} color are borrowed from speech recognition, but have never been considered for securing SRSs.}
    \resizebox{0.48\textwidth}{!}{%
    \begin{tabular}{c|c|c|c|c}
    \hline
      & {\bf Name} & {\bf Parameter} & {\bf D} & {\bf R} \\ \hline
   \multirow{4}{*}{\rotatebox{270}{\begin{tabular}[c]{@{}c@{}}{\bf Time} \\  {\bf Domain}\end{tabular}}}
     & {\bf Quantization (QT)}~\cite{yang2018characterizing} &    $q$: quantized factor   & \xmark & \xmark \\ \cline{2-5}
     & \textcolor{blue}{{\bf Audio Turbulence (AT)}~\cite{yuan2018commandersong}} & $snr$: signal-to-noise ratio  & \cmark & \cmark \\ \cline{2-5}
     & {\bf Average Smoothing (AS)}~\cite{du2020sirenattack} & $k$: kernel size & \cmark & \xmark \\ \cline{2-5}
     & {\bf Median Smoothing (MS)}~\cite{yang2018characterizing} & $k$: kernel size& \cmark & \xmark \\ \hline
   \multirow{3}{*}{\rotatebox{270}{\begin{tabular}[c]{@{}c@{}}{\bf Freq.} \\  {\bf Domain}\end{tabular}}}
     &{\bf Down Sampling (DS)}~\cite{yuan2018commandersong,yang2018characterizing} & $\tau$: downsampling  frequency  & \cmark & \xmark \\ \cline{2-5}
     & \textcolor{blue}{{\bf Low Pass Filter (LPF)}~\cite{kwon2019poster}} & $f_s$: stopband edge  frequency & \cmark & \xmark \\ \cline{2-5}
     & \textcolor{blue}{{\bf Band Pass Filter (BPF)}~\cite{rajaratnam2018speech}} & \begin{tabular}[c]{@{}c@{}}$f_{pl},f_{pu}$: passband edge frequency \\ $f_{sl},f_{su}$: stopband edge frequency \end{tabular} & \cmark & \xmark \\ \hline
    \multirow{7}{*}{\rotatebox{270}{\begin{tabular}[c]{@{}c@{}}{\bf Speech} \\  {\bf Compression}\end{tabular}}}
      & \textcolor{blue}{{\bf OPUS}~\cite{vos2013voice}} & $b_o$: compression  bitrate  & \xmark & \xmark \\ \cline{2-5}
      & \textcolor{blue}{{\bf SPEEX}~\cite{valin2016speex}} & $b_s$: compression  bitrate& \xmark & \xmark \\ \cline{2-5}
      & \textcolor{blue}{{\bf AMR}~\cite{ekudden1999adaptive}} & $b_r$: compression  bitrate & \xmark & \xmark \\ \cline{2-5}
      & \textcolor{blue}{{\bf AAC-V}~\cite{bosi1997iso}} & $q_c$: quality & \xmark & \xmark \\ \cline{2-5}
      & \textcolor{blue}{{\bf AAC-C}~\cite{bosi1997iso}} & $b_c$: compression  bitrate & \xmark & \xmark \\ \cline{2-5}
      & \textcolor{blue}{{\bf MP3-V}~\cite{hacker2000mp3}} & $q_m$: quality  & \xmark & \xmark \\ \cline{2-5}
      & \textcolor{blue}{{\bf MP3-C}~\cite{hacker2000mp3}} & $b_m$: compression  bitrate & \xmark & \xmark \\ \hline
     {\bf Ours} & {\bf \defensename (\defensenameabbr)} &
     \begin{tabular}[c]{@{}c@{}}$cl_m$: cluster method \\ $cl_r$: cluster ratio \end{tabular}& \cmark & \cmark \\ \hline
    \end{tabular}
    }
    \label{tab:input-transform}
    \vspace*{-3mm}
\end{table}

\subsubsection{Input Transformation}\label{sec:input-transformation}
Input transformations mitigate adversarial examples by pre-processing the inputs before feeding them to the model.
However, we are not aware of any transformations dedicated for securing SR and existing transformations for securing speech/speaker recognition adopt the transformations that were originally used for securing image recognition.
Thus, 
we study various input transformations that may be effective for securing SR.
We group them into time/frequency domain,  and speech
compression, listed in Table~\ref{tab:input-transform},
where the third column shows tunable parameters, 
and the last two columns indicate if a method is differentiable and randomized. 



\smallskip
\noindent {\bf Time domain}. We consider
four time domain transformations:
Quantization (QT)~\cite{yang2018characterizing},
Audio Turbulence (AT)~\cite{yuan2018commandersong},
Average Smoothing (AS)~\cite{du2020sirenattack}
and Median Smoothing (MS)~\cite{yang2018characterizing}.
QT rounds the amplitude of each sample point of a voice to the nearest integer multiple of a factor $q$.
AT assumes that an adversarial perturbation is sensitive to the noise, so it adds random noise to an input voice to disrupt the perturbation.
The magnitude of the noise is adjusted by $snr=\frac{P_I}{P_n}$ where $P_I$ (resp. $P_n$) is the power of input voice (resp. random noise).
AS mitigates adversarial examples by applying mean smooth to the waveform of the input voice. A mean smooth with kernel size $k$ (must be odd) replaces each element $x_k$ with the \emph{mean} value of its $k$ neighbors. 
MS is similar to AS except that it replaces a voice element $x_k$ with the \emph{median} value of its $k$ neighbors.
QT is non-differentiable due to the round operation, while the others are differentiable.
AT is a randomized transformation while the others are deterministic.


%
%

\smallskip
\noindent {\bf Frequency domain}.
We consider three frequency domain transformations:
Down Sampling (DS)~\cite{yang2018characterizing},
Low Pass Filter (LPF)~\cite{kwon2019poster}
and Band Pass Filter (BPF)~\cite{rajaratnam2018speech}.
DS, called Audio Squeezing in~\cite{yuan2018commandersong}, down-samples voices and applies signal
recovery to disrupt perturbations.
The down-sample frequency is determined by the ratio, denoted by $\tau$, between the new and original sampling frequencies.
LPF assumes that human speeches are within relatively lower frequencies than adversarial perturbation, and applies a low-pass filter to remove the high-frequent perturbations.
A low-pass filter has two  parameters: the edge frequencies of the pass and stop band, denoted by $f_p$ and $f_s$, respectively.
Since LBF fails to remove perturbations whose frequencies are lower than that of human speeches,
BPF combines LPF with a high-pass filter. BPF has four parameters: lower and upper edge frequencies of pass band (denoted by $f_{pl}$ and $f_{pu}$),  lower and upper cutoff frequencies of stop band (denoted by $f_{sl}$ and $f_{su}$). These transformations are differentiable and deterministic.

\smallskip\noindent {\bf Speech compression~\cite{das2018adagio}\cite{rajaratnam2018speech}\cite{andronic2020mp3}}. Based on psychoacoustic principle,
speech compression aims to suppress the redundant information within a speech to improve storage or transmission efficiency.
When imperceptible adversarial perturbation is redundant information, it can be eliminated by speech compression.
Speech compression achieves the aforementioned purpose by reducing the bit rate. We investigate 7 standard speech compression techniques, which are grouped into two categories: Constant Bit Rate (CBR) and Variable Bit Rate (VBR). The former uses a fixed bit rate and the latter exploits dynamic bit rate schedule controlled by the quality parameter.
For CBR, we consider OPUS~\cite{vos2013voice}, SPEEX~\cite{valin2016speex}, Adaptive Multi-Rate Codec~\cite{ekudden1999adaptive} (AMR), Advanced Audio Coding~\cite{bosi1997iso} (AAC-C) and MP3~\cite{hacker2000mp3} (MP3-C). For VBR, we consider AAC (AAC-V) and MP3 (MP3-V).
These transformations are non-differentiable and deterministic.

%

\subsubsection{Defense Metrics}\label{sec:defense-metric}
An effective defense should not only improves resistance to adversarial examples,
but also sacrifices accuracy on benign examples as less as possible.
These requirements are measured by accuracy on adversarial examples $A_a$ and
accuracy on benign examples $A_b$, respectively.
We also use R1 score, defined as $R1=\frac{2\times A_b\times A_a}{A_b+A_a}$~\cite{rajaratnam2018speech},
to assign equal importance to both  $A_b$ and $A_a$.

\subsection{Adaptive Attacks}\label{sec:adaptive-attack-2-defense}
Adaptive attacks~\cite{CW17a,tramer2020adaptive} are specific attacks
designed to circumvent a given defense.
The parameter $\kappa$ in CW and the number of iterations in PGD (see above) are positively correlated with the strength of adversarial examples, thus
have been used to circumvent defenses in image recognition~\cite{zhao2021attack,tramer2020adaptive}.
Below, we introduce three more techniques for adaptive attacks from image recognition:
Backward Pass Differentiable Approximation (BPDA), Expectation over Transformation (EOT)
and Natural Evolution Strategy (NES),
to circumvent non-differentiable, randomized transformations
and estimate gradients in block-box setting, respectively.

\smallskip
\noindent {\bf BPDA}~\cite{athalye2018obfuscated}.
To defeat gradient-based attacks, one common solution is to make the system non-differentiable, e.g., by adding
some non-differentiable components into the system. BPDA was proposed to circumvent such defenses.
Consider a system $f(x)=f^n\circ \cdots \circ f^{k} \circ \cdots \circ f^{1}(x)$ consisting of $n$ consecutive components
$f^i$. Suppose $f^{k}(\cdot)$ is non-differentiable, then gradient-based attacks that leverage backpropagation become ineffective, as
$\nabla_{x}f^{k}(x)$ is unavailable or uninformative.
BPDA solves this problem by replacing the backward pass of $f^{k}(\cdot)$ with one of a differentiable function $g(\cdot)$, i.e., approximating $\nabla_x f^{k}(x)$ with $\nabla_x g(x)$.
In SR, there are several non-differentiable input transformations (cf. Table~\ref{tab:input-transform}),
when deployed to mitigate adversarial examples, the component $f^{1}(\cdot)$ becomes non-differentiable.
Thus, BPDA might be useful to circumvent such defenses.

%

\smallskip
\noindent {\bf EOT}~\cite{athalye2018synthesizing}.
Randomization is a general technique to mitigate adversarial attacks by
introducing some randomized transformations into the system.
Randomized transformations behaves differently during the process of adversarial example generation and inference, thus a successfully crafted adversarial example may become ineffective at inference time~\cite{xie2017mitigating}. EOT was proposed for the construction of
adversarial examples that remain adversarial over a chosen transformation distribution~\cite{athalye2018synthesizing}
and has been used to circumvent randomized transformations~\cite{athalye2018obfuscated}.
In SR, there are randomized transformations (cf. Table~\ref{tab:input-transform}).
When they are deployed in SR, EOT might be useful to circumvent them,
where the transformation distribution is approximated by sampling $r$ times.

\smallskip
\noindent {\bf NES}~\cite{wierstra2014natural}.
Both BPDA and EOT are not suitable for mount black-box adaptive attacks.
Thus, we provide NES for estimating gradients in black-box attacks.
Specifically, it first creates $m$ noisy examples by adding Gaussian noises onto an example.
Then, the values of the loss function of $m$ examples are obtained by querying the model, which are finally exploited to approximate the gradient.
NES has already been used in the black-box attack FAKEBOB
which is able to achieve high attack success rate.

\section{Our Approach: \defensename}\label{sec:our-approach}
Apart from the above input transformations,
in this work, we propose a novel type of \emph{feature-level} transformations dedicated to SRSs, called \defensename.


\subsection{Motivation}
Due to the success of input transformations (e.g., JPEG compression~\cite{GuoRCM18} and local smoothing~\cite{Xu0Q18})
for mitigating adversarial images, some input transformations (e.g., MP3 compression~\cite{andronic2020mp3} and median smoothing~\cite{yang2018characterizing}) have been proposed to
mitigating adversarial voices~\cite{das2018adagio}. However, existing transformations for adversarial voices overlook the internal difference between image and voice recognitions. Thus,
transformations are \emph{only} applied to the input waveform signals, and \emph{feature-level} transformations have never been considered before.
%


For state-of-the-art neural network based image recognition, an image is directly fed to a system without feature engineering.
In other words, feature engineering is left to be handled by the internal neurons of the network.
Due to the time-varying non-stationary property of voices, voices are not resilient enough to noises and other variation,
and waveform signals themselves cannot effectively represent speaker characteristics~\cite{voice-feature-review,abs-1305-1145}.
Hence, to achieve better feature representative capacity and system performance~\cite{xiao2016speech}, SR often relies on feature engineering to extract hand-crafted features from voice waveforms (cf. Figure~\ref{fig:typical-SRSs}), such as speech spectrogram~\cite{hannun2014deep}\cite{amodei2016deep}, Filter-bank~\cite{FilterBanks}\cite{li2017deep}, MFCC~\cite{muda2010voice}, and PLP~\cite{hermansky1990perceptual}.
When an adversarial perturbation is crafted in the voice waveform,
it will also be propagated to the features during inference time.

Based on the above observation, we investigate feature-level transformations to mitigate adversarial examples in SR.
The design of our approach
is motivated by the following
questions:
(Q1) What kind of features can be transformed?
(Q2) How to transform features?
(Q3) How effective are they?



\begin{figure}[t]
    \centering
    \includegraphics[width=0.5\textwidth]{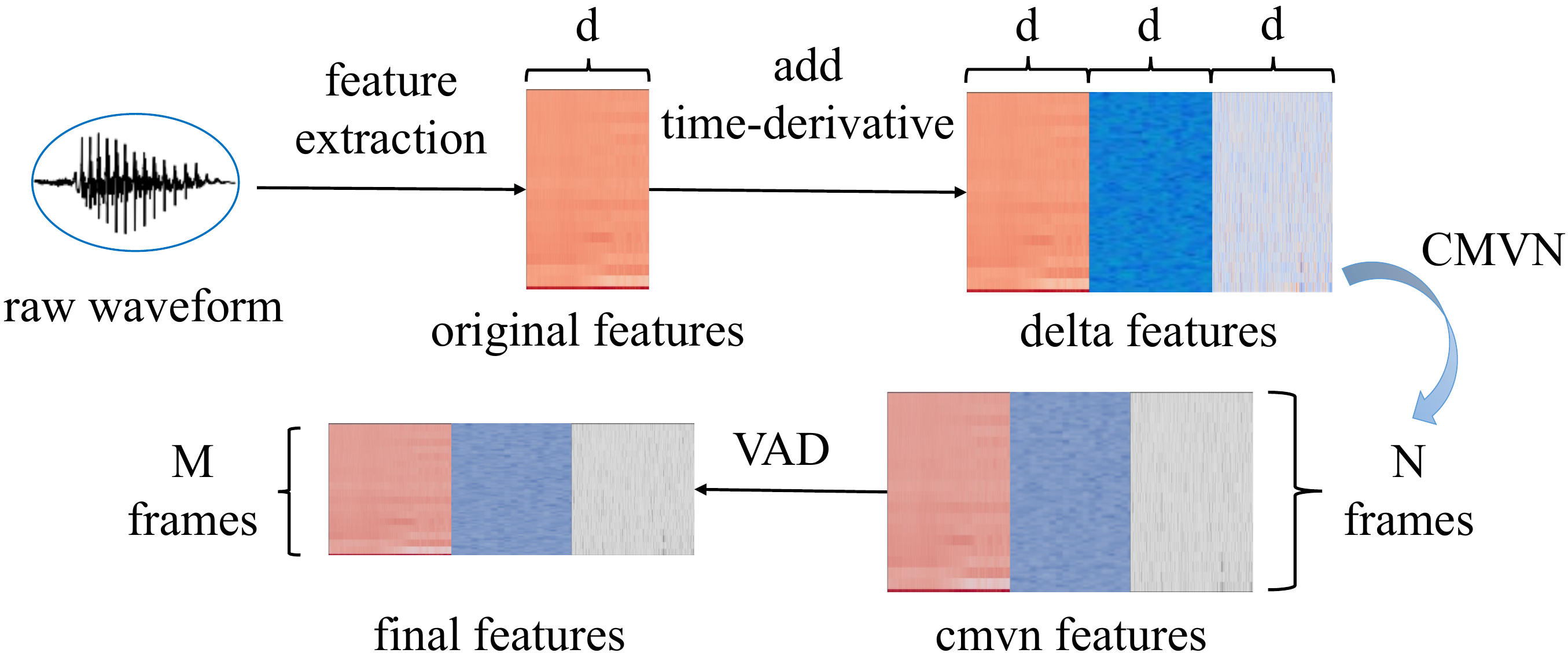}\vspace*{-3mm}
    \caption{A typical flow of feature processing.}
    \label{fig:different-features}\vspace*{-3mm}
\end{figure}

\subsection{Design}\label{sec:our-approach-implementation}
In this section, we present our solution addressing Q1 and Q2, while Q3 is studied in Section~\ref{sec:evaldefense}.

To address Q1, we have to understand what kind of features are used in SR.
Figure~\ref{fig:different-features} shows a typical flow of feature processing in SR.
First, the \emph{original features} (e.g., MFCC or PLP) are extracted from an input raw waveform.
Next, to capture temporal information, time-derivative features including first-order and second-order time derivatives~\cite{xiao2016speech}
are 
successively
extracted from and added into the original features, leading to the \emph{delta features}.
At the third step, cepstral mean and variance normalization (CMVN)~\cite{viikki1998cepstral} is applied to the delta features
which reduces channel and reverberation effects, resulting in \emph{cmvn features}.
Finally, voice activity detection (VAD)~\cite{sohn1999statistical} is utilized to remove the unvoiced frames from the cmvn features and obtain the \emph{final features}.
Therefore, these four types of features could be transformed, which may disrupt adversarial perturbations at feature level.

To address Q2, a straightforward idea is to directly extend the existing transformations from input waveforms to features.
However, such extension is \emph{not} trivial due to the following reasons.
(1) All the transformations mentioned in Section~\ref{sec:input-transformation} work on voice waveforms, each of which
is a vector, where the value at each index denotes the magnitude at the corresponding sample point.
While features of an input voice
are represented by a matrix,
one row of features per frame.
This difference prevents frequency domain transformations (i.e., DS, LPF, and BPF) and speech compressions from being extended to feature level.
(2) The mapping from waveforms to features is not linear,
and a small perturbation in the input voice may result in a large
perturbation at the feature level.
This difference refuses time domain transformations (i.e., QT, AT, AS and MS)
which assumed that perturbations are small and/or sensitive to noises.

Alternatively, inspired by speech compressions, we propose, \defensename (FC), a feature compression approach
to disrupt perturbations at feature level. For a feature matrix $M$ with $N$ frames
and each frame consists of $d$ features, we regard the matrix $M$ as $N$ data points in a $d$-dimensional space and
partition $N$ data points into $K$ clusters for a given parameter $K<N$.
Then, all data points in one cluster are represented by a representative vector.
Finally, $K$ representative vectors are combined to form the new feature matrix $M'$.
We denote the ratio between $N$ and $K$ by $cl_r=\frac{K}{N}$.

To partition $N$ data points into $K$ clusters, various clustering methods such as kmeans~\cite{hartigan1979algorithm}, soft-kmeans~\cite{mackay2003information} and Gaussian Mixture Model~\cite{fraley1998many}
could be leveraged. Currently, our platform supports kmeans and its variant warped-kmeans~\cite{LeivaV13}.
We leave other cluster methods as future work.
Compared to kmeans, warped-kmeans preserves the temporal dependency of the data by imposing some constraints on the partition operation, thus is more suitable to cluster sequential data.
For both kmeans and warped-kmeans, the representative vector of a cluster is the average of all data points within this cluster.
Furthermore, both kmeans and warped-kmeans are randomized methods, thus our approach is randomized.

Our approach could be applied to any types of features mentioned above, i.e., original, delta, cmvn and final features.
We use \defensename-origin (\defensenameabbr-o), \defensename-delta (\defensenameabbr-d), \defensename-cmvn (\defensenameabbr-c), \defensename-final (\defensenameabbr-f) to denote
these four concrete feature compression methods.

\section{\platformname Platform}\label{sec:design-implement}
To advance future research on adversarial examples in SR,
in this section, we present the design and implementation of a SECurity analysis platform for SR, named \platformname~\cite{SEC4SR}.
Our platform is designed to be modular, flexible and extensible
so that new models, datasets,
attacks and defenses can be easily integrated into the platform to comprehensively and systematically evaluate
their performance.

\begin{figure}[t]
    \centering
    \includegraphics[width=0.5\textwidth]{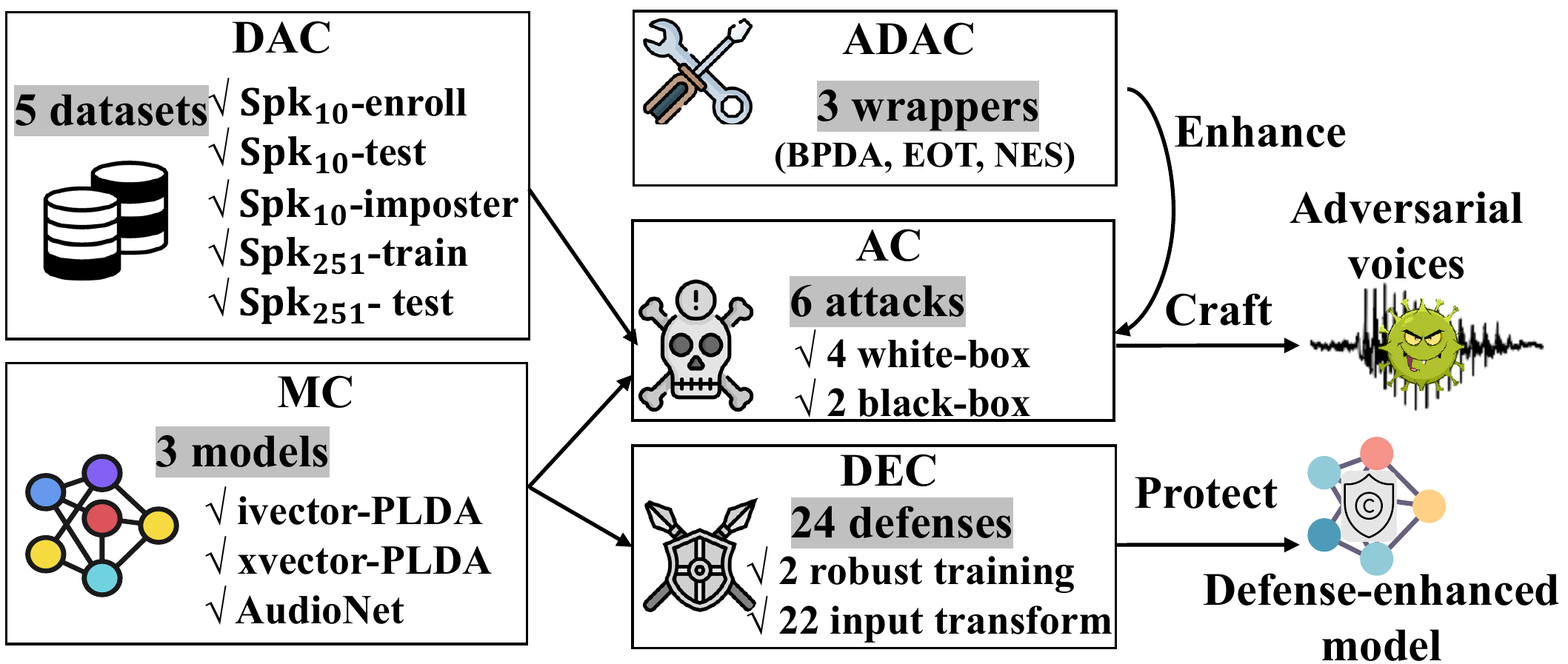}\vspace*{-3mm}
    \caption{Overview of our platform \platformname.}
    \label{fig:overview-platform}\vspace*{-3mm}
\end{figure}

The overview of our platform is shown in Figure~\ref{fig:overview-platform}, consisting of
the following five main components.


\smallskip

\smallskip\noindent {\bf Model component (MC).}
Adversarial attacks and defenses in SR have attracted increasing attentions~\cite{chen2019real,du2020sirenattack,ASGBWYST19,li2020adversarial,xie2021real,abs-1801-03339,jati2021adversarial,AWBPT20}.
However, previous works benchmark their attacks and defenses on different SRSs,
varying in front-end features (e.g., MFCC and PLP), model architectures, recognition tasks, training algorithms, 
and back-end scoring methods.
This hampers researchers to comprehensively and systematically evaluate the proposed methods.

To tackle this problem, \platformname provides 3 mainstream SRSs including
ivector-PLDA~\cite{kaldi-ivector-plda} and xvector-PLDA~\cite{kaldi-xvector-plda} that are pre-trained models of the popular open-source platform KALDI having 10k stars and 4.4k forks on GitHub~\cite{kaldi},
and a one-dimension convolution neural network based model AudioNet that was proposed in \cite{becker2018interpreting} and studied in \cite{jati2021adversarial}.
These models cover all the recognition tasks.
Details of these models are summarized in Table~\ref{tab:MC}.

\smallskip\noindent {\bf Dataset component (DAC).}
Open-access voice datasets are provided for training and testing of speaker or speech recognition,
but not for evaluating adversarial attacks and defenses.
The latter often requires different types of voice sets for different recognition tasks.
For instance, enrollment speakers are not required for the CSI-NE task, but are required for the other tasks (e.g., CSI-E, SV, and OSI).
Imposters are not required for the CSI-E and CSI-NE tasks, but are required for the SV and OSI tasks.
{Since there is no uniform set up of voice sets for evaluating adversarial attacks or defenses,
previous studies \textit{randomly} choose different subsets of voices from different voice datasets~\cite{chen2019real, shamsabadi2021foolhd, WangGX20, LiZXZ0M020}.}
This hinders comprehensive and systematic evaluation of adversarial attacks and defenses.

To solve this problem, \platformname provides 5 datasets: Spk$_{10}$-enroll, Spk$_{10}$-test, Spk$_{10}$-imposter, Spk$_{251}$-train, and Spk$_{251}$-test,
covering the required datasets for all the recognition tasks (i.e., CSI-NE, CSI-E, SV, and OSI).
The datasets are summarized in Table~\ref{tab:DAC}.

\begin{table}[t]
    \centering
     \caption{Models in MC, where T/D denote traditional/deep model, US/S stand for unsupervised/supervised learning.}
    \resizebox{0.45\textwidth}{!}{
    \begin{tabular}{c|c|c|c}
    \hline
     & {\bf ivector-PLDA} & {\bf xvector-PLDA} & {\bf AudioNet} \\ \hline
      \begin{tabular}[c]{@{}c@{}}{\bf Embedding type}\end{tabular} & T & D & D \\ \hline
      {\bf Feature type} & MFCC & MFCC & MFCC \\ \hline
      {\bf Add delta} & \cmark & \xmark & \xmark \\ \hline
      {\bf Add acceleration} & \xmark & \cmark & \xmark \\ \hline
      {\bf Apply CMVN} & \cmark & \cmark & \xmark \\ \hline
      {\bf Apply VAD} & \cmark & \cmark & \xmark \\ \hline
      {\bf \#Feature dim} & 72 & 30 & 32 \\ \hline
      {\bf Training algorithm} & US & S & S \\ \hline
      {\bf Scoring method} & PLDA & PLDA & COSS \\ \hline
    \end{tabular}
    }
    \label{tab:MC} \vspace*{-2mm}
\end{table}
\begin{table}[t]
    \centering
\setlength\tabcolsep{2pt}
    \caption{Voice datasets in DAC, where x-y(z) indicates that the minimal, maximal and average length of voices are x, y, z.}
    \resizebox{0.48\textwidth}{!}{
    \begin{tabular}{c|c|c|c|c|c}
    \hline
         & {\bf Spk$_{10}$-enroll} & {\bf Spk$_{10}$-test} & {\bf Spk$_{10}$-imposter} & {\bf Spk$_{251}$-train} & {\bf Spk$_{251}$-test} \\ \hline
         {\bf Task} & \multicolumn{3}{c|}{CSI-E/SV/OSI } & \multicolumn{2}{c}{CSI-NE} \\ \hline  
         {\bf \#Speakers} & 10(5M,5F) & 10(5M,5F) & 10(5M,5F) &
        {\begin{tabular}[c]{@{}c@{}} 251\\(126M,125F)\end{tabular}}
         &
        {\begin{tabular}[c]{@{}c@{}} 251\\(126M,125F)\end{tabular}}
         \\ \hline
         {\bf \#Voices} & 10$\times$10 & 100$\times$10 & 100$\times$10 & 25652 & 2887 \\ \hline
         {\bf Length} & 3--21s (7.2s) & 1--15s(4.3s) & 1--17s(8.33s) & 1--24s (12.3s) & 1--19s (11.7s) \\ \hline
    \end{tabular}
    }
    \label{tab:DAC} \vspace*{-2mm}
\end{table}


Spk$_{10}$-enroll consists of 10 speakers (5 males and 5 females), 10 voices per speaker.
The speakers are randomly selected from the ``test-other'' and ``dev-other'' subsets of the popular dataset Librispeech~\cite{panayotov2015librispeech,LiZJXZWM020,zhang2021attack,jati2021adversarial,chen2019real}.
For each speaker, we select the top-10 longest voices in order to have better enrollment embedding~\cite{ParkYKKA17, app9183697}. The voices in Spk$_{10}$-enroll are used for speaker enrollment
of the CSI-E, SV, and OSI tasks.
%
Spk$_{10}$-test (resp. Spk$_{10}$-imposter) consists 10 speakers (5 males and 5 females), 100 randomly selected voices per speaker.
Spk$_{10}$-test has the same speakers as Spk$_{10}$-enroll, but distinct voices.
The speakers in Spk$_{10}$-imposter are randomly chosen from the ``train-other-500" subset of Librispeech, but are different from the speakers in Spk$_{10}$-test.
Both Spk$_{10}$-test and Spk$_{10}$-imposter can be used for adversarial attacks on the CSI-E, SV, and OSI tasks,
where the speakers of Spk$_{10}$-test have been enrolled, but
the speakers of Spk$_{10}$-imposter have not been enrolled.


Both Spk$_{251}$-train and Spk$_{251}$-test
are taken from the ``train-clean-100'' subset of Librispeech,
each of which has the same 251 speakers (126 males and 125 females).
Following \cite{jati2021adversarial}, for each speaker, 90\% of his/her voices are added into Spk$_{251}$-train,
and the remaining 10\% are added into Spk$_{251}$-test.
Spk$_{251}$-train can be used to train background
models while Spk$_{251}$-test can be used for adversarial attacks
on the CSI-NE task.
Since there are no overlapping speakers among Spk$_{251}$-train, Spk$_{10}$-test, and Spk$_{10}$-imposter,
if one prefers to attack the CSI-NE task using the voices
of the speakers that are not in Spk$_{251}$-train,
voices in Spk$_{10}$-enroll, Spk$_{10}$-test and Spk$_{10}$-imposter could be used.
%
%

\smallskip\noindent {\bf Attack Component (AC).}
The main function of AC is to explore vulnerabilities of SRSs by launching adversarial attacks.
AC implements the representative and state-of-the-art attacks mentioned in Section~\ref{sec:review-attacks},
namely, 4 white-box attacks: FGSM, PGD and CW$_{\bm{\infty}}$, CW$_2$; and 2 black-box attacks: FAKEBOB and SirenAttack.
Note that CW$_{\bm{\infty}}$ is implemented by adopting the loss function of CW  but optimizing by PGD, the same as \cite{madry2017towards} and \cite{dong2020adversarial},
in order to improve the attack efficiency.

\smallskip\noindent {\bf Defense Component (DEC).}
The main function of DEC is to protect SRSs and improve their resistance to adversarial examples.
DEC implements all defense methods mentioned in Section~\ref{sec:review-defenses}
and Section~\ref{sec:our-approach}, including robust training and input-/feature-level transformations.
Specifically, for robust training, we implement both FGSM and PGD based adversarial training.
For input transformation, we implement 14 input transformations and 8 novel feature transformations.

\smallskip\noindent {\bf Adaptive Attack Component (ADAC).}
ADAC is mainly used to implement adaptive attacks when the adversary aims to circumvent a chosen defense.
Currently, ADAC includes both BPDA, EOT and NES (cf. Section~\ref{sec:adaptive-attack-2-defense}).
They are implemented as standalone wrappers so that they can be easily plugged into attacks
to mount adaptive attacks. 

We strongly encourage researchers to
add new models, datasets, attacks and defenses into our platform,
and  comprehensively and
systematically evaluate them using our platform.
We expect our platform can advance future research on adversarial examples in SR.

\section{Evaluation}\label{sec:evaluation}
In this section, we first evaluate the effectiveness and stealthiness of
adversarial attacks implemented in \platformname, then study the effectiveness of defenses
against non-adaptive and adaptive attacks.
Throughout the evaluation, we limit the 
perturbation budget 
$\epsilon$ to 0.002 for $L_\infty$ attacks, the same as \cite{chen2019real}\cite{jati2021adversarial}.
Note that CW$_2$ minimizes adversarial perturbations in the loss function, and hence does not have any limitations.
To avoid fake adversarial examples due to the discretization problem~\cite{bu2019taking}, namely,  adversarial examples become benign after transformed back into concrete voices,
we evaluate adversarial voices after storing them back into the 16-bit PCM form.

Due to massive experiments, we only target the ivector-PLDA and AudioNet models (cf. Section~\ref{sec:design-implement})
for the CSI task (CSI-E or CSI-NE). 
The results on the SV and OSI tasks could be similar, as demonstrated in \cite{chen2019real}.
We conduct  experiments on a machine with an Intel Xeon E5-2697 v2 2.70GHz CPU, 376GiB memory, and a GeForce RTX 2080Ti GPU.

\begin{table}[t]
    \centering
    \caption{Effectiveness and stealthiness of untargeted attacks.}
\resizebox{0.45\textwidth}{!}{
    \begin{tabular}{c|c|c|c|c|c}
    \hline
      \multicolumn{2}{c|}{\multirow{2}{*}{\bf Attack}} & {\bf Effectiveness} & \multicolumn{3}{c}{\bf Stealthiness} \\ \cline{3-6}
       \multicolumn{2}{c|}{} & {\bf ASR} & \bm{$L_2$} & {\bf SNR} & {\bf PESQ} \\ \hline
       \multicolumn{2}{c|}{\bf FGSM} & 57.7\% & 0.537 & 28.53 & 2.23 \\ \hline
       \multirow{6}{*}{\bf {PGD-$\mathbf{x}$}} & $\mathbf{x}${\bf =10} & 100\% & 0.330 & 32.77 & 2.85 \\ \cline{2-6}
       & $\mathbf{x}${\bf =20} & 100\% & 0.378 & 31.57 & 2.72 \\ \cline{2-6}
       & $\mathbf{x}${\bf =30} & 100\% & 0.385 & 31.42 & 2.70 \\ \cline{2-6}
       & $\mathbf{x}${\bf =40} & 100\% & 0.384 & 31.45 & 2.71 \\ \cline{2-6}
       & $\mathbf{x}${\bf =50} & 100\% & 0.390 & 31.31 & 2.69 \\ \cline{2-6}
       & $\mathbf{x}${\bf =100} & 100\% & 0.391 & 31.29 & 2.70 \\ \hline
       \multicolumn{2}{c|}{\bf CW\bm{$\infty$}-0} & 100\% & 0.378 & 31.57 & 2.72 \\ \hline
       \multirow{5}{*}{\bf CW$_2$-$\kappa$} & $\mathbf{\kappa}${\bf =0} & 93.5\% & 0.041 & 52.99 & 4.24 \\\cline{2-6}
       & $\mathbf{\kappa}${\bf =5} & 100\% & 0.058 & 49.73 & 4.10 \\ \cline{2-6}
       & $\mathbf{\kappa}${\bf =10} & 100\% & 0.078 & 47.09 & 3.95\\ \cline{2-6}
    & $\mathbf{\kappa}${\bf =20} & 100\% & 0.131 & 42.14 & 3.60 \\ \cline{2-6}
      & $\mathbf{\kappa}${\bf =50} & 100\% & 0.476 & 30.44 & 2.46 \\ \hline
      \multirow{2}{*}{\bf FAKEBOB}&\#Iter=1000 & 99.7\% & 0.397 & 31.12  & 2.67  \\ \cline{2-6}
     &\#Iter=200& 80.2\% & 0.383 & 31.40 & 2.71 \\ \hline
    \end{tabular}}
    \label{tab:evaluate-atatck} \vspace*{-2mm}
\end{table}

\subsection{Evaluation of Attacks}\label{sec:evaluate-attack}
\noindent{\bf Experimental Setup}.
We consider the ivector-PLDA model for the CSI-E task which is enrolled with 10 speakers using Spk$_{10}$-enroll. 
We use Spk$_{10}$-test to test the model, resulting in 99.8\% accuracy on benign examples.
We will also use Spk$_{10}$-test to generate adversarial examples.
The target label of targeted attack for each voice is randomly chosen among the labels except the ground truth.

We use 10, 20, 30, 40, 50 and 100 steps for PGD, and 30 steps for CW$_{\bm{\infty}}$,
with step size $\alpha=\frac{\epsilon}{5}=0.0004$.
The step size of FGSM is $\epsilon=0.002$.
For CW$_2$, we use 9 binary search steps to minimize adversarial perturbations, run 900-9000 iterations to converge, and vary the parameter $\kappa$ from 0, 5, 10, 20 to 50.
For FAKEBOB, we set the iteration limit (denoted by \#Iter) to 1000 and samples\_per\_draw $m$ of NES to $50$, the same as~\cite{chen2019real}.
We set $\kappa=0.5$ for FAKEBOB so that adversarial voices remain adversarial after stored
back into the 16-bit PCM form.
We do not study  SirenAttack as it is less effective than FAKEBOB~\cite{chen2019real}.

\begin{table}[t]
    \centering
    \caption{Effectiveness and stealthiness of targeted attacks.}
    \resizebox{0.45\textwidth}{!}{
    \begin{tabular}{c|c|c|c|c|c}
    \hline
      \multicolumn{2}{c|}{\multirow{2}{*}{\bf Attack}} & {\bf Effectiveness} & \multicolumn{3}{c}{\bf Stealthiness} \\ \cline{3-6}
       \multicolumn{2}{c|}{} & {\bf ASR} & \bm{$L_2$} & {\bf SNR} & {\bf PESQ} \\ \hline
       \multicolumn{2}{c|}{\bf FGSM} & 40.2\% & 0.537 & 28.53 & 2.23 \\ \hline
       \multirow{6}{*}{\bf {PGD-$\mathbf{x}$}} & $\mathbf{x}${\bf =10} & 99.7\% & 0.317 & 33.12 & 2.87 \\ \cline{2-6}
       & $\mathbf{x}${\bf =20} & 99.7\% & 0.356 & 32.10 & 2.77 \\ \cline{2-6}
       & $\mathbf{x}${\bf =30} & 99.9\% & 0.368 & 31.82 & 2.75 \\ \cline{2-6}
       & $\mathbf{x}${\bf =40} & 99.9\% & 0.373 & 31.69 & 2.74 \\ \cline{2-6}
       & $\mathbf{x}${\bf =50} & 100\% & 0.376 & 31.63 & 2.73 \\ \cline{2-6}
       & $\mathbf{x}${\bf =100} & 100\% & 0.380 & 31.53 & 2.72 \\ \hline
       \multicolumn{2}{c|}{\bf CW\bm{$\infty$}-0} & 100\% & 0.369 & 31.79 & 2.74 \\ \hline
       \multirow{5}{*}{\bf CW$_2$-$\kappa$}  & $\mathbf{\kappa}${\bf =0} & 96.9\% & 0.083 & 46.93 & 3.92 \\ \cline{2-6}
       & $\mathbf{\kappa}${\bf =5} & 100\% & 0.105 & 44.53 & 3.77 \\ \cline{2-6}
       & $\mathbf{\kappa}${\bf =10} & 100\% & 0.139 & 41.79 & 3.57 \\ \cline{2-6}
    & $\mathbf{\kappa}${\bf =20} &  100\% & 0.290 & 35.03 & 2.94 \\ \cline{2-6}
      & $\mathbf{\kappa}${\bf =50} &  100\% &  1.714 & 18.91 & 1.51 \\ \hline
      \multirow{2}{*}{\bf FAKEBOB}&\#Iter=1000 & 95.2\% & 0.383 & 31.44 & 2.70  \\ \cline{2-6}
     &\#Iter=200& 51.5\% & 0.378 & 31.60 & 2.72 \\ \hline
    \end{tabular}
    }
    \label{tab:evaluate-atatck-target} \vspace*{-2mm}
\end{table}
{Table~\ref{tab:evaluate-atatck} and Table~\ref{tab:evaluate-atatck-target} show the results of untargeted and target attacks, respectively.}

\smallskip\noindent {\bf Effectiveness.}
We can observe from Table~\ref{tab:evaluate-atatck} that
FGSM is less effective than others, with only $57.7\%$ ASR. This is because FGSM is a single-step attack.
PGD, CW$_{\bm{\infty}}$ and FAKEBOB are the most powerful attacks whose ASR is close to 100\% ASR.
When $\kappa=0$, the ASR of CW$_2$ is 93.5\%.
With the increase of $\kappa$, CW$_2$ achieves 100\% ASR.

From Table~\ref{tab:evaluate-atatck-target}, we can observe that
FGSM, PGD-10, PGD-20, PGD-30, PGD-40, and FAKEBOB become slightly less effective
for targeted attack, while the others have the same ASR
for both targeted and untargeted attacks.

We remark that CW$_2$-0 achieved 100\% ASR before storing the adversarial examples back into the 16-bit PCM form.
This means that CW$_2$-0 suffers from the discretization problem~\cite{bu2019taking}.
This is because  CW$_2$ seeks for minimal adversarial perturbations which are too small to be easily disrupted when $\kappa=0$.
This problem is avoided by increasing $\kappa$.

We also report the results of FAKEBOB with 200 iteration limit.
It becomes 19.5\% less effective for untargeted attack and 43.7\% less effective for targeted attack.
FAKEBOB with 200 iteration limit will be used to evaluate defenses
against untargeted attacks in consideration of the experiment effort.

\smallskip\noindent {\bf Stealthiness.}
From Table~\ref{tab:evaluate-atatck}, we observe that FAKEBOB, PGD and CW$_{\bm{\infty}}$  achieve comparable
results in terms of $L_2$ norm, SNR, and PESQ.
Unsurprisingly, adversarial voices crafted by FGSM are more perceptible than others.
When $\kappa\leq 20$, CW$_2$ achieves the best imperceptibility with at least 42 dB SNR and 3.6 PESQ.
This is because CW$_2$ uses a binary search to minimize perturbations. However, with the increase of $\kappa$, CW$_2$ becomes significantly more perceptible.

From Table~\ref{tab:evaluate-atatck-target}, we can observe that
FGSM, PGD-10, PGD-20, PGD-30, PGD-40, and FAKEBOB for targeted attack
achieve similar stealthiness as for untargeted attack.
But CW$_2$ results in larger $L_2$ distance, smaller SNR and PESQ for targeted attack  than for untargeted attack,
indicating that targeted attack introduces larger distortions.

Overall, targeted attack is more challenging (i.e., lower ASR and more perceptible) than untargeted attack.

\subsection{Evaluation of Defenses}\label{sec:evaldefense}

We first evaluate transformations against non-adaptive attacks, where
adversarial examples are crafted on the original model without any transformations but are fed to the model with a transformation.
Then we consider adaptive attacks where the adversary knows the transformation and  crafts
adversarial examples on the model with the transformation.
Finally, we study adaptive attacks on adversarially trained (AdvT) models with some transformation.
To avoid bias when a randomized transformation is involved,
we report average results after testing 10 times.
We only consider untargeted attacks which are more challenging to be defeated than targeted attacks~\cite{athalye2018obfuscated}.

\subsubsection{Transformations against Non-Adaptive Attacks}\label{sec:non-adaptive-attack}
\noindent{\bf Experimental setup}.
We use the same setup as in Section~\ref{sec:evaluate-attack}, except
that transformations are involved when evaluating adversarial examples.
Though the ivector-PLDA model is pre-trained without any transformations,
it still produces sufficient accuracy on benign examples (cf. the third column in Table~\ref{tab:evaluate-defense-non-adaptive}).
Thus, we do not re-train the pre-trained ivector-PLDA model using voices after transformations.
As each transformation contains at least one tunable parameter
which may affect effectiveness, we tune parameters and choose the best ones according to their R1 scores for the remaining experiments.
Details are given in Appendix~\ref{sec:parameter}.

\begin{table*}
\caption{Results of transformations against non-adaptive attacks, where BLA denotes black-box attack, k (resp. wk) denotes kmeans (resp. warped-kmeans).
The top-3 highest/lowest results are highlighted in \textcolor{blue}{blue}/\textcolor{red}{red} color except for Baseline where no defense is deployed. The accuracy $A_a$ used for computing
R1 Score is the average of all the attacks.}
\label{tab:evaluate-defense-non-adaptive}
\resizebox{\textwidth}{!}{%
\begin{tabular}{c|c|c||c|c|c|c|c|c|c|c||c|c|c|c|c||c}
\hline
\multirow{2}{*}{\bf Defense} &  {\bf R1} & \multirow{2}{*}{$\mathbf{A_b}$} &  \multicolumn{8}{c||}{\bf ${A_a}$ of L$_\infty$ white-box attacks} &\multicolumn{5}{c||}{\bf ${A_a}$ of L$_2$ white-box attacks} &  {\bf ${A_a}$ of L$_\infty$ BLA}\\ \cline{4-17}
  & {\bf Score} &  &  {\bf FGSM} & {\bf PGD-10}& {\bf PGD-20} & {\bf PGD-30}& {\bf PGD-40} & {\bf PGD-50} & {\bf PGD-100}& {\bf CW\bm{$_\infty$}-0} & {\bf CW$_2$-0} & {\bf CW$_2$-5} &{\bf CW$_2$-10} & {\bf CW$_2$-20}&{\bf  CW$_2$-50}& {\bf FAKEBOB} \\ \hline \hline
{\bf Baseline} & 9.3\% & 99.8\% & 42.3\% & 0\% & 0\% & 0\% & 0\% & 0\% & 0\% & 0\% & 6.5\% & 0\% & 0\% & 0\% & 0\% & 19.8\% \\ \hline\hline
\textbf{QT} & \textcolor{blue}{\bf 77.7\%} & \textcolor{red}{\bf 86.8\%} & \textcolor{blue}{\bf 76.8\%} & \textcolor{blue}{\bf 61.2\%} & \textcolor{blue}{\bf 55.4\%} & \textcolor{blue}{\bf 56.6\%} & \textcolor{blue}{\bf 62.5\%} & \textcolor{blue}{\bf 59.8\%} & \textcolor{blue}{\bf 67.2\%} & \textcolor{blue}{\bf 60.2\%} & 86.8\% & 86.4\% & 86.2\% & \textcolor{blue}{\bf 84.9\%} & \textcolor{blue}{\bf 49.9\%} & \textcolor{blue}{\bf 91.3\%} \\ \hline
\textbf{AT} & \textcolor{blue}{\bf 85.8\%} & {89.2\%} & \textcolor{blue}{\bf 82.9\%} & \textcolor{blue}{\bf 77.8\%} & \textcolor{blue}{\bf 75.9\%} & \textcolor{blue}{\bf 75.6\%} & \textcolor{blue}{\bf 78.5\%} & \textcolor{blue}{\bf 76.6\%} & \textcolor{blue}{\bf 81.2\%} & \textcolor{blue}{\bf 78.4\%} & 89.1\% & 89.1\% & \textcolor{blue}{\bf 89.2\%} & \textcolor{blue}{\bf 88.9\%} & \textcolor{blue}{\bf 78.5\%} & \textcolor{blue}{\bf 95.4\%} \\ \hline
\textbf{AS} & 41.0\% & 98.1\% & \textcolor{red}{\bf 46.0\%} &  \textcolor{red}{\bf 0.0\%} & \textcolor{red}{\bf 0.0\%} & \textcolor{red}{\bf 0.0\%} & \textcolor{red}{\bf 0.0\%} &\textcolor{red}{\bf 0.0\%} & \textcolor{red}{\bf 0.0\%} & \textcolor{red}{\bf 0.0\%} & \textcolor{blue}{\bf 96.8\%} & 87.4\% & 65.5\% & \textcolor{red}{\bf 20.1\%} & \textcolor{red}{\bf 0.0\%} & \textcolor{red}{\bf 47.5\%} \\ \hline
\textbf{MS} & 55.6\% & \textcolor{red}{\bf 83.9\%} & 65.6\% & \textcolor{blue}{\bf 21.3\%} & \textcolor{blue}{\bf 17.1\%} & \textcolor{blue}{\bf 17.3\%} & \textcolor{blue}{\bf 22.1\%} & \textcolor{blue}{\bf 18.3\%} & \textcolor{blue}{\bf 24.5\%} & \textcolor{blue}{\bf 20.1\%} & \textcolor{red}{\bf 77.1\%} & 73.2\% & 68.8\% & 57.9\% & 26.9\% & 71.5\% \\ \hline \hline
\textbf{DS} & 41.1\% & 91.8\% & 57.2\% & 0.3\% & 0.2\% & 0.2\% & 0.2\% & 0.1\% & 0.2\% & 0.2\% & \textcolor{red}{\bf 77.2\%} & \textcolor{red}{\bf 68.1\%} & 59.9\% & 39.3\% & 0.7\% & 67.3\% \\ \hline
\textbf{LPF} & 40.0\% & 96.9\% & 59.8\% & \textcolor{red}{\bf 0.0\%} & \textcolor{red}{\bf 0.0\%} & \textcolor{red}{\bf 0.0\%} & \textcolor{red}{\bf 0.0\%}&\textcolor{red}{\bf 0.0\%} &\textcolor{red}{\bf 0.0\%} &\textcolor{red}{\bf 0.0\%}& 84.6\% & 71.7\% & 59.7\% & 22.2\% & \textcolor{red}{\bf 0.0\%} & 54.3\% \\ \hline
\textbf{BPF} & \textcolor{red}{\bf 37.6\%} & 91.0\% & 51.4\% & \textcolor{red}{\bf 0.0\%} & \textcolor{red}{\bf 0.0\%} & \textcolor{red}{\bf 0.0\%} & \textcolor{red}{\bf 0.0\%}& \textcolor{red}{\bf 0.0\%} & 0.1\% & \textcolor{red}{\bf 0.0\%} & \textcolor{red}{\bf 79.0\%} & 68.5\% & \textcolor{red}{\bf 52.9\%} & 21.2\% & 0.2\% & 58.5\% \\ \hline \hline
\textbf{OPUS} & 59.4\% & \textcolor{red}{\bf 88.6\%} & 67.9\% & 17.4\% & 14.1\% & 15.0\% & 17.9\% & 17.1\% & 23.3\% & 18.2\% & 84.0\% & 81.0\% & 78.8\% & 71.8\% & 31.5\% & 87.5\% \\ \hline
\textbf{SPEEX} & 56.4\% &  93.8\% & \textcolor{blue}{\bf 71.8\%} & 7.2\% & 6.6\% & 7.9\% & 11.9\% & 10.6\% & 21.8\% & 10.8\% & 88.1\% & 84.0\% & 77.4\% & 59.6\% & 18.3\% & 87.9\% \\ \hline
\textbf{AMR} & 59.0\% & 96.8\% & 67.4\% & 6.4\% & 7.0\% & 7.7\% & 11.0\% & 8.1\% & 15.9\% & 9.4\% & 94.8\% & \textcolor{blue}{\bf 92.3\%} & \textcolor{blue}{\bf 88.6\%} & 67.2\% & 24.6\% & \textcolor{blue}{\bf 94.2\%} \\ \hline
\textbf{AAC-V} & \textcolor{red}{\bf 26.6\%} & \textcolor{blue}{\bf 99.8\%} & \textcolor{red}{\bf 47.1\%} & \textcolor{red}{\bf 0.0\%}  & \textcolor{red}{\bf 0.0\%} & \textcolor{red}{\bf 0.0\%}  &\textcolor{red}{\bf 0.0\%}  &\textcolor{red}{\bf 0.0\%}  & \textcolor{red}{\bf 0.0\%}  & \textcolor{red}{\bf 0.0\%}  & 89.7\% & \textcolor{red}{\bf 37.3\%} & \textcolor{red}{\bf 5.9\%} & \textcolor{red}{\bf 0.0\%}  & \textcolor{red}{\bf 0.0\%} & \textcolor{red}{\bf 34.6\%} \\ \hline
\textbf{AAC-C} & 47.7\% & 92.7\% & 64.2\% & 2.8\% & 2.3\% & 1.8\% & 2.5\% & 2.4\% & 2.7\% & 1.9\% & 83.6\% & 78.5\% & 71.8\% & 51.1\% & 8.1\% & 76.4\% \\ \hline
\textbf{MP3-V} & \textcolor{red}{\bf 23.2\%} & \textcolor{blue}{\bf 99.6\%} & \textcolor{red}{\bf 48.0\%} & \textcolor{red}{\bf 0.0\%} & \textcolor{red}{\bf 0.0\%} & \textcolor{red}{\bf 0.0\%} & \textcolor{red}{\bf 0.0\%} & \textcolor{red}{\bf 0.0\%} & \textcolor{red}{\bf 0.0\%} & \textcolor{red}{\bf 0.0\%} & 87.4\% & \textcolor{red}{\bf 15.9\%} & \textcolor{red}{\bf 0.3\%} & \textcolor{red}{\bf 0.0\%} & \textcolor{red}{\bf 0.0\%} & \textcolor{red}{\bf 32.0\%} \\ \hline
 \textbf{MP3-C} & 42.8\% & 96.4\% & 53.1\% &\textcolor{red}{\bf 0.0\%} & \textcolor{red}{\bf 0.0\%}  & 0.1\% & 0.1\% &\textcolor{red}{\bf 0.0\%} & \textcolor{red}{\bf 0.0\%}  & 0.1\% & 87.6\% & 79.0\% & 63.9\% & 29.3\% & 0.4\% & 71.1\% \\ \hline \hline
\textbf{FC-o(k)} & \textcolor{blue}{\bf 61.6\%} & 94.0\% & 70.4\% & 16.3\% & 13.8\% & 13.0\% & 17.0\% & 12.7\% & 20.8\% & 14.4\% & 91.4\% & 86.5\% & 83.4\% & \textcolor{blue}{\bf 74.0\%} & \textcolor{blue}{\bf 42.0\%} & 85.5\% \\ \hline
\textbf{FC-d(k)} &  53.7\% & \textcolor{blue}{\bf 99.4\%} & 70.5\% & 0.2\% &\textcolor{red}{\bf 0.0\%}  & 0.2\% & 0.9\% & 0.3\% & 1.1\% & 0.7\% & \textcolor{blue}{\bf 97.1\%} & \textcolor{blue}{\bf 94.1\%} & \textcolor{blue}{\bf 87.3\%} & 62.8\% & 14.7\% & 85.6\% \\ \hline
\textbf{FC-c(k)} & 51.7\% & 98.8\% & 68.8\% & \textcolor{red}{\bf 0.0\%}  & 0.2\% & 0.1\% & 0.1\% & 0.1\% & 0.5\% & 0.1\% &  \textcolor{blue}{\bf 96.3\%} &  \textcolor{blue}{\bf 91.0\%} & 82.1\% & 55.0\% & 11.2\% & 84.0\% \\ \hline
\textbf{FC-f(k)} & 50.4\% & 98.2\% & 67.1\% & 0.3\% & 0.3\% & 0.5\% & 0.3\% & 0.4\% & 0.9\% & 0.2\% & 93.4\% & 86.6\% & 78.7\% & 51.2\% & 10.8\% & 83.6\% \\ \hline
\textbf{FC-o(wk)} & 54.0\% & 96.7\% & 66.6\% & 3.9\% & 3.5\% & 3.7\% & 4.2\% & 4.0\% & 6.5\% & 3.3\% & 91.3\% & 84.4\% & 77.5\% & 58.5\% & 26.8\% & 89.6\% \\ \hline
\textbf{FC-d(wk)} & 54.4\% & 98.2\% & 70.2\% & 1.7\% & 1.1\% & 1.1\% & 3.0\% & 1.8\% & 3.5\% & 3.2\% & 93.9\% & 88.3\% & 82.9\% & 64.0\% & 23.4\% & 88.1\% \\ \hline
\textbf{FC-c(wk)} & 52.7\% & 98.0\% & 68.3\% & 1.4\% & 0.7\% & 0.7\% & 2.4\% & 1.3\% & 2.5\% & 1.9\% & 93.0\% & 87.1\% & 79.4\% & 58.6\% & 20.1\% & 87.6\% \\ \hline
\textbf{FC-f(wk)} & 53.1\% & 97.6\% & 68.5\% & 2.0\% & 1.2\% & 0.8\% & 3.0\% & 1.5\% & 3.0\% & 2.1\% & 91.6\% & 85.7\% & 79.5\% & 60.4\% & 22.1\% & 88.7\% \\ \hline
\end{tabular}%
}\vspace*{-2mm}
\end{table*}

The results are shown in Table~\ref{tab:evaluate-defense-non-adaptive}, where Baseline
means no defense is deployed. In general, the accuracy on benign
and adversarial examples significantly
vary with transformations and attacks.
The results provide many findings, including but not limited to the following.

\smallskip\noindent {\bf The side effect on benign examples.}
Most transformations slightly degrade accuracy on benign examples, but the degradation varies.
The accuracy degradation reflects the degree of distortions induced by each transformation.
Among all the transformations,
QT, AT, MS, and OPUS cause the greatest accuracy degradation ($>10\%$),
indicating that they add more distortions.
AAC-V, MP3-V and 
\defensenameabbr-d(k)
almost have no side effects, reduced only
$0\%$, $0.2\%$ and $0.4\%$ accuracy, respectively.
Comparing MP3-V over MP3-C (resp. AAC-V over AAC-C),
we found that dynamic bit rate based speech compressions
have less side effects, as they preserve the better quality of voices.
Among all the feature-level transformations, we found that
\defensenameabbr-d outperforms the others, indicating
that
\defensenameabbr
has less effects on the delta
features than the others.

\begin{tcolorbox}[size=title,opacityfill=0.1,breakable]
\textbf{Findings 1.}
The transformations vary in accuracy degradation.
Input transformations with dynamic bit rate and delta feature transformation have the least side effects,
but QT, AT, MS, and OPUS
have the greatest side effects.
%
\end{tcolorbox}

\noindent {\bf Resilience to $L_\infty$ adversarial examples.}
On adversarial examples crafted by $L_\infty$ attacks except for FGSM and FAKEBOB,
most transformations are not very effective, in particular, AS, DS, LPF, BPF, AAC-V, AAC-C, MP3-V,
MP3-C, FC-d, FC-c and FC-f are almost completely ineffective.
AT and QT are the two most promising transformations, which respectively improve more than 40\% and 34\% accuracy on adversarial examples regardless of the attack.
However, recall that AT and QT also result in the greatest accuracy degradation on benign examples.
This means that
transformations are often double-edged swords.

On adversarial examples crafted by FGSM and FAKEBOB,
we found that all the transformations are more effective.
This is because FGSM is a single-step attack and FAKEBOB adopts an early-stop strategy, so adversarial examples crafted by FGSM and FAKEBOB
are weak (i.e., close to the decision boundary).
In contrast, PGD and CW$_{\bm{\infty}}$ continue searching for
strong adversarial examples (i.e., far from the decision boundary)
even if an adversarial example has been found.
We also found that strong adversarial examples do not necessarily
have larger distortions than the weak ones according to the results in Table~\ref{tab:evaluate-atatck},
where PGD and CW$_{\bm{\infty}}$ have similar distortions with FAKEBOB, but smaller than FGSM.

\begin{tcolorbox}[size=title,opacityfill=0.1,breakable]
\textbf{Findings 2.}
In general,
the lower quality a transformation preserves, the more effective
it is against $L_\infty$ adversarial examples.
QT and AT are the two most effective  transformations against $L_\infty$ adversarial examples,
while AS, LPF, AAC-V and MP3-V are the least effective.
%
\end{tcolorbox}

\noindent {\bf Resilience to $L_2$ adversarial examples.}
All the transformations are effective on $L_2$ adversarial examples when $\kappa$ is small.
With the increase of the parameter $\kappa$, the transformations become less effective,
as increasing $\kappa$ improves the strength of adversarial examples at the cost of distortion.
This is consistent with the results on $L_\infty$ adversarial examples between PGD/CW\bm{$_\infty$} and FGSM/FAKEBOB.
When 
$\kappa< 10$, 
FC-d(k) and FC-c(k)
outperform the others, indicating
that they are effective on weak $L_2$  adversarial examples.
In contrast, QT and AT outperform the others when 
$\kappa\geq 10$,
indicating that QT and AT are more effective on strong adversarial examples.

\begin{tcolorbox}[size=title,opacityfill=0.1,breakable]
\textbf{Findings 3.}
The transformations are effective on $L_2$ adversarial examples, but
become less effective with the increase of
$\kappa$.
FC-d(k) and FC-c(k)  
are the two most effective transformation against weak $L_2$ adversarial examples.
%
\end{tcolorbox}

\smallskip
\noindent {\bf Effectiveness of transformations.}
Though effective transformations against adversarial examples degrade accuracy  on benign examples,
compared to Baseline, all transformations are effective in terms of R1 score.
The best one (i.e., AT) improves R1 score by 76.5\% and the worst one (i.e., MP3-V) improves it by 13.9\%.
This is because the accuracy improvements on adversarial examples  in general
are often larger than the accuracy degradation on benign examples.

Among the feature-level transformations, we can observe that
\defensenameabbr-o and \defensenameabbr-d often significantly outperform.
This is because transformation on preceding features also affects succeeding features, which amplifies the effect of the transformation.
Between two clustering algorithms kmeans and warped-kmeans,
the effectiveness varies with attacks
and in general they are almost comparable.
In terms of R1 score, FC-o with kmeans, i.e., FC-o(k), ranks the first place.

 \begin{tcolorbox}[size=title,opacityfill=0.1,breakable]
 \textbf{Findings 4.}
 Though QT, AT and FC-o(k) degrade accuracy on benign examples, they
 are the three most effective transformations against non-adaptive attacks.
 \end{tcolorbox}

We also found that: the strength of adversarial examples crafted by CW$_2$ with increase of $\kappa$
remains monotonic after the transformation,
but the strength of adversarial examples crafted by PGD with increase of  \#Steps
becomes non-monotonic after the transformation.
More details on the number of steps in PGD and $\kappa$ in CW$_2$ refer to Appendix~\ref{sec:moredetails}.

\begin{table*}
    \centering\setlength\tabcolsep{2pt}
    \caption{Results ($A_a$ / L2 / SNR / PESQ) of transformations against adaptive attacks, where BLA denotes black-box attack. The accuracy highlighted in \textcolor{red}{red} indicates that an adaptive attack is less effective than its non-adaptive version.
    The rows highlighted with \textcolor{gray!100}{gray} (resp. \textcolor{green!80}{green}) color indicate that the transformations are non-differentiable (resp. randomized).}
    \resizebox{1.0\textwidth}{!}{
    \begin{tabular}{c||c|c|c|c||c|c||c}
    \hline
    \multirow{2}{*}{\bf Defense} & \multicolumn{4}{c||}{\bf L$_\infty$ white-box attacks} & \multicolumn{2}{c||}{\bf L$_2$ white-box attacks} &  {\bf L$_\infty$ BLA} \\ \cline{2-8}
     & {\bf FGSM} & {\bf PGD-10} &
     {\bf PGD-100} & {\bf CW$_{\bm{\infty}}$-0} & {\bf CW$_2$-0} & {\bf CW$_2$-2} & {\bf FAKEBOB} \\   \hline \hline
      \rowcolor{gray!40} {\bf QT} & 18.6\%/ 0.537 / 28.53 / 2.44 & 0\% / 0.407 / 30.96 / 2.74 & 0\% / 0.441 / 30.26 / 2.70 & 0\% / 0.434 / 30.40 / 2.71 &  14.6\% / 0.154 / 46.81 /  3.86 & 0\% / 0.198 / 44.04 / 3.71  & 40.1\% / 0.393 / 31.43 / 2.70 \\ \hline
     \rowcolor{green!40} {\bf AT} &  18.7\% / 0.537 / 28.53 / 2.64 & 4.3\% / 0.461 / 29.74 / 2.86& 1.8\% / 0.512 / 28.89 / 2.79 & 2.0\% / 0.503 / 29.04 / 2.81 &  64.4\% / 0.462 / 37.47 / 3.03 & \begin{tabular}[c]{@{}c@{}}26.2\% / 0.563 / 35.45 / 2.88 \\ ($\kappa$=50): 0\% / 1.391 / 20.71 / 1.70 \end{tabular} & {\textcolor{red}{96.67\%}} / 0.508 / 29.01 / 2.78  \\ \hline
      {\bf AS} & 31.5\% / 0.537 / 28.53 / 2.33 & - & - & - & 19.0\% / 0.059 / 49.70 / 4.16& 0\% / 0.067 / 48.49 / 4.11 & 14.5\% / 0.386 / 31.35 / 2.70 \\ \hline
    {\bf MS} & 1.6\% / 0.423 / 30.63 / 2.51 & 0\% / 0.279 / 34.23 / 3.09  & 0\% / 0.402 / 31.03 / 2.71 & 0\% / 0.387 / 31.36 / 2.73 & 4.7\% / 0.018 / 61.76 / 4.45& - & 0.3 \% / 0.421 / 30.71 / 2.63 \\ \hline \hline
   {\bf DS} & 24.2\%/ 0.537 / 28.53 / 2.41 & - & - & - & 18.2\% / 0.033 / 57.28 / 4.35 & 0\% / 0.041 / 55.02 / 4.29  & 15.0\% / 0.383 / 31.43 / 2.71 \\ \hline
       {\bf LPF} & 32.6\% / 0.537 / 28.53 / 2.38 & - & - & - & 20.2\% / 0.034 / 55.34 / 4.35 & 0\% / 0.041 / 53.46 / 4.29 & 18.8\% / 0.380 / 31.51 / 2.72 \\ \hline
     {\bf BPF} & 26.4\% / 0.537 / 28.53 / 2.35 & - & - & - & 17.3\% / 0.030 / 57.98 / 4.37 & 0\% / 0.036 / 55.99 / 4.31 & 12.3\% / 0.398 / 31.25 / 2.69 \\ \hline \hline
      \rowcolor{gray!40} {\bf OPUS} & \textcolor{red}{89.1\%} / 0.537 / 28.53 / 2.33 & \textcolor{red}{86.8\%} / 0.287 / 34.70 / 2.91 & \textcolor{red}{84.4\%} / 0.324 / 32.82 / 2.77 &  \textcolor{red}{85.5\%} / 0.318 / 33.30 / 2.81 & 25.1\% / 6.786 / 20.97 / 1.89  & 0\% / 8.933 / 15.94 / 1.71 & 82.3\% / 0.410 / 30.90 / 2.65 \\ \hline
      \rowcolor{gray!40} {\bf SPEEX} & \textcolor{red}{89.7\%} / 0.537 / 28.53 / 2.28 & \textcolor{red}{80.6\%} / 0.352 / 31.82 / 2.72 &  \textcolor{red}{75.4\%} / 0.415 / 30.51 / 2.64 & \textcolor{red}{75.7\%} / 0.407 / 30.77 / 2.65 & 1.9\% / 2.745 /  24.33 / 1.92  & - & \textcolor{red}{89.4\%} / 0.410 / 30.89 / 2.65 \\ \hline
    \rowcolor{gray!40} {\bf AMR} & \textcolor{red}{90.4\%} / 0.537 / 28.53 / 2.27 & \textcolor{red}{73.2\%} / 0.361 / 30.39 / 2.78 & \textcolor{red}{63.4\%} / 0.424 / 29.41 / 2.64 & \textcolor{red}{65.8\%} / 0.413 / 29.48 / 2.66 & 2.1\% / 2.816 / 24.30 / 1.96 & - & 92.0\% / 0.407 / 30.95 / 2.65 \\ \hline
      \rowcolor{gray!40} {\bf AAC-V} &  \textcolor{red}{51.9\%} / 0.537 / 28.53 / 2.24 &  0\% / 0.342 / 32.46 / 2.83 & 0\% / 0.405 / 31.48  / 2.69 & 0\% / 0.391 / 31.55 / 2.71 & 2.3\% / 0.070 / 48.96 / 4.06 & - & \textcolor{red}{44.9\%} / 0.377 / 31.60 / 2.72 \\ \hline
      \rowcolor{gray!40} {\bf AAC-C} & \textcolor{red}{88.8\%} / 0.537 / 28.53 / 2.33 & \textcolor{red}{43.2\%}  / 0.324 / 32.51 / 2.70 & \textcolor{red}{6.2\%} / 0.344 / 32.26 / 2.75 & \textcolor{red}{12.0\%} / 0.340 / 32.26 / 2.74 & 19.9\% / 0.967 / 32.67 / 2.59 & 0\% / 1.161 / 29.23 / 2.36 & 23.1\% / 0 .413 / 30.76 / 2.64 \\ \hline
      \rowcolor{gray!40} {\bf MP3-V} & \textcolor{red}{52.2\%} / 0.537 / 28.53 / 2.24 & 0\% / 0.339 / 32.53 / 2.84 & 0\% / 0.356 / 31.50 / 2.72 & 0\% / 0.348 / 31.53 / 2.75  &2.4\% / 0.060 / 49.95 / 4.12 & - & \textcolor{red}{46.4\%} / 0.376 / 31.62 / 2.73 \\ \hline
    \rowcolor{gray!40} {\bf MP3-C} & \textcolor{red}{89.4\%} / 0.537 / 28.53 / 2.33 & \textcolor{red}{10.2\%} / 0.302 / 33.43 / 2.88 & \textcolor{red}{0.9\%} / 0.346 / 32.34 / 2.77 & \textcolor{red}{1.9\%} / 0.343 / 32.39 / 2.78 & 15.5\% / 0.630 / 34.70 / 2.88& 0\% / 0.770 / 31.11 / 2.64 & 54.2\% / 0.405 / 30.98 / 2.66 \\ \hline \hline
    \rowcolor{green!40} {\bf \defensenameabbr-o(k)} & 54.1\% / 0.537 / 28.53 / 2.23 &  0\% / 0.373 / 31.73 / 2.74  & 0\% / 0.413 / 30.50 / 2.66 & 0\% / 0.392 / 30.83 / 2.72 &90.4\% / 0.074 / 56.20 / 4.14  & \begin{tabular}[c]{@{}c@{}}88.0\% / 0.095 / 53.54 / 4.05 \\ ($\kappa$=50): 1.2\% / 2.435 / 18.38 / 1.57 \end{tabular} & {\textcolor{red}{92.17\%}} / 0.478 / 30.02 / 2.47 \\ \hline
    \end{tabular}
    }
    \label{tab:evaluate-defense-adaptive-iv}
\end{table*}

\subsubsection{Transformations against Adaptive Attacks}\label{sec:adaptive-attack}
\noindent{\bf Experimental setup}.
To evaluate the effectiveness of the transformations against adaptive attacks, we use the same setup
as in Section~\ref{sec:non-adaptive-attack}, except for the following.
We only consider  adaptive attacks derived from a subset of representative attacks (FGSM, PGD-10, PGD-100, CW$_{\bm{\infty}}$-0, CW$_2$-0, CW$_2$-2, CW$_2$-50 and FAKEBOB).
For adaptive attacks derived from FGSM, CW$_2$-0 and FAKEBOB, we consider all the transformations, as they are effective on adversarial examples crafted by their non-adaptive versions
but the effectiveness varies.
For adaptive attacks derived from PGD-10, PGD-100, CW$_{\bm{\infty}}$-0,
we do not consider AS, DS, LPF and BPF, as they are differentiable, deterministic, and almost completely ineffective on adversarial examples crafted by (non-adaptive) PGD-10, PGD-100 and CW$_{\bm{\infty}}$-0 attacks.
CW$_2$-2 (resp. CW$_2$-50) is considered only when a transformation is effective (i.e., at least 5\% accuracy)  on adversarial examples crafted by CW$_2$-0 (resp. CW$_2$-2).
We do not consider all the
combinations of attacks and transformations,
as the current experiments
already require substantial effort.

To derive adaptive white-box attacks,
we exploit EOT with $r=50$ to circumvent randomized transformations 
except for CW$_2$,
and exploit BPDA to circumvent non-differentiable transformations 
by replacing a transformation with the identity function in the backward pass
as done
in \cite{tramer2020adaptive, athalye2018obfuscated, yang2019me}.
For CW$_2$, instead of using EOT to handle randomized transformations
which is computationally expensive,
we increase the parameter $\kappa$ which is sufficient 
according to our experiments.
Note that the black-box attack FAKEBOB is regarded as an adaptive attack for simplifying representation, but it is the same as the non-adaptive one.

The results are shown in Table~\ref{tab:evaluate-defense-adaptive-iv}.
Overall, the effectiveness varies with transformations and attacks.
Below, we compare the results with those obtained using non-adaptive attacks (i.e., Table~\ref{tab:evaluate-defense-non-adaptive}), by distinguishing if
the transformations are differentiable or not.

\smallskip\noindent{\bf Non-differentiable transformations} (\textcolor{gray!100}{gray} color in Table~\ref{tab:evaluate-defense-adaptive-iv}).
We can observe that QT becomes less effective, indicating both BPDA and NES are able to circumvent QT.
However, except for AAC-V and MP3-V which achieve similar results as in Table~\ref{tab:evaluate-defense-non-adaptive}, all the speech compressions (OPUS, SPEEX, AMR, AAC-C and MP3-C)
become more effective on the adversarial examples crafted by the white-box  attacks, indicating that BPDA is not able to circumvent them.
Indeed, (1) OPUS, SPEEX, AMR, AAC-C and MP3-C achieve higher accuracy on the adversarial examples crafted by FGSM, PGD and CW$_{\bm{\infty}}$-0
when compared with the results in Table~\ref{tab:evaluate-defense-non-adaptive}.
(2) Though BPDA can reduce the accuracy on the adversarial examples crafted by CW$_2$-0 and CW$_2$-2,
much more distortions are introduced when compared with the results in Table~\ref{tab:evaluate-atatck}.
(We remark that CW$_2$ does have any perturbation thresholds,
while the others do have. Thus, adaptive CW$_2$ attacks still achieve high ASR at the cost of distortion.)
To understand why BPDA has different effectiveness,
we checked the quality of approximating non-differentiable transformations by the identity function.
We found that QT, AAC-V and MP3-V are much closer to the identity function than the others (cf. Appendix~\ref{sec:evaapptrans}).
This means that more accurate approximation functions are
required to circumvent the other speech compressions.
We leave this as future work.

On the adversarial examples crafted by FAKEBOB,
the speech compressions except for
AAC-V and MP3-V achieve similar results as those obtained by non-adaptive attacks. The slight difference may be due to the random nature of NES.
We notice that AAC-V and MP3-V become more effective ($>10\%$ improvements over the results in Table~\ref{tab:evaluate-defense-non-adaptive}).
We suspect that
it is because the gradients estimated by NES for AAC-V and MP3-V
are not informative enough,
due to the variable bit rate of AAC-V and MP3-V.

\begin{tcolorbox}[size=title,opacityfill=0.1,breakable]
\textbf{Findings 5.}
BPDA and NES are able to circumvent QT, but fail to circumvent
other non-differentiable transformations.
More accurate approximation functions are required for BPDA
to circumvent speech compressions, otherwise it is better to launch
non-adaptive PGD attacks or CW$_2$ attacks at the cost of distortion.
\end{tcolorbox}

\smallskip\noindent{\bf Differentiable transformations}.
Almost all the differentiable transformations become less effective, indicating
that they can be easily circumvented by using EOT or NES, or increasing the parameter $\kappa$.
However, AT and FC-o(k) become more effective on the adversarial examples
crafted by FAKEBOB.
This is because NES fails to estimate accurate gradients of
the randomized transformations AT and FC-o(k).

\begin{tcolorbox}[size=title,opacityfill=0.1,breakable]
\textbf{Findings 6.}
Differentiable transformations become less effective against the white-box adaptive attacks,
but randomized transformations become more effective against the black-box adaptive attack.
\end{tcolorbox}

\begin{table*}[h]
    \centering
    \caption{Results ($A_a$ / L2 / SNR / PESQ) on {\bf Standard}, {\bf Vanilla  AdvT}, and {\bf AdvT+X}. The top-1 is in highlighted in \textcolor{blue}{blue} color excluding the baseline Standard.}
    \resizebox{1.0\textwidth}{!}{
    \begin{tabular}{c||c||c|c|c|c||c||c}
    \hline
    \multirow{2}{*}{} & \multirow{2}{*}{$\mathbf{A_b}$} & \multicolumn{4}{c||}{{\bf L$_\infty$ white-box attacks}} & {\bf L$_2$ white-box attacks} & {\bf L$_\infty$ BLA} \\ \cline{3-8}
     & & {\bf FGSM} & {\bf PGD-10} & {\bf PGD-100} & {\bf CW$_{\bm{\infty}}$-0} & {\bf CW$_2$-1} & {\bf FAKEBOB} \\ \hline \hline
     {\bf Standard} & {99.06\%} & 19.61\% / 0.868 / 28.94 / 2.42 & 0\% / 0.447 / 34.63 / 3.31 & 0\% / 0.591 / 32.22 / 3.0 & 0\% / 0.542 / 32.96 / 3.10 & 0\% / 0.046 / 55.87 / 4.47 & 0.35\% / 0.650 / 31.53 / 2.89 \\ \hline \hline
     {\bf Vanilla AdvT} & 95.67\% & 75.20\% / 0.726 / 30.41 / 3.04 & 58.19\% / 0.608 / 31.55 / 3.56 & 53.83\% / 0.675 / 30.64 / 3.43 & 56.39\% / 0.660 / 30.84 / 3.46 & 0\% / 1.297 / 29.10 / 2.83 & 85.63\% / 0.680 / 31.09 / 2.84 \\ \hline \hline
     {\bf AdvT+QT} & 95.74\% & 88.19\% / 0.716 / 30.54 / 2.93 & 72.12\% / 0.569 / 32.19 / 3.28 & 64.08\% / 0.680 / 30.66 / 3.06 & 67.86\% / 0.649 / 31.00 / 3.11 & 0.94\% / 1.240 / 32.10 / 2.75 & 79.84\% / 0.651 / 31.47 / 2.88 \\ \hline
     {\bf AdvT+AT} & 95.57\% & 67.24\% / 0.866 / 28.96 / 3.31 & 59.79\%  / 0.767 / 29.40 / 3.44 & 57.96\% / 0.806 / 28.94 / 3.30 & 58.69\% / 0.800 / 29.00 / 3.32 & 5.90\% / 1.318 / 29.16 / 2.89 & 94.69\% / 0.708 / 30.71 / 2.79 \\ \hline
     {\bf AdvT+AS} & 93.59\% & 82.72\% / 0.728 / 30.39 / 2.92 & 53.83\% / 0.572 / 32.14 / 3.50 & 43.12\% / 0.657 / 31.00 / 3.31 & 47.49\% / 0.635 / 31.24 / 3.36 & 0\% / 1.267 / 31.56 / 2.83 & 83.55\% / 0.664 / 31.27 / 2.87 \\ \hline
     {\bf AdvT+MS} & 92.76\% & 65.85\% / 0.582 / 32.35 / 3.20 & 49.77\% / 0.494 / 33.38 / 3.61 & 44.13\% / 0.572 / 32.12 / 3.38 & 47.56\% / 0.547 / 32.46 / 3.45 & 0\% /  0.958 / 32.46 / 2.90 & 76.38\% / 0.654 / 31.45 / 2.89 \\ \hline \hline
     {\bf AdvT+DS} & 95.32\% & 70.14\% / 0.737 / 30.28 / 3.20 & 51.44\% / 0.610 / 31.81 / 3.58 & 44.06\% / 0.687 / 30.77 / 3.37 & 47.59\% / 0.665 / 31.02 / 3.44 & 0\% / 1.187 / 30.16 / 3.06 & 79.91\% / 0.647 / 31.55 / 2.90 \\ \hline \hline

     {\bf AdvT+\defensenameabbr-o(k)} & \textcolor{blue}{\bf 97.81\%} & \textcolor{blue}{\bf 92.58\%} / 0.868 / 28.94 / 2.44 & \textcolor{blue}{\bf 86.25\%} / 0.570 / 31.64 / 3.11 & \textcolor{blue}{\bf 71.13\%} / 0.674 / 30.42 / 3.18 & \textcolor{blue}{\bf 73.76\%} / 0.668 / 30.59 / 3.16 &
     \textcolor{blue}{\bf 76.47\%} / 2.131 / 26.86 / 2.16 &
     \textcolor{blue}{\bf 98.08\%} / 0.688 / 30.98 / 2.82\\ \hline

    \end{tabular}
    }
    \label{tab:evaluate-defense-adaptive-cnn}
\end{table*}

\subsubsection{Transformations+AdvT against Adaptive Attacks}\label{sec:advTadaptive-attack}
\noindent{\bf Experimental setup}.
Since ivector-PLDA cannot be adversarially trained due to unsupervised learning
and xvector-PLDA is too complicated (nearly 4.4 million trainable parameters) to be adversarially trained,
we use AudioNet for adversarial training.
We train AudioNet for the CSI-NE task.
The training and testing datasets are 251Spks-train and 251Spks-test, respectively.
The training uses a minibatch of size 128 for 300 epoches,
Cross-Entropy Loss as the objective function,
and Adam~\cite{kingma2014adam} with default hyper-parameters to optimize trainable parameters.
To avoid overfitting, we add uniform noises with budget $\epsilon'=0.002$ to benign examples for each minibatch.
The model is denoted by {\bf Standard}.

For adversarial training, we use PGD with 10 steps (i.e., PGD-10) to generate adversarial examples,
with $0.5$ adversarial example ratio, the same as \cite{jati2021adversarial}.
The others are the same as for {\bf Standard}.
The model is denoted by {\bf Vanilla AdvT}.

We also explore the combinations of adversarial training with transformations.
For a chosen transformation {\bf X}, we implement it as a proper layer
in AudioNet. Note that this layer does not have any trainable parameters, similar to the ReLU activation layer~\cite{eckle2019comparison}.
The resulting network is adversarially trained the same as above,
except that BPDA is adopted for non-differentiable transformations
and EOT with $r=10$ is adopted for  randomized transformations.
The resulting model is denoted by {\bf AdvT+X}.
We do not consider speech compressions, LPF and BPF,
as BPDA is not effective for estimating gradients of speech compressions,
and the accuracy of the resulting model with LPF/BPF
is extreme low on both training (i.e., 24.10\%/23.65\%) and test set (i.e., 2.04\%/2.25\%).

The adaptive attacks are derived from
FGSM, PGD-10, PGD-100, CW$\mathbf{_\infty}$-0, CW$_2$-1 and FAKEBOB, similar 
with
Section~\ref{sec:adaptive-attack}.
To increase attack power,
the samples\_per\_draw $m$ of FAKEBOB is increased to 300 to estimate more precise gradients,
and the batch size $r$ of EOT is increased to 300 to compute more precise
transformation distributions.

\smallskip\noindent{\bf Results}.
The results are reported in Table~\ref{tab:evaluate-defense-adaptive-cnn}.
We can observe that sole adversarial training (i.e., {\bf Vanilla AdvT}) is effective for defeating adversarial examples compared over {\bf Standard}, at the cost of the accuracy on benign examples (i.e., $A_b$ reduces from 99.06\% to 95.67\%).
Adversarial training either significantly improves the accuracy by more than 53\% on adversarial examples crafted by $L_\infty$ attacks,
or significantly amplifies the distortions of adversarial examples crafted by CW$_2$-1 (more than 28 times in terms of L$_2$ distance).

Although the sole adversarial training is effective compared over {\bf Standard},
the combination of adversarial training with a transformation does not necessarily bring the best of both worlds, 
which also exists in image recognition~\cite{tramer2020adaptive}.
For instance, {\bf AdvT+AT}, {\bf AdvT+AS}, {\bf AdvT+MS} and {\bf AdvT+DS} slightly degrade accuracy on benign examples
compared over the sole adversarial training (i.e., {\bf Vanilla AdvT}).
Both {\bf AdvT+QT} and {\bf AdvT+AT} improve the accuracy on adversarial examples crafted by
all the attacks except for FGSM and FAKEBOB, but {\bf AdvT+AS}, {\bf AdvT+MS} and {\bf AdvT+DS} do not.
{\bf AdvT+QT} (resp. {\bf AdvT+AT}) reduces the accuracy on the adversarial examples crafted by FAKEBOB (resp. FGSM), compared over {\bf Vanilla AdvT}.

Surprisingly, we found that adversarial training combined with FC-o(k), i.e., AdvT+FC-o(k), is very effective.
It improves the accuracy on both adversarial and benign examples (compared with {\bf Vanilla AdvT}),
which definitely brings the best of both worlds.

\begin{figure}[t]
    \centering
    \includegraphics[width=0.45\textwidth]{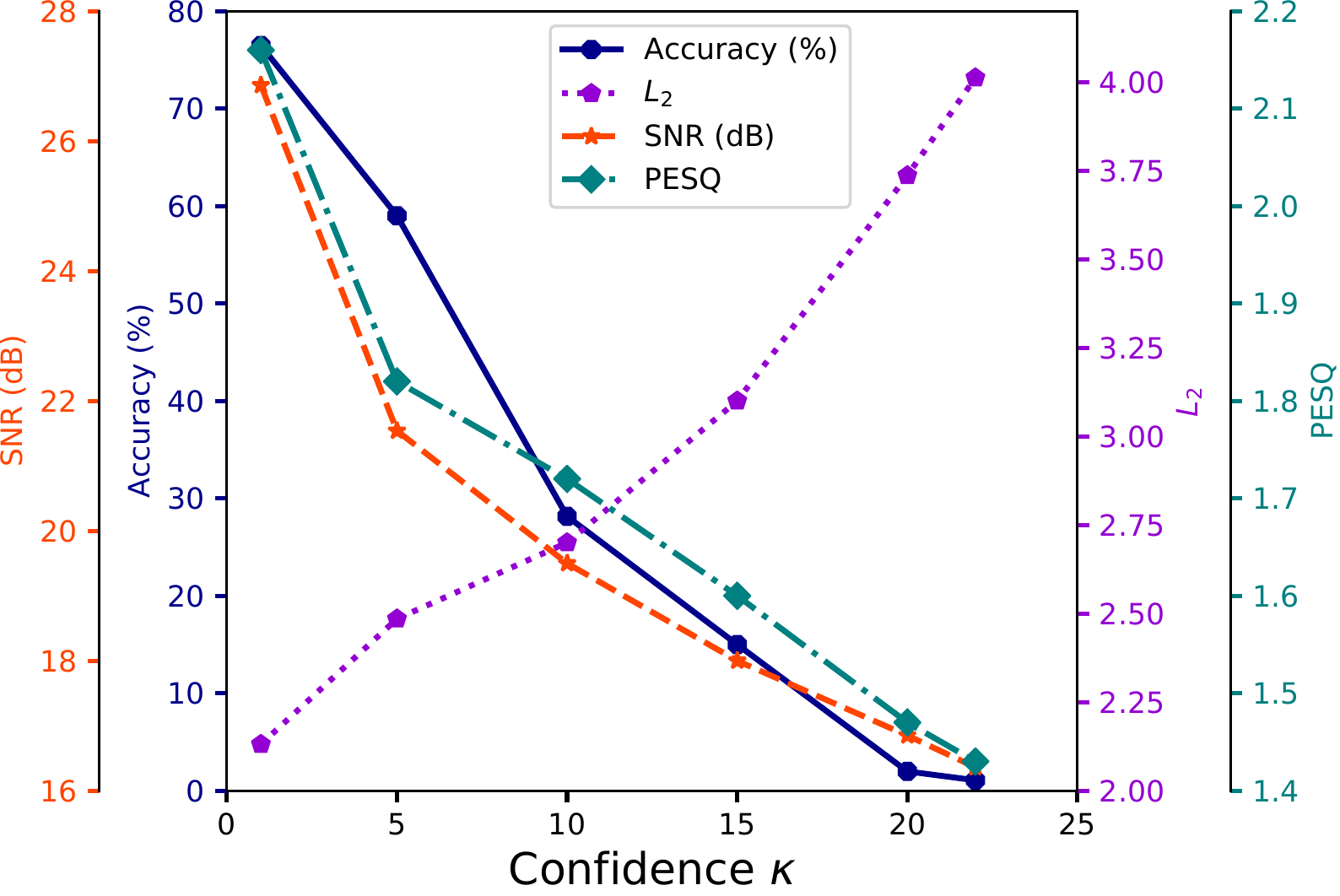}
    \caption{The accuracy and distortion on adversarial examples with increase of $\kappa$ of CW$_2$ using AdvT+\defensenameabbr-o(k).}
    \label{fig:AdvT-FF-CW2-kappa}
\end{figure}

To thoroughly evaluate the effectiveness of AdvT+FC-o(k), we launch
CW$_2$ attacks by increasing the parameter $\kappa$ from 1, 5, 10, 15, 20 and 22, as the
increasing of $\kappa$ makes CW$_2$ more powerful (cf. Table~\ref{tab:evaluate-defense-non-adaptive} and Table~\ref{tab:evaluate-defense-adaptive-iv}).
Although the accuracy on the adversarial examples decreases with increase of
$\kappa$, we found that the distortion also increases.
For instance, when $\kappa=22$, the ASR is close to 100\%, but the $L_2$ distance (resp. SNR and PESQ) is
4.0 (resp. 16.36 dB and 1.43), which is 86 times larger (resp. 3 times smaller and 2 times smaller)
than that of {\bf Standard}, meaning that the adversarial examples become significantly more perceptible. This demonstrates the
effectiveness of AdvT+FC-o(k) against powerful attacks.

\begin{tcolorbox}[size=title,opacityfill=0.1,breakable]
\textbf{Findings 7.}
Combined with adversarial training, AdvT+FC-o(k) is the unique one that is effective against all the attacks.
\end{tcolorbox}

\section{Related Work}\label{sec:relate}
Adversarial attacks and defenses in speech and speaker recognitions have attracted intensive attention.
Though the modern speech recognition and speaker recognition systems are very
similar to each other, they perform different tasks and differ at the last stage of the processing~\cite{AWBPT20,chen2021sok,chen2019real}.
Thus, in this section, we do not discuss adversarial attacks and defenses for speech recognition~\cite{Carlini018,carlini2016hidden,qin2019imperceptible,yuan2018commandersong,ijcai2019-741,247642,LiW00020,TaoriKCV19,alzantot2018did,yang2018characterizing} (cf.~\cite{AWBPT20,chen2021sok} for survey).

There are other voice attacks against SRSs,
such as poisoning attacks~\cite{ASGBWYST19}, hidden voice attacks~\cite{AbdullahGPTBW19}, and spoofing attacks~\cite{hautamaki2013vectors,shirvanian2019quantifying,mukhopadhyay2015all,shirvanian2014wiretapping,shirvanian2018short}.
These attacks have different attack goals and scenarios from adversarial attacks~\cite{chen2019real}.

Below, we discuss adversarial attacks and defenses in SR
and security analysis platforms on adversarial attacks.

\smallskip\noindent {\bf Adversarial attacks.}
Existing adversarial attacks in SR can be roughly classified into white-box attacks~\cite{DBLP:journals/corr/abs-1711-03280, abs-1801-03339, li2020adversarial,jati2021adversarial, zhang2021attack, xie2021real, LiZJXZWM020, xie2020enabling,WangGX20, shamsabadi2021foolhd}
and black-box attacks~\cite{chen2019real,du2020sirenattack}.

FGSM was adopted to attack the CSI-NE task~\cite{DBLP:journals/corr/abs-1711-03280}
and the SV task~\cite{abs-1801-03339,li2020adversarial}.
Zhang at al. used PGD to attack the CSI-NE task~\cite{zhang2021attack}.
Jati at al. attacked the CSI-NE task by leveraging FGSM, PGD, CW$\mathbf{_\infty}$, and CW$_2$~\cite{jati2021adversarial}.
These attacks vary in tasks, models, and datasets. They also did not consider any defenses except that \cite{jati2021adversarial} considered adversarial training.
Thus, it is difficult to compare the effectiveness of these attacks due to a lack of a uniform platform.
This work aims to fill this gap by providing the platform \platformname, which covers all these white-box attacks, and furthermore their adaptive versions.

Besides, there are also some specific white-box attacks, aimed at
crafting universal perturbations~\cite{xie2021real,LiZJXZWM020,xie2020enabling}
and improving the imperceptibility of adversarial voices~\cite{WangGX20, shamsabadi2021foolhd}.
These attacks considered either the CSI-E or CSI-NE task, but no defense was considered.
We do not incorporate these methods into our platform \platformname, as all of them are not publicly available and non-trivial to reproduce.

As mentioned in Section~\ref{sec:blackboxattacks},
FAKEBOB~\cite{chen2019real} and SirenAttack~\cite{du2020sirenattack}
are two black-box attacks on SR systems. FAKEBOB considered
CSI-E, SV, and OSI tasks while
SirenAttack considered CSI-NE only.
Both FAKEBOB and SirenAttack are implemented in our platform \platformname.

\smallskip\noindent {\bf Adversarial defenses and detection.}
Robust training was proposed to mitigate adversarial voices in SR.
\cite{du2020sirenattack,du2020sirenattack,jati2021adversarial} showed that adversarial training can improve resistance to adversarial attacks.
\cite{jati2021adversarial} also proposed another training technique
which adds a regularization term using Lipschitz smoothness to the objective function for model training.
This training technique performs better than FGSM adversarial training, but worse
than PGD adversarial training. This motivated us to evaluate PGD adversarial training in this work.

FAKEBOB~\cite{chen2019real} and SirenAttack~\cite{du2020sirenattack} respectively evaluated their attacks against
(QT, MS and DS) and (DS and AS) transformations.
Their results are similar to ours. The evaluation reported in this work is more
systematic and comprehensive, covering both non-adaptive and adaptive attacks,
22 input-level and feature-level transformations.

To our knowledge, there does not exist any detection approaches dedicated for SR.
FAKEBOB~\cite{chen2019real} evaluated temporal dependency in SR, which
was originally proposed to detect adversarial examples in speech recognition~\cite{yang2018characterizing}.
However, it was shown that temporal dependency is ineffective in SR,
since adversarial voices in SR do not alter the transcription, thus the temporal dependency is preserved.

\smallskip\noindent {\bf Security analysis platforms.}
To systematically and comprehensively evaluate recognition models,
adversarial attacks and defenses, various platforms have been proposed in the literature
such as Cleverhans~\cite{papernot2016technical}, Foolbox~\cite{rauber2017foolbox}, AdvBox~\cite{goodman2020advbox}, ART~\cite{nicolae2018adversarial}, advertorch~\cite{ding2019advertorch}, ARES~\cite{DDongFYPSXZ20}, DEEPSEC~\cite{LingJZWWLW19}, FenceBox~\cite{qiu2020fencebox}, and DeepRobust~\cite{LJXT20}.
These platforms advanced the research on adversarial
examples.

However,
these platforms cannot be directly adopted in SR,
due to the following reasons:
(1) The input transformations (e.g., image compression and random cropping) provided by these platforms target images rather than voices, thus cannot be used to defending against adversarial voices in SR.
(2) The pre-trained models provided by these platforms are not designed for SR, thus may not perform well in SR.
(3) No voice datasets are provided by these platforms, while existing voice datasets
are provided for training and testing of speaker or speech
recognition, but not for evaluating
 adversarial attacks and defenses.
(4) The distortion metrics provided by these platforms are designed for measuring the similarity between original and adversarial images,
while human listening perception is different from human visual perception. 

To our knowledge, \platformname is the \textit{first} platform specifically designed for evaluating adversarial attacks and defenses in SR.
\platformname overcomes the aforementioned limitations by
(1) introducing diverse voice-dedicated transformations for defending against adversarial voices,
(2) providing ready-to-use pre-trained SRSs and standardized datasets covering different recognition tasks,
(3) featuring several distortion metrics which are closely related to human listening perception,
and (4) implementing several representative and state-of-the art white-box and black-box attacks
which can be configured for mounting both non-adaptive and adaptive attacks.

\section{Conclusion}
We reported the design and implementation
of \platformname, which incorporates various adversarial attacks (white-box, black-box, non-adaptive and adaptive) with attack metrics,
and diverse defense solutions (adversarial training,  speech compressions, time-domain/frequency-domain/feature-level transformations) with defense metrics.
Our feature-level transformation is the most effective one when combined with adversarial training.
\platformname is the first platform that supports
systematic, comprehensive, extensible evaluation of adversarial attacks and defenses in SR.
Using \platformname, we conduct extensive evaluation of adversarial attacks and defenses, resulting in a set of interesting and insightful findings.
We envision that \platformname is able to facilitate research on adversarial examples in SR.

\appendix

\section{Tuning the Parameters of Transformations}\label{sec:parameter}
To tune the parameters of the transformations,
we vary the parameters as shown in Table~\ref{tab:range-optimal}
and conduct all the attacks mentioned in Section~\ref{sec:evaluate-attack}.

\begin{figure*}
    \centering

    \subfigure[QT]{
    \begin{minipage}[t]{0.23\textwidth}
    \includegraphics[width=1.0\textwidth]{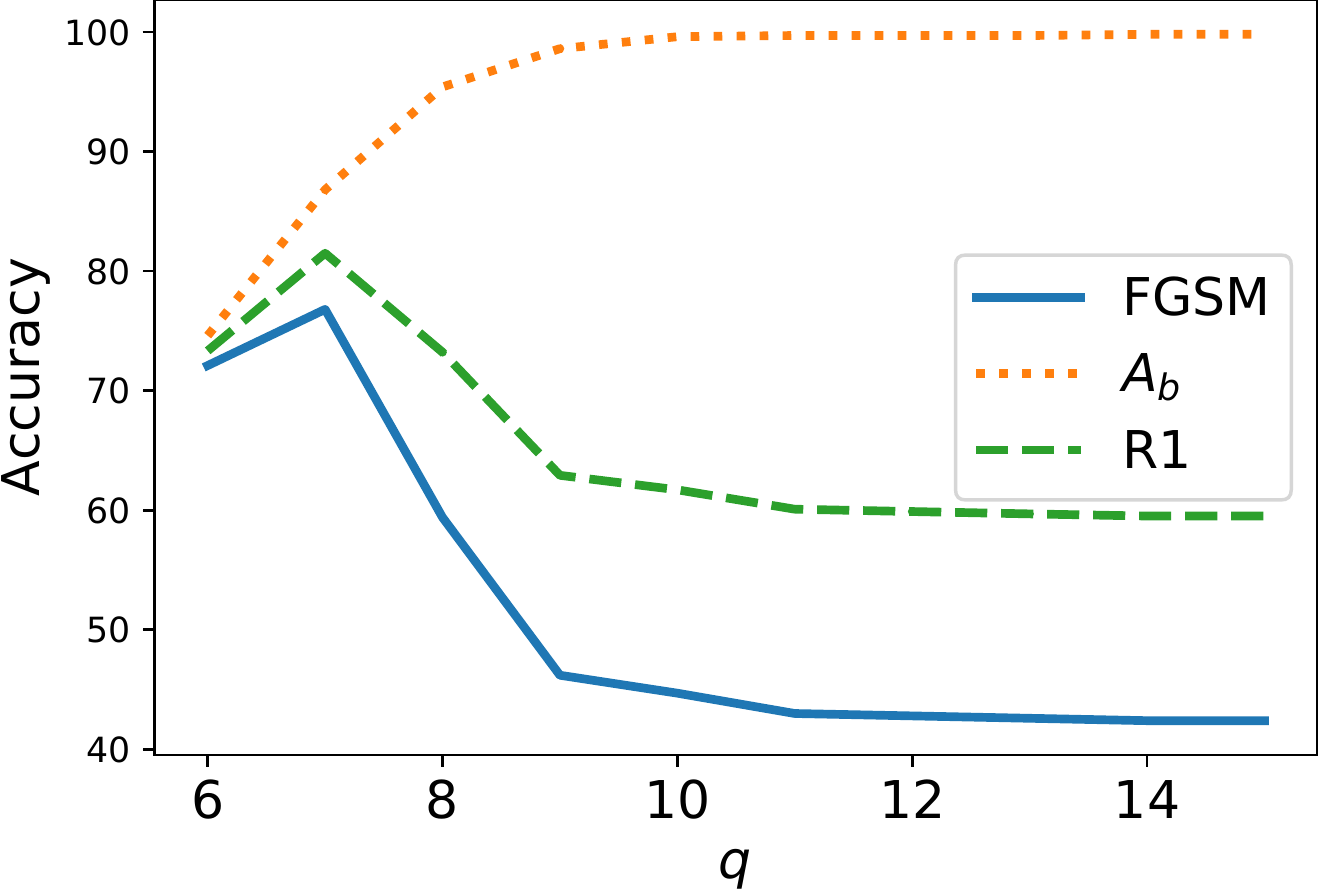}
    \end{minipage}
    \begin{minipage}[t]{0.23\textwidth}
    \includegraphics[width=1.0\textwidth]{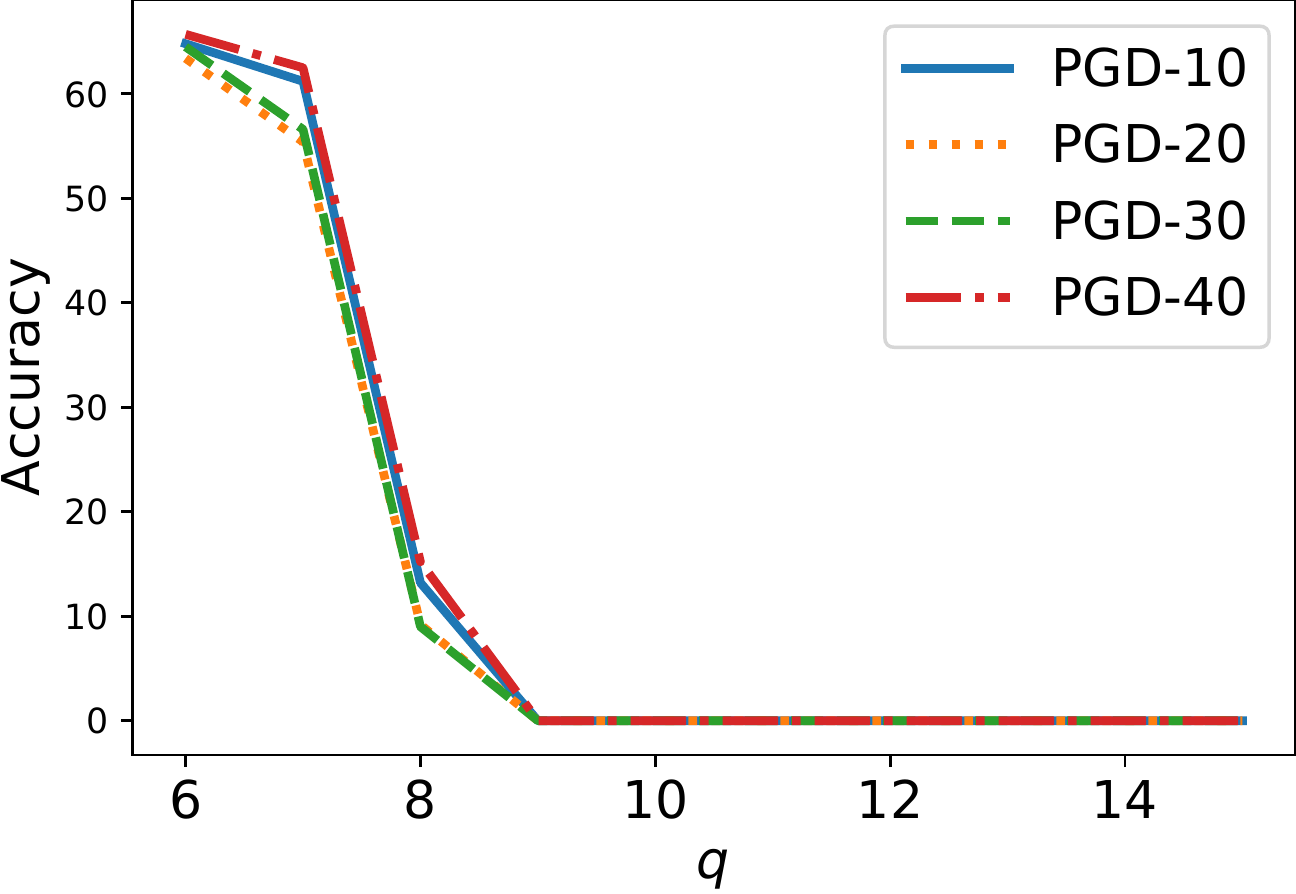}
    \end{minipage}
    \begin{minipage}[t]{0.23\textwidth}
    \includegraphics[width=1.0\textwidth]{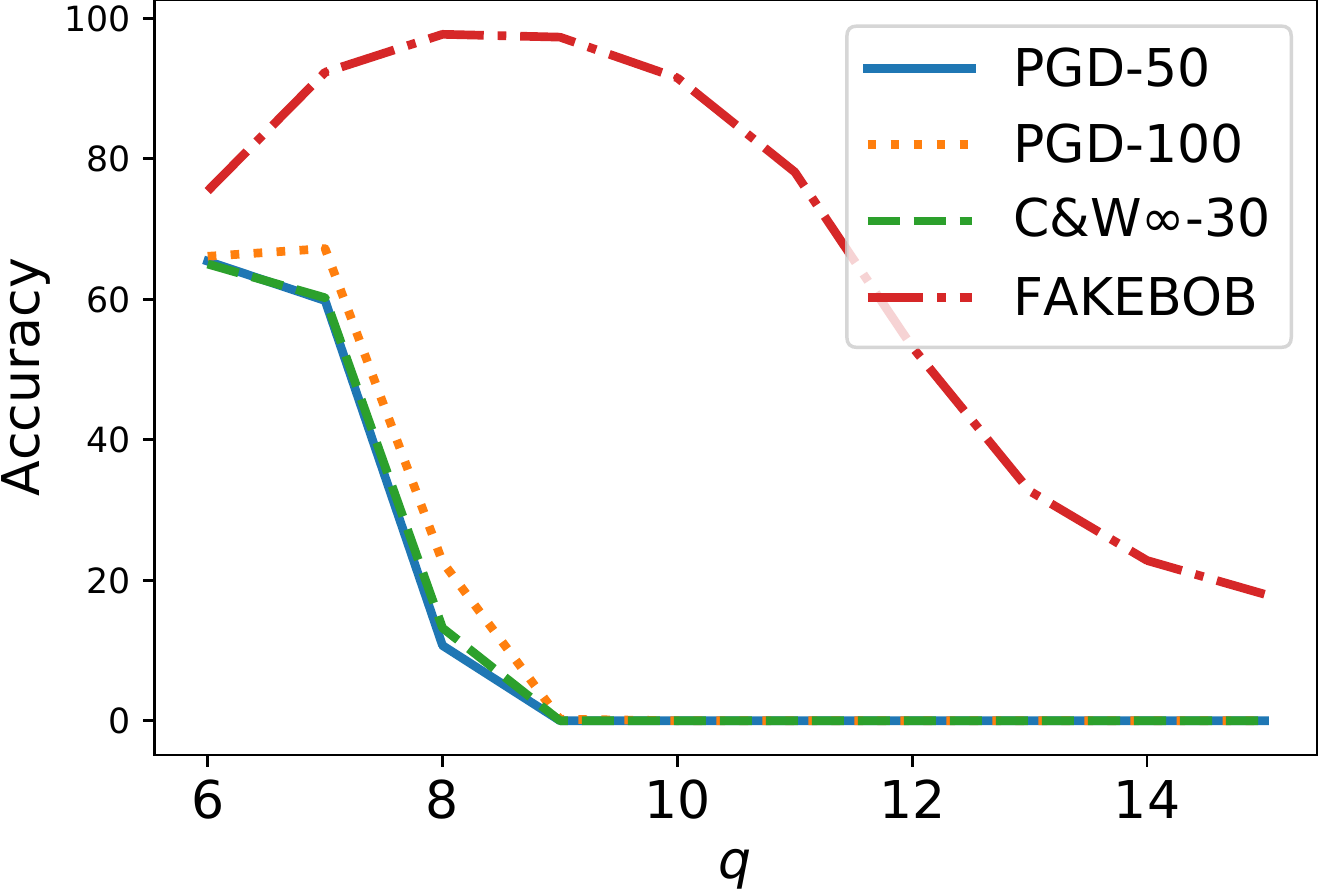}
    \end{minipage}
    \begin{minipage}[t]{0.23\textwidth}
    \includegraphics[width=1.0\textwidth]{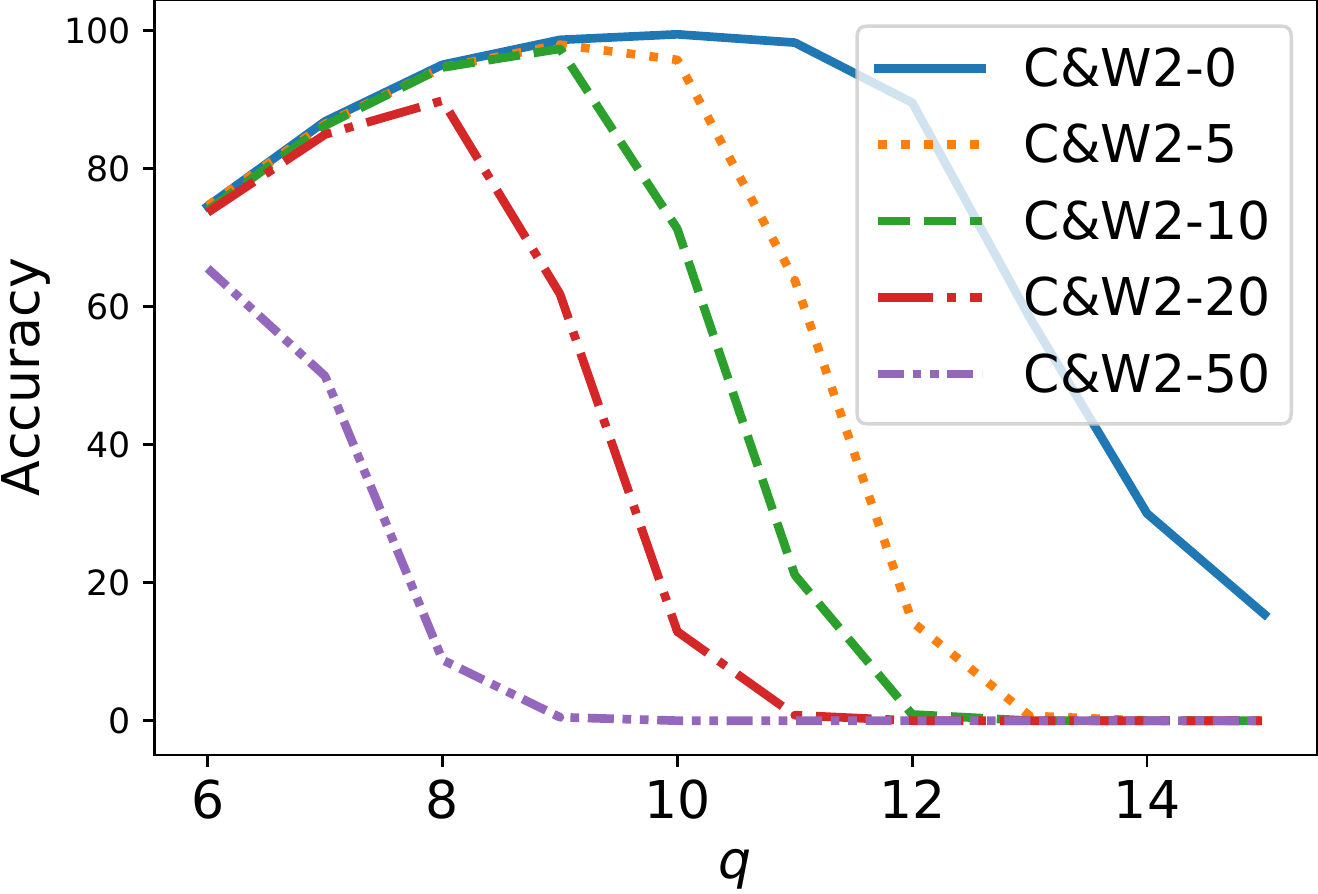}
    \end{minipage}
    }

    \subfigure[DS]{
    \begin{minipage}[t]{0.23\textwidth}
    \includegraphics[width=1.0\textwidth]{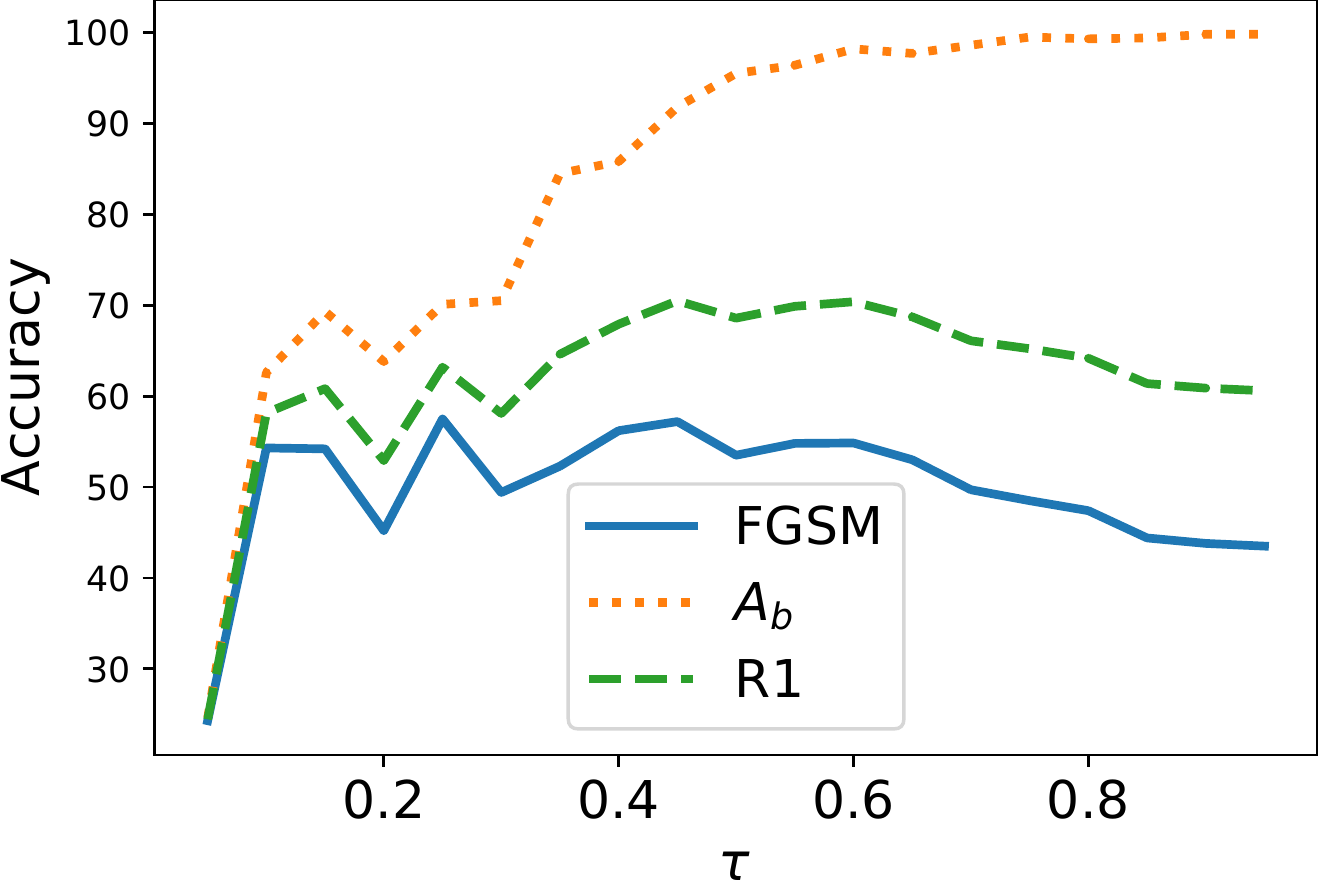}
    \end{minipage}
    \begin{minipage}[t]{0.23\textwidth}
    \includegraphics[width=1.0\textwidth]{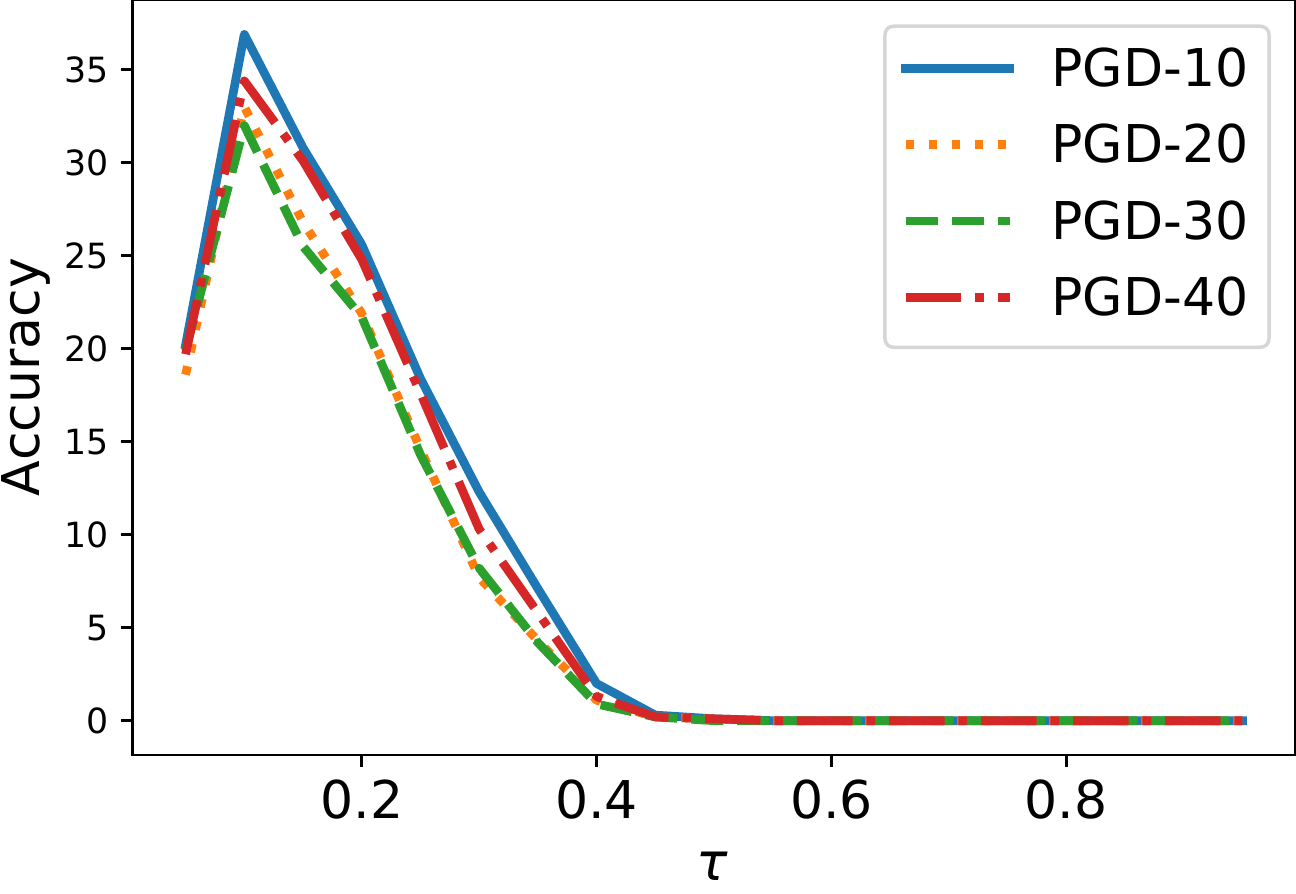}
    \end{minipage}
    \begin{minipage}[t]{0.23\textwidth}
    \includegraphics[width=1.0\textwidth]{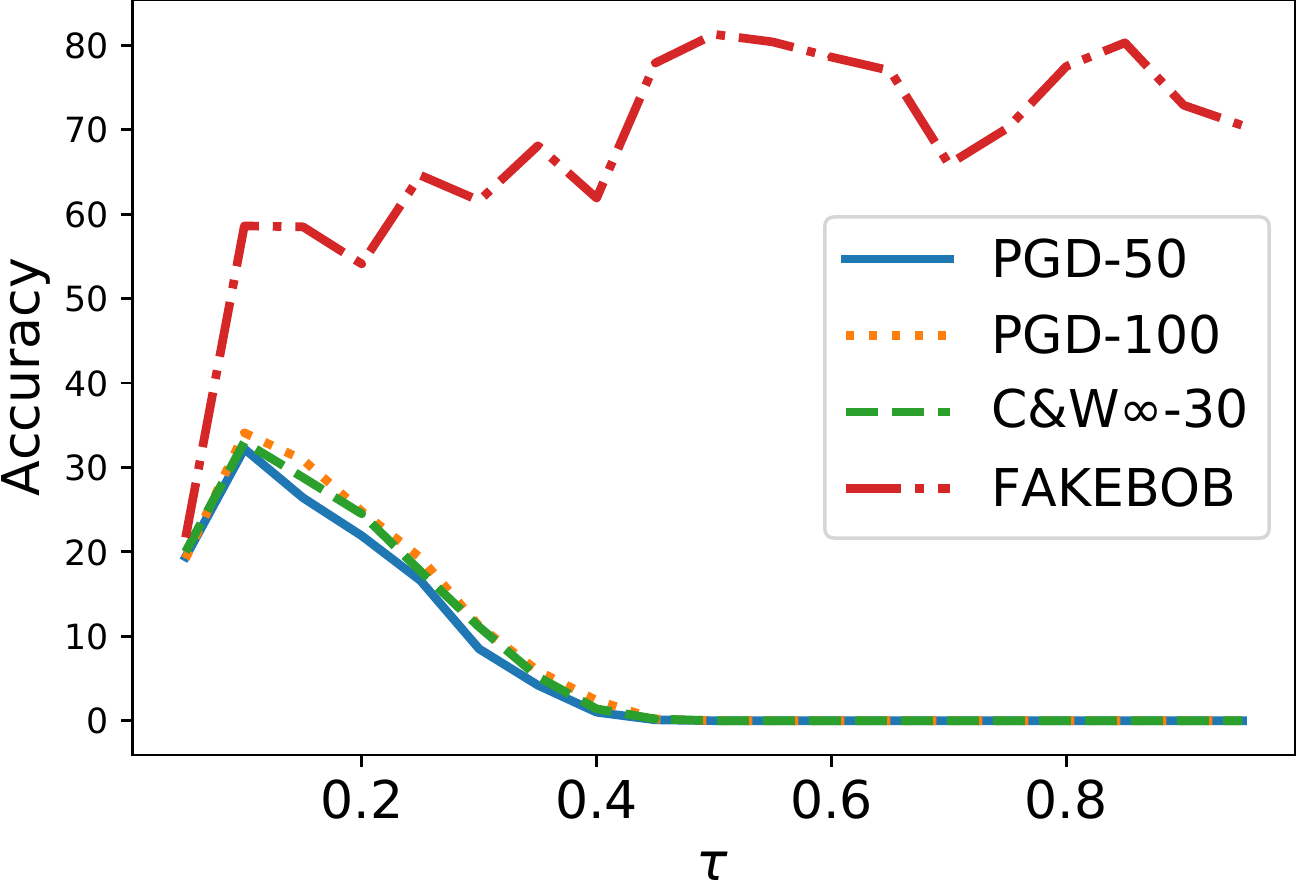}
    \end{minipage}
    \begin{minipage}[t]{0.23\textwidth}
    \includegraphics[width=1.0\textwidth]{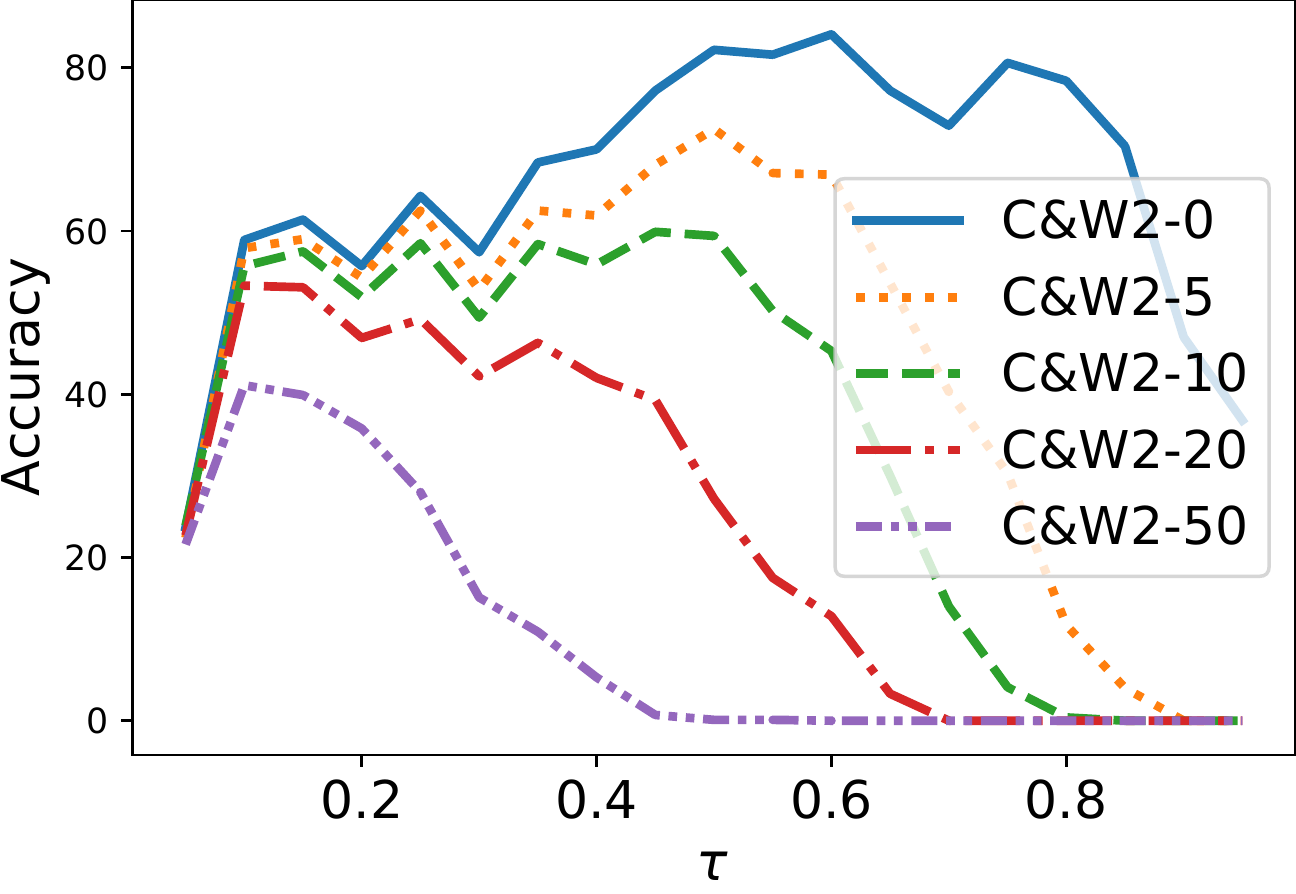}
    \end{minipage}
    }

    \subfigure[AT]{
    \begin{minipage}[t]{0.23\textwidth}
    \includegraphics[width=1.0\textwidth]{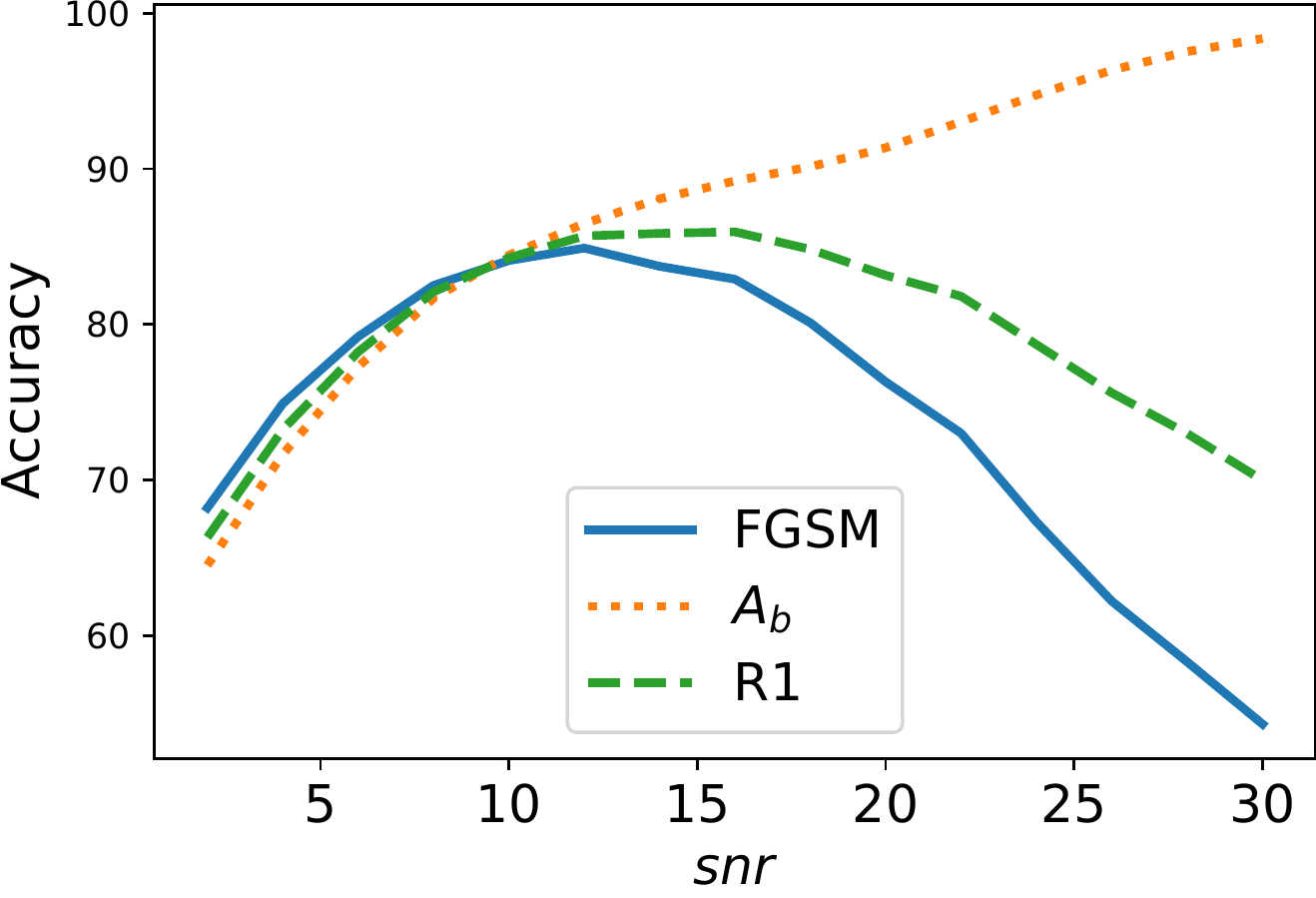}
    \end{minipage}
    \begin{minipage}[t]{0.23\textwidth}
    \includegraphics[width=1.0\textwidth]{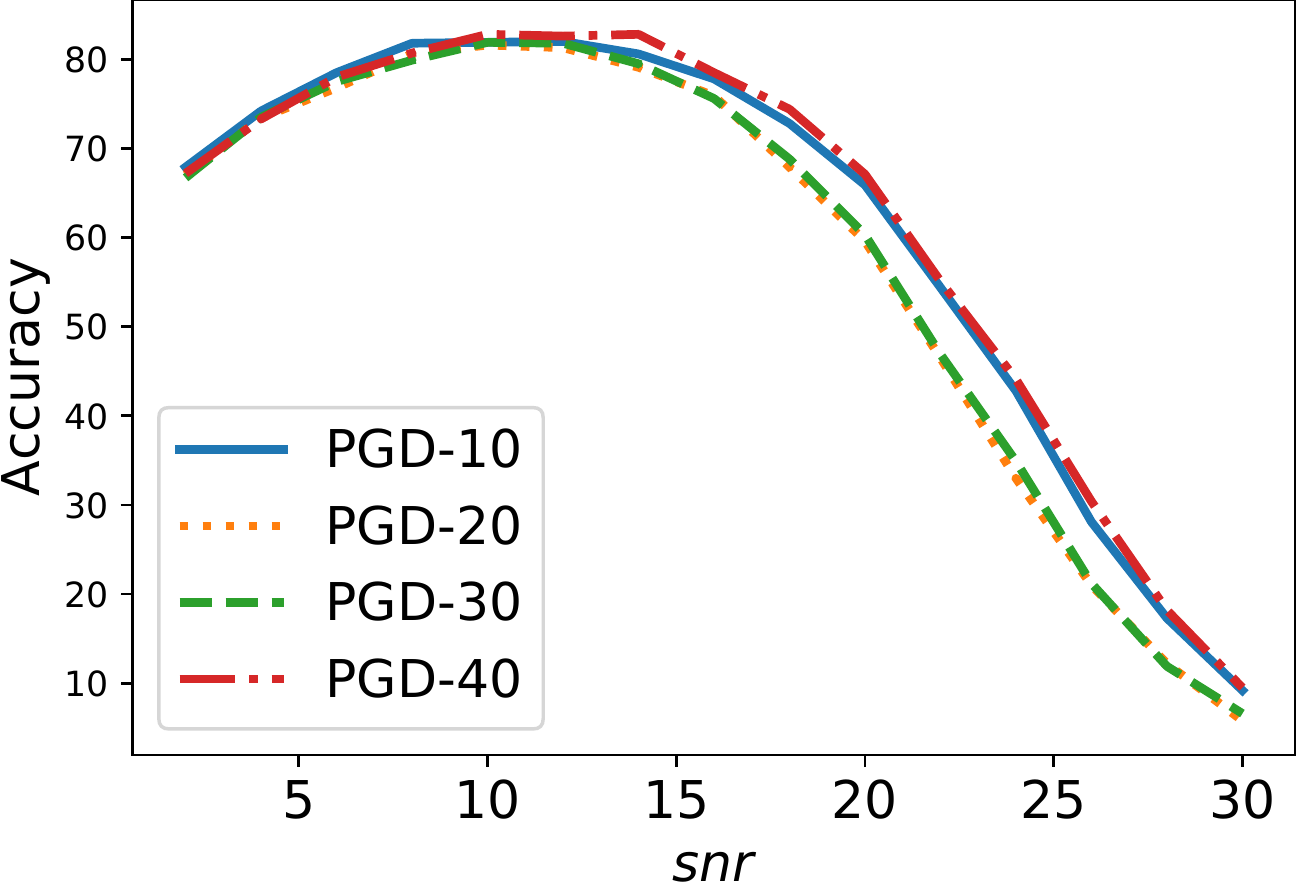}
    \end{minipage}
    \begin{minipage}[t]{0.23\textwidth}
    \includegraphics[width=1.0\textwidth]{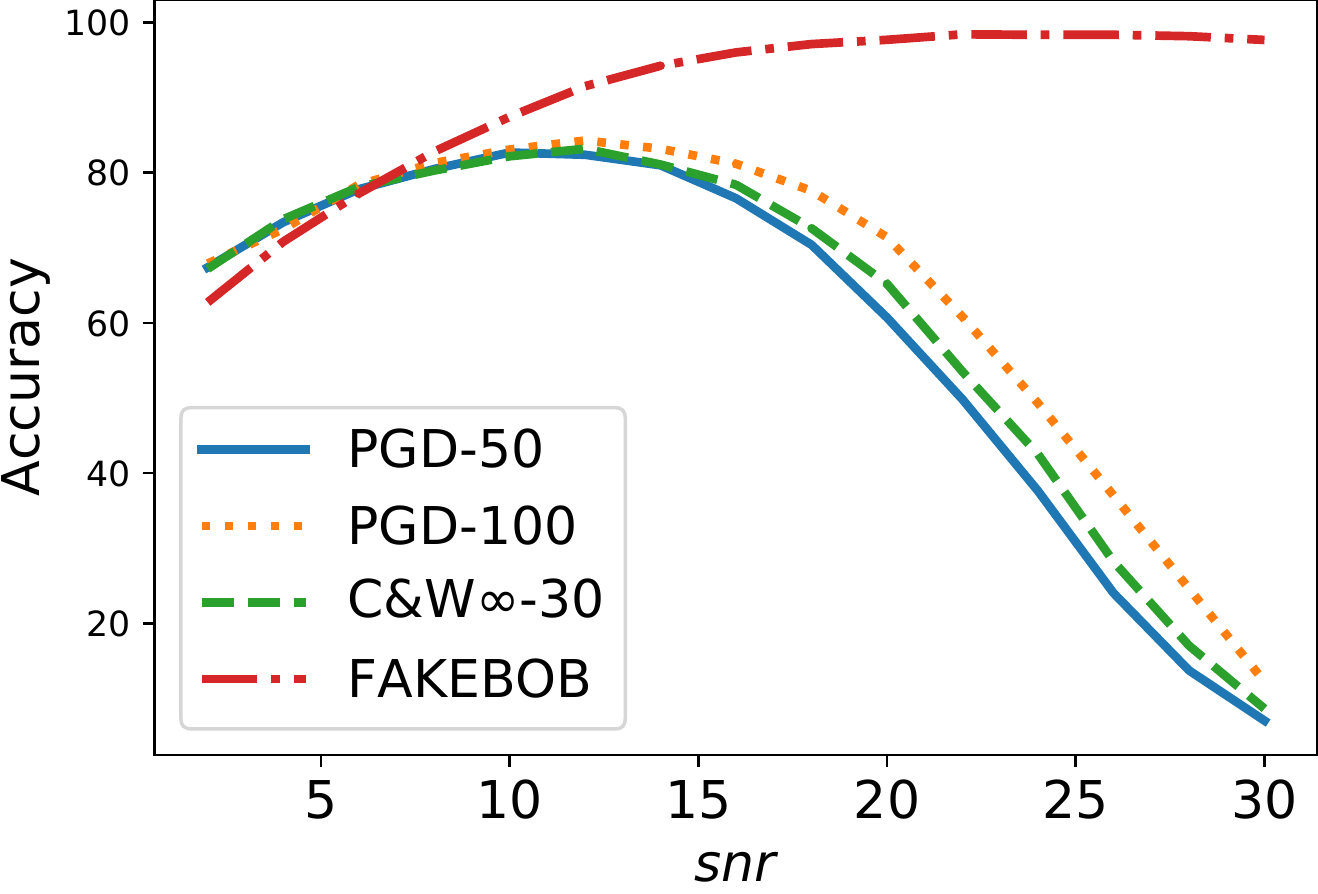}
    \end{minipage}
    \begin{minipage}[t]{0.23\textwidth}
    \includegraphics[width=1.0\textwidth]{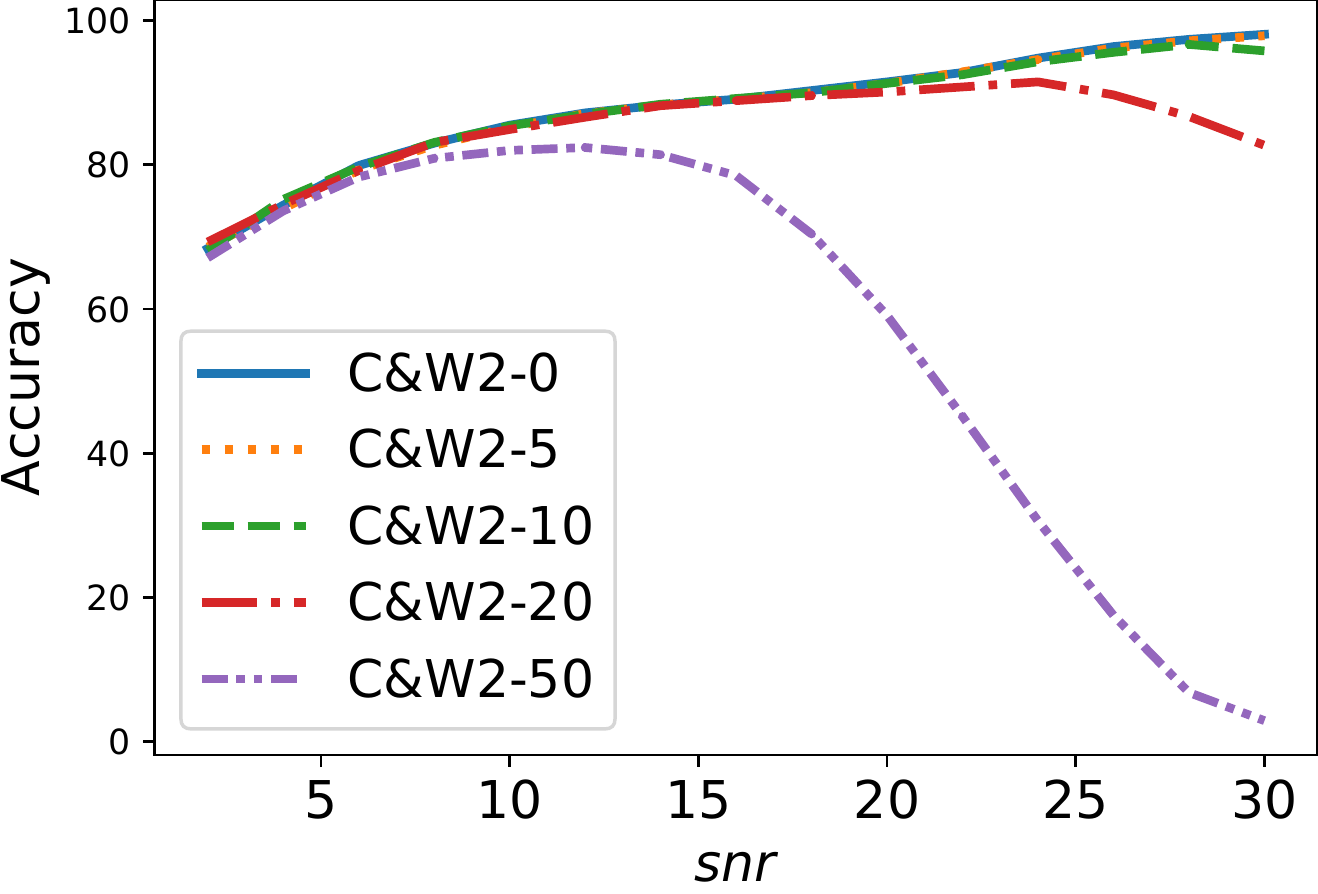}
    \end{minipage}
    }

    \subfigure[AS]{
    \begin{minipage}[t]{0.23\textwidth}
    \includegraphics[width=1.0\textwidth]{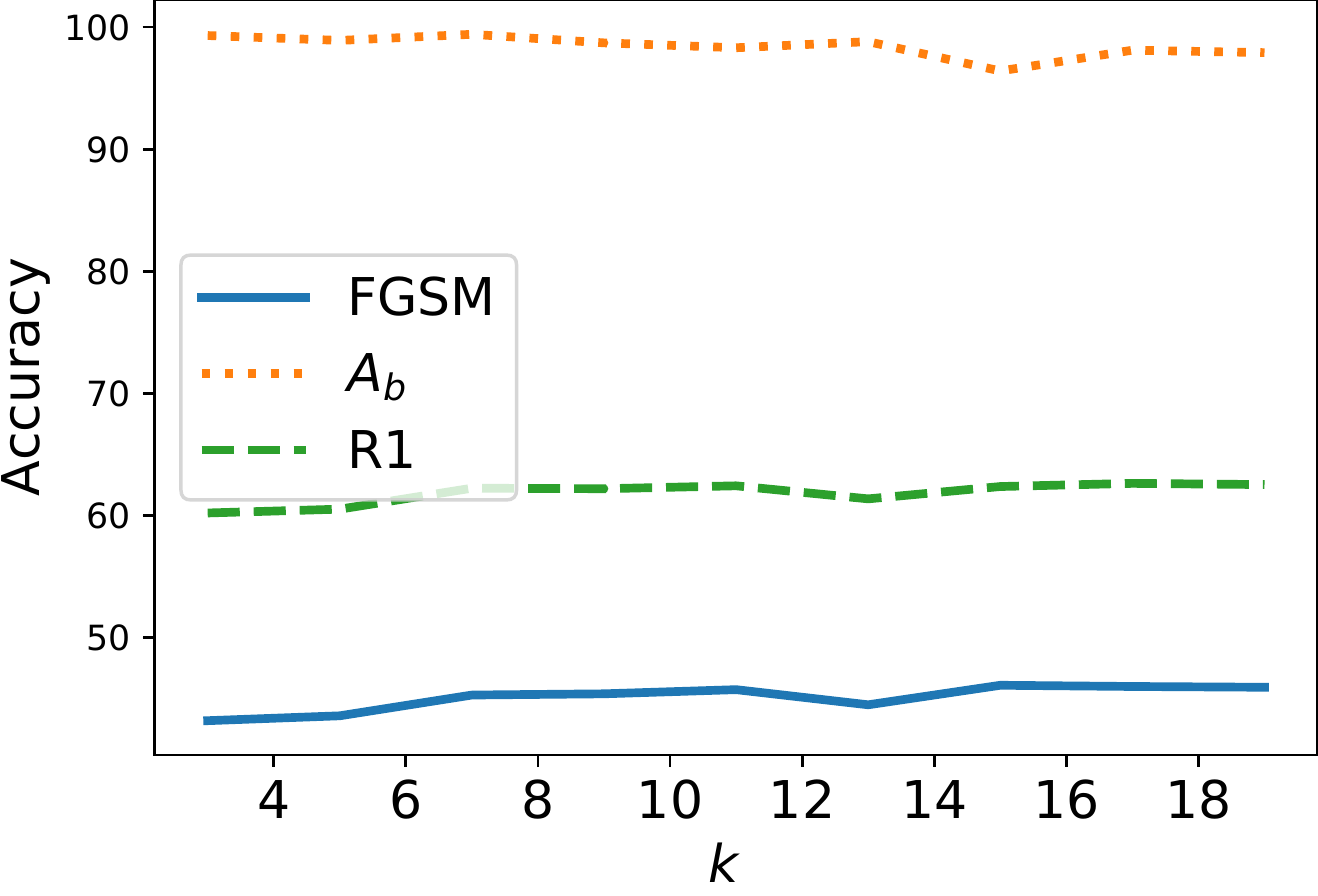}
    \end{minipage}
    \begin{minipage}[t]{0.23\textwidth}
    \includegraphics[width=1.0\textwidth]{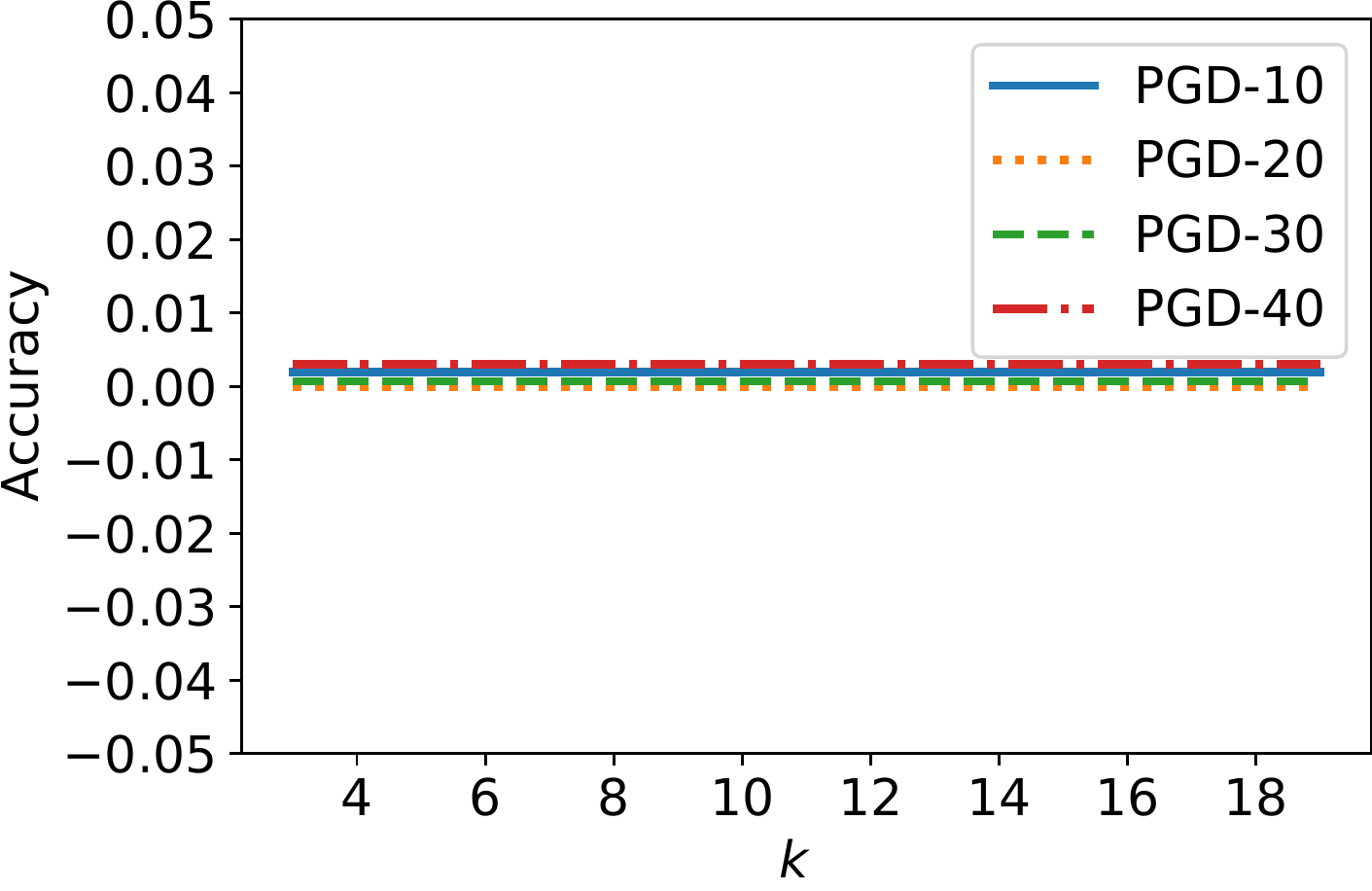}
    \end{minipage}
    \begin{minipage}[t]{0.23\textwidth}
    \includegraphics[width=1.0\textwidth]{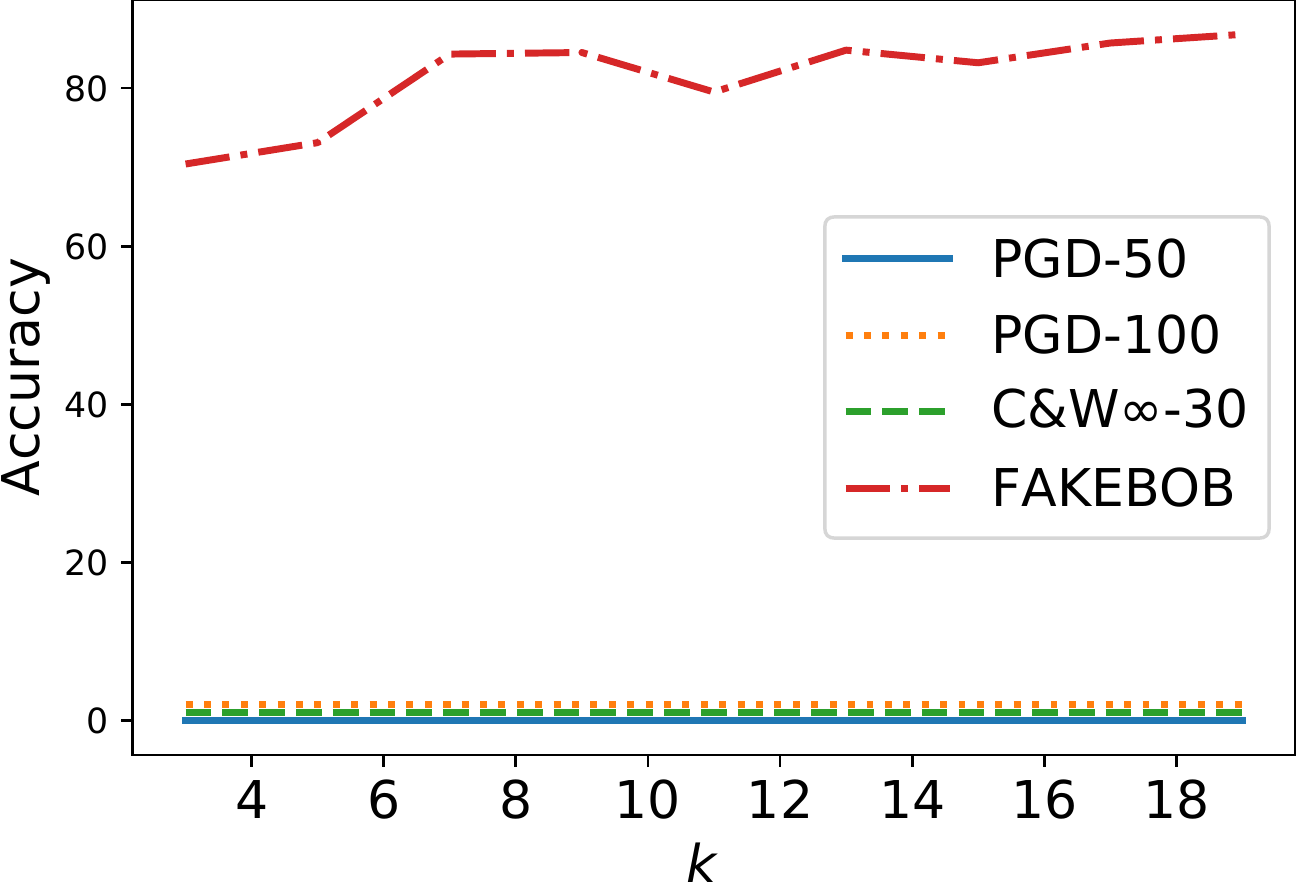}
    \end{minipage}
    \begin{minipage}[t]{0.23\textwidth}
    \includegraphics[width=1.0\textwidth]{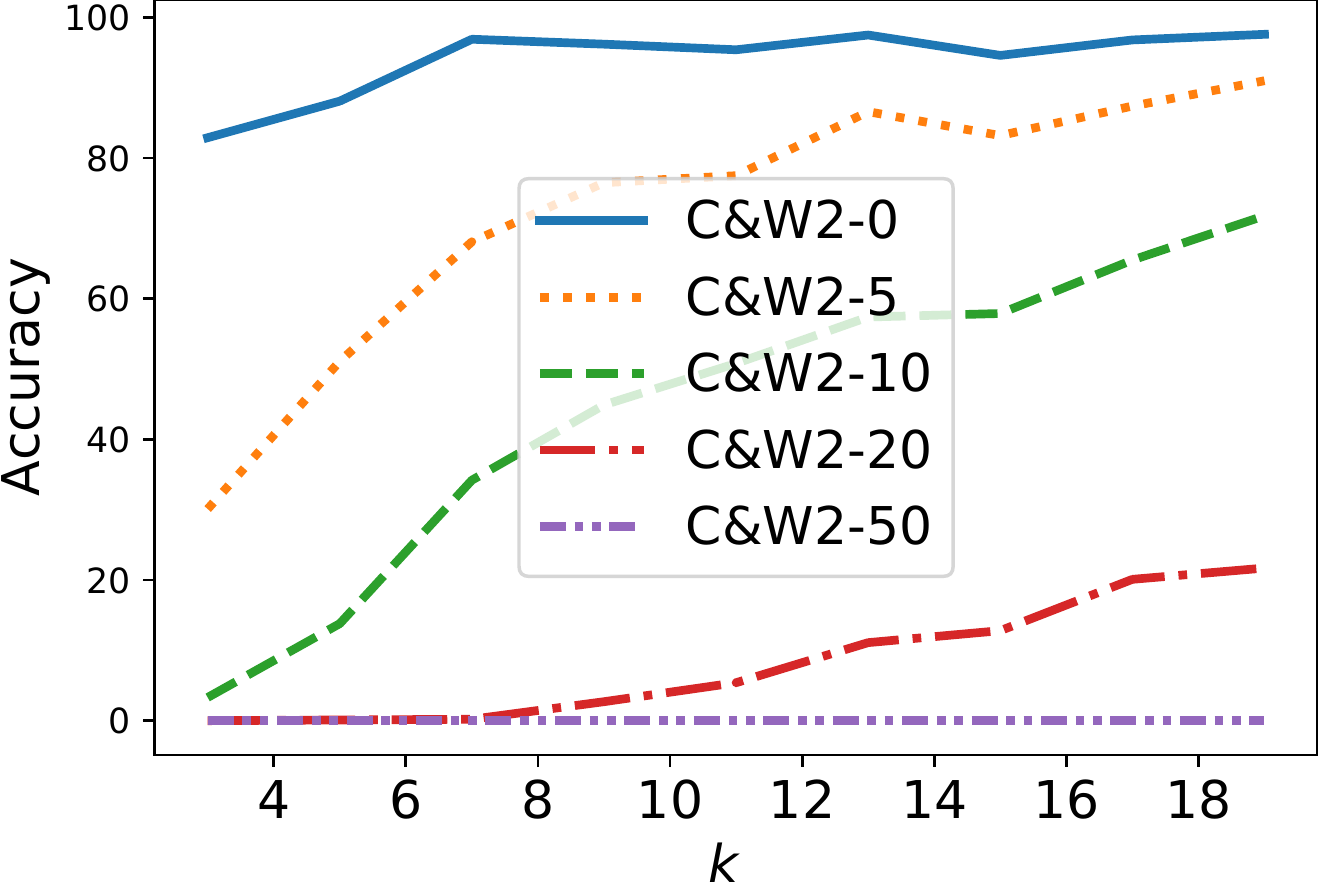}
    \end{minipage}
    }

    \subfigure[MS]{
    \begin{minipage}[t]{0.23\textwidth}
    \includegraphics[width=1.0\textwidth]{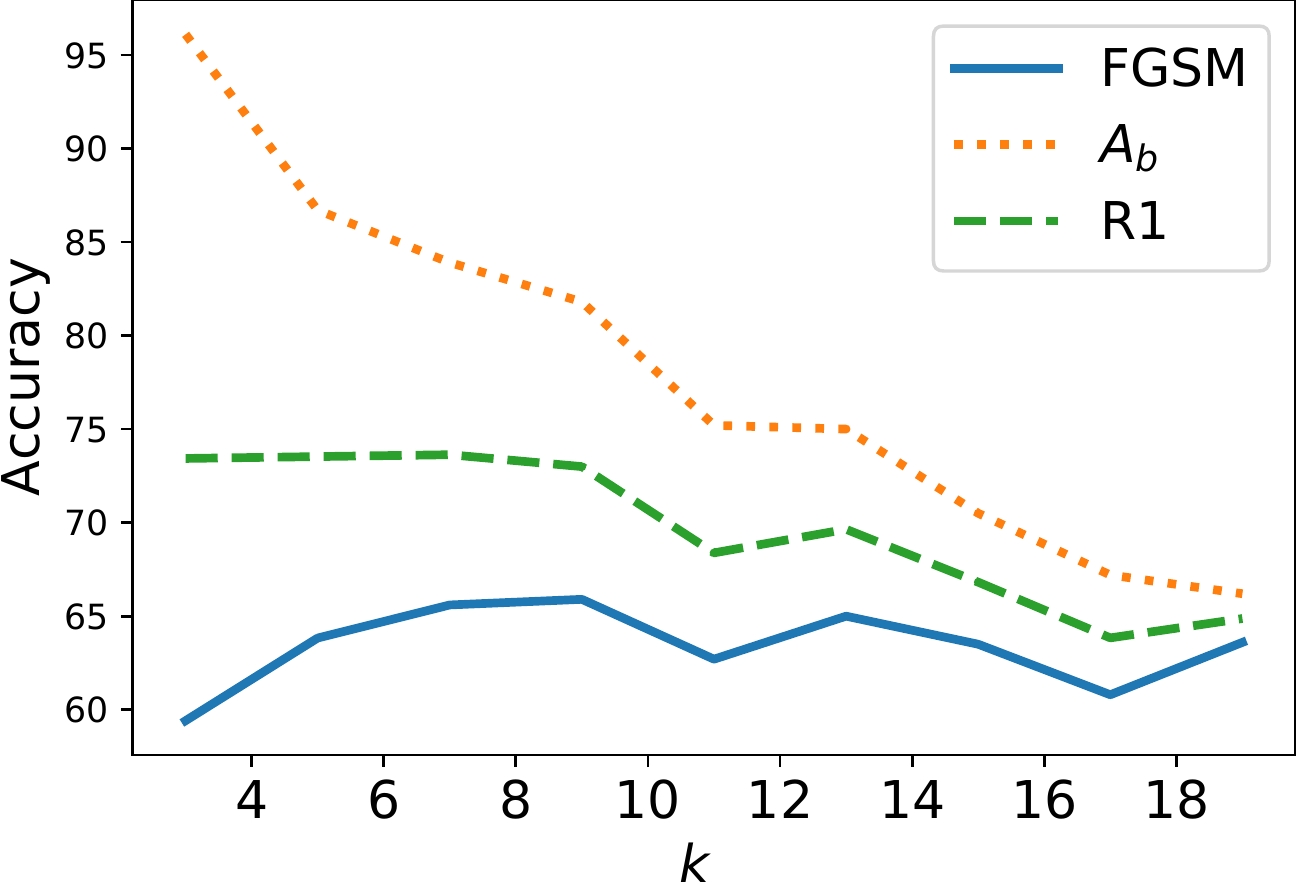}
    \end{minipage}
    \begin{minipage}[t]{0.23\textwidth}
    \includegraphics[width=1.0\textwidth]{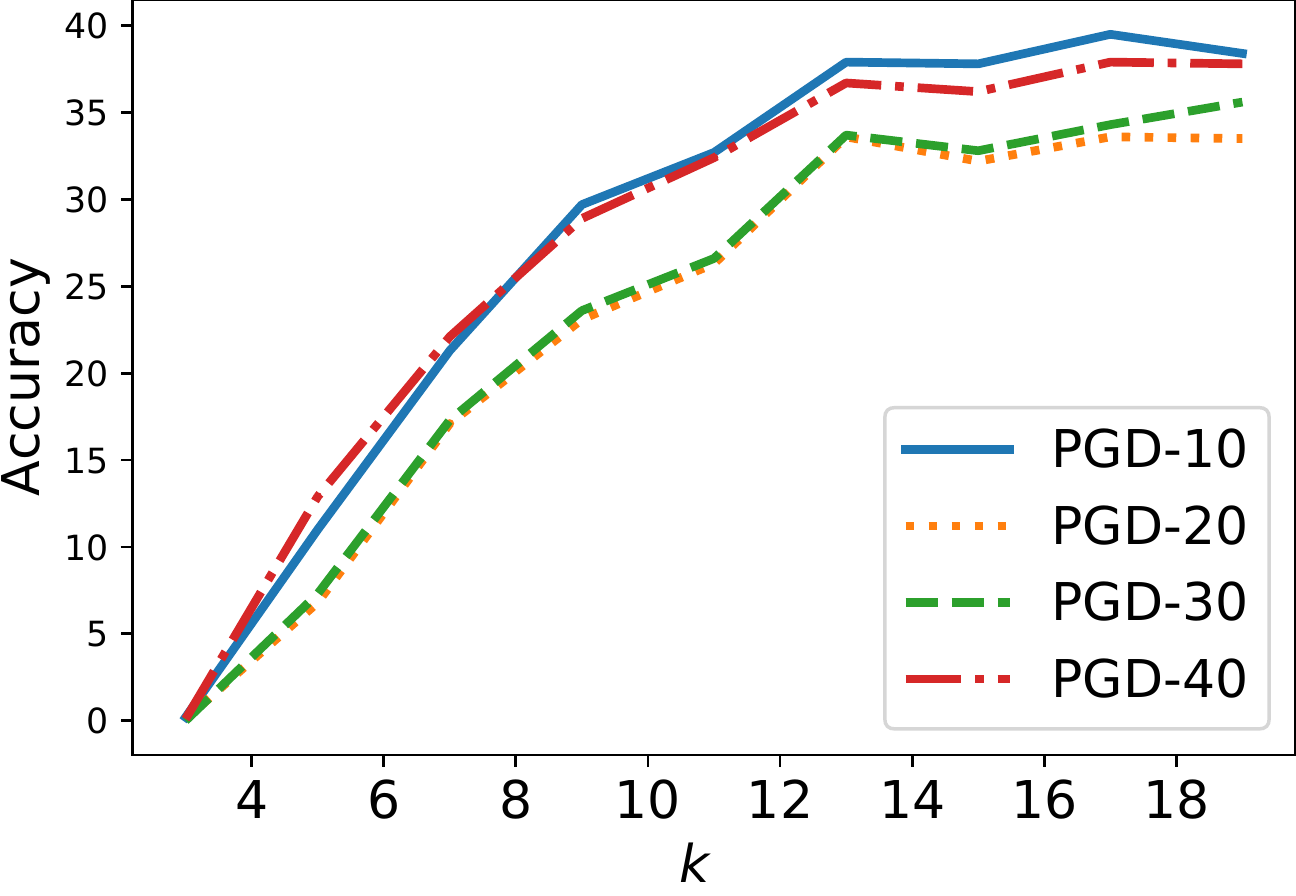}
    \end{minipage}
    \begin{minipage}[t]{0.23\textwidth}
    \includegraphics[width=1.0\textwidth]{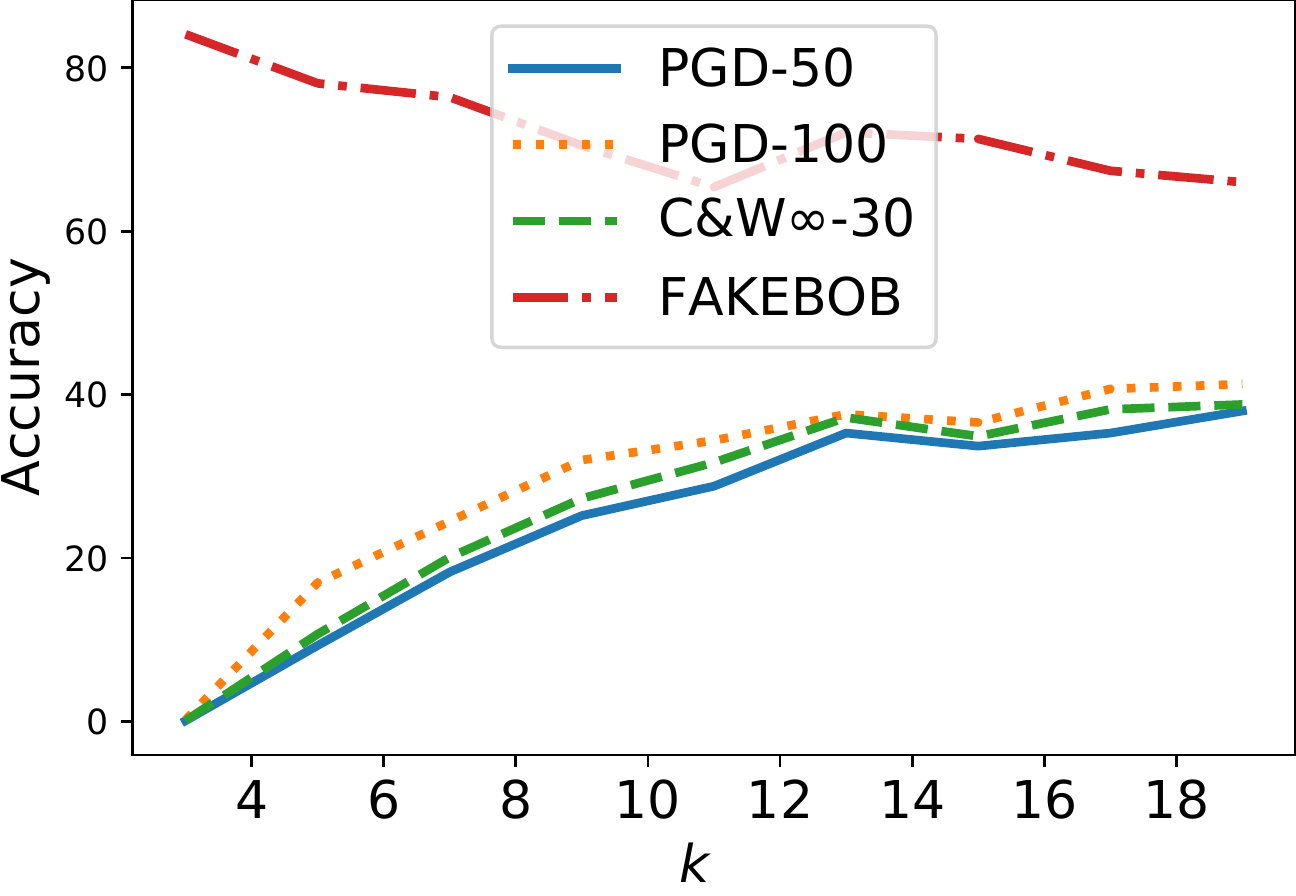}
    \end{minipage}
    \begin{minipage}[t]{0.23\textwidth}
    \includegraphics[width=1.0\textwidth]{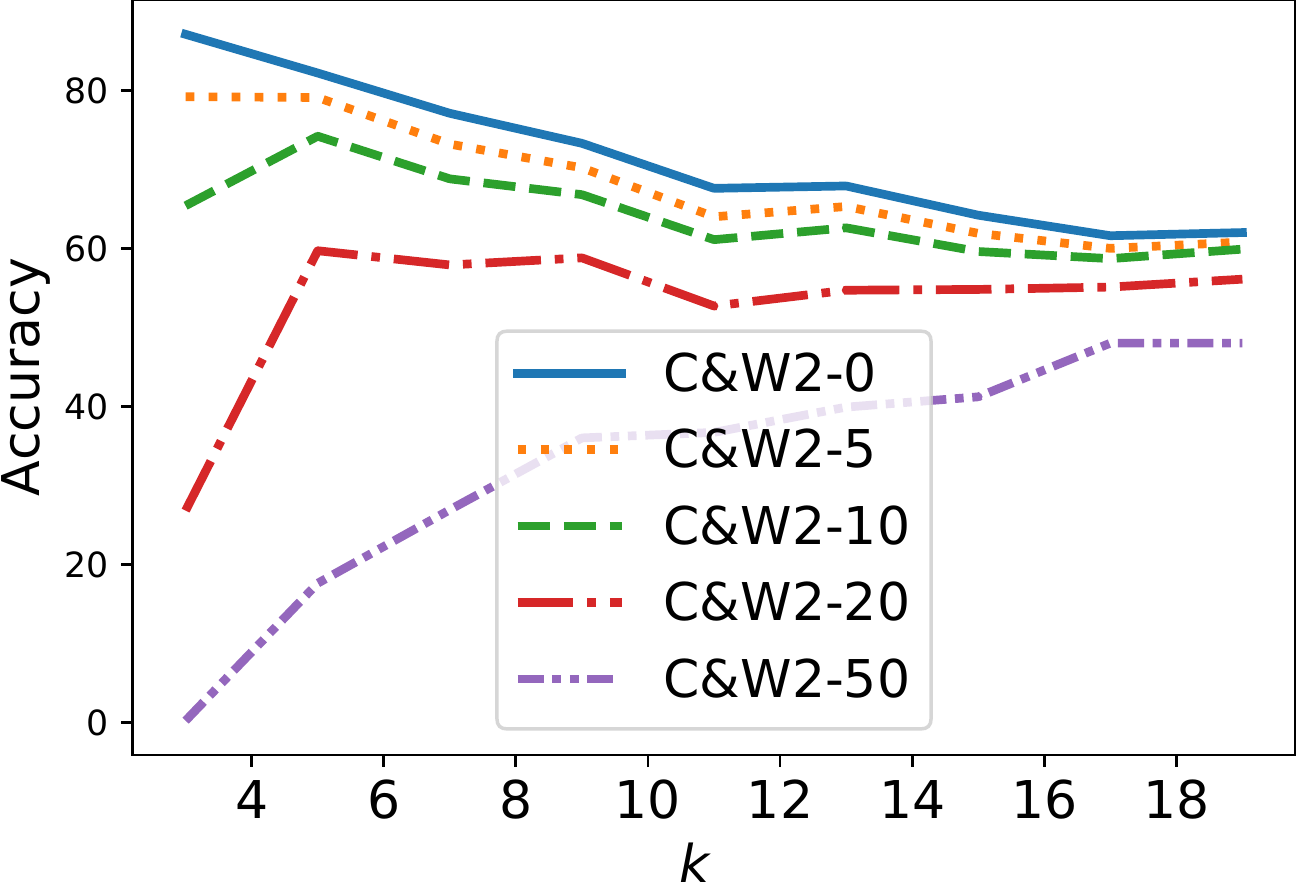}
    \end{minipage}
    }

    \subfigure[LPF]{
    \begin{minipage}[t]{0.23\textwidth}
    \includegraphics[width=1.0\textwidth]{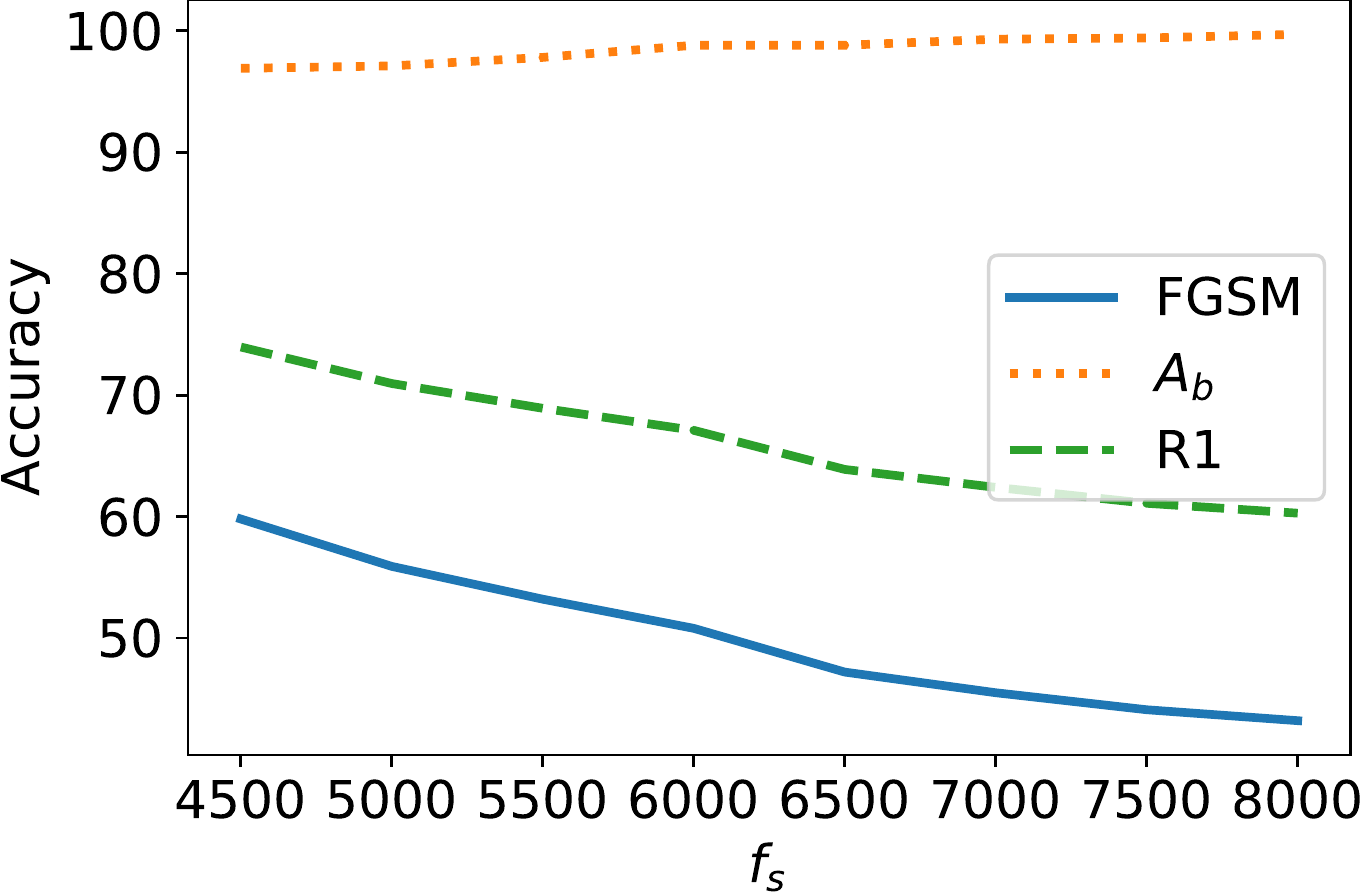}
    \end{minipage}
    \begin{minipage}[t]{0.23\textwidth}
    \includegraphics[width=1.0\textwidth]{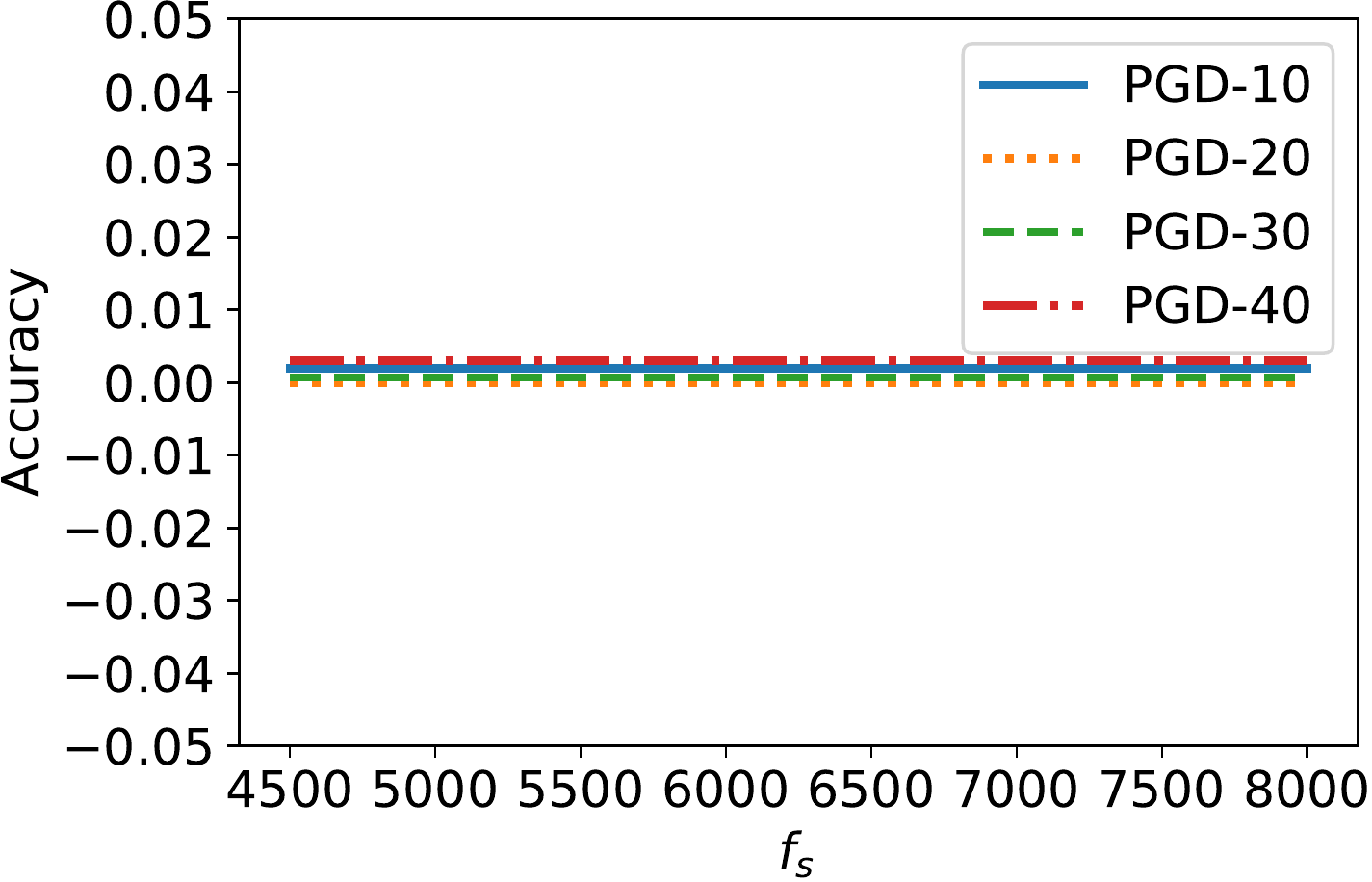}
    \end{minipage}
    \begin{minipage}[t]{0.23\textwidth}
    \includegraphics[width=1.0\textwidth]{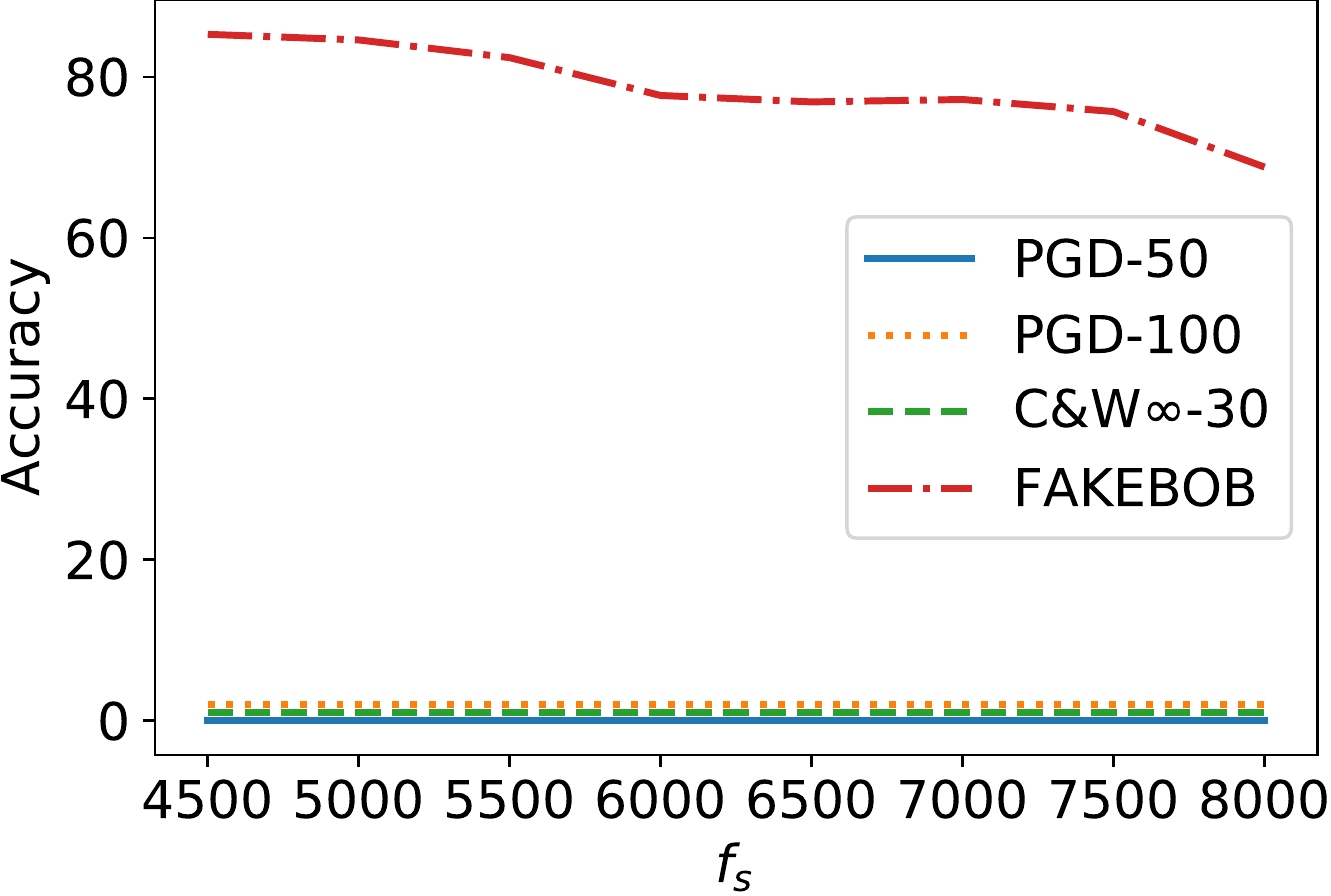}
    \end{minipage}
    \begin{minipage}[t]{0.23\textwidth}
    \includegraphics[width=1.0\textwidth]{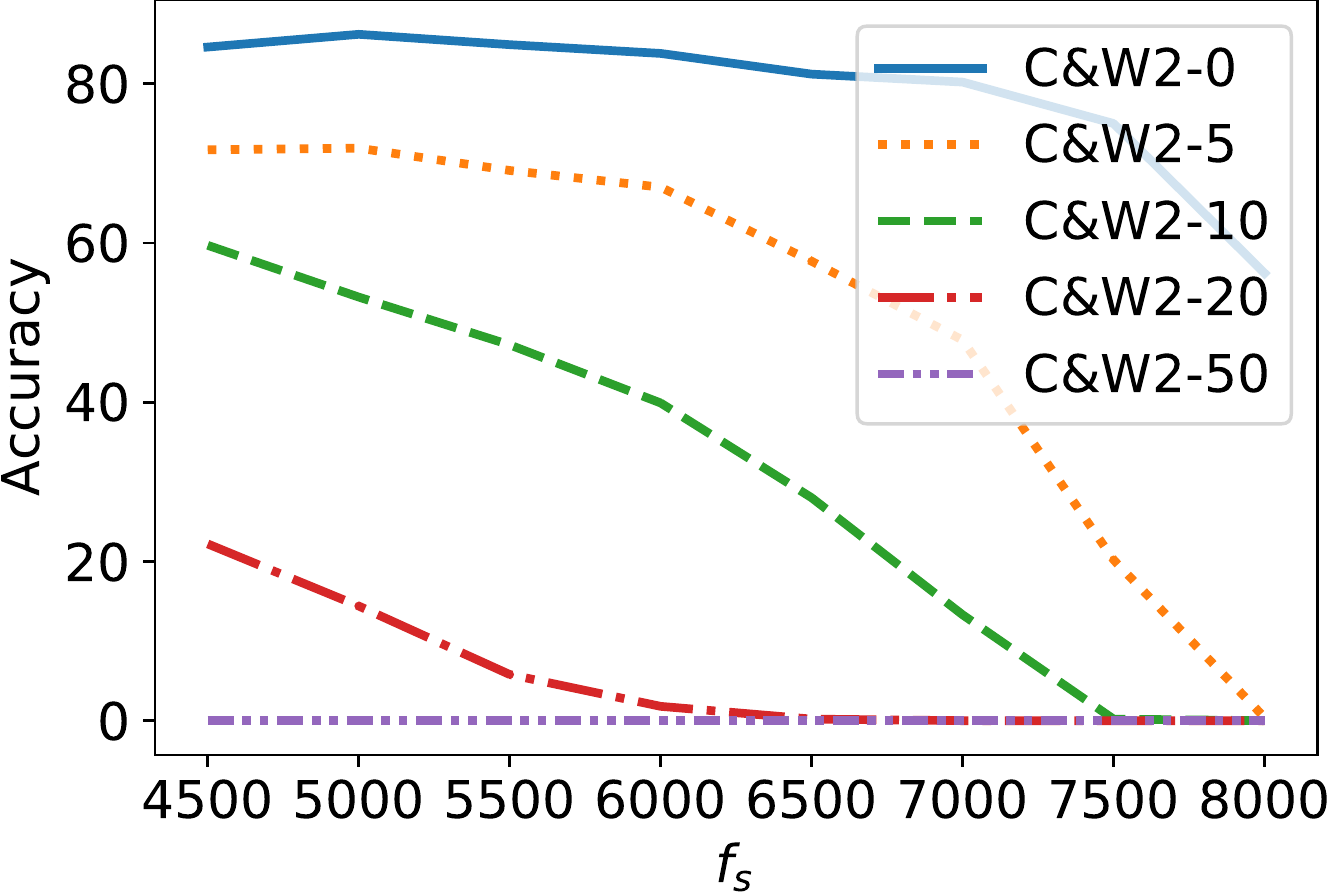}
    \end{minipage}
    }


    \caption{The performance of input transformations.}
    \label{fig:parameter-1}
\end{figure*}

\begin{figure*}
    \centering

    \subfigure[BPF]{
    \begin{minipage}[t]{0.23\textwidth}
    \includegraphics[width=1.0\textwidth]{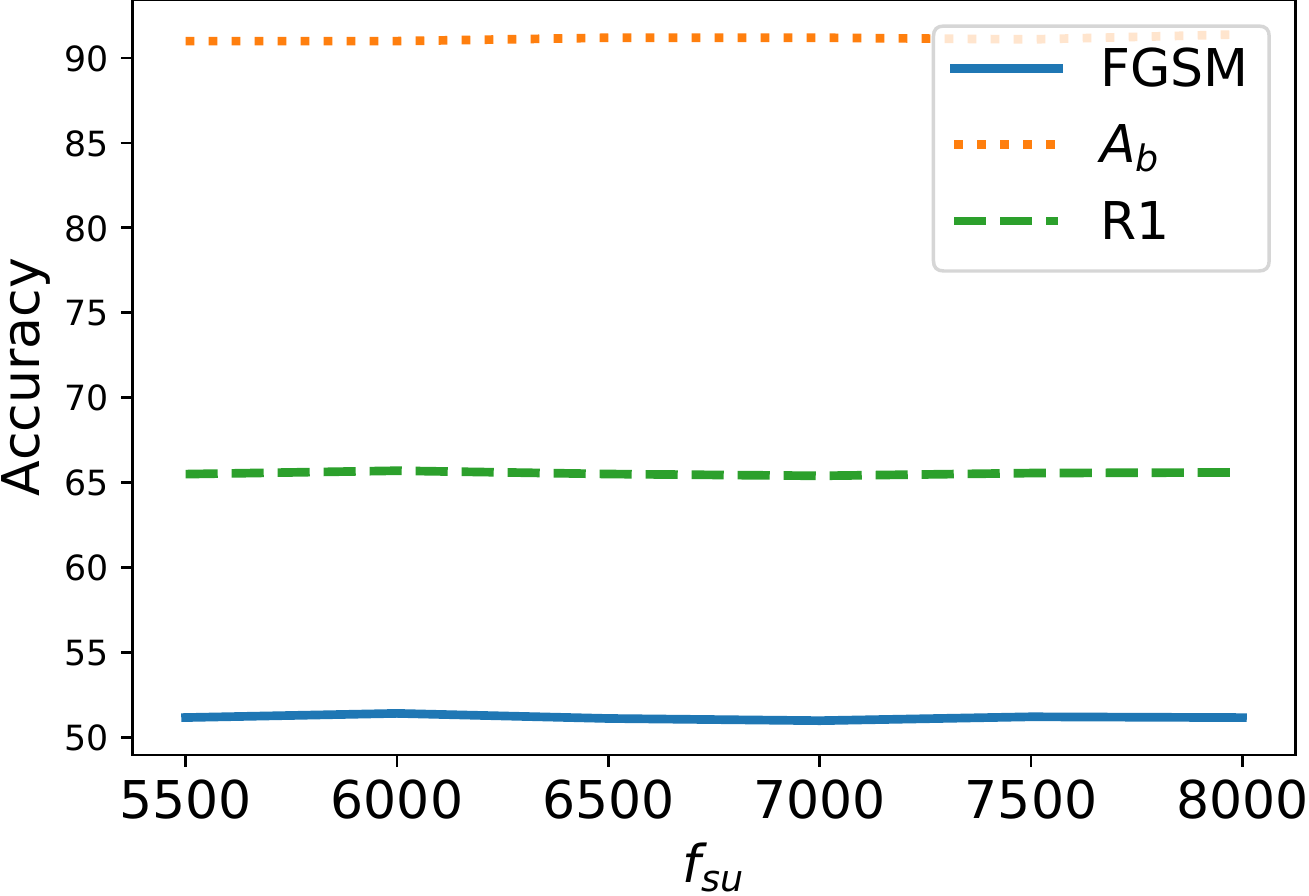}
    \end{minipage}
    \begin{minipage}[t]{0.23\textwidth}
    \includegraphics[width=1.0\textwidth]{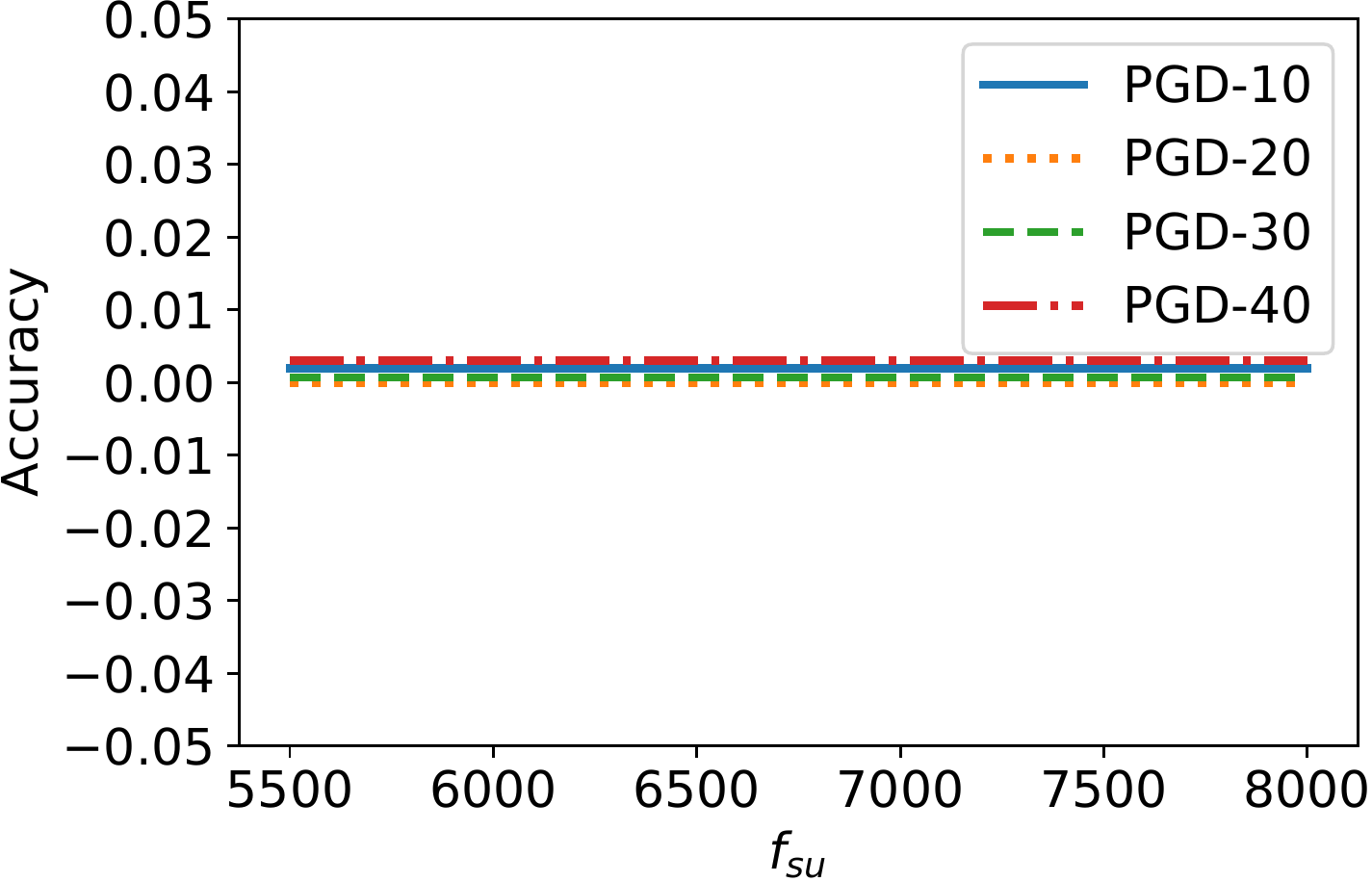}
    \end{minipage}
    \begin{minipage}[t]{0.23\textwidth}
    \includegraphics[width=1.0\textwidth]{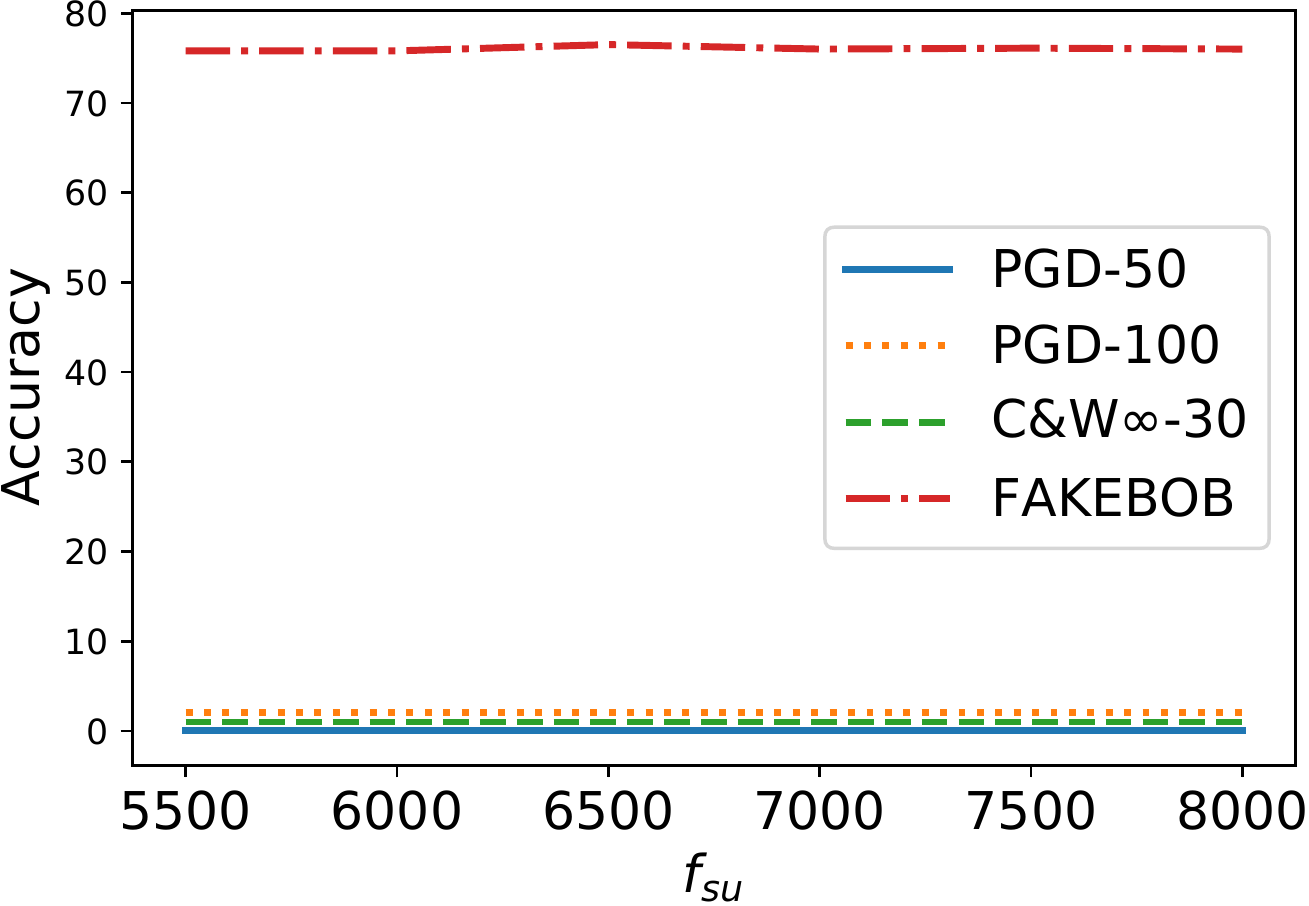}
    \end{minipage}
    \begin{minipage}[t]{0.23\textwidth}
    \includegraphics[width=1.0\textwidth]{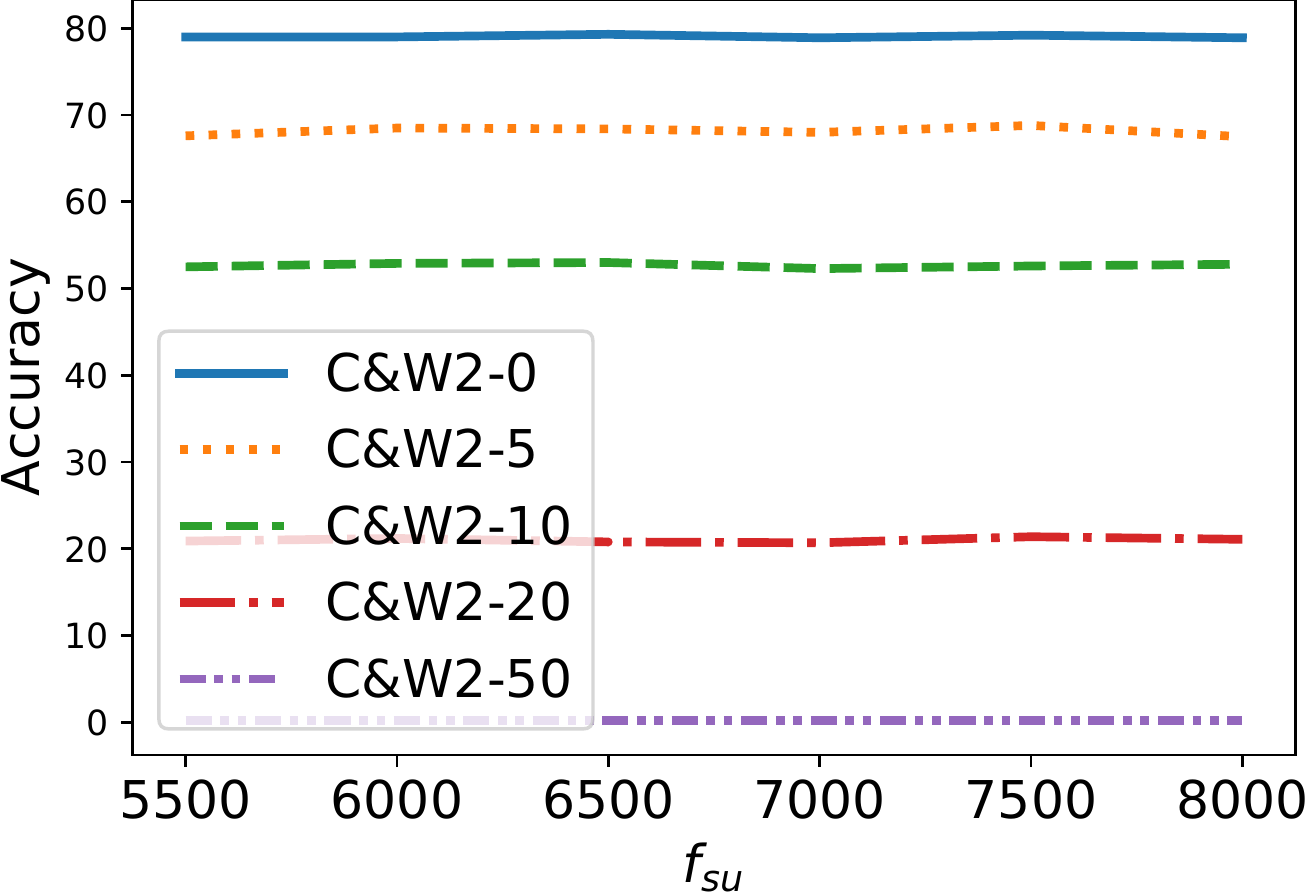}
    \end{minipage}
    }

    \subfigure[OPUS]{
    \begin{minipage}[t]{0.23\textwidth}
    \includegraphics[width=1.0\textwidth]{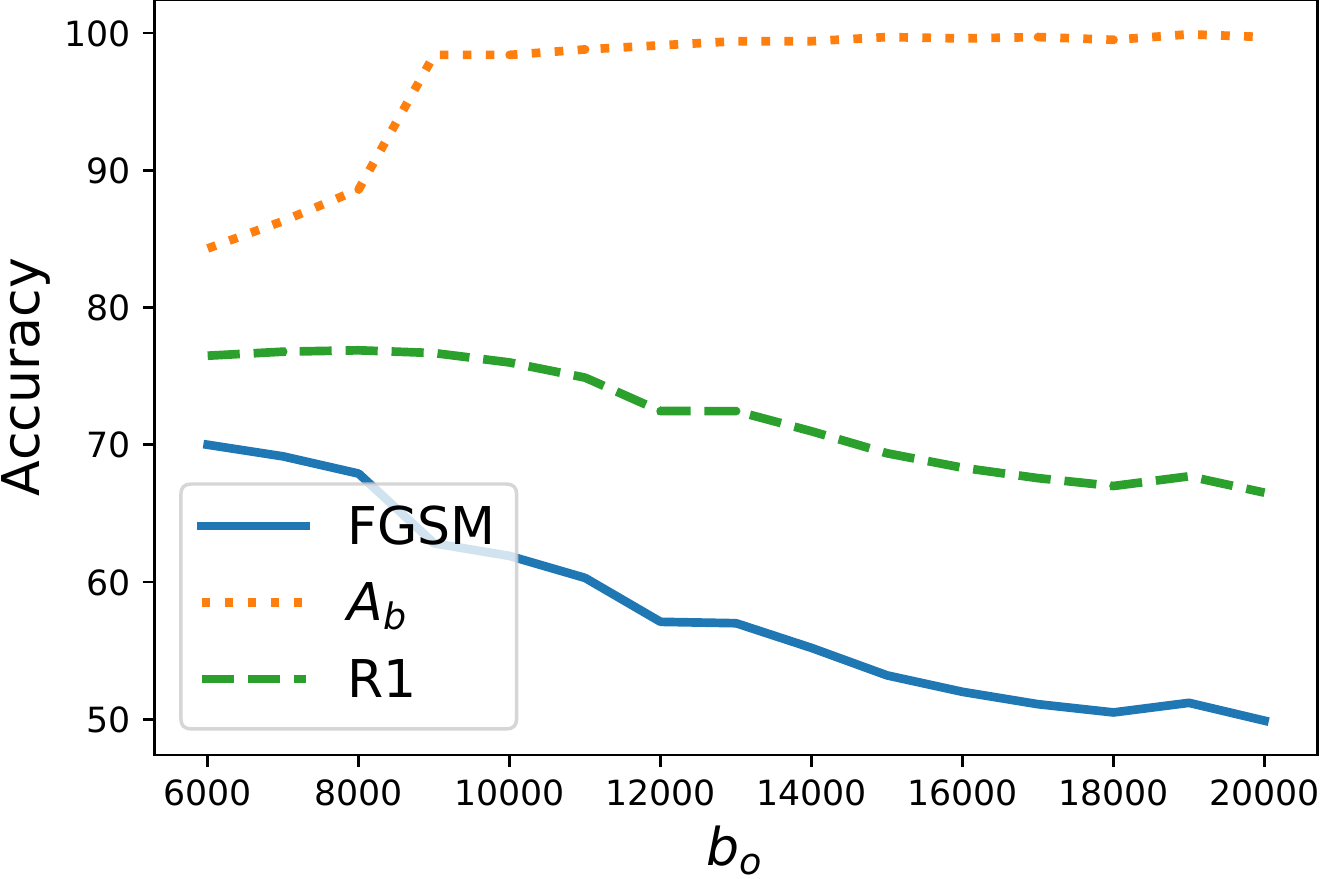}
    \end{minipage}
    \begin{minipage}[t]{0.23\textwidth}
    \includegraphics[width=1.0\textwidth]{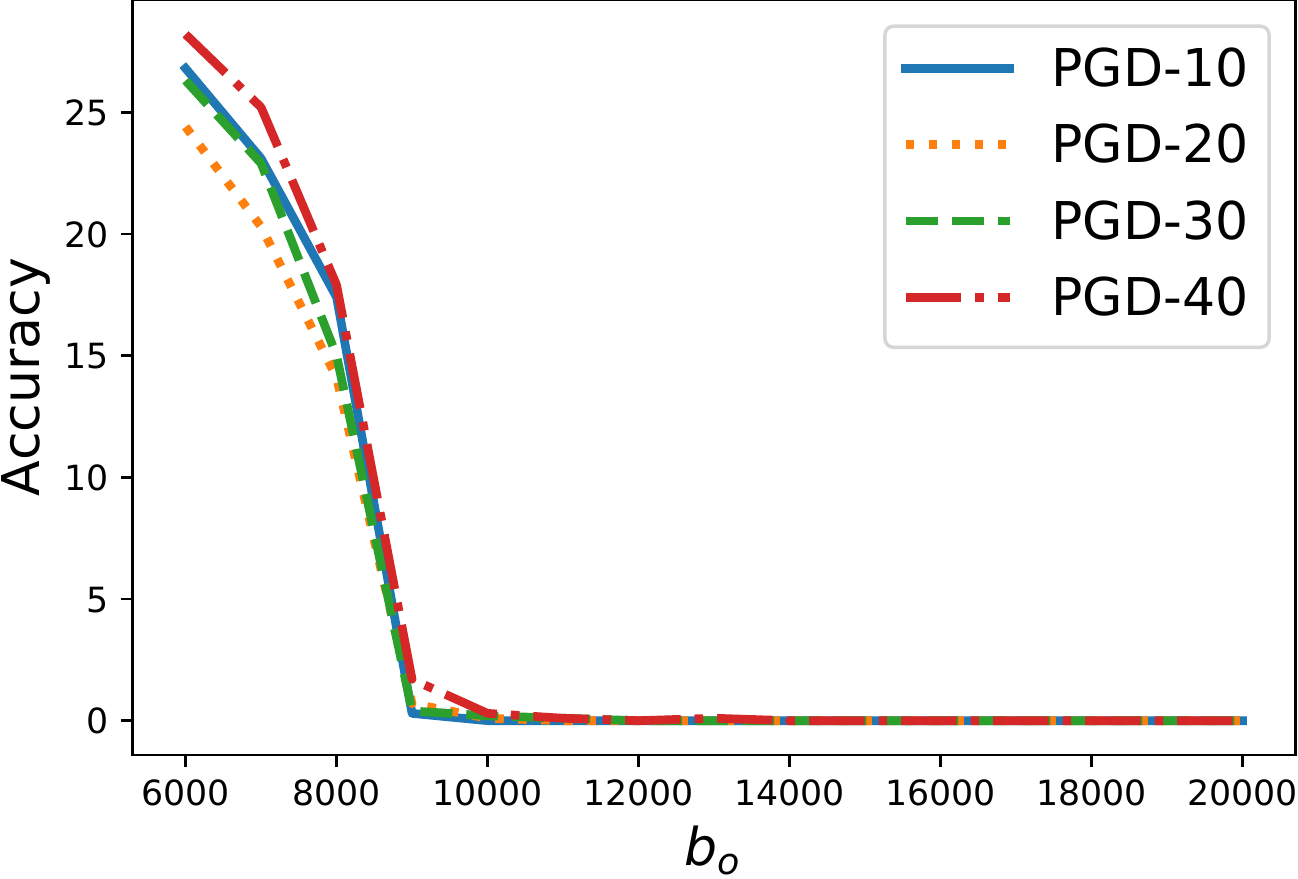}
    \end{minipage}
    \begin{minipage}[t]{0.23\textwidth}
    \includegraphics[width=1.0\textwidth]{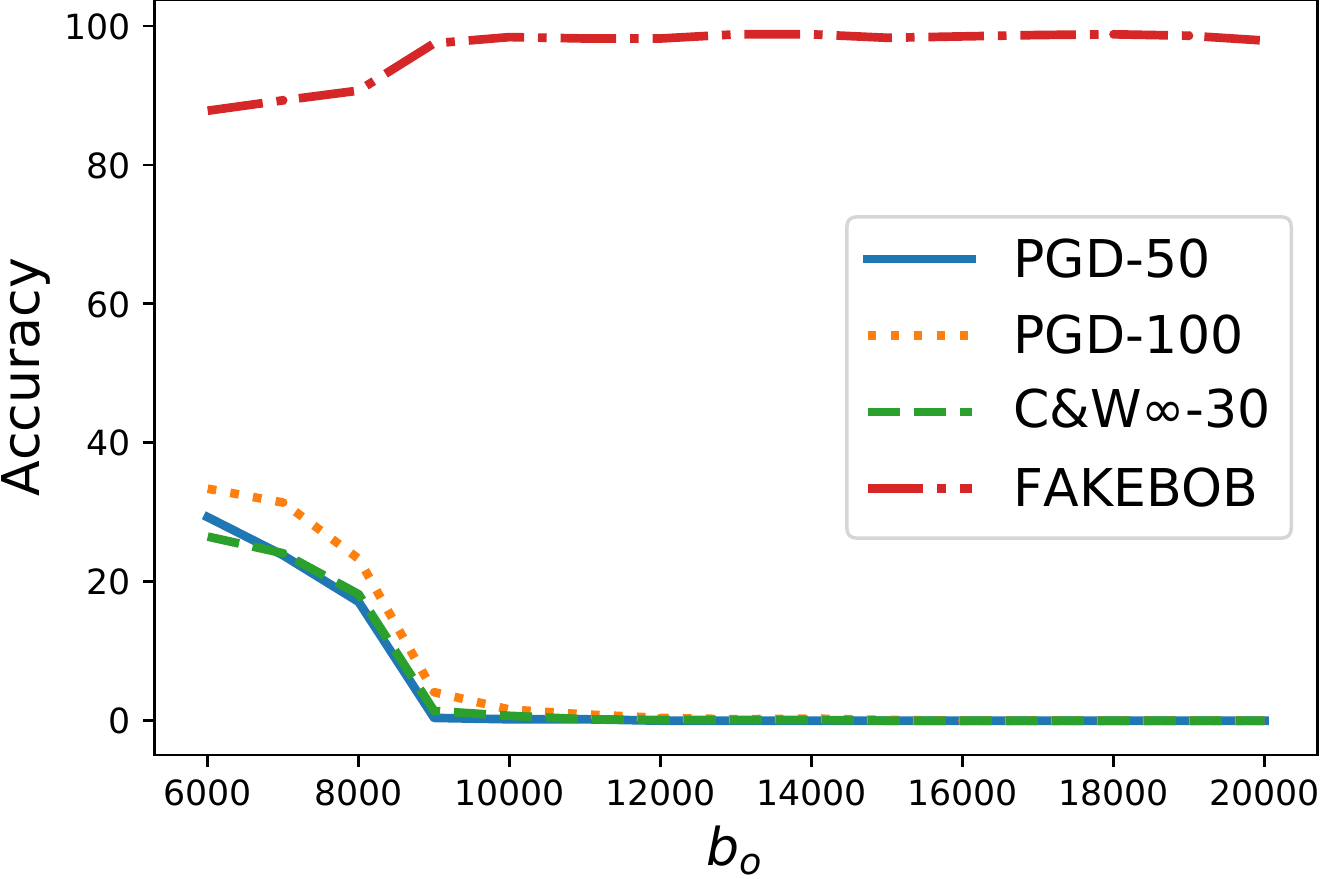}
    \end{minipage}
    \begin{minipage}[t]{0.23\textwidth}
    \includegraphics[width=1.0\textwidth]{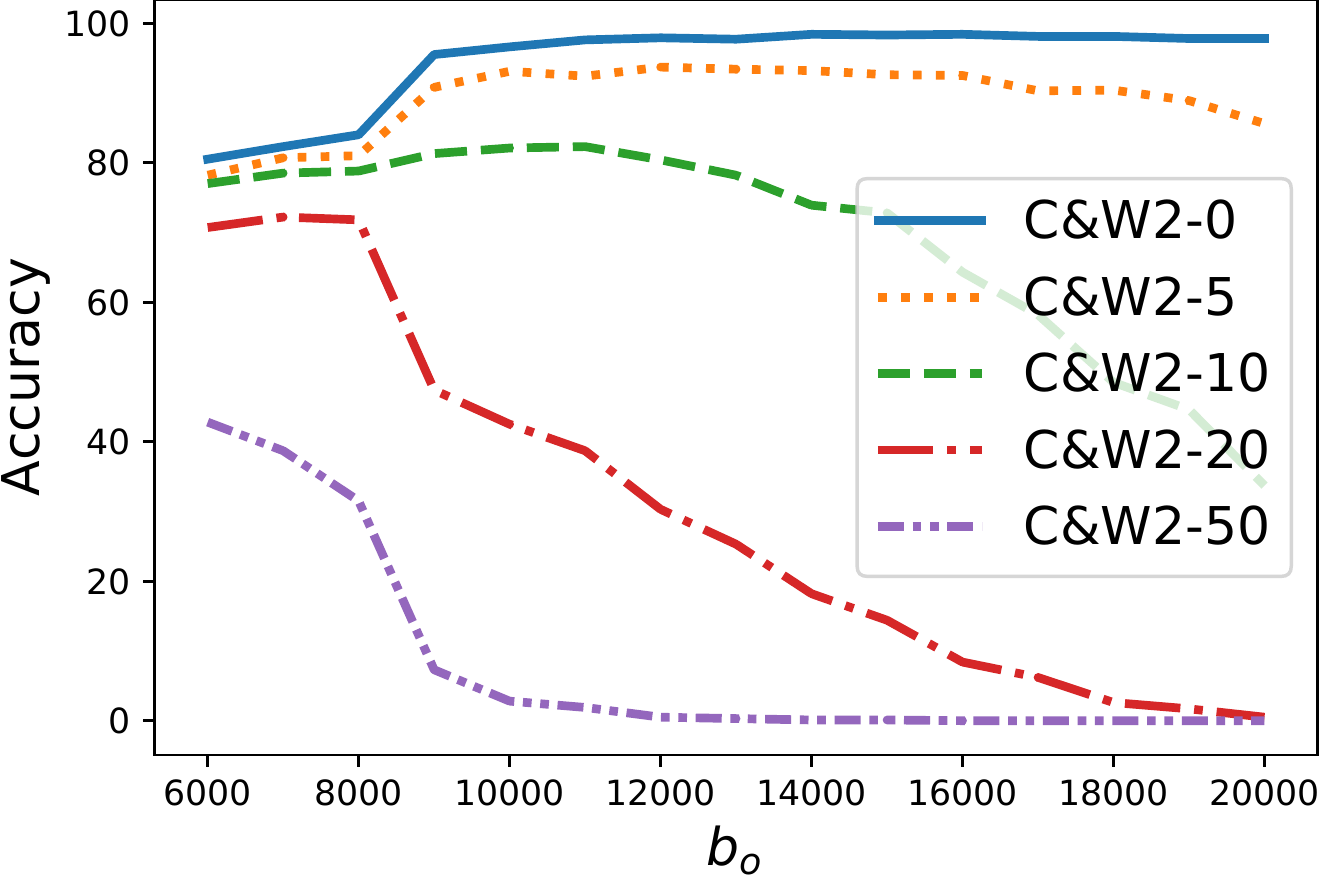}
    \end{minipage}
    }

    \subfigure[SPEEX]{
    \begin{minipage}[t]{0.23\textwidth}
    \includegraphics[width=1.0\textwidth]{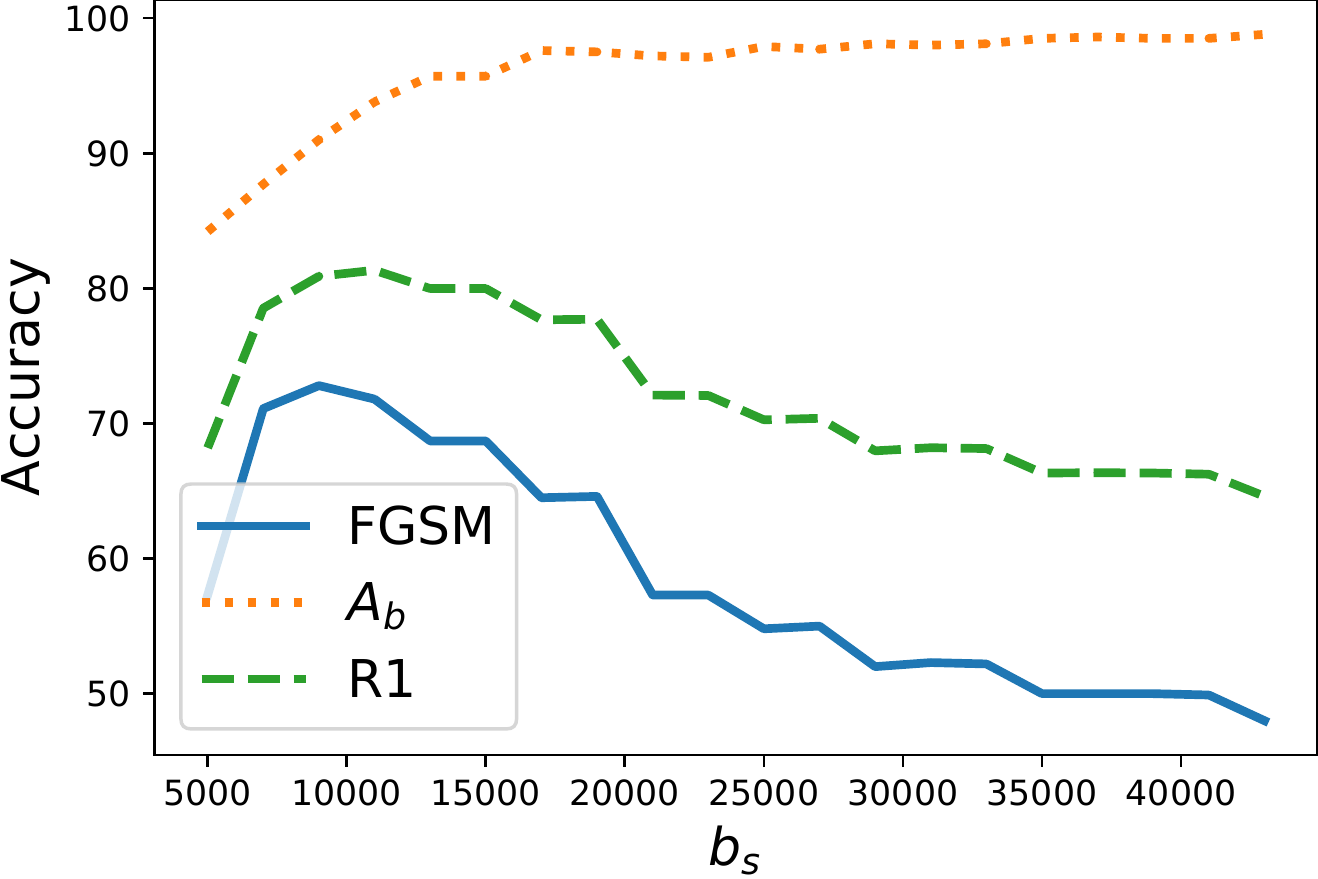}
    \end{minipage}
    \begin{minipage}[t]{0.23\textwidth}
    \includegraphics[width=1.0\textwidth]{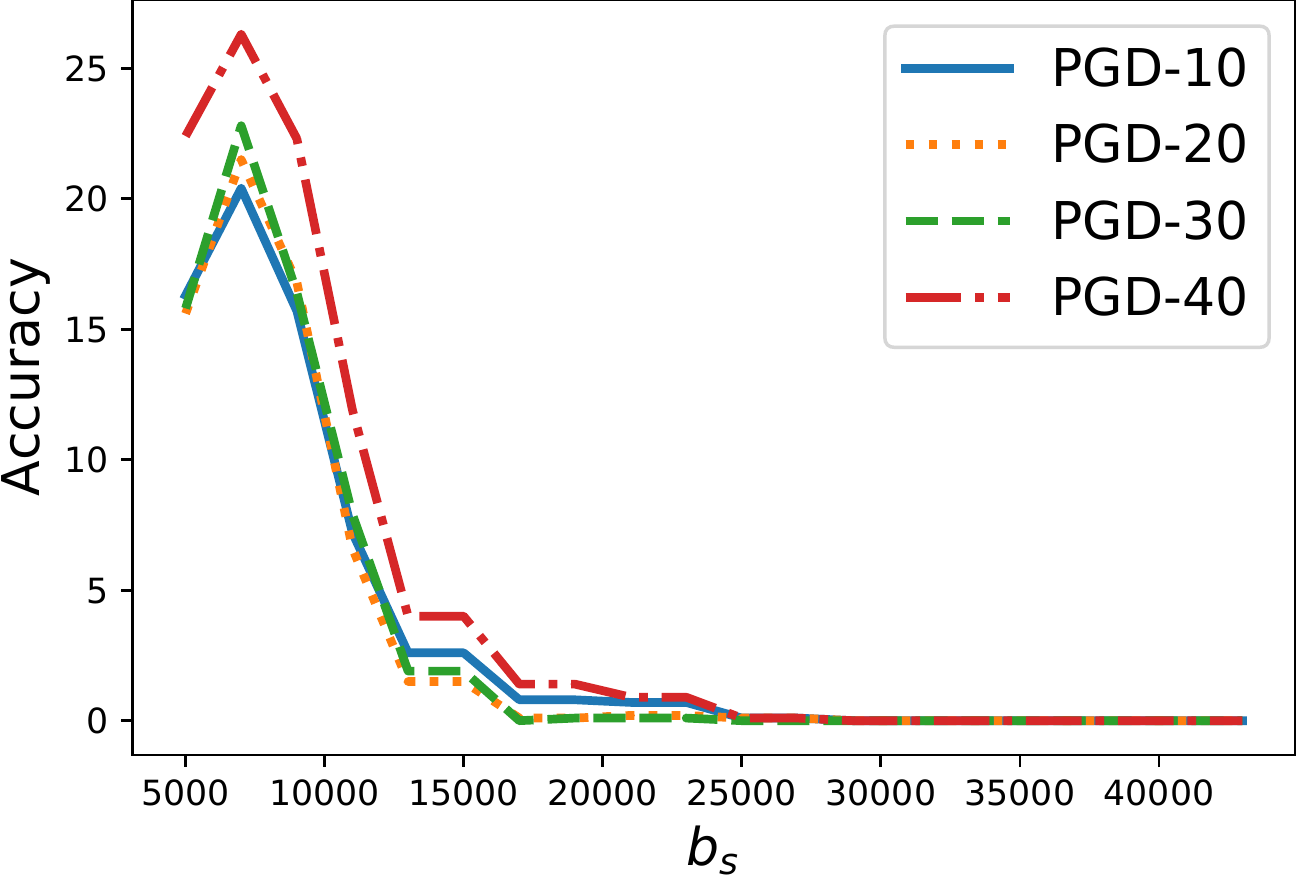}
    \end{minipage}
    \begin{minipage}[t]{0.23\textwidth}
    \includegraphics[width=1.0\textwidth]{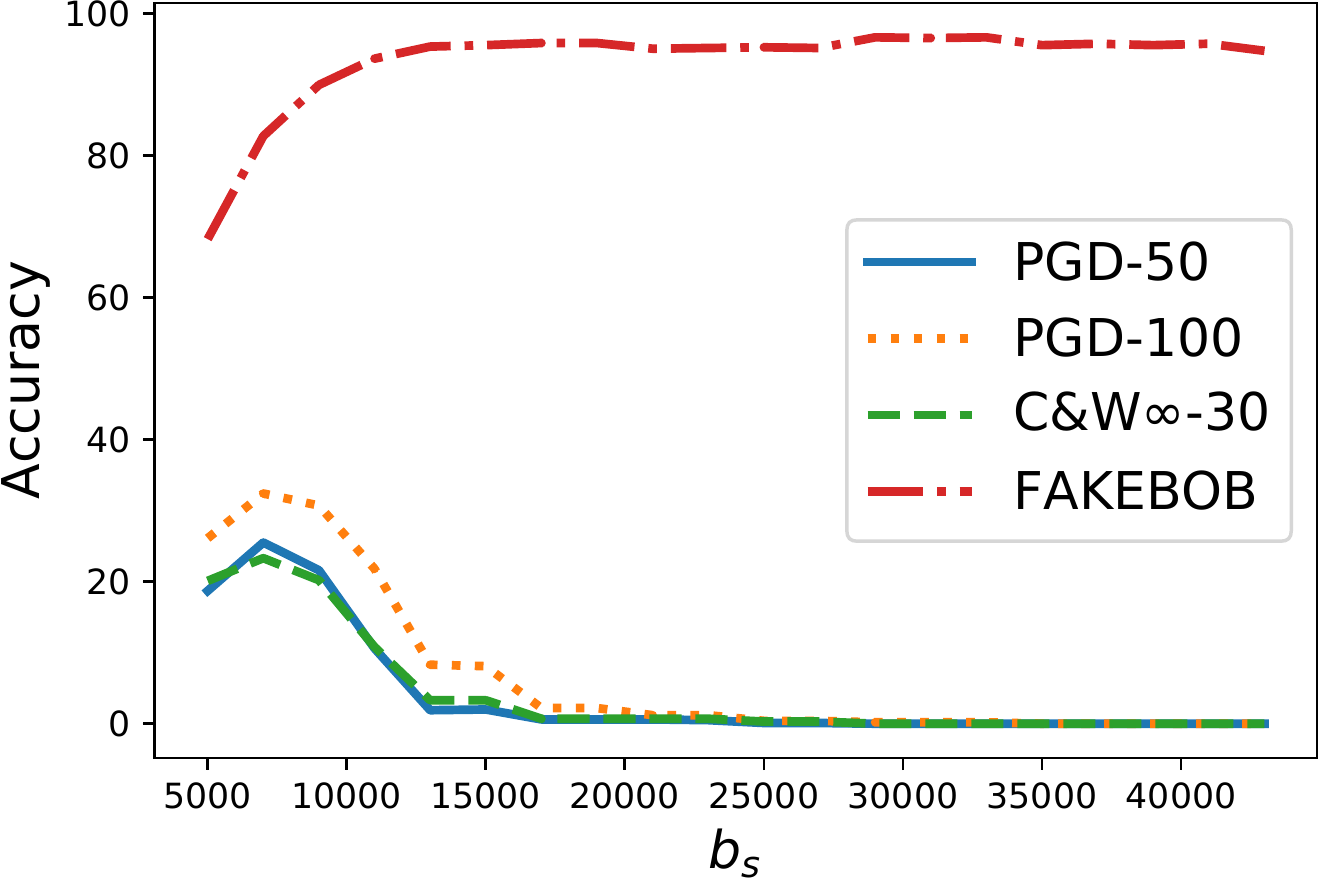}
    \end{minipage}
    \begin{minipage}[t]{0.23\textwidth}
    \includegraphics[width=1.0\textwidth]{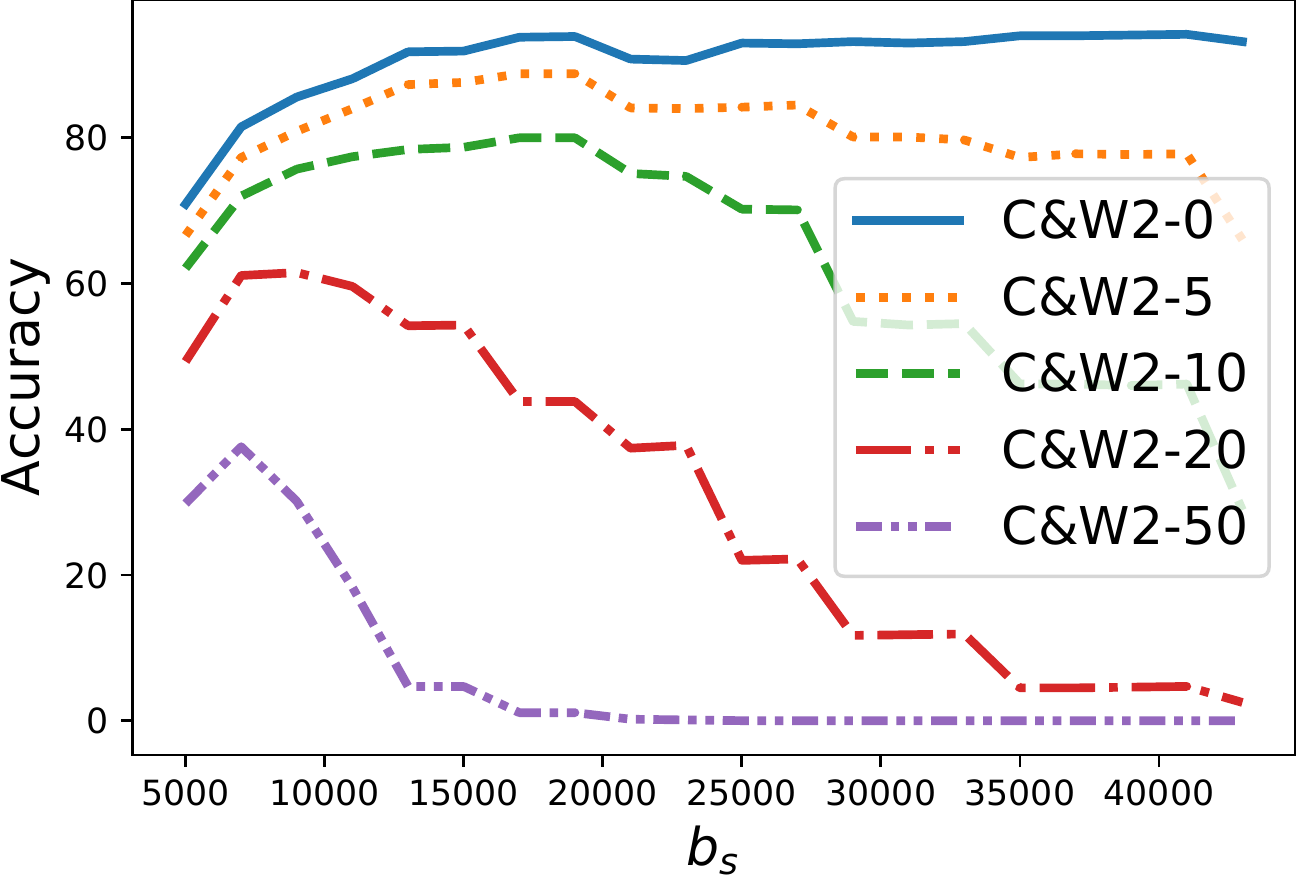}
    \end{minipage}
    }

    \subfigure[AMR]{
    \begin{minipage}[t]{0.23\textwidth}
    \includegraphics[width=1.0\textwidth]{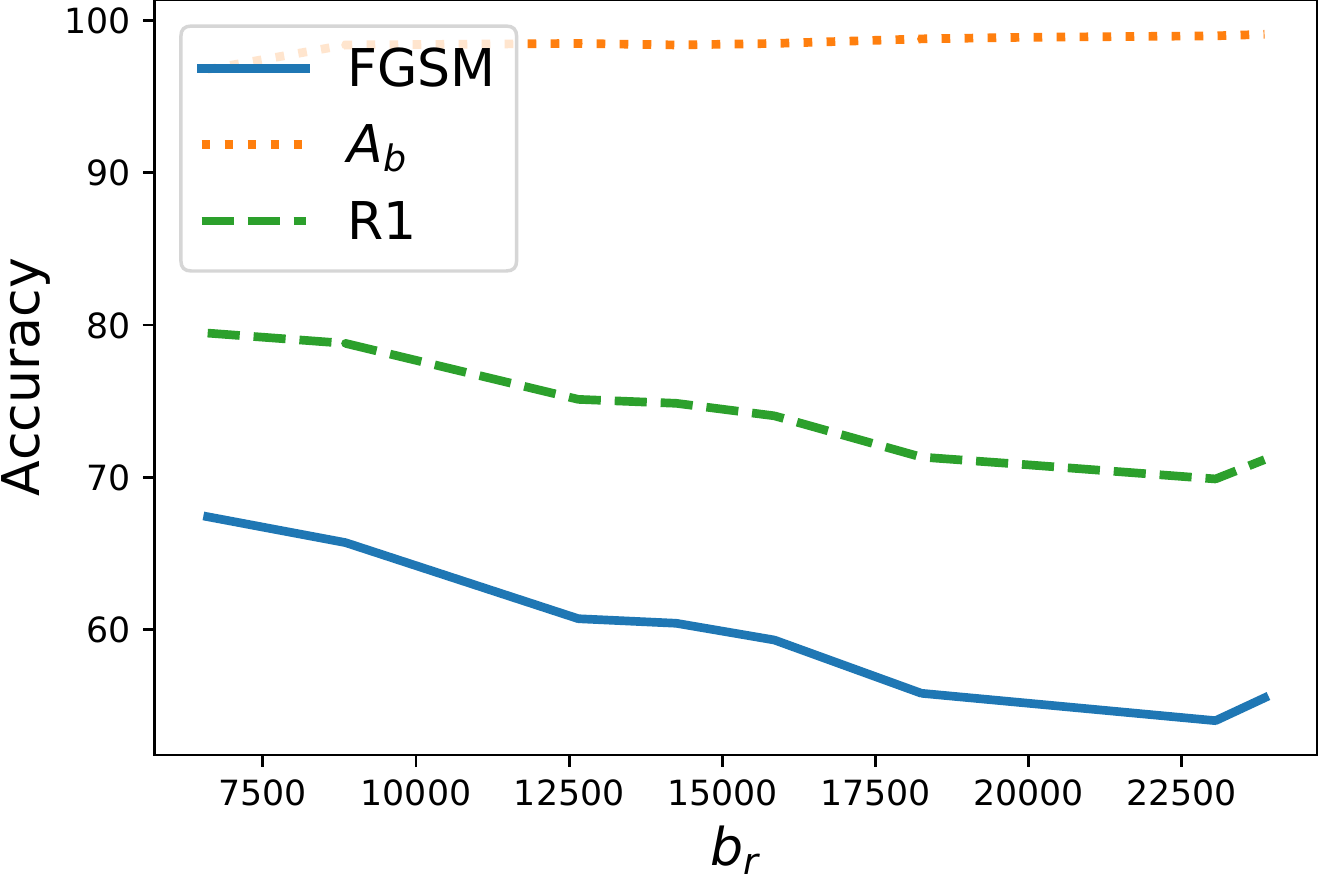}
    \end{minipage}
    \begin{minipage}[t]{0.23\textwidth}
    \includegraphics[width=1.0\textwidth]{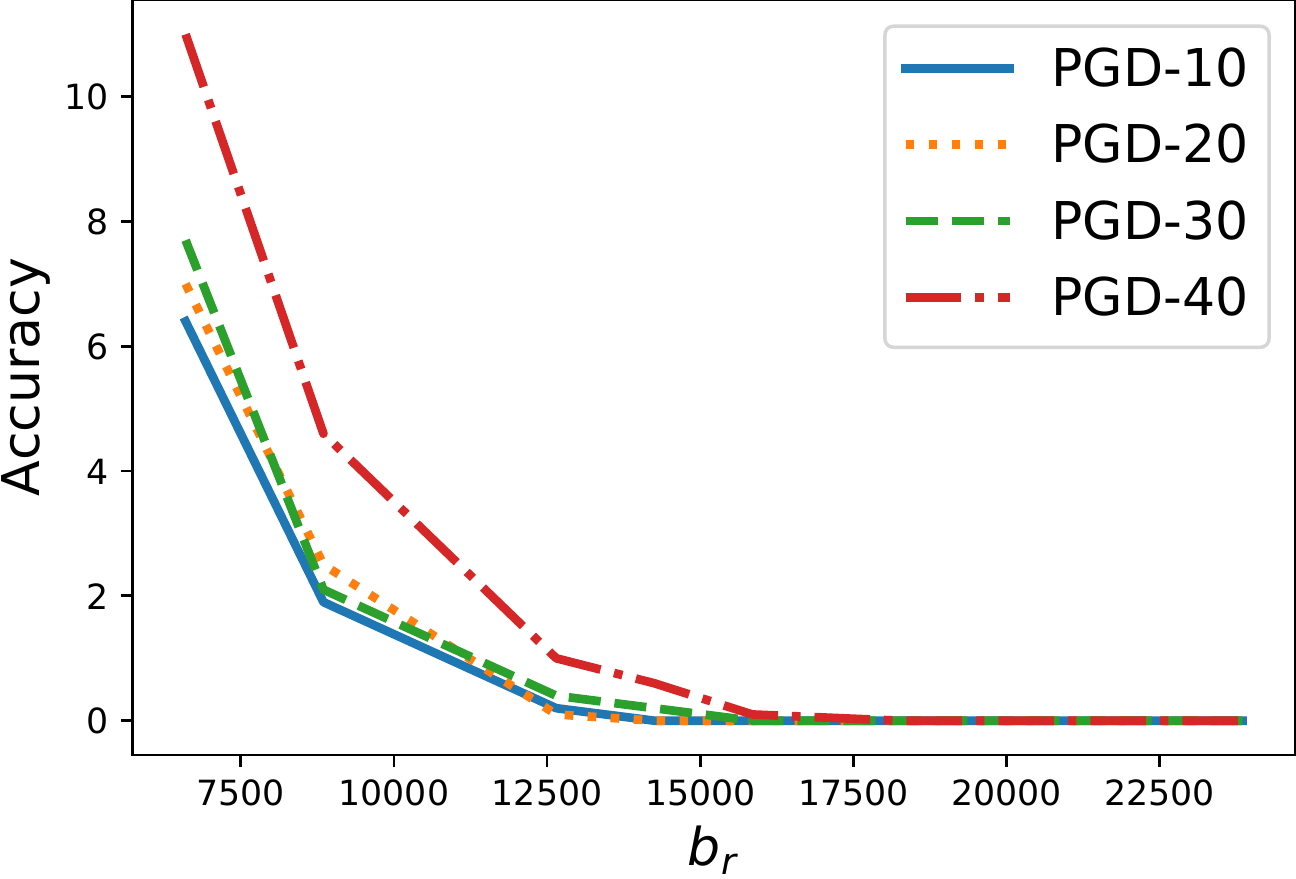}
    \end{minipage}
    \begin{minipage}[t]{0.23\textwidth}
    \includegraphics[width=1.0\textwidth]{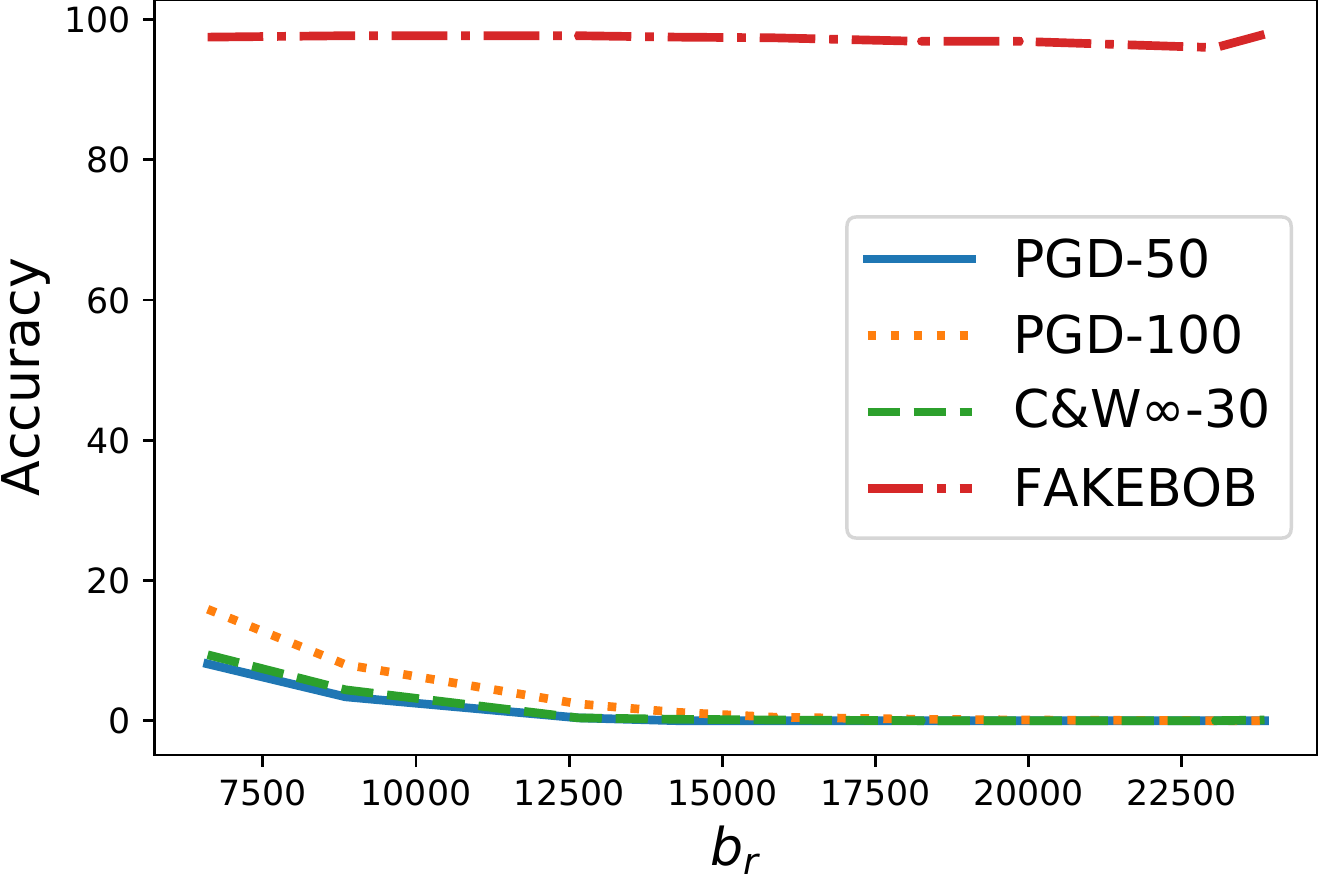}
    \end{minipage}
    \begin{minipage}[t]{0.23\textwidth}
    \includegraphics[width=1.0\textwidth]{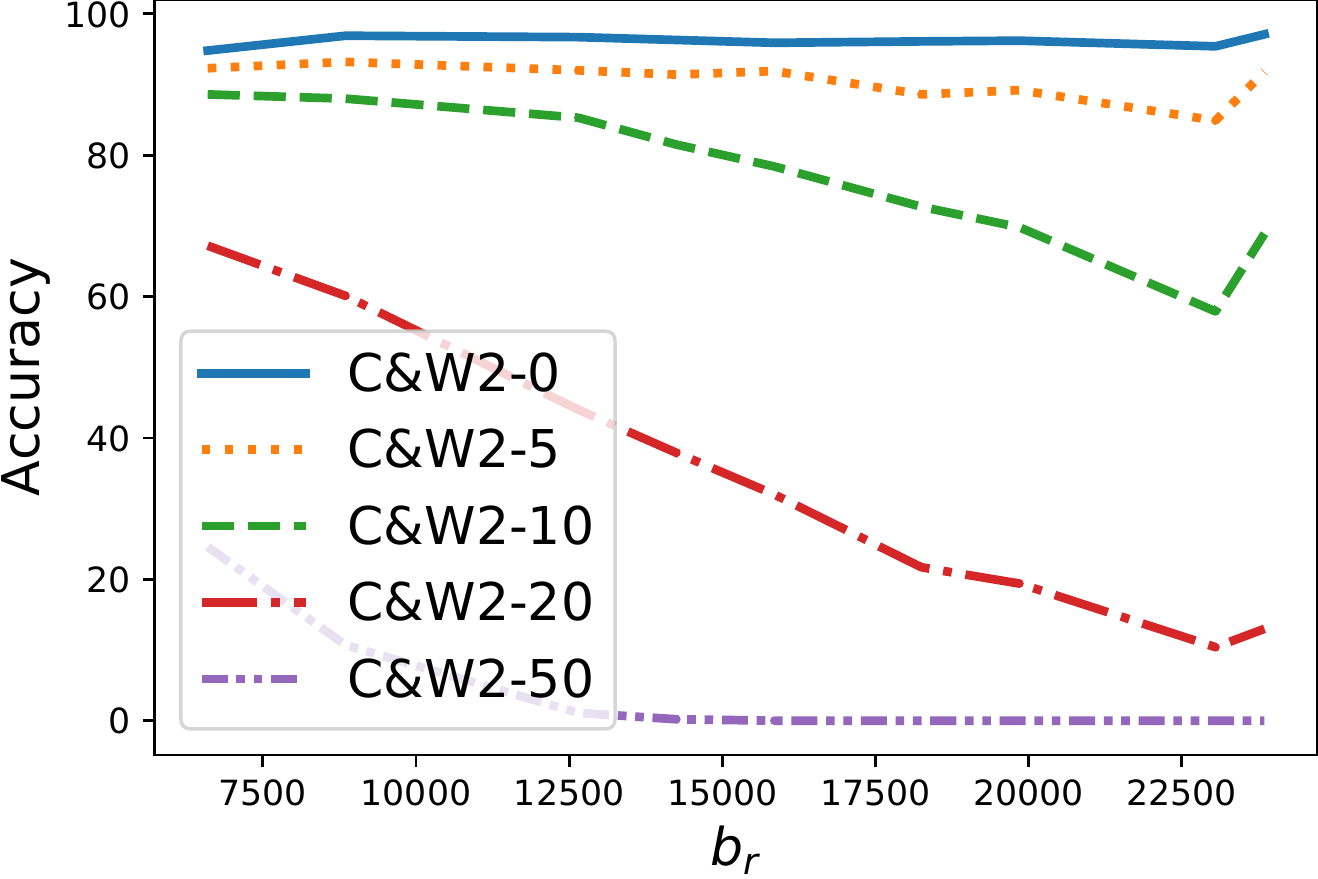}
    \end{minipage}
    }

     \subfigure[AAC-V]{
    \begin{minipage}[t]{0.23\textwidth}
    \includegraphics[width=1.0\textwidth]{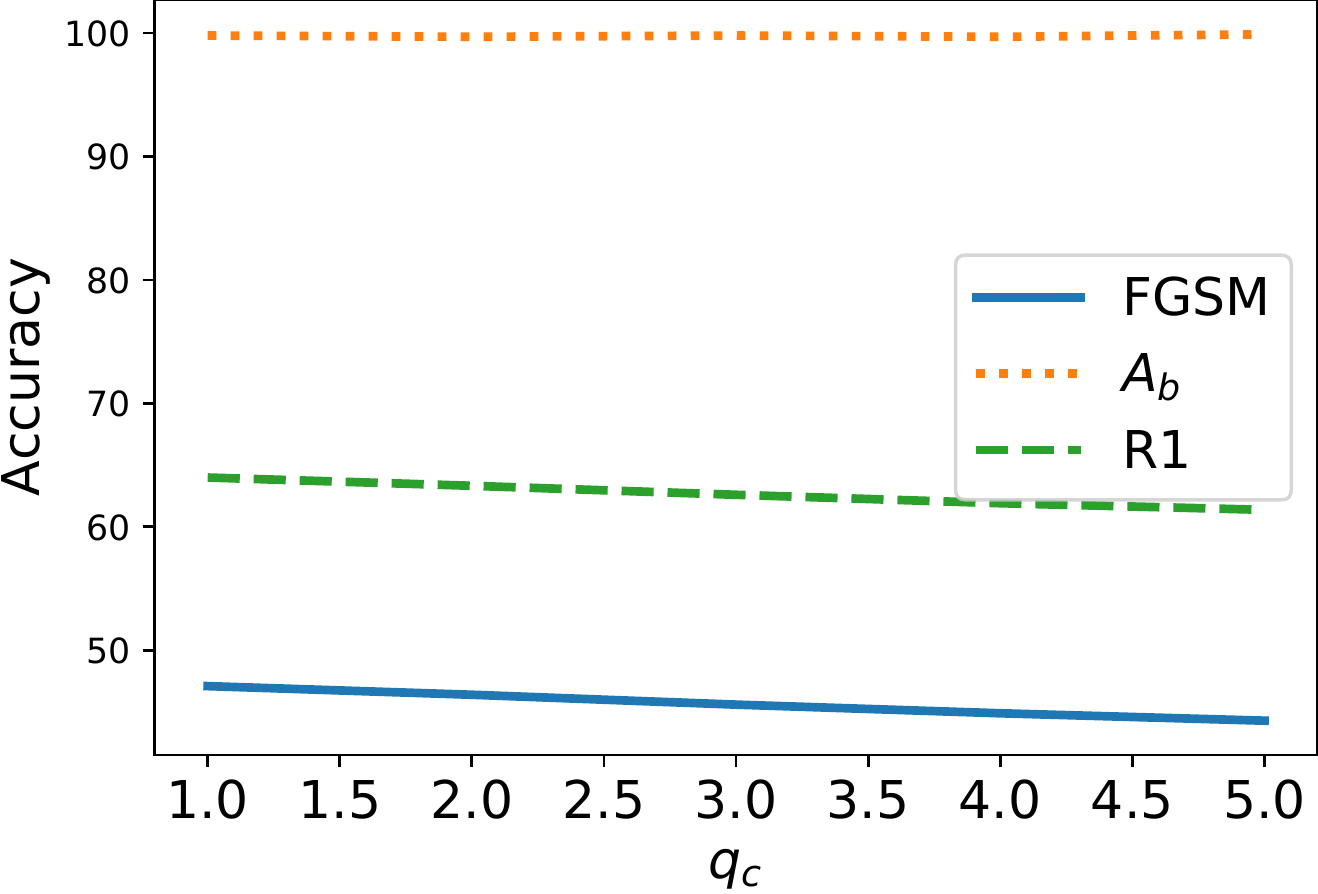}
    \end{minipage}
    \begin{minipage}[t]{0.23\textwidth}
    \includegraphics[width=1.0\textwidth]{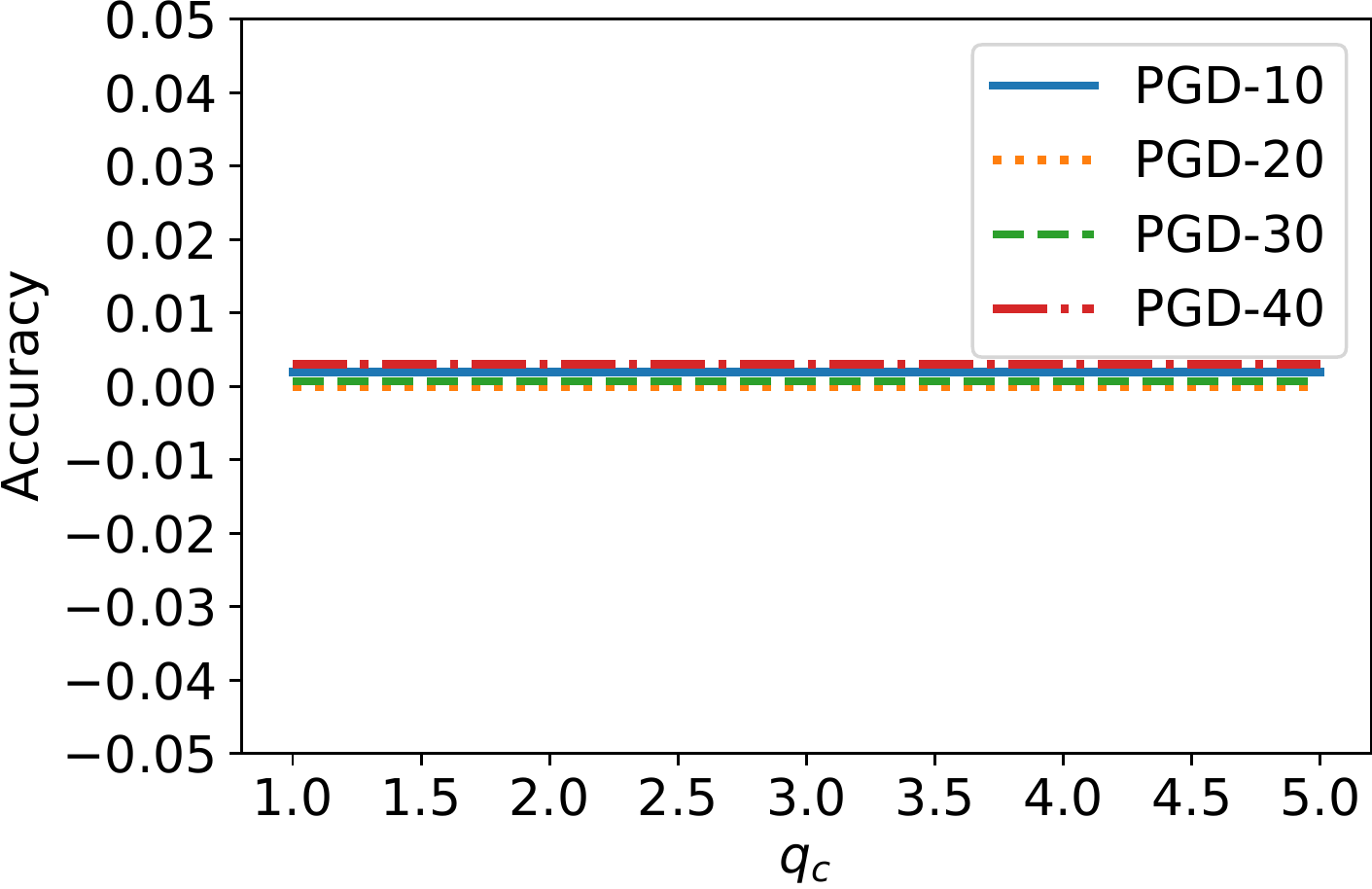}
    \end{minipage}
    \begin{minipage}[t]{0.23\textwidth}
    \includegraphics[width=1.0\textwidth]{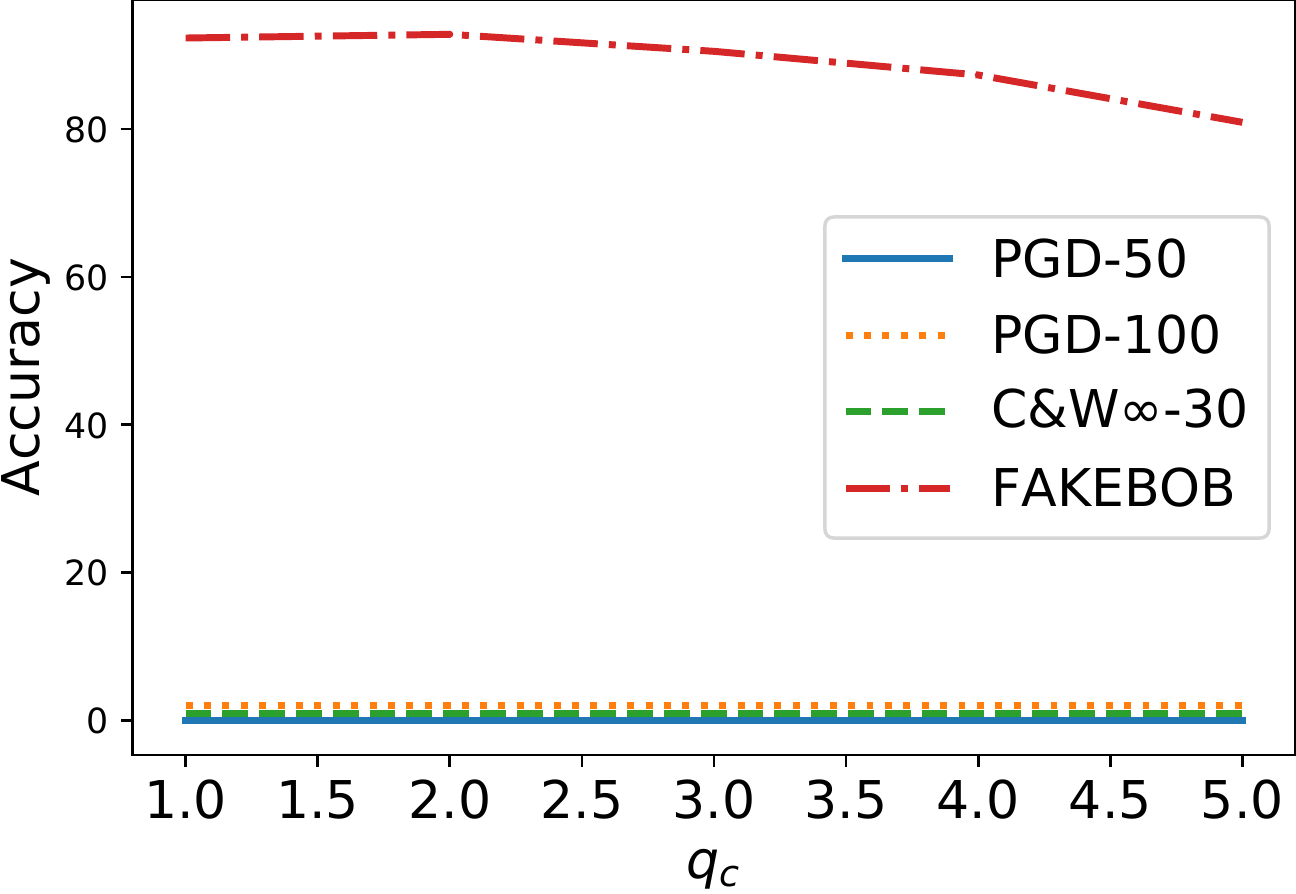}
    \end{minipage}
    \begin{minipage}[t]{0.23\textwidth}
    \includegraphics[width=1.0\textwidth]{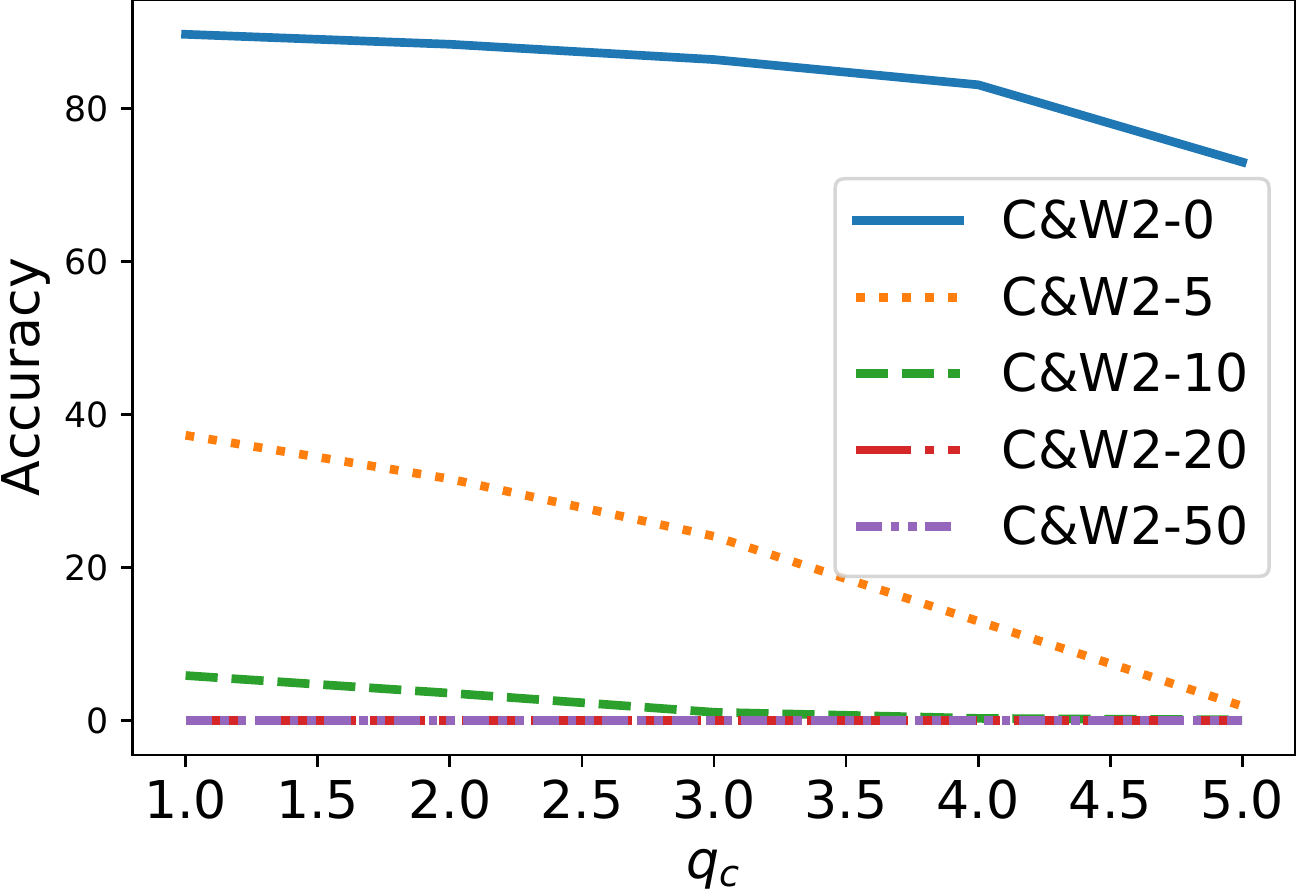}
    \end{minipage}
    }

    \subfigure[AAC-C]{
    \begin{minipage}[t]{0.23\textwidth}
    \includegraphics[width=1.0\textwidth]{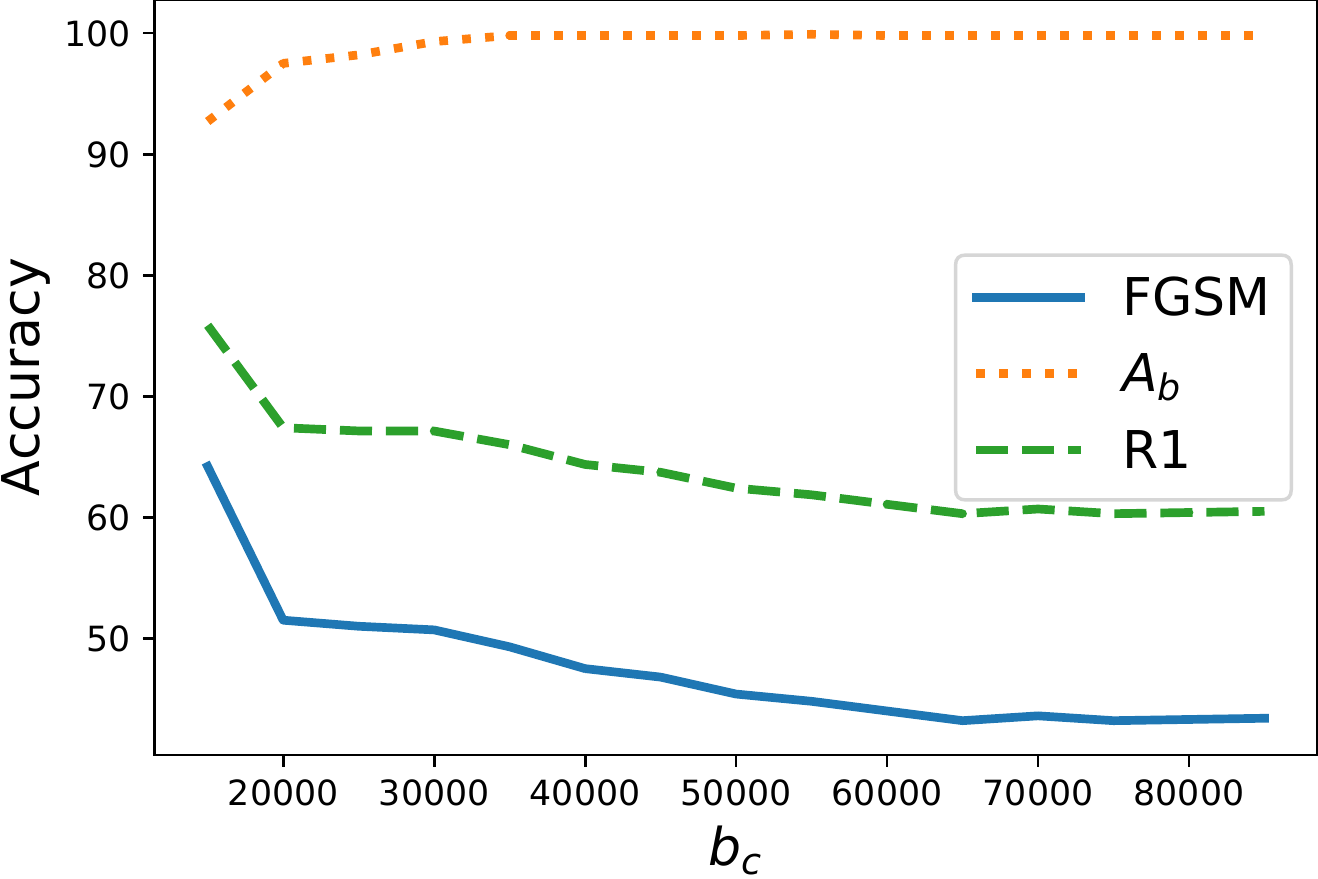}
    \end{minipage}
    \begin{minipage}[t]{0.23\textwidth}
    \includegraphics[width=1.0\textwidth]{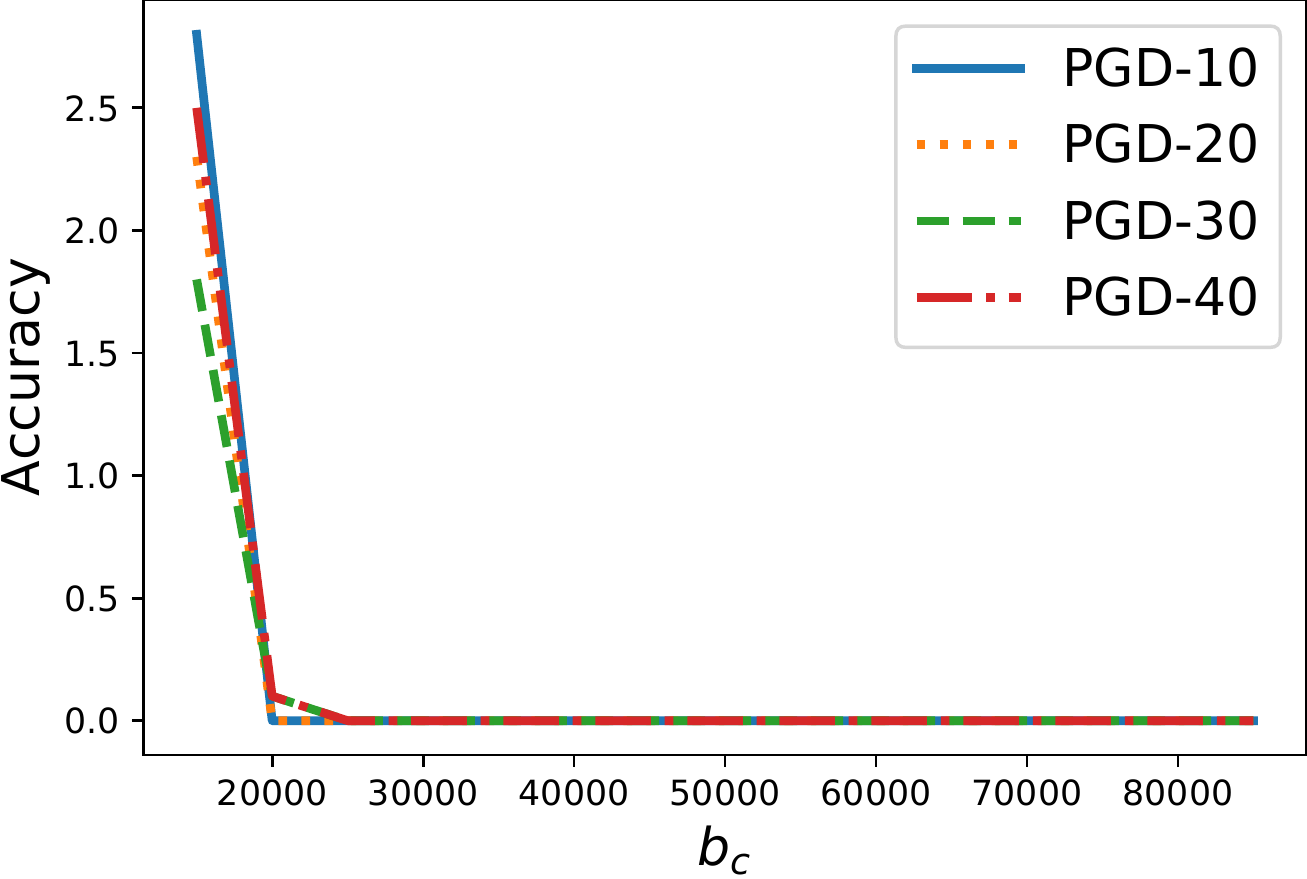}
    \end{minipage}
    \begin{minipage}[t]{0.23\textwidth}
    \includegraphics[width=1.0\textwidth]{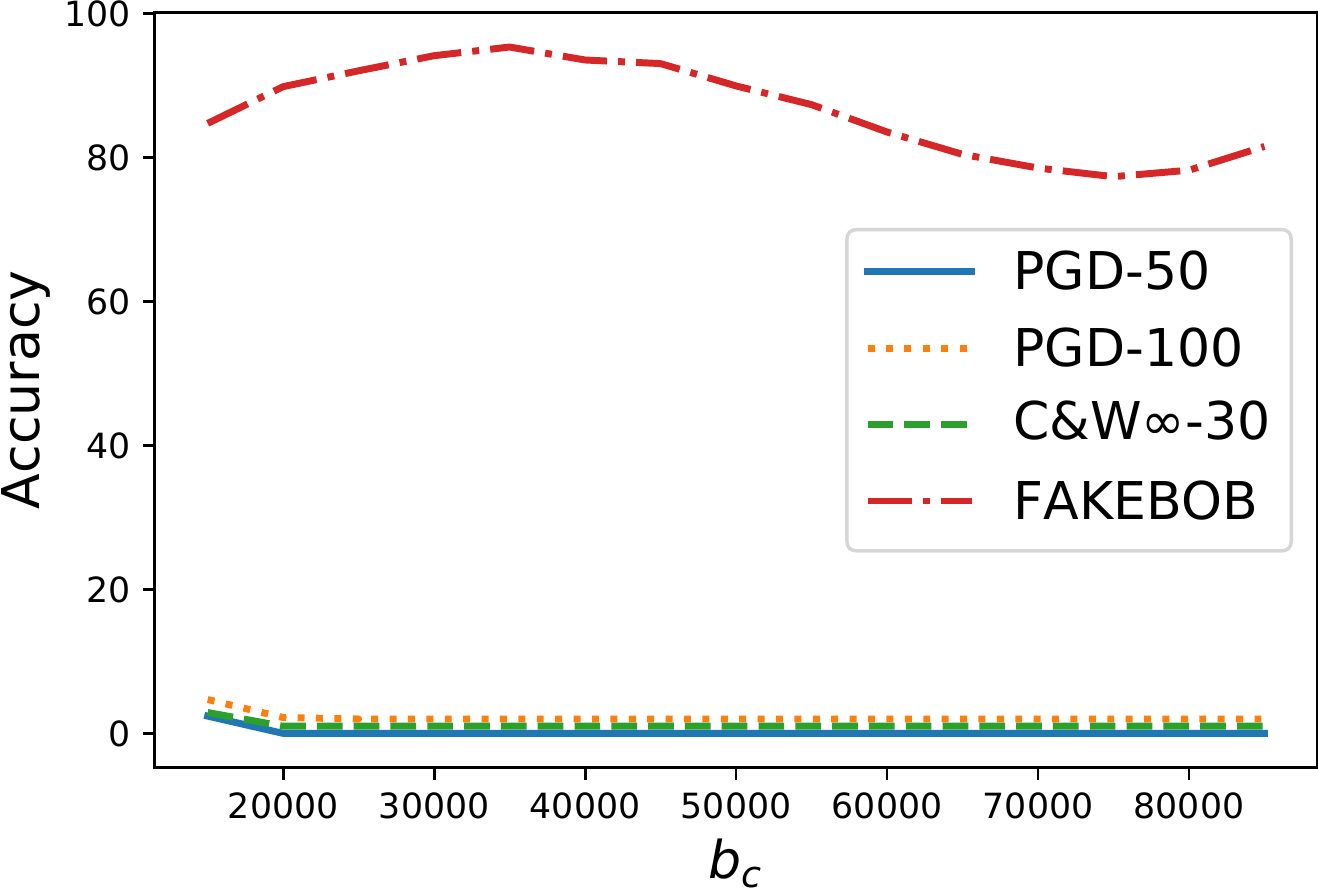}
    \end{minipage}
    \begin{minipage}[t]{0.23\textwidth}
    \includegraphics[width=1.0\textwidth]{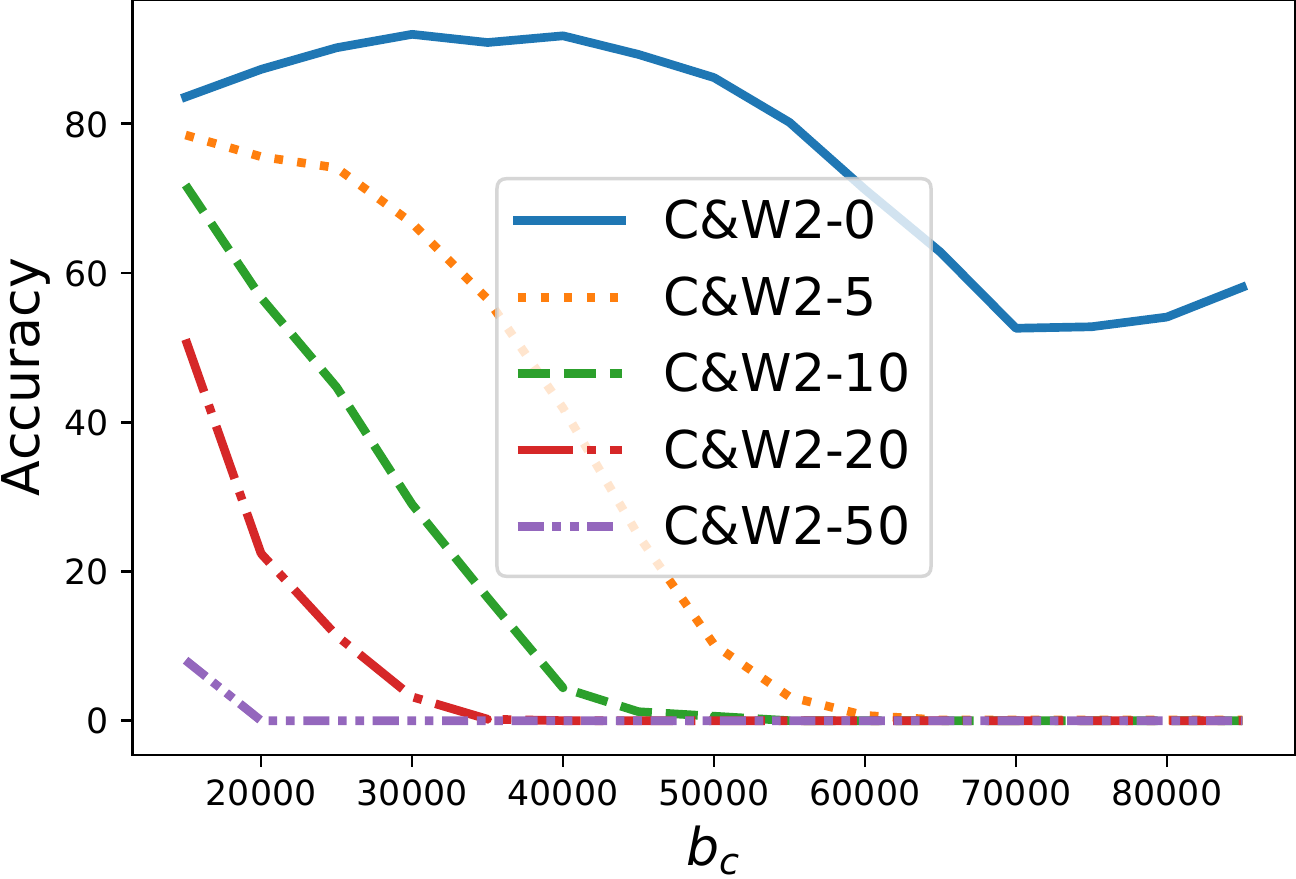}
    \end{minipage}
    }

     \caption{The performance of input transformations. For better visualization, we fix $f_{sl}=150$ Hz of BPF and shows how its performance varies with $f_{su}$.}
    \label{fig:parameter-2}
\end{figure*}

\begin{figure*}
    \centering

    \subfigure[MP3-V]{
    \begin{minipage}[t]{0.23\textwidth}
    \includegraphics[width=1.0\textwidth]{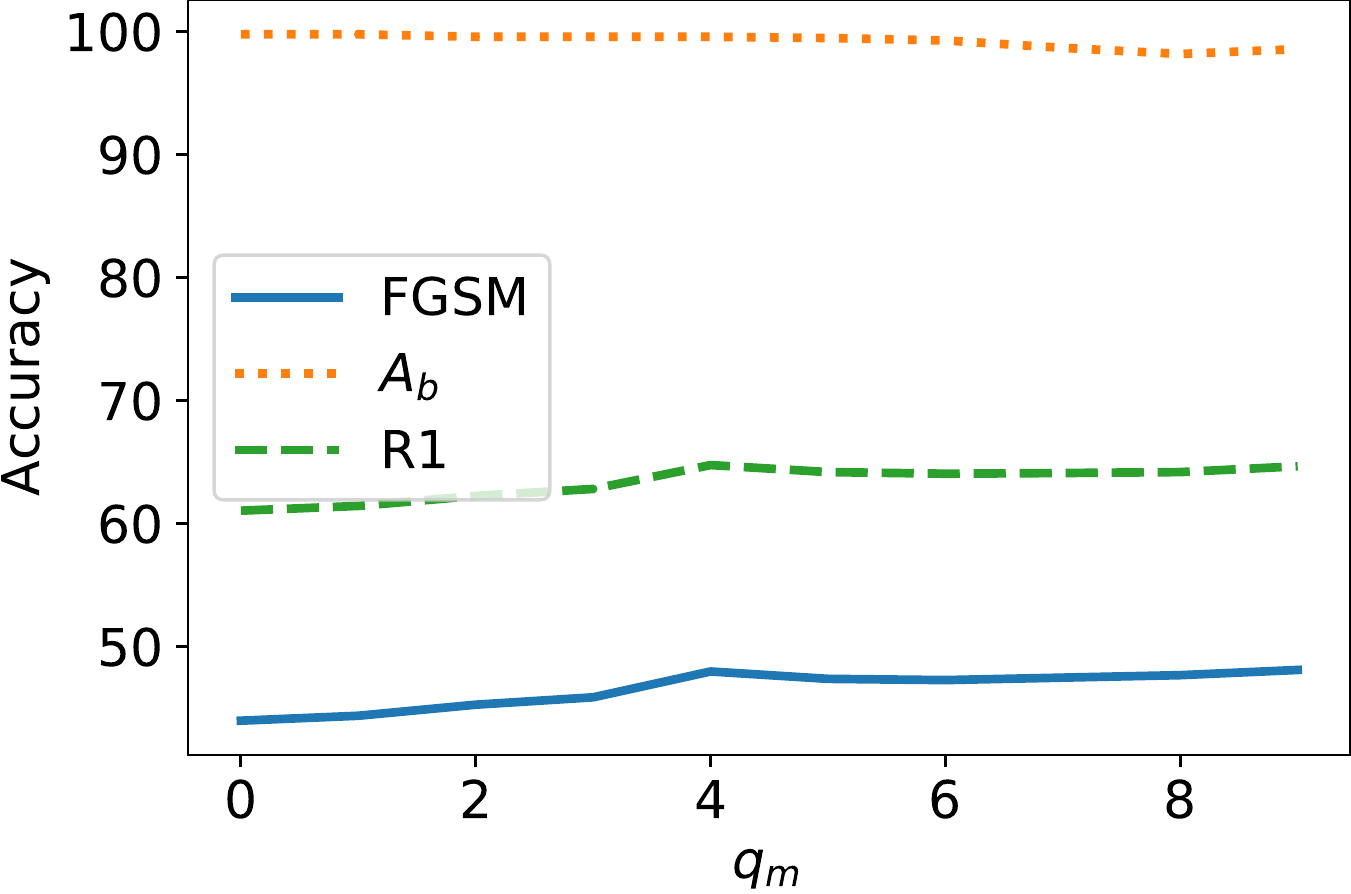}
    \end{minipage}
    \begin{minipage}[t]{0.23\textwidth}
    \includegraphics[width=1.0\textwidth]{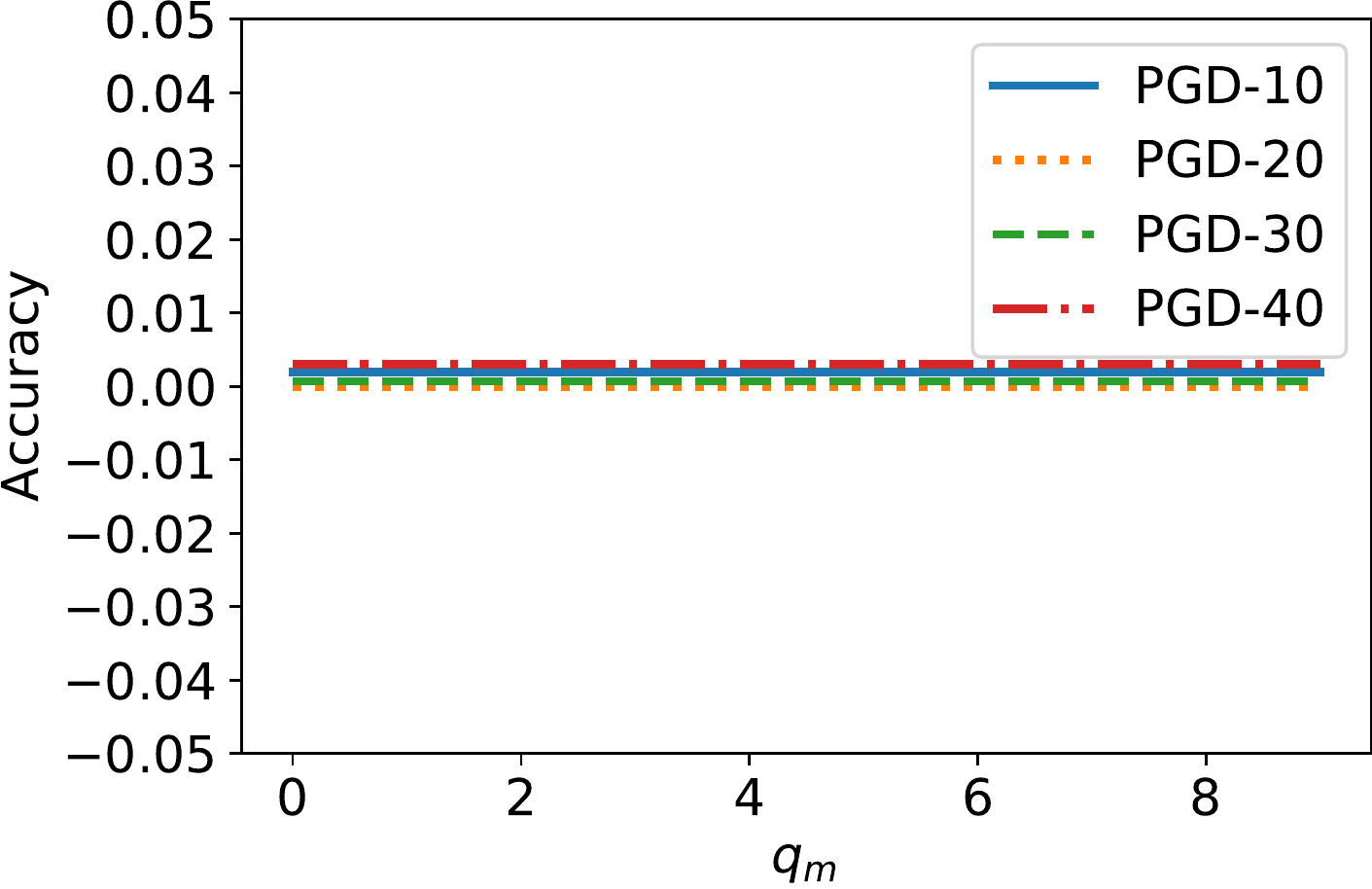}
    \end{minipage}
    \begin{minipage}[t]{0.23\textwidth}
    \includegraphics[width=1.0\textwidth]{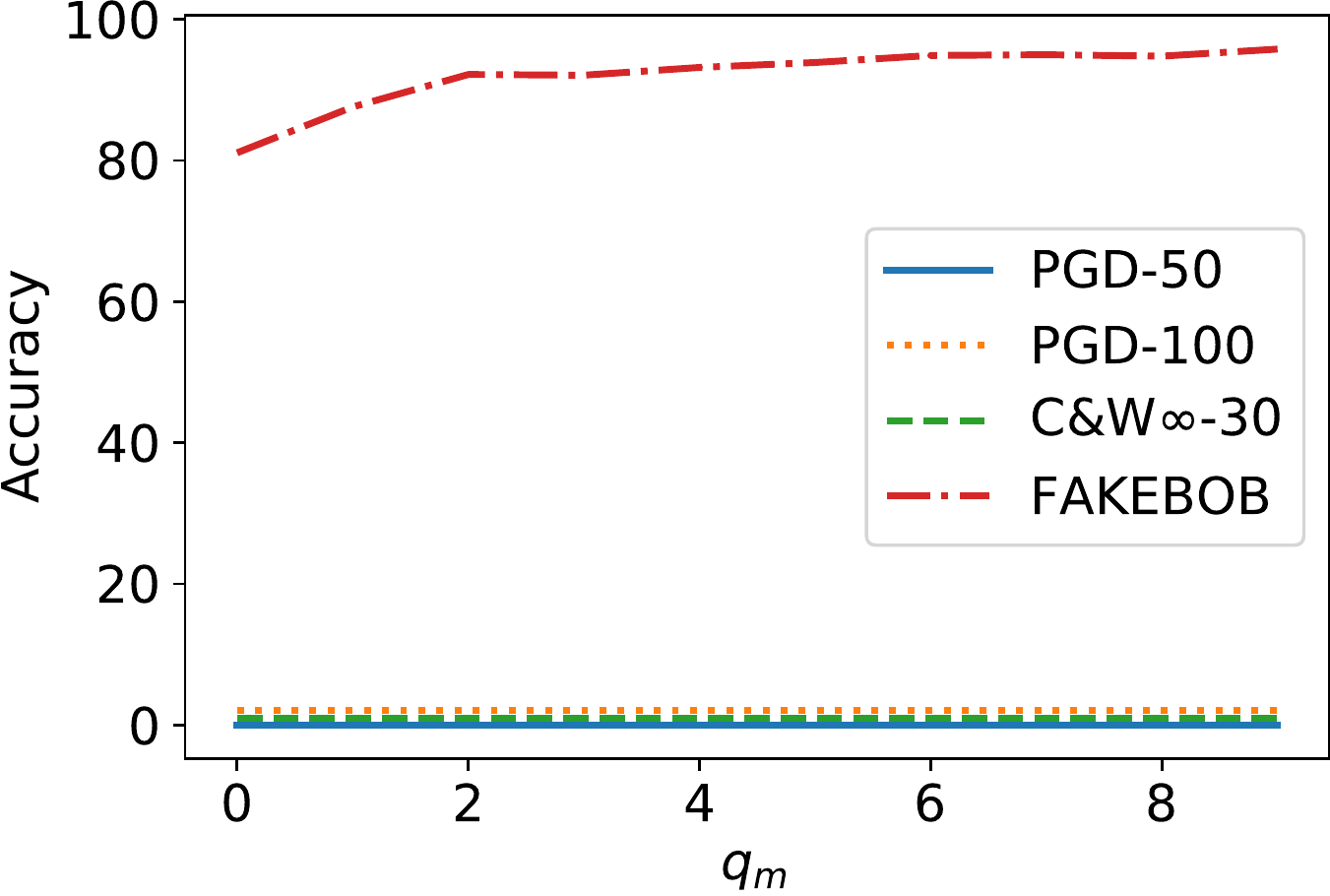}
    \end{minipage}
    \begin{minipage}[t]{0.23\textwidth}
    \includegraphics[width=1.0\textwidth]{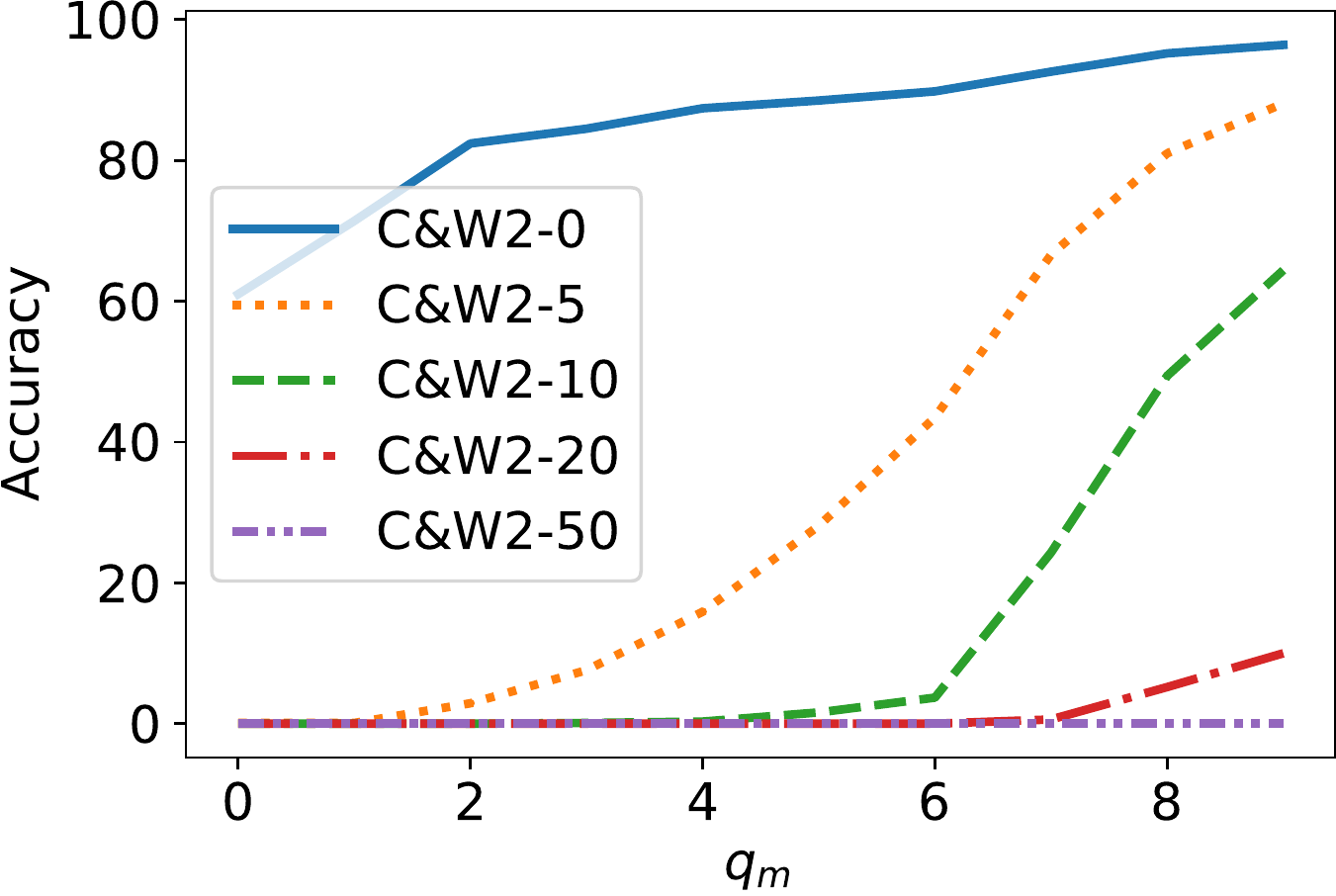}
    \end{minipage}
    }


    \subfigure[MP3-C]{
    \begin{minipage}[t]{0.23\textwidth}
    \includegraphics[width=1.0\textwidth]{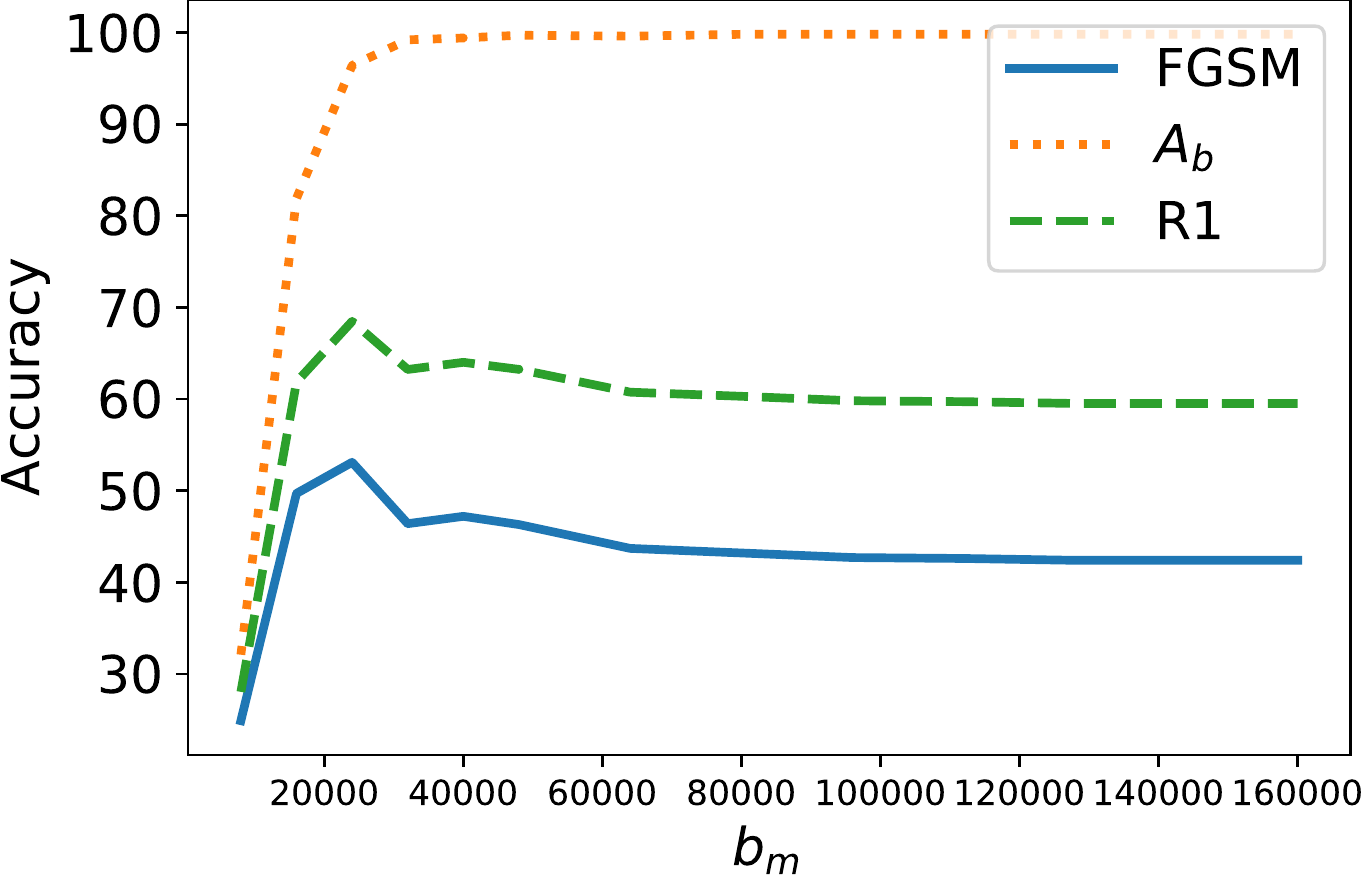}
    \end{minipage}
    \begin{minipage}[t]{0.23\textwidth}
    \includegraphics[width=1.0\textwidth]{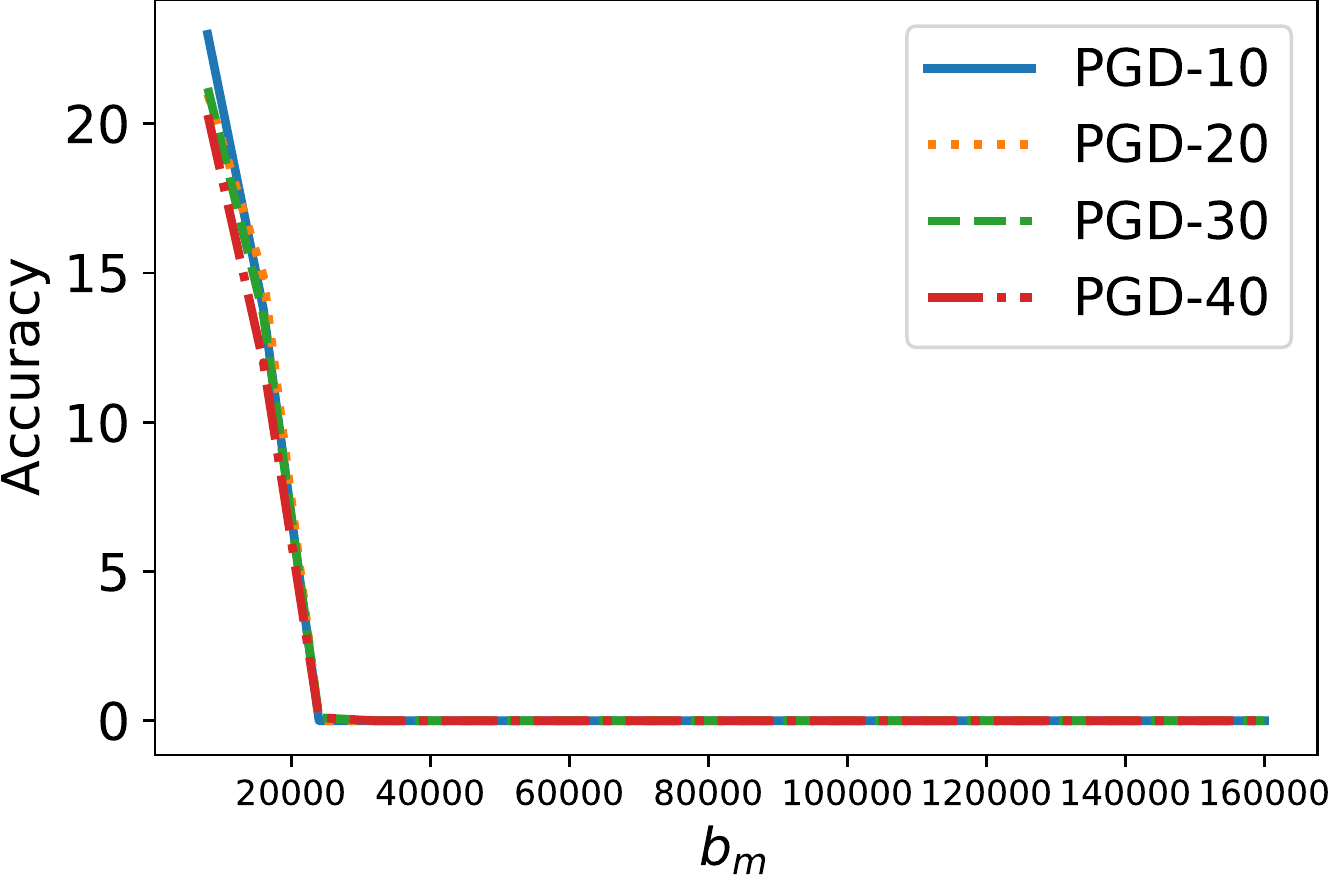}
    \end{minipage}
    \begin{minipage}[t]{0.23\textwidth}
    \includegraphics[width=1.0\textwidth]{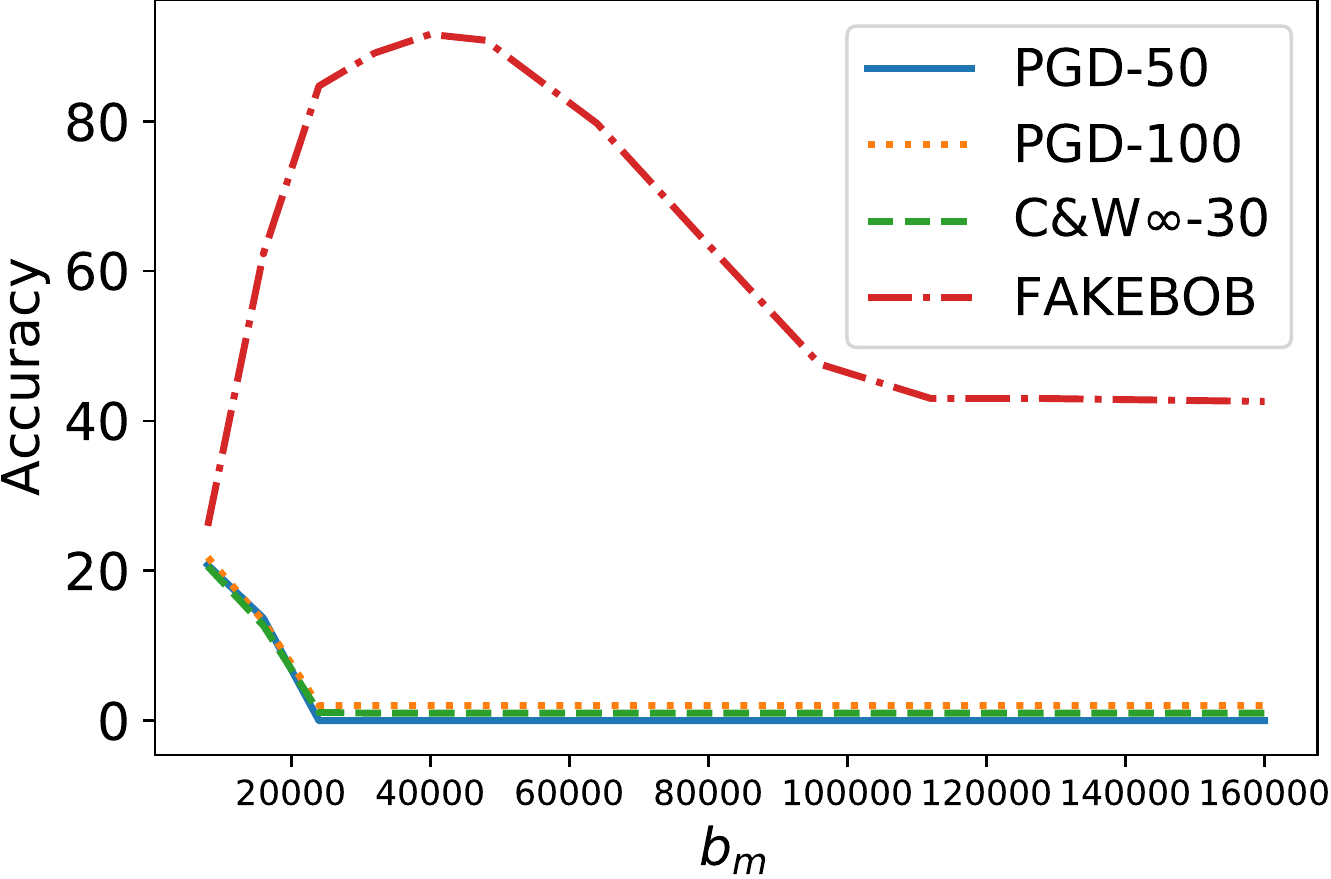}
    \end{minipage}
    \begin{minipage}[t]{0.23\textwidth}
    \includegraphics[width=1.0\textwidth]{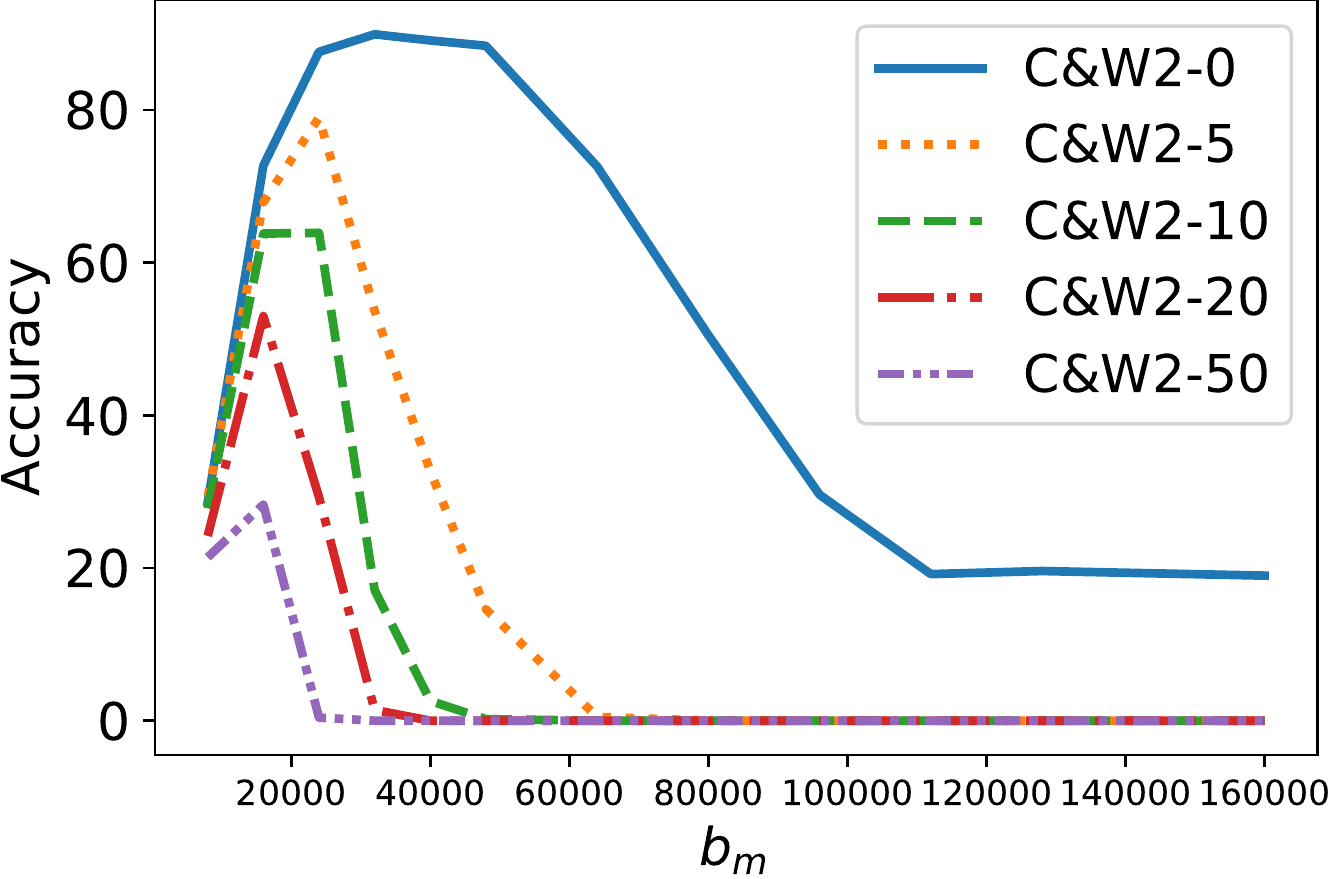}
    \end{minipage}
    }

    \subfigure[\defensenameabbr-o]{
    \begin{minipage}[t]{0.23\textwidth}
    \includegraphics[width=1.0\textwidth]{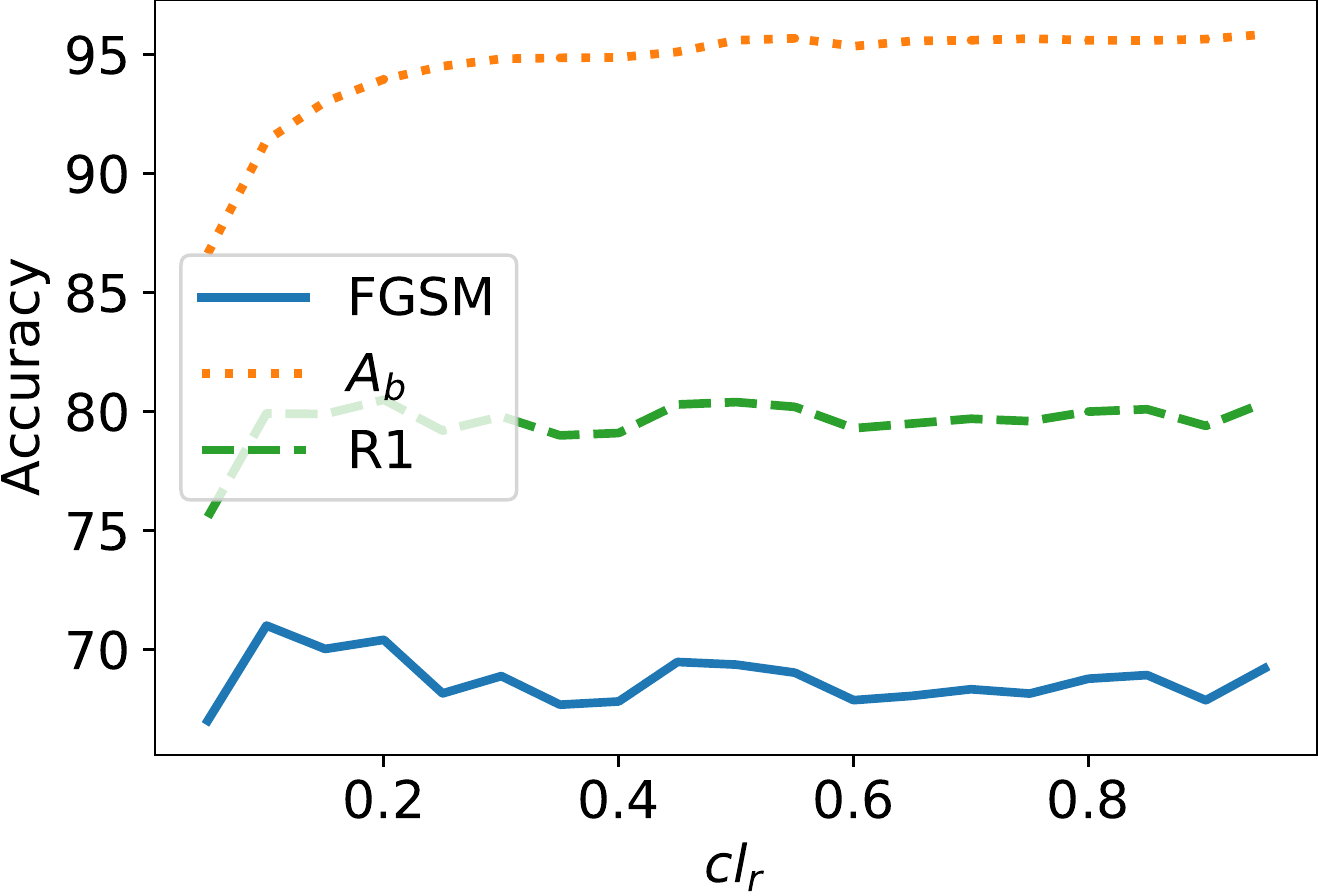}
    \end{minipage}
    \begin{minipage}[t]{0.23\textwidth}
    \includegraphics[width=1.0\textwidth]{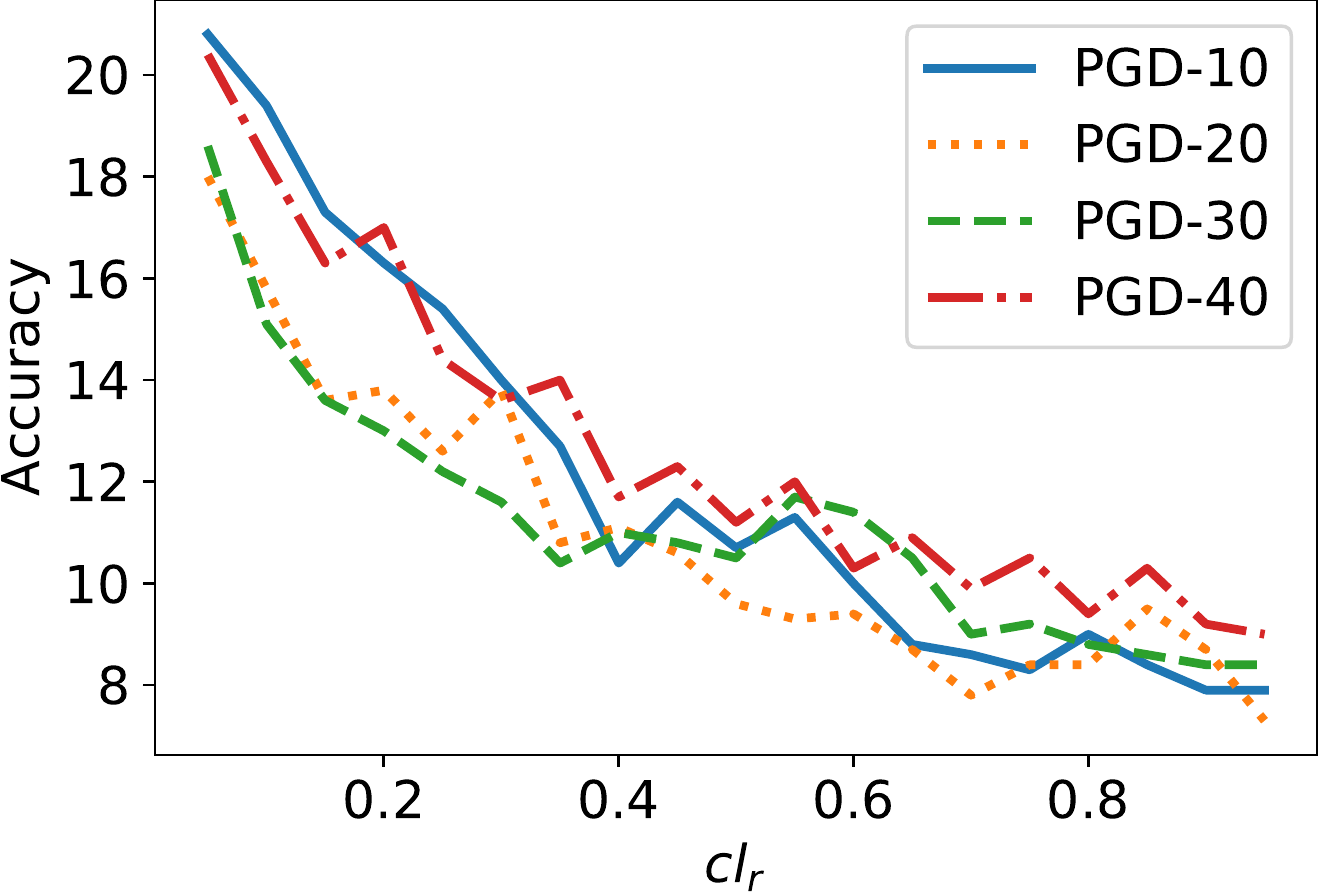}
    \end{minipage}
    \begin{minipage}[t]{0.23\textwidth}
    \includegraphics[width=1.0\textwidth]{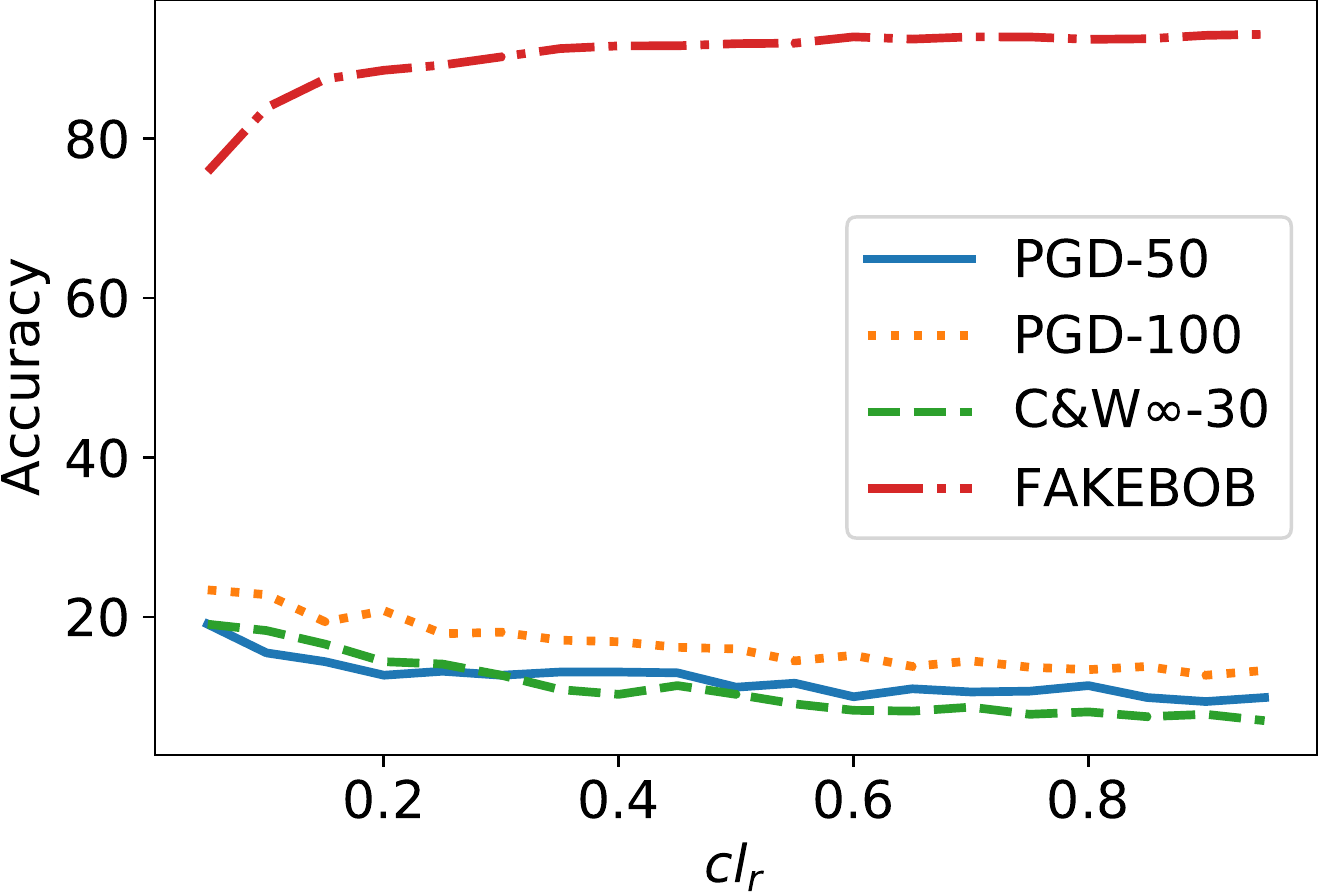}
    \end{minipage}
    \begin{minipage}[t]{0.23\textwidth}
    \includegraphics[width=1.0\textwidth]{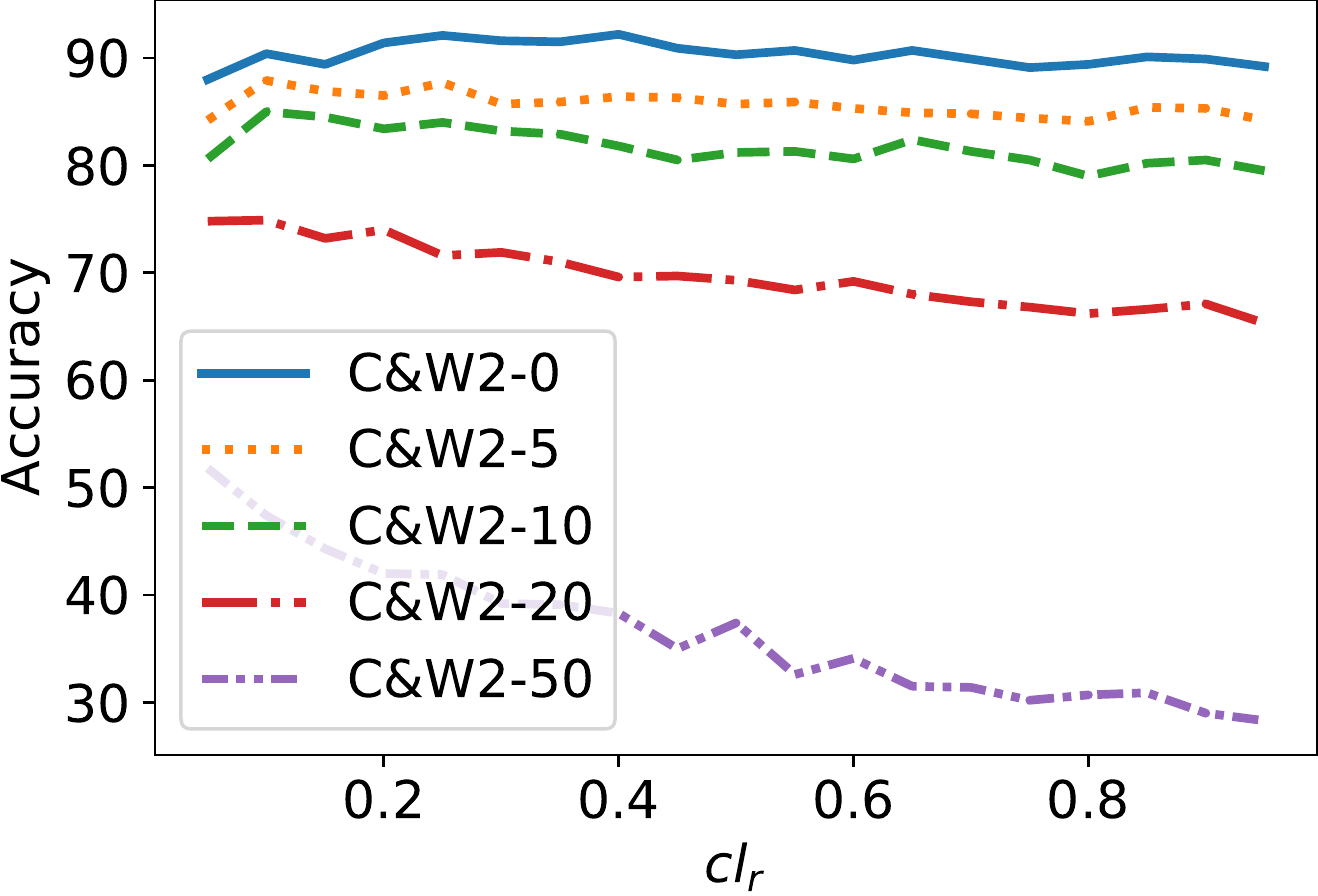}
    \end{minipage}
    \label{fig:p-ff-o}
    }

     \subfigure[\defensenameabbr-d]{
    \begin{minipage}[t]{0.23\textwidth}
    \includegraphics[width=1.0\textwidth]{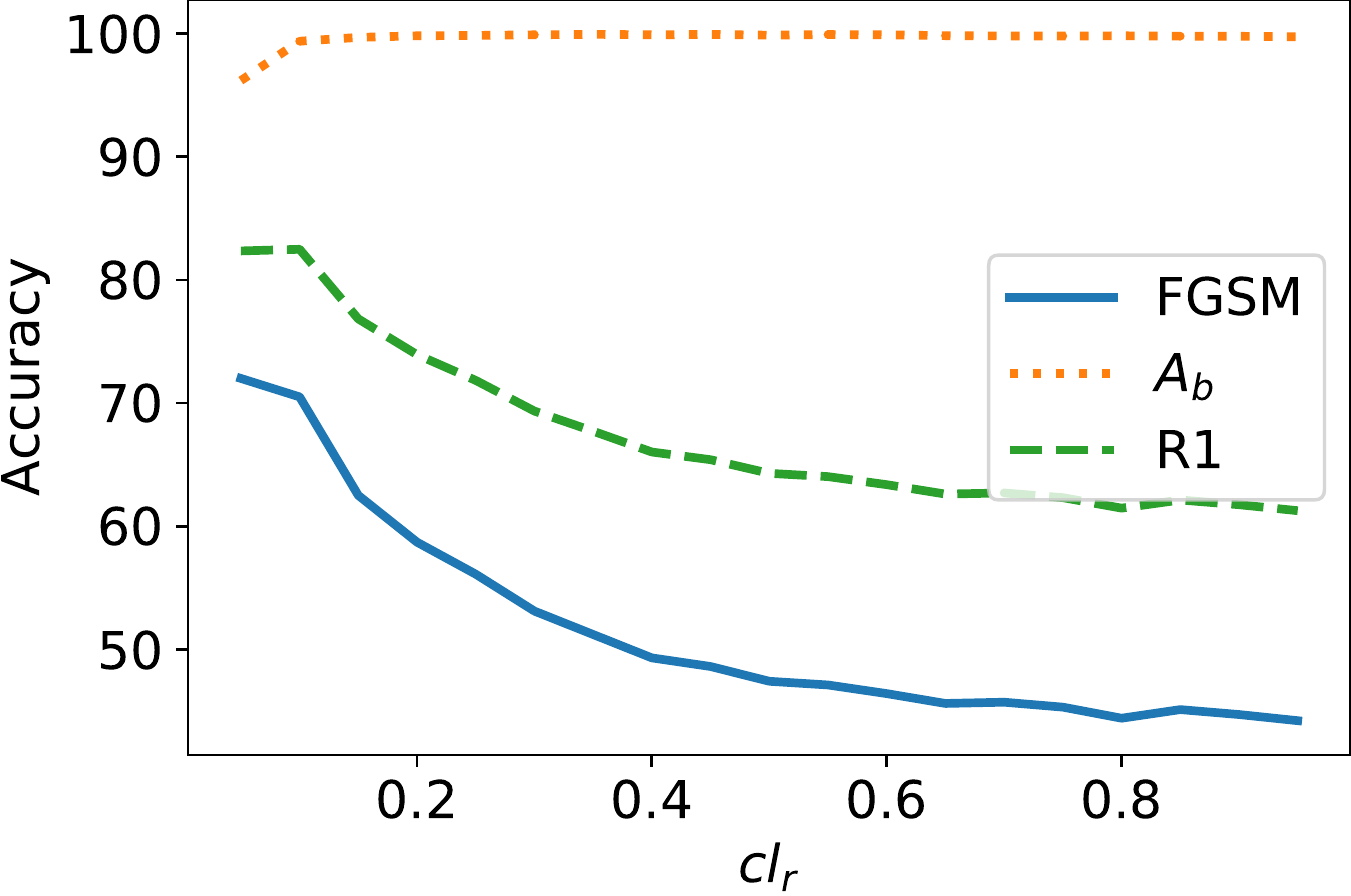}
    \end{minipage}
    \begin{minipage}[t]{0.23\textwidth}
    \includegraphics[width=1.0\textwidth]{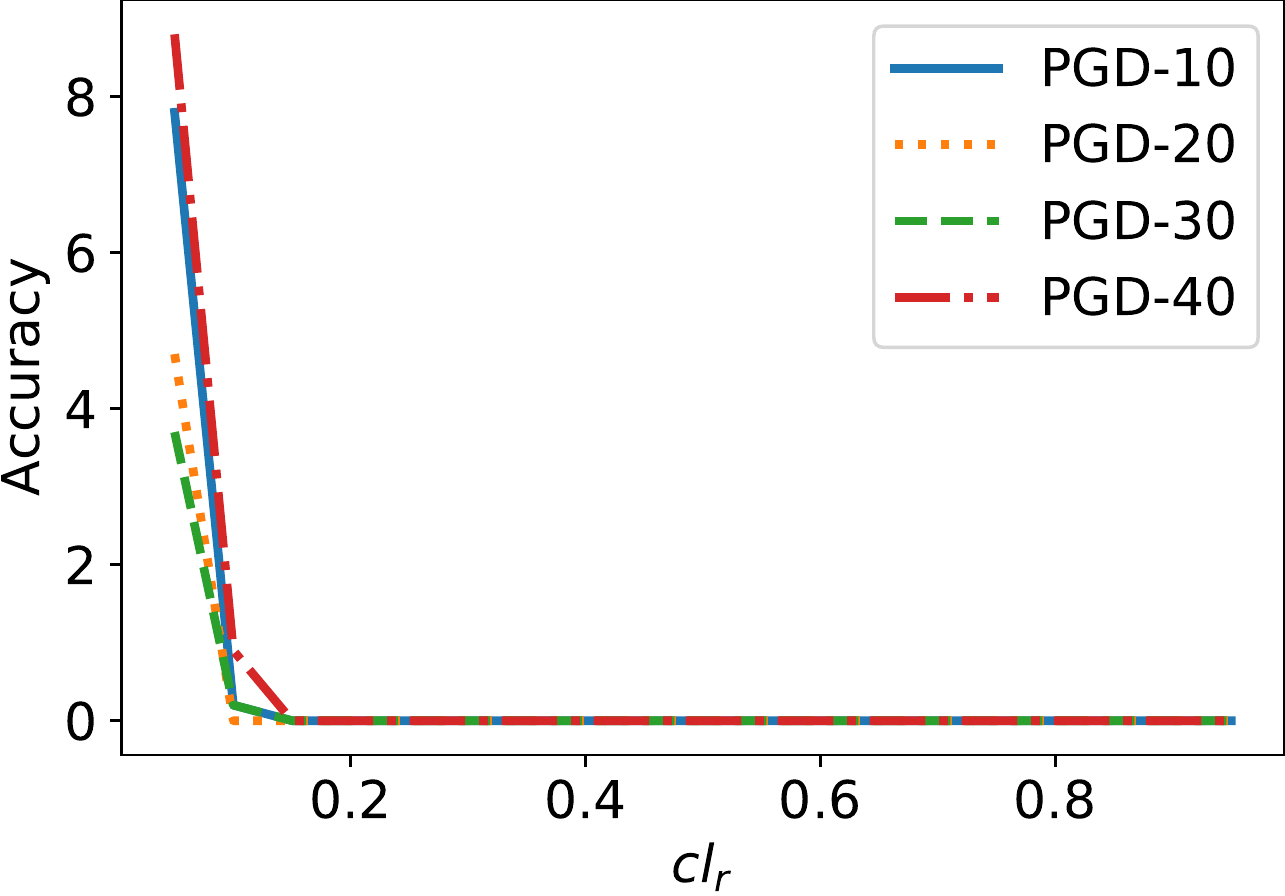}
    \end{minipage}
    \begin{minipage}[t]{0.23\textwidth}
    \includegraphics[width=1.0\textwidth]{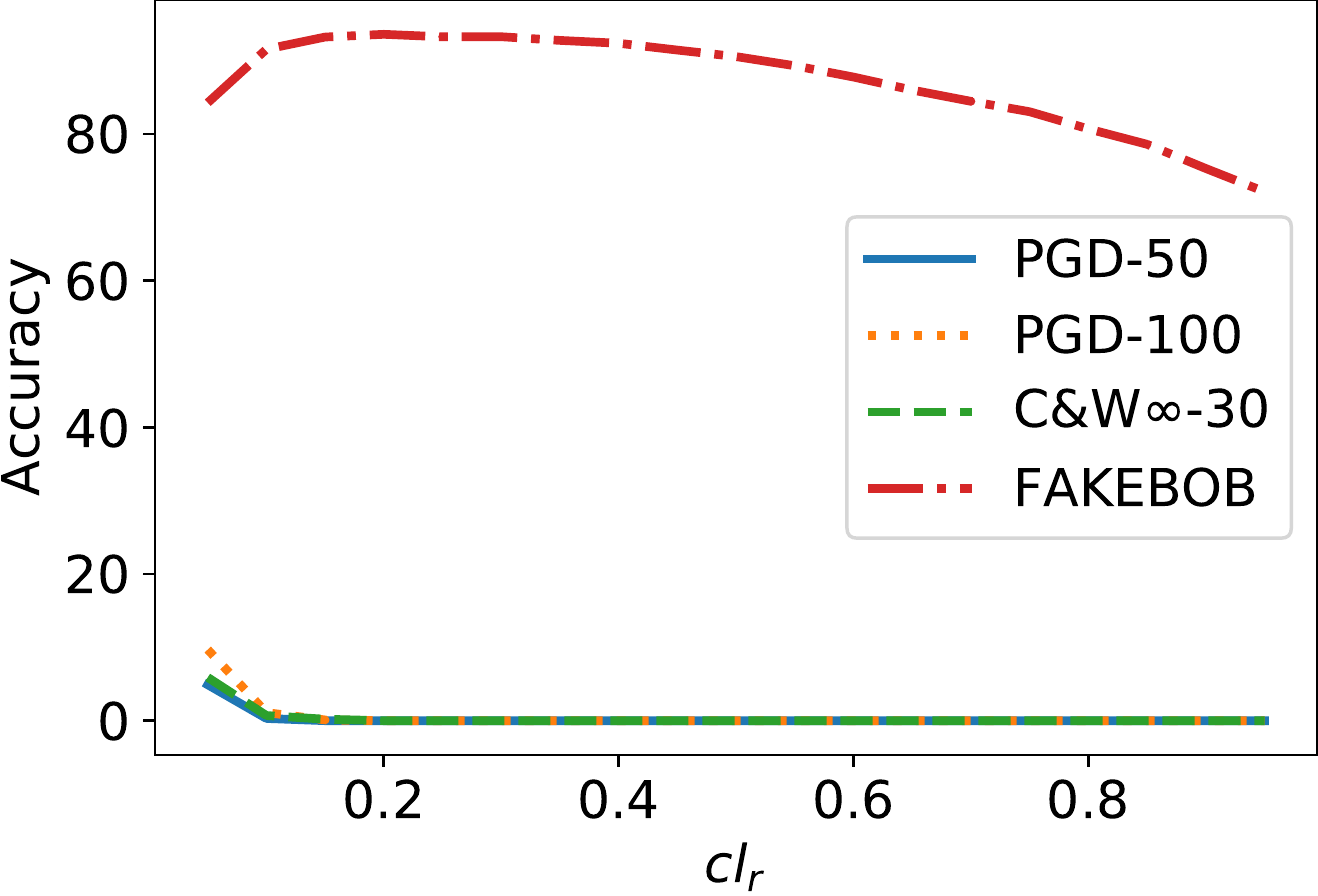}
    \end{minipage}
    \begin{minipage}[t]{0.23\textwidth}
    \includegraphics[width=1.0\textwidth]{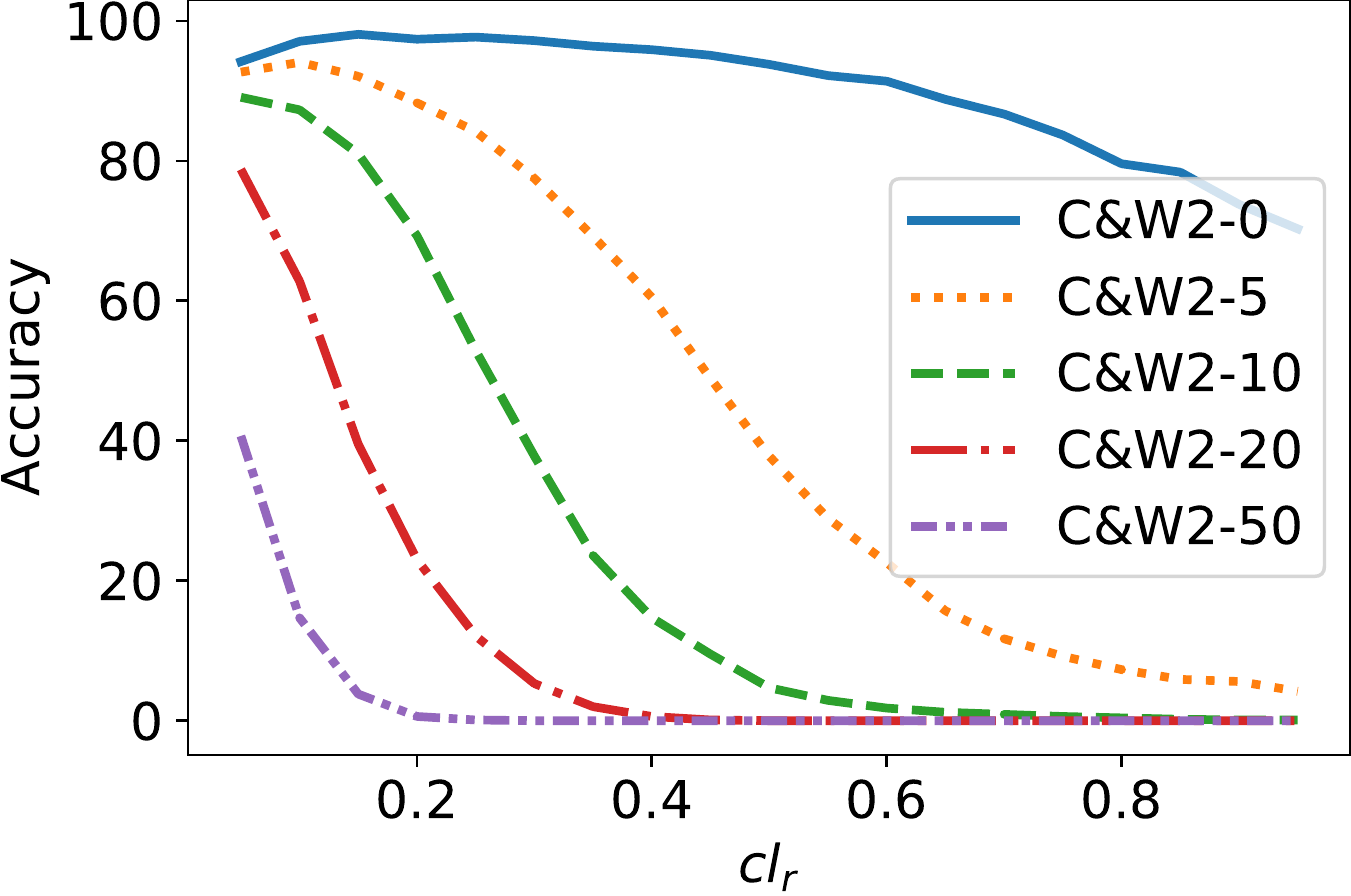}
    \end{minipage}
    \label{fig:p-ff-d}
    }

     \subfigure[\defensenameabbr-c]{
    \begin{minipage}[t]{0.23\textwidth}
    \includegraphics[width=1.0\textwidth]{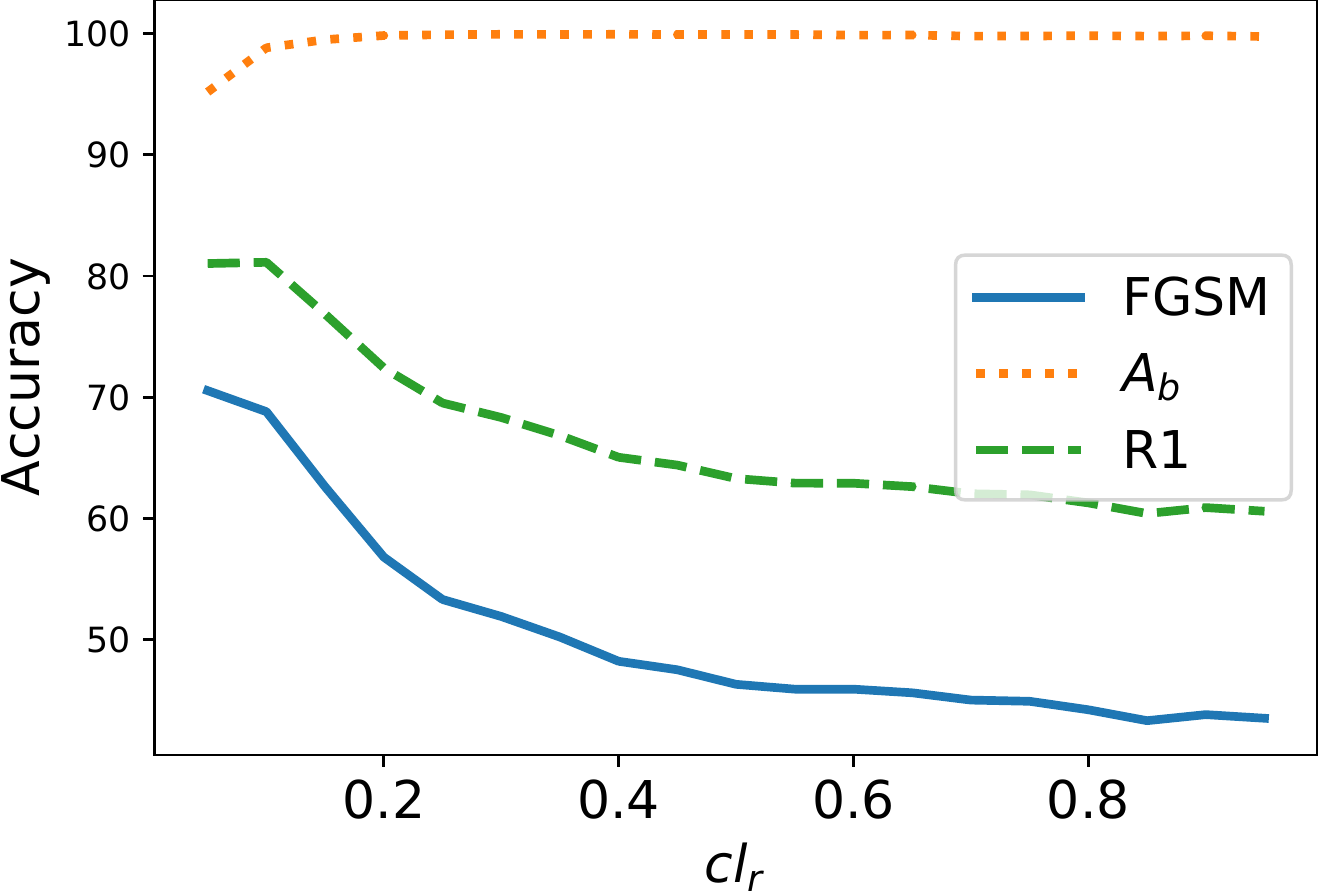}
    \end{minipage}
    \begin{minipage}[t]{0.23\textwidth}
    \includegraphics[width=1.0\textwidth]{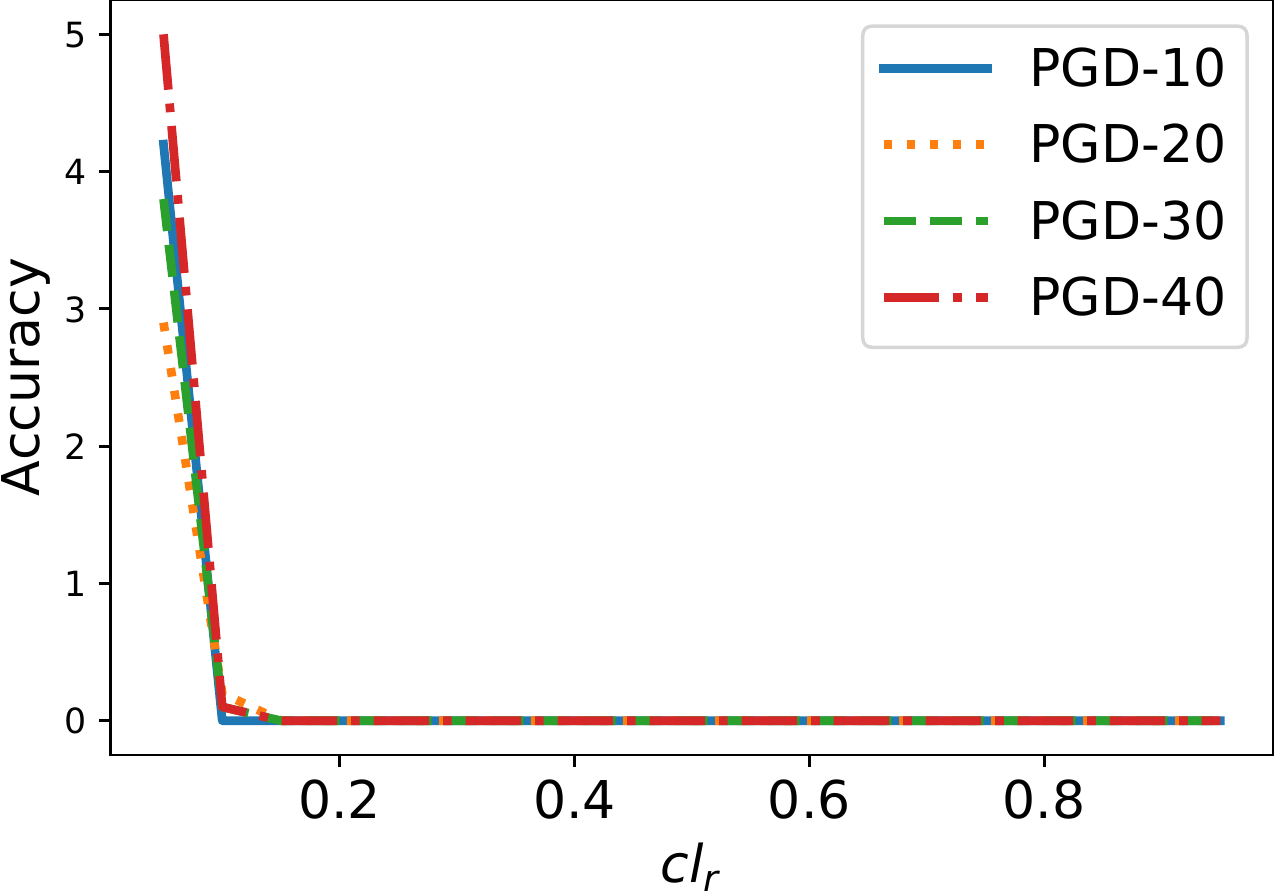}
    \end{minipage}
    \begin{minipage}[t]{0.23\textwidth}
    \includegraphics[width=1.0\textwidth]{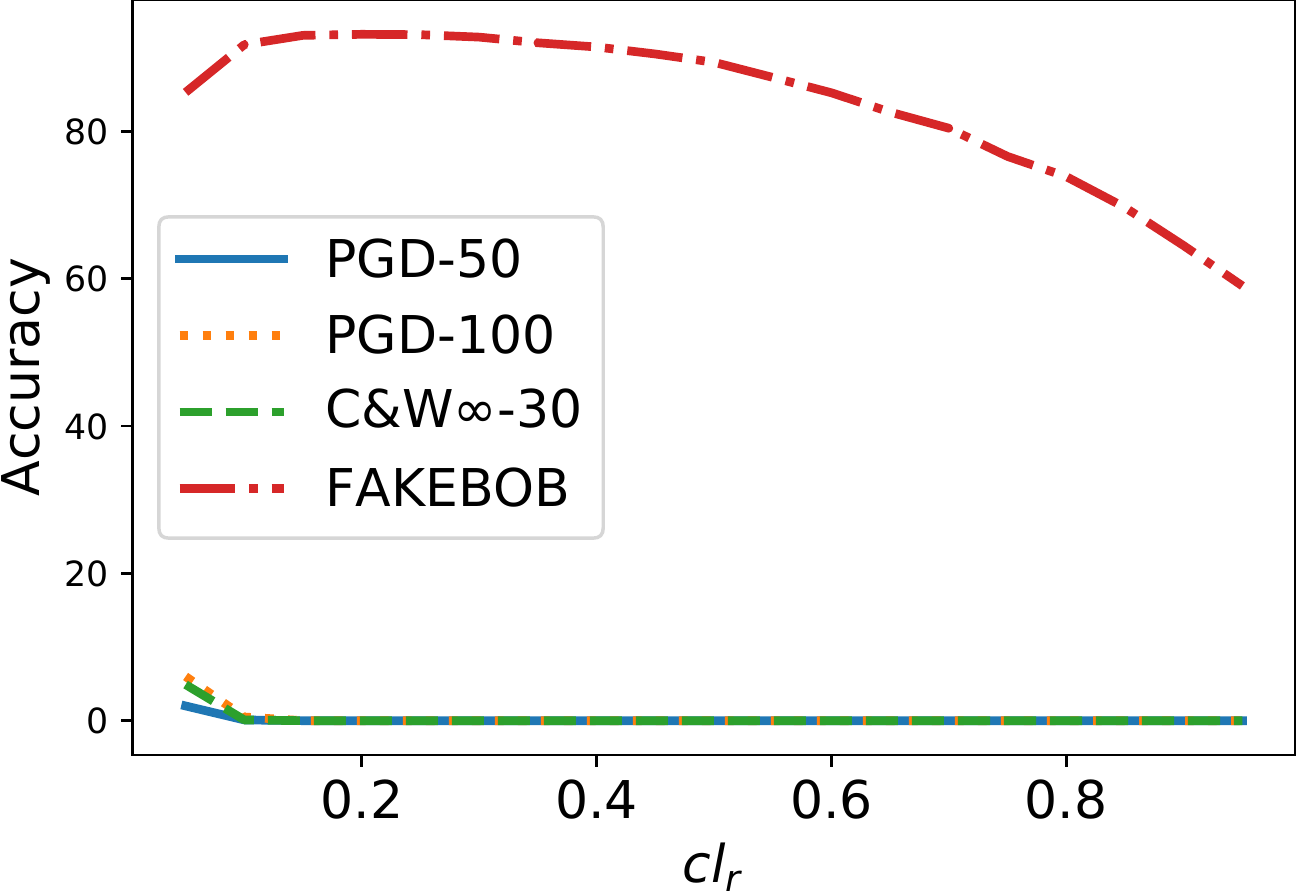}
    \end{minipage}
    \begin{minipage}[t]{0.23\textwidth}
    \includegraphics[width=1.0\textwidth]{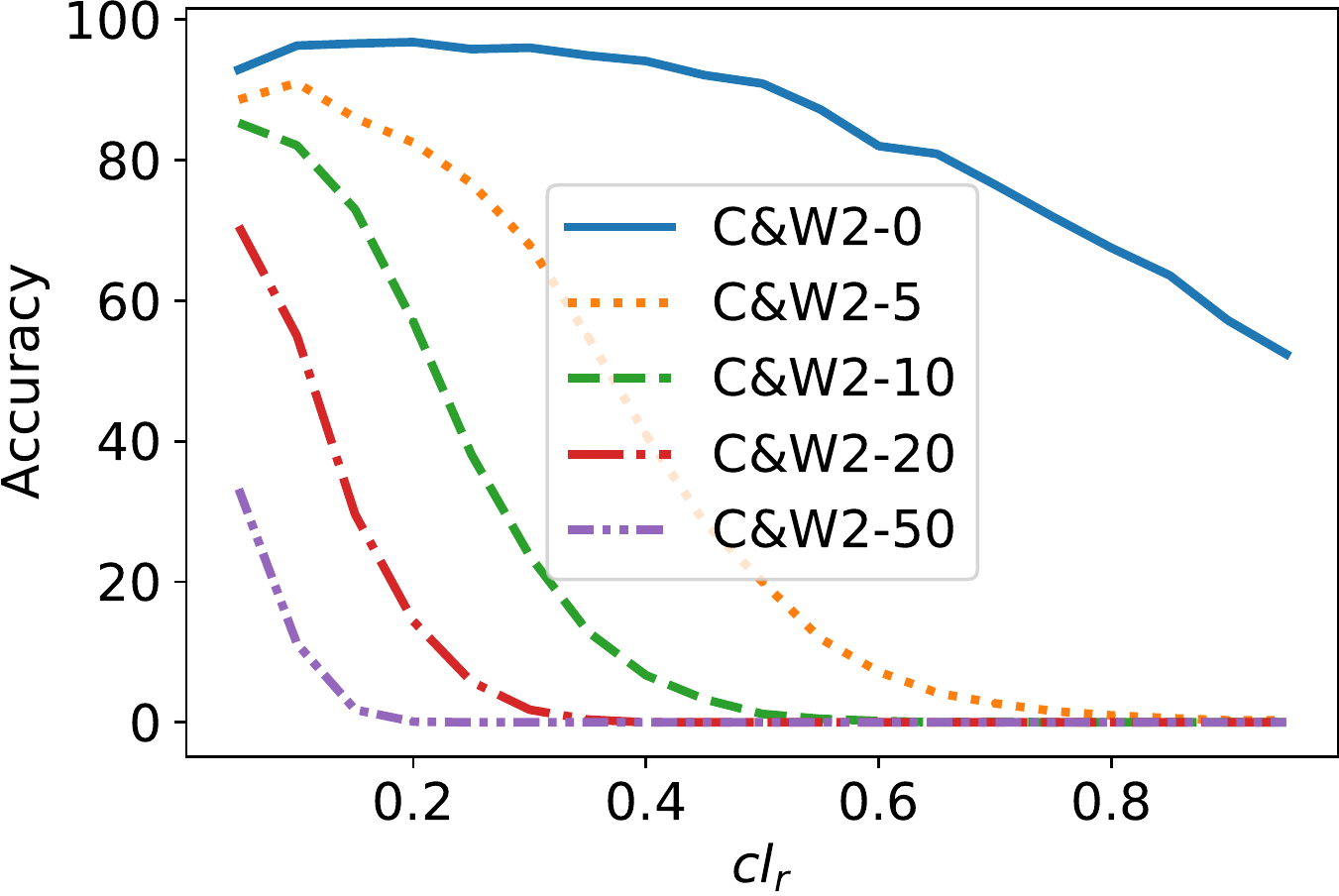}
    \end{minipage}
    \label{fig:p-ff-c}
    }

    \subfigure[\defensenameabbr-f]{
    \begin{minipage}[t]{0.23\textwidth}
    \includegraphics[width=1.0\textwidth]{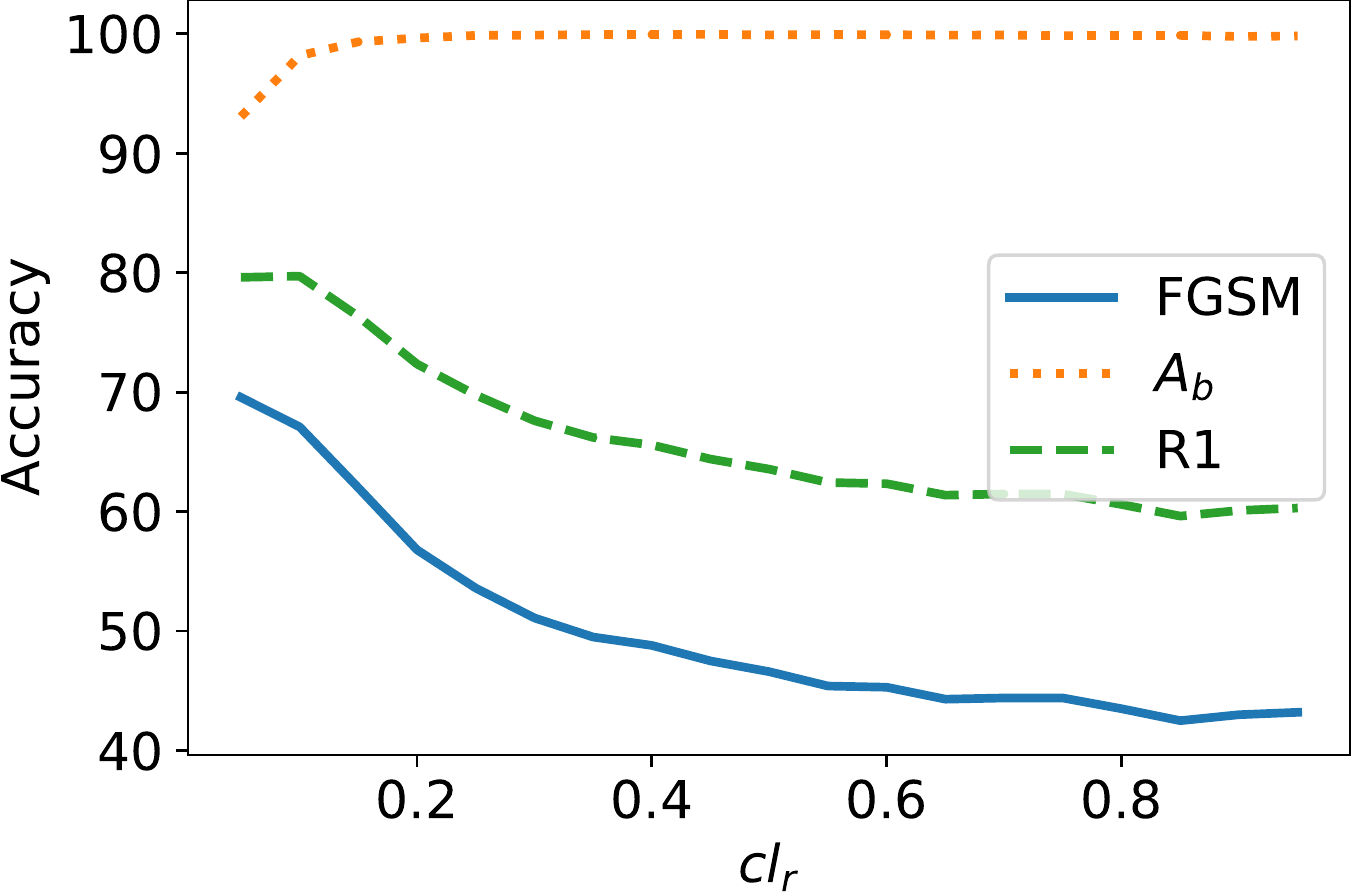}
    \end{minipage}
    \begin{minipage}[t]{0.23\textwidth}
    \includegraphics[width=1.0\textwidth]{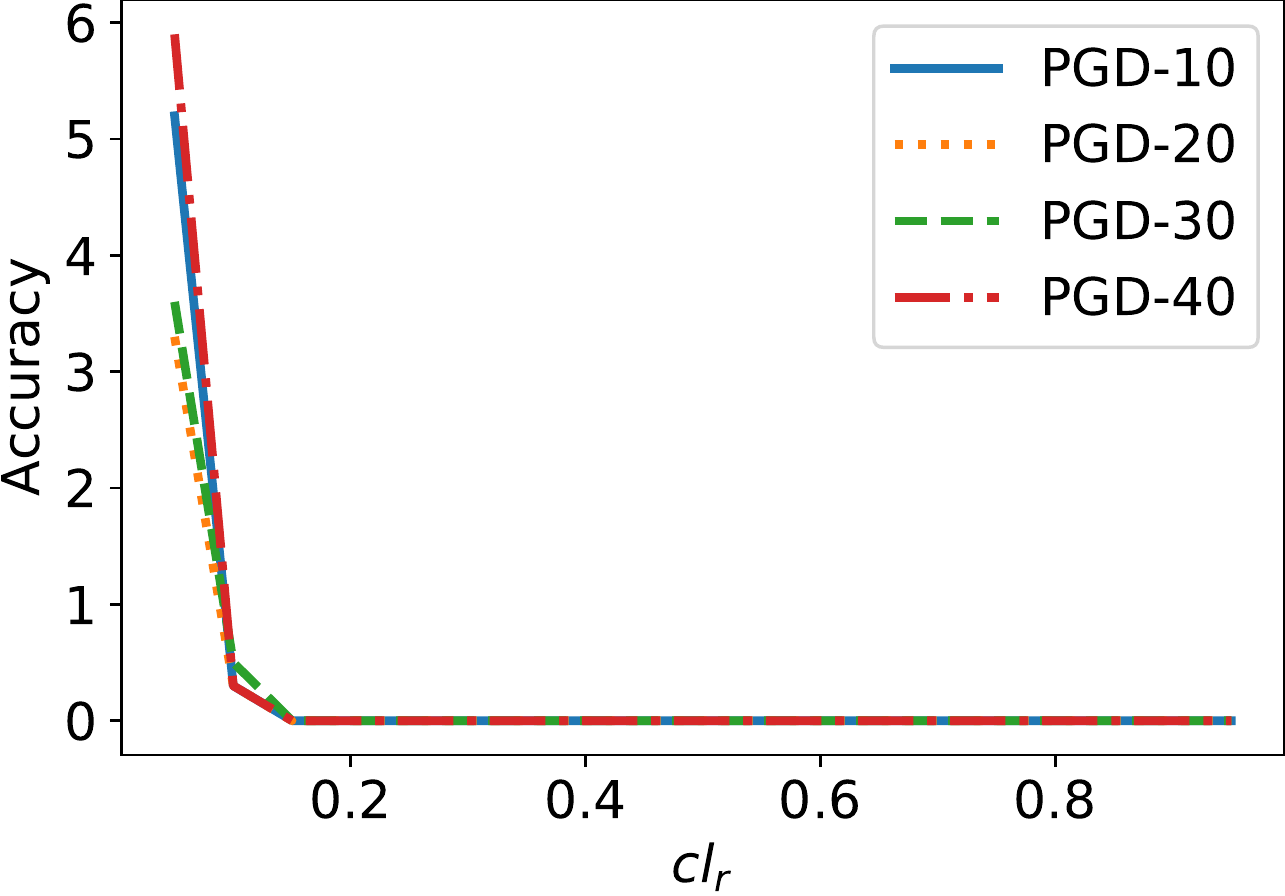}
    \end{minipage}
    \begin{minipage}[t]{0.23\textwidth}
    \includegraphics[width=1.0\textwidth]{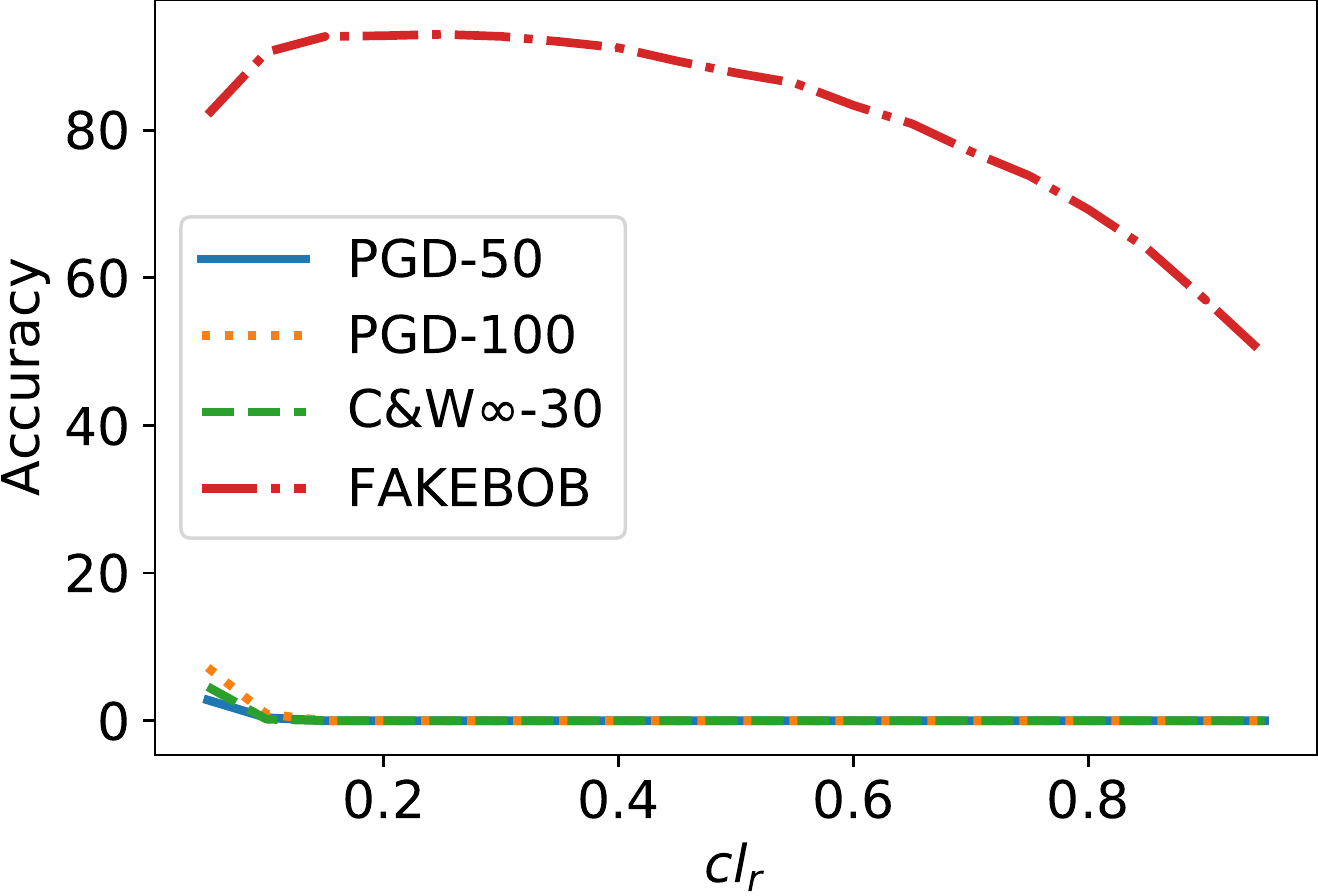}
    \end{minipage}
    \begin{minipage}[t]{0.23\textwidth}
    \includegraphics[width=1.0\textwidth]{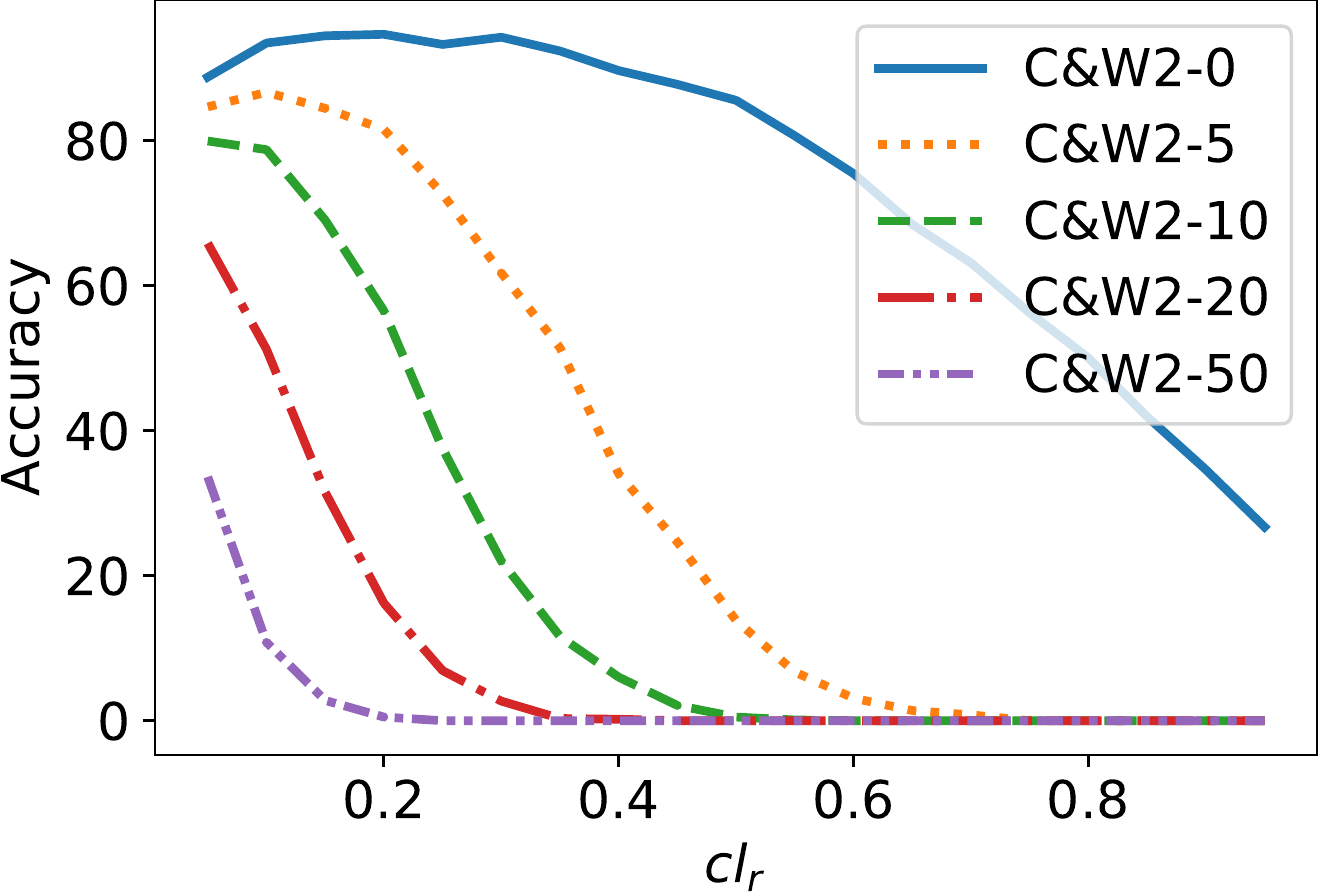}
    \end{minipage}
    \label{fig:p-ff-f}
    }

    \caption{The performance of input transformations and feature transformations.}
    \label{fig:parameter-3}
\end{figure*}

The results are depicted as curves  in Figures~\ref{fig:parameter-1}-\ref{fig:parameter-3}.
We choose the optimal parameters according to the R1 scores on FGSM, as R1 score assigns equal importance to both the accuracy on benign examples
and the accuracy on adversarial examples.
We consider FGSM as it is the weakest one among all the attacks, as shown in Section~\ref{sec:evaluate-attack},
and a good parameter should provide strong resilience to the weakest attack.
Although these optimal parameters may not be the optimal ones against the other attacks,
they are still very promising.

\begin{table}[t]
    \centering
     \caption{The ranges and optimal values for parameters of transformations.}
    \resizebox{0.48\textwidth}{!}{
    \begin{tabular}{c|c|c}
    \hline
         {\bf Method (Parameter)} &  {\bf Range} & {\bf Optimal} \\ \hline
        {\bf QT ($q$)} & 128, 256, 512, 1024 & 512 \\ \hline
        {\bf DS ($\tau$)} & 0.05 to 0.95, step 0.05 & 0.45 \\ \hline
        {\bf AT ($snr$)} & 2 to 20 dB, step 2 dB & 16 dB \\ \hline
        {\bf AS ($k$)} & 3 to 21, step 2 & 17 \\ \hline
        {\bf MS ($k$)} & 3 to 21, step 2 & 7 \\ \hline
        {\bf LPF ($f_p$, $f_s$)} & \begin{tabular}[c]{@{}c@{}}$f_p$: 4000 Hz \\ $f_s$: 4500 to 8000 Hz, step 500 Hz\end{tabular} & $f_s$=4500 Hz \\ \hline
        {\bf BPF ($f_{pl}$, $f_{pu}$, $f_{sl}$, $f_{su}$)} & \begin{tabular}[c]{@{}c@{}}$f_{pl}$: 300 Hz \\ $f_{pu}$: 4000 Hz \\ $f_{sl}$: 50 to 200 Hz, step 50 Hz \\ $f_{su}$: 5000 Hz to 8000 Hz, step 500 Hz \end{tabular} & \begin{tabular}[c]{@{}c@{}}$f_{sl}$=150 Hz \\ $f_{su}$=6000 Hz \end{tabular} \\ \hline
        {\bf OPUS ($b_o$)} & 6-20 kbps, step 1 kbps & 8 kbps \\ \hline
        {\bf SPEEX ($b_s$)} & 4-44 kbps, step 2 kbps & 11 kbps \\ \hline
        {\bf AMR ($b_r$)} & \begin{tabular}[c]{@{}c@{}}6.6, 8.85, 12.65, 14.25, 15.85 \\ 18.25, 19.85, 23.05, 23.85 kbps \end{tabular} & 6.6 kbps \\ \hline
        {\bf AAC-V ($q_c$)} & 1-5, step 1 & 1 \\ \hline
        {\bf AAC-C ($b_c$)} & 15-85 kbps, step 5 kbps & 15 kbps \\ \hline
        {\bf MP3-V ($q_m$)} & 0-9, step 1 & 4 \\ \hline
        {\bf MP3-C ($b_m$)} & \begin{tabular}[c]{@{}c@{}}8, 16, 24, 32, 40, 48, \\ 64, 80, 96, 112, 128, 160 kbps \end{tabular} & 24 kbps \\ \hline
        {\bf \defensenameabbr ($cl_m$, $cl_r$)} & \begin{tabular}[c]{@{}c@{}} $cl_{m}$: kmeans/warped-kmeans \\ $cl_{r}$: 0.05 to 0.95, step 0.05 \end{tabular} & \begin{tabular}[c]{@{}c@{}} \defensenameabbr-o (k): $cl_{r}$=0.2 \\ \defensenameabbr-o (wk): $cl_{r}$=0.35 \\ \defensenameabbr-d: $cl_{r}$=0.1 \\ \defensenameabbr-c: $cl_{r}$=0.1 \\ \defensenameabbr-f: $cl_{r}$=0.1 \end{tabular} \\ \hline
    \end{tabular}
    }
    \label{tab:range-optimal}
    \vspace*{-3mm}
\end{table}

\begin{figure}[t]
    \centering
    \includegraphics[width=0.35\textwidth]{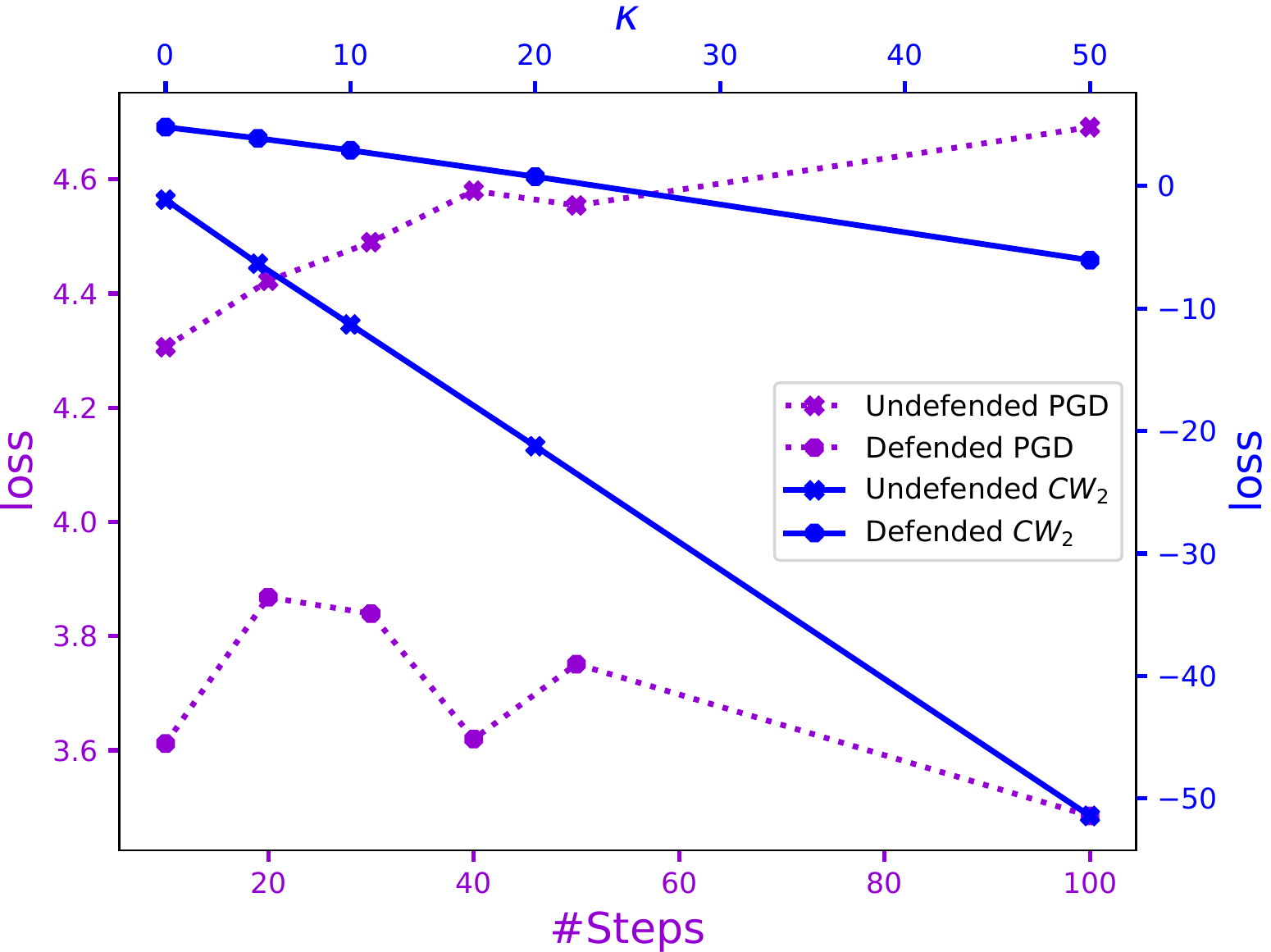}\vspace*{-3mm}
    \caption{Loss values of adversarial examples of PGD and CW$_2$ on the model without/with the MS transformation.}
    \label{fig:loss-step-kappa}\vspace*{-3mm}
\end{figure}

\begin{figure}[t]
    \centering
    \subfigure[QT, 0.92]{
    \includegraphics[width=0.22\textwidth]{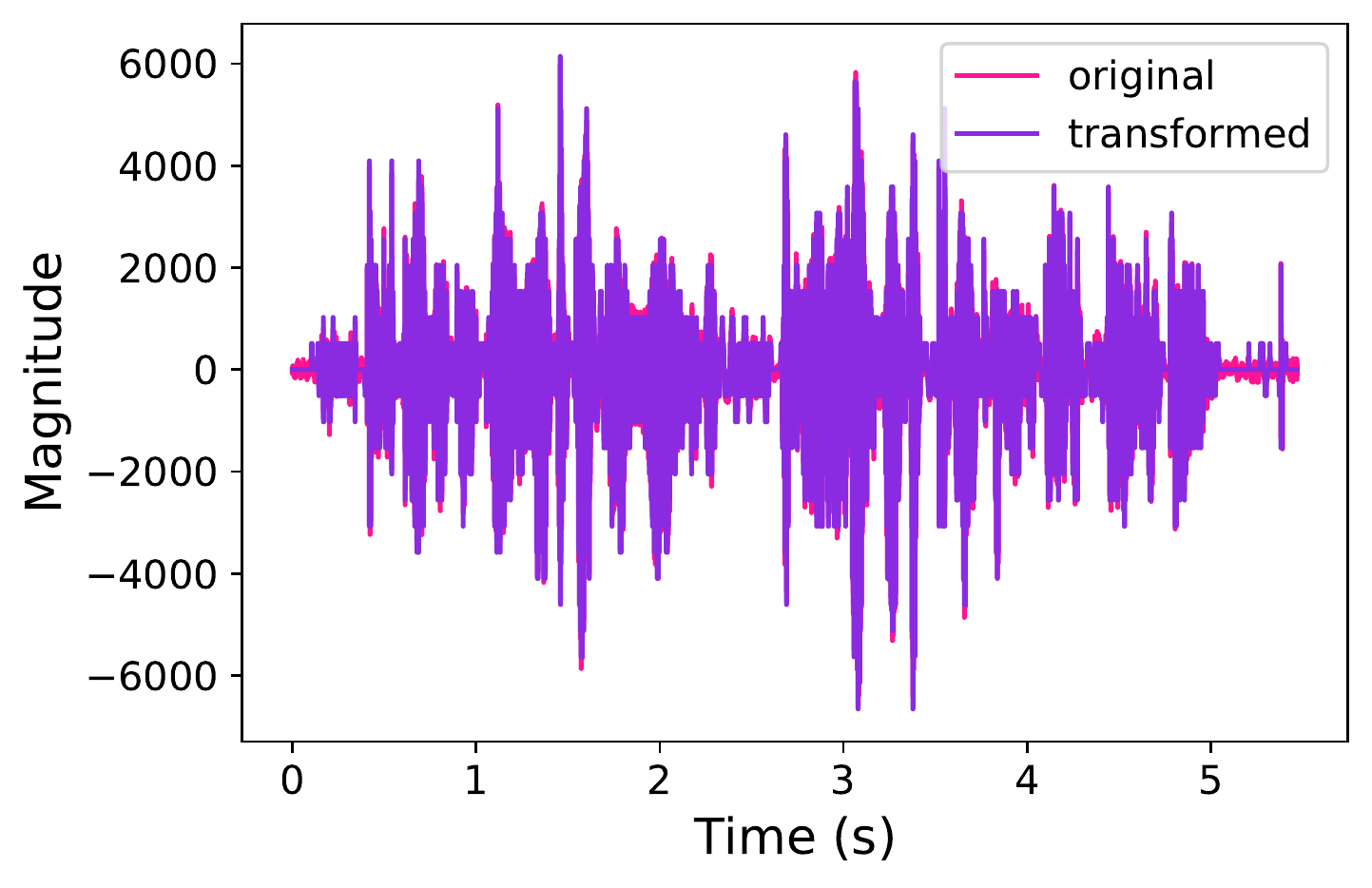}
    }
    \subfigure[OPUS, 15.24]{
    \includegraphics[width=0.22\textwidth]{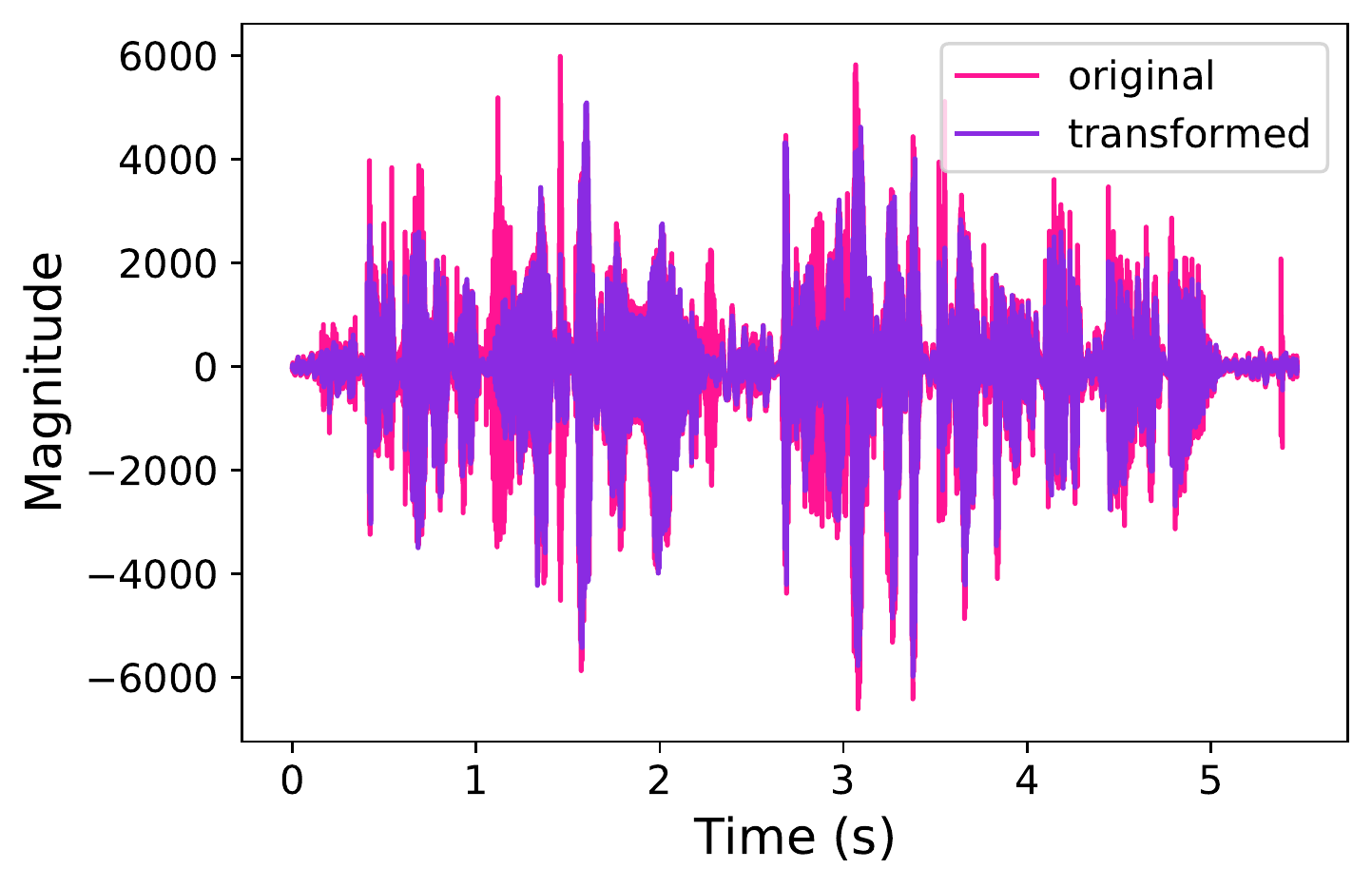}
    }\quad
    \subfigure[SPEEX, 29.57]{
    \includegraphics[width=0.22\textwidth]{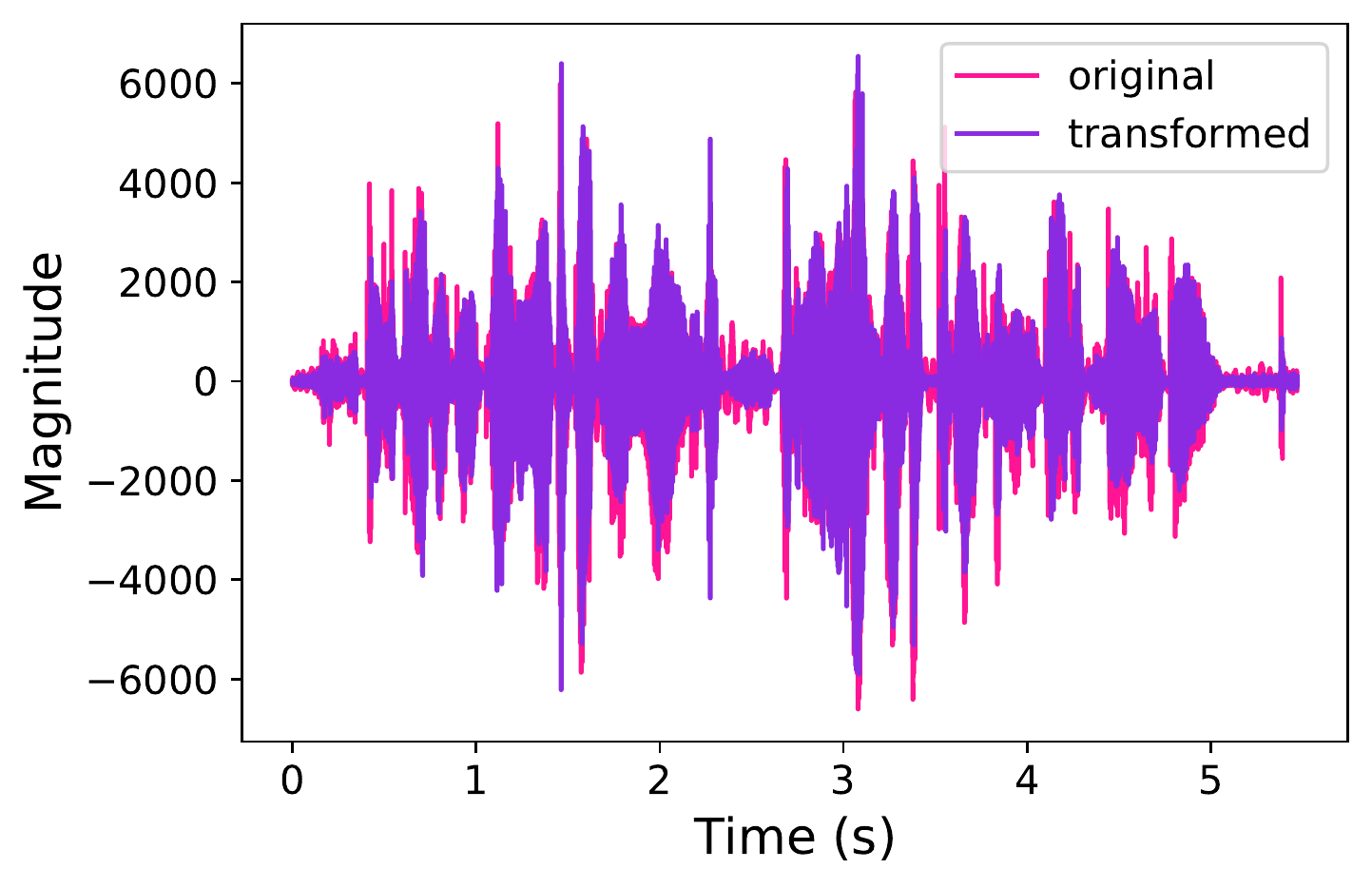}
    }
    \subfigure[AMR, 23.68]{
    \includegraphics[width=0.22\textwidth]{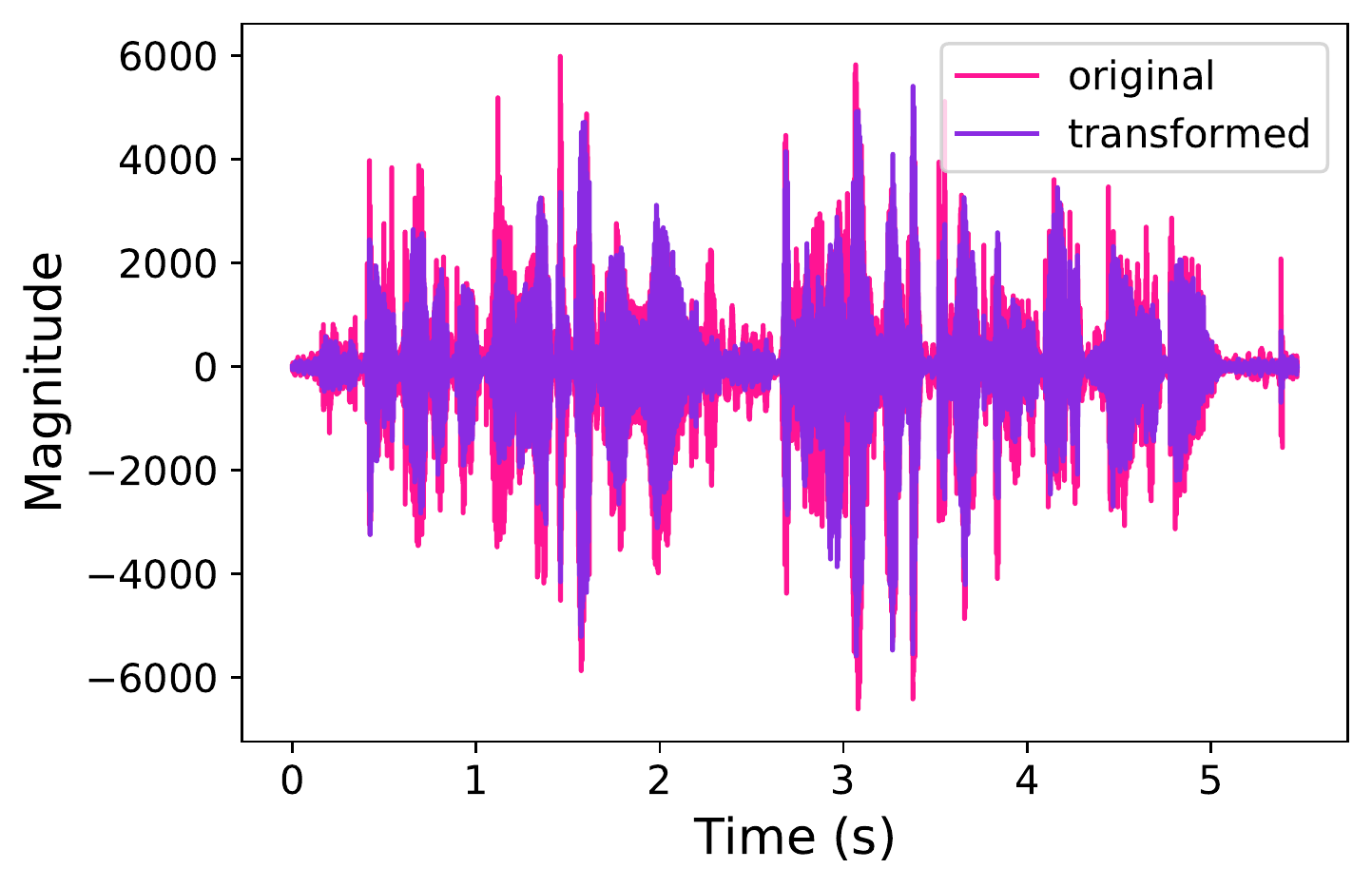}
    }
    \quad
   \subfigure[AAC-V, 3.87]{
    \includegraphics[width=0.22\textwidth]{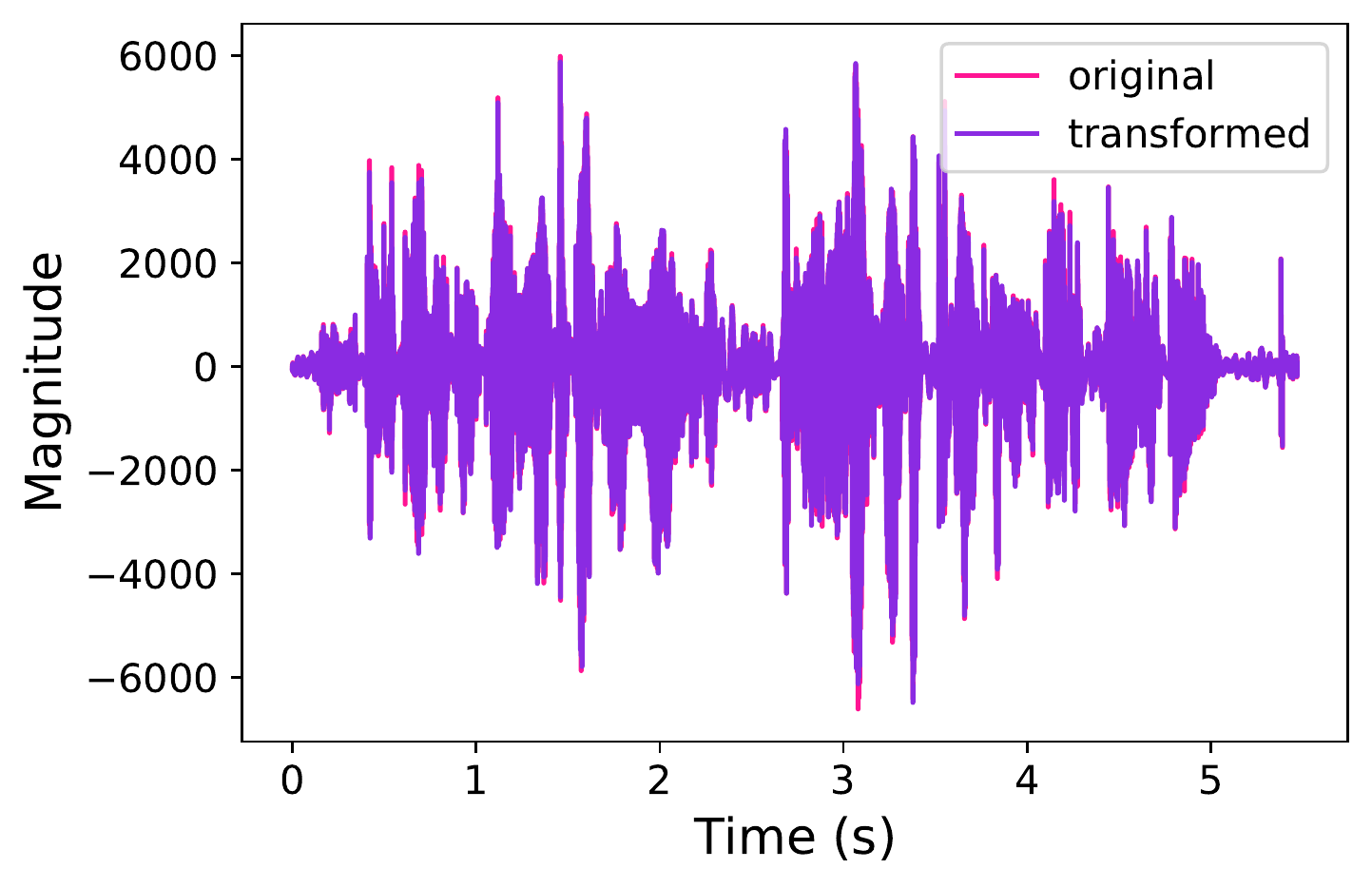}
    }
    \subfigure[AAC-C, 6.92]{
    \includegraphics[width=0.22\textwidth]{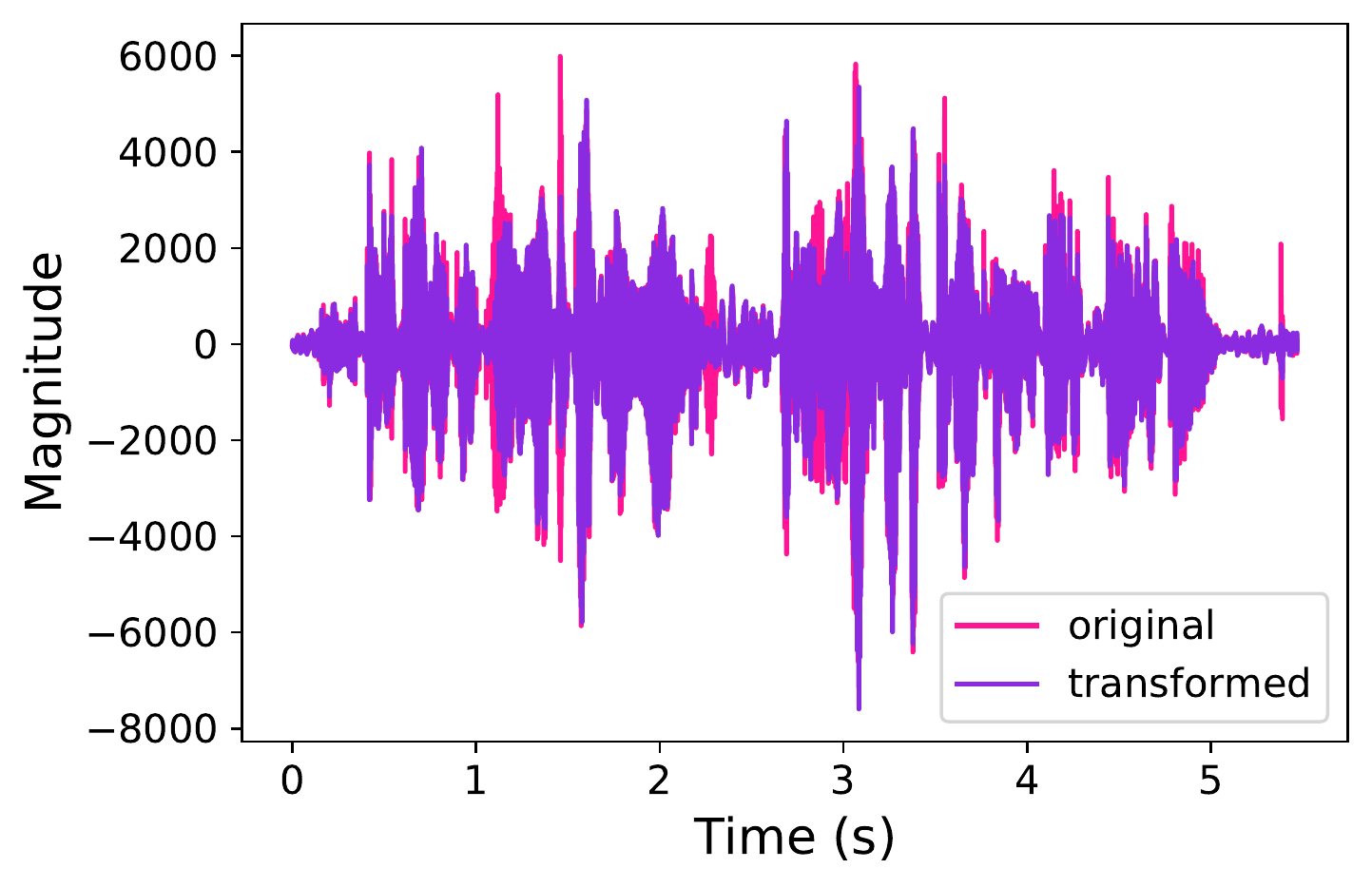}
    }\quad
    \subfigure[MP3-V, 4.19]{
    \includegraphics[width=0.22\textwidth]{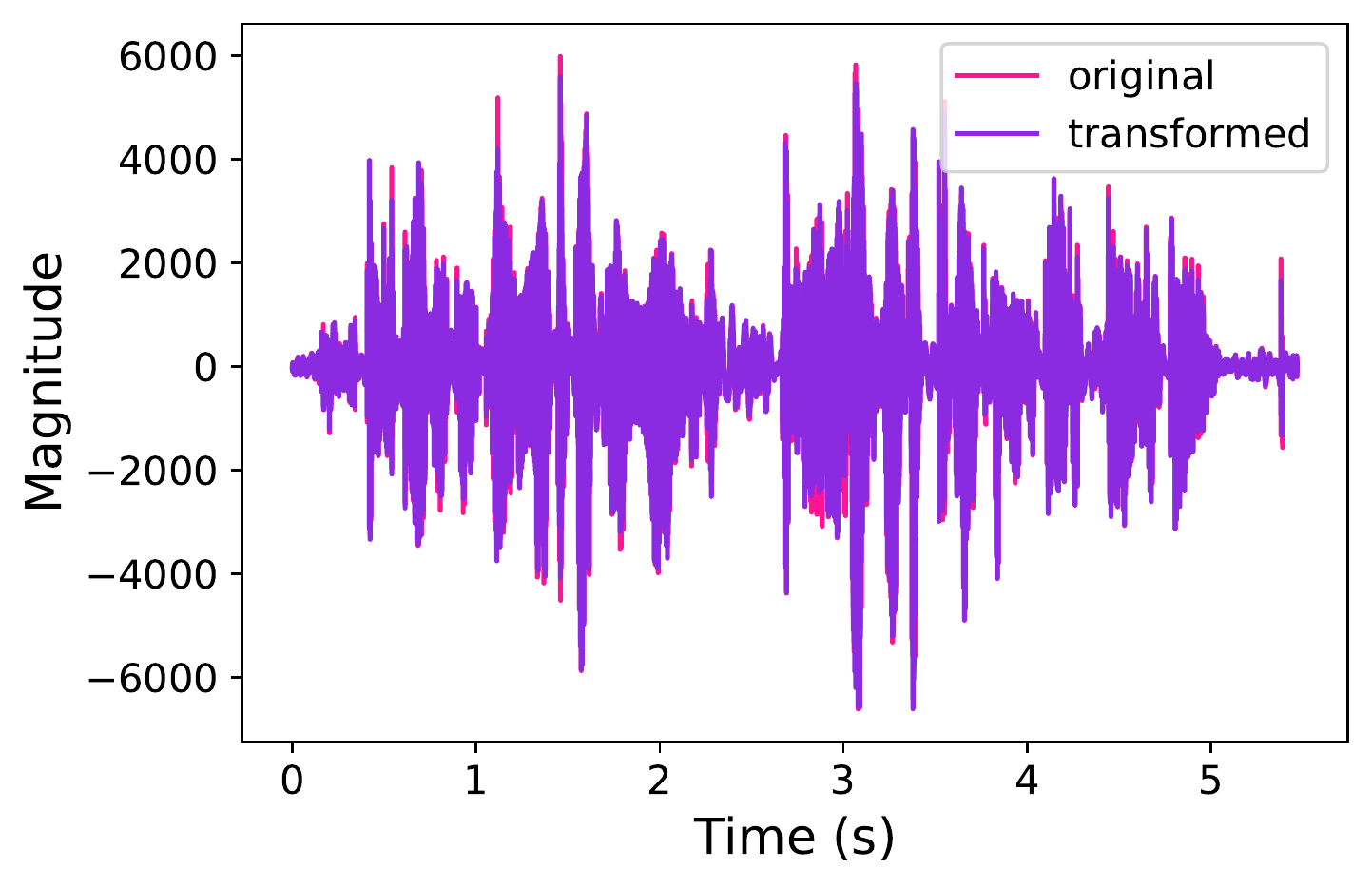}
    }
    \subfigure[MP3-C, 5.92]{
    \includegraphics[width=0.22\textwidth]{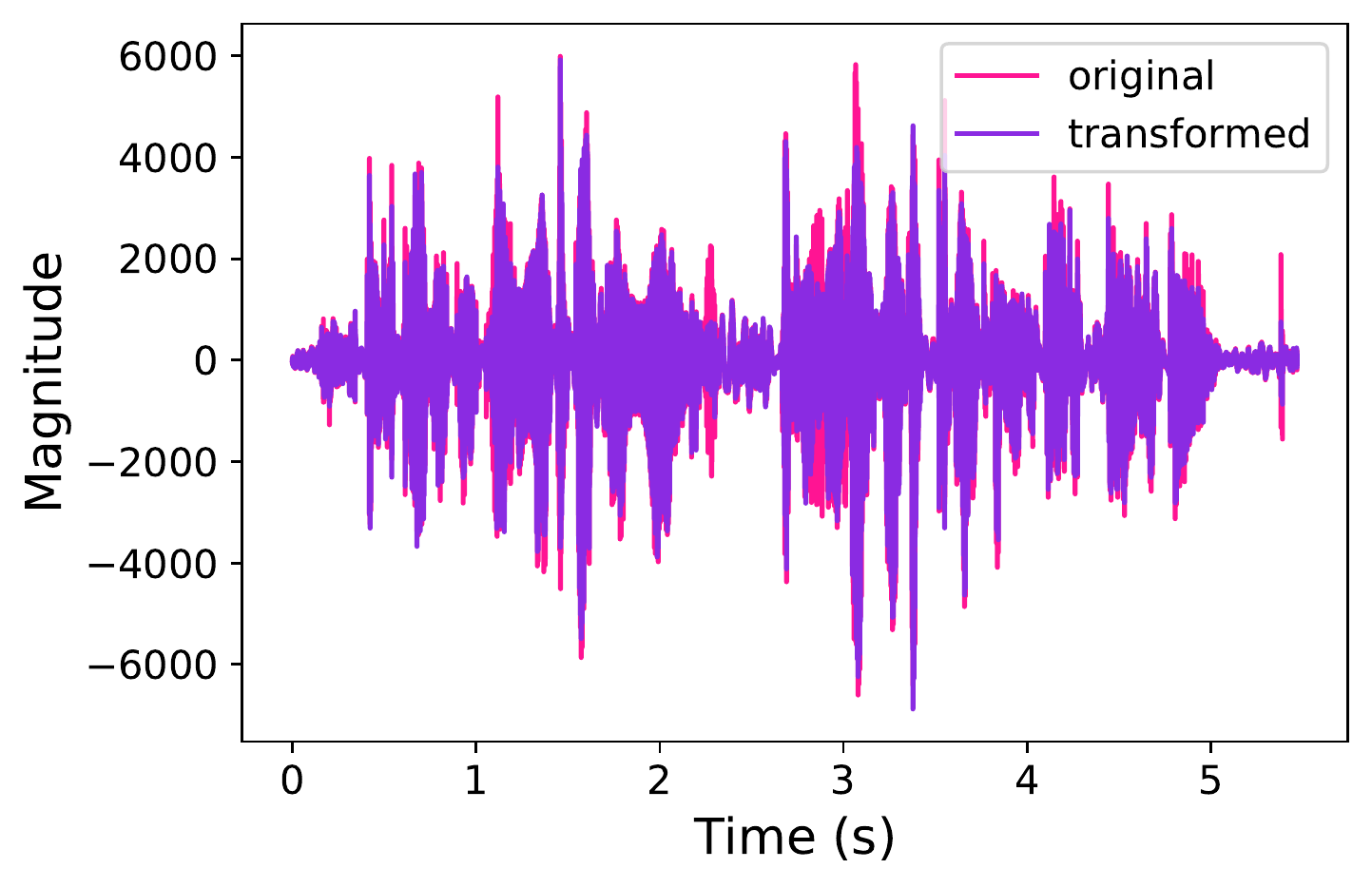}
    }
\vspace*{-2mm}
    \caption{The visualization of an original voice and transformed voice by different input transformations. The average $L_2$ distance between original and transformed voices is listed right of the method name.}
    \label{fig:BPDA-differ}\vspace*{-3mm}
\end{figure}

\section{More Details of Transformations against Non-Adaptive Attacks}\label{sec:moredetails}

\smallskip
\noindent {\bf The number of steps in PGD vs $\kappa$ in CW$_2$.}
As mentioned previously, with the increase of $\kappa$ in CW$_2$, the strength and distortion of adversarial examples increase,
and the effectiveness of the transformations decreases.
In contrast, although the strength of adversarial examples also increases with the number of steps (denoted by \#Steps) in PGD,
the distortion of adversarial examples are almost the same (cf. Table~\ref{tab:evaluate-atatck}) and the effectiveness of the transformations does not decrease monotonically.
For instance, consider the transformation MS, the accuracy $A_a$
decreases from 21.3\% to 17.1\% when \#Steps increases from 10 to 20,
but $A_a$ increases from 17.1\% to 24.5\% when \#Steps increases 20 to 100.

To understand the above gap, we study
whether strong adversarial examples remain strong after the transformation MS.
Figure~\ref{fig:loss-step-kappa} shows the loss values of the adversarial examples crafted by PGD and CW$_2$ on
the model without/with MS by varying
\#Steps and $\kappa$, where the larger the loss of PGD (resp. the smaller  the loss of CW$_2$) is, the stronger the adversarial examples are.
We can observe that: the strength of adversarial examples crafted by CW$_2$ with increase of $\kappa$
remains monotonic after the transformation,
but the strength of adversarial examples crafted by PGD with increase of  \#Steps
becomes non-monotonic after the transformation.
This may be because
CW$_2$ introduces larger distortions with increase of $\kappa$,
but PGD does not introduce larger distortions with increase of \#Steps.


\begin{tcolorbox}[size=title,opacityfill=0.1,breakable]
\textbf{Findings.}
Without having access to a transformation,
the transformation may be still effective against strong adversarial examples (e.g., PGD).
\end{tcolorbox}

\section{Approximation of Non-differentiable Transformations by the Identity Function}
\label{sec:evaapptrans}
To measure how accurate it is to substitute a non-differentiable transformation with
the identity function, we compute the average $L_2$ distance between the original voices and voices after the transformation.
The results are shown in Figure~\ref{fig:BPDA-differ}, where the $L_2$ distance is given in the caption of each sub-figure,
and the curves in each sub-figure are the waveform of a random chosen voice and the voice after transformation.


From Figure~\ref{fig:BPDA-differ}, we can observe that the $L_2$ distance of QT, AAC-V and MP3-V is much smaller than that of OPUS, SPEEX, AMR, AAC-C and MP3-C,
indicating that QT, AAC-V and MP3-V are much closer to the identity function.
We can also observe that the difference between original voice and the voice after transformation by CBR speech compressions  are more significant than QT and VBR speech compressions (i.e., AAC-V and MP3-V).
In conclusion, it seems that it suffices to replace QT and VBR speech compressions with the identity function in the backward pass,
but more accurate approximation functions or more advanced adaptive attacks than BPDA are required to circumvent other speech compressions.

\end{document}